%% file: tesi.tex
\documentclass[12pt,twoside,openright]{report}
\usepackage{color}
\usepackage{cite}
\usepackage{a4wide}
\usepackage{latexsym}
\usepackage{fancyhed}
\usepackage{longtable}
\usepackage{amsmath}
\usepackage{amssymb}
\usepackage{bbm}
\usepackage{feynarts}
\usepackage{slashed}
\usepackage{sola_cites}
\usepackage{sbl_tsn}
\usepackage{sbl_th}
\usepackage{sbl_figs}
\usepackage{import}
\usepackage{diagrames}
\newcommand{\clearemptydoublepage}{
  \newpage{\thispagestyle{empty}\cleardoublepage}}
\renewcommand{\title}{Flavor changing neutral decay effects in
  models with two Higgs boson doublets: Applications to LHC Physics.}

\begin{document}
\begin{fmffile}{fmftesi}
\pagenumbering{roman}
\include{portada}              
\clearemptydoublepage
\include{greetings}            
\clearemptydoublepage
\input{headings_toc}

\tableofcontents               
\clearemptydoublepage
\pagenumbering{arabic}         
\input{headings}

\include{introduccio}             
\clearemptydoublepage
\include{2hdm}
\clearemptydoublepage
\include{mssm}
\clearemptydoublepage
\import{tch2HDM/}{tch2HDM}
\clearemptydoublepage
\import{htc2HDM/}{htc2HDM}
\clearemptydoublepage
\import{hbsSUSY_br/}{hbsSUSY_br}
\clearemptydoublepage
\import{hbsSUSY_prod/}{hbsSUSY_prod}
\clearemptydoublepage
\include{conclusions}             
\clearemptydoublepage
\appendix                                                       
\include{func_npunts}               
\clearemptydoublepage
\include{diag_matrix}               
\clearemptydoublepage
\end{fmffile}

\addcontentsline{toc}{chapter}{\listfigurename}
\listoffigures
\clearemptydoublepage
\addcontentsline{toc}{chapter}{\listtablename}
\listoftables
\clearemptydoublepage
\addcontentsline{toc}{chapter}{\bibname}
\providecommand{\href}[2]{#2}
\bibliography{bib}          

\end{document}

%% file: portada.tex
\begin{titlepage}
\begin{center}
    \includegraphics[width=7cm]{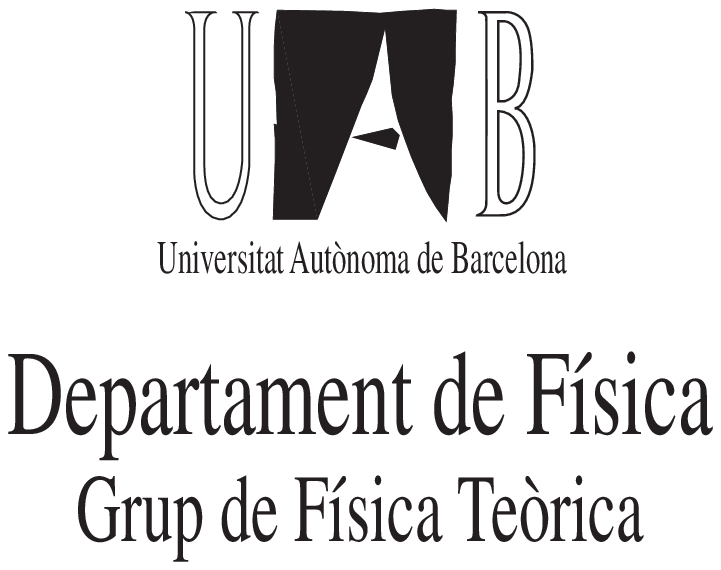}
    \vspace{1cm}
\end{center}
    \begin{center}
        \rule[0.15cm]{\textwidth}{0.1cm}
        {\huge{\bf Flavor changing neutral decay}}\\[0.2cm]

        {\huge{\bf effects in models with}}\\[0.3cm]

        {\huge{\bf two Higgs boson doublets:}}\\[0.2cm]

        {\huge{\bf Applications to LHC Physics.}}
        \rule[0.15cm]{\textwidth}{0.1cm}
        {\large
          Efectes de deca{\"\i}ments amb canvi de sabor neutres \\
          en models amb dos doblets de bosons de Higgs:\\
          Aplicacions a la F{\'\i}sica de LHC.\\
        }
        \rule[0.15cm]{0.5\textwidth}{0.05cm}

        \vspace{0.8cm}
        {\Large Santi B{\'e}jar Latonda}
        \vspace{0.8cm}

        Universitat Aut{\`o}noma de Barcelona

        Grup de F{\'\i}sica Te{\`o}rica

        Institut de F{\'\i}sica d'Altes Energies

        2005/2006

        \vspace{0.5cm}
        {\large Mem{\`o}ria presentada per a optar al grau de\\
          Doctor en Ci{\`e}ncies F{\'\i}siques
          \vspace{.5cm}\\
          Director:\\
          Dr. Joan Sol{\`a} i Peracaula}
    \end{center}
\end{titlepage}


%% file: greetings.tex
\begin{titlepage}
    \vspace*{1cm}

    Vull agrair a en Joan Sol{\`a} la seva gran dedicaci{\'o}, sense el qual aquest
    treball no hauria estat possible. Per la seva confian{\c c}a en el meu saber
    fer, tant en la feina com en el temps de compilar el resultat final.  Ha
    estat comprensiu i un gran guia en els moments de dubte.  La seva visi{\'o}
    sobre quins temes eren interessants estudiar de entre tots els camins que
    hi ha en aquest camp i entre els que ben segur m'hauria perdut.

    A en Jaume Guasch, un company infatigable i constant, disposat a resoldre
    dubtes i a prendre un caf{\`e} quan ho necessitava. A m{\'e}s li he d'agrair que
    m'acoll{\'\i}s com a un amic al seu despatx tant de temps sense queixar-se. I per tot el
    treball que hem fet junts.

    Al companys del Grup de F{\'\i}sica Te{\`o}rica els hi he d'agrair l'entorn que
    m'ha perm{\'e}s seguir treballant en el camp que m'agrada i no haver perdut
    l'entusiasme. El treballar dia a dia amb aquests f{\'\i}sics ha estat molt
    enriquidor.  Las seves xerrades sobre f{\'\i}sica, sobre els f{\'\i}sics, temes
    mundans i no tant mundans m'han ajudat a seguir treballant amb il·lusi{\'o} a
    la f{\'\i}sica.

    Al Grup de F{\'\i}sica Te{\`o}rica per haver-me acollit proporcionat tots
    els medis necessaris per poder treballar.

    A la M$^a$ del Carmen per escoltar i intentar entendre les meves
    divagacions, les meves explicacions, tot i que la son {\'e}s m{\'e}s forta de
    vegades. Espero haver-te transm{\`e}s una miqueta de la meva passi{\'o} per la
    f{\'\i}sica.

    A la fam{\'\i}lia i amics per comprendre que el ser f{\'\i}sic te{\`o}ric no {\'e}s nom{\'e}s
    una excusa per estar assegut dient que est{\`a}s meditant sobre part{\'\i}cules
    fonamentals, o l'origen de l'univers, en lloc de parar la taula.

    \vspace*{\fill}
    \noindent
    This Thesis has been written using Free Software.

    \noindent
    The \LaTeX$2_\varepsilon$\  Typesetting system.

    \noindent
    Feynman graphs using \texttt{feyn}\textsf{MF}.

    \noindent
    Plots using Xmgr and Grace plotting tools.

    \noindent
    GNU Emacs. 

    \noindent
    Running in a GNU/Linux system.

\end{titlepage}


%% file: headings_toc.tex

\addtolength{\headheight}{0.5cm}
\addtolength{\textheight}{-0.5cm}

\lhead[\fancyplain{}{}]{ \fancyplain{}{}   }
\rhead[ \fancyplain{}{} ]{ \fancyplain{}{}   }
\chead[ \fancyplain{}{}  ]{\fancyplain{}{}}
\cfoot{\fancyplain{\thepage}{\thepage}}

\pagestyle{fancyplain}

%% file: headings.tex

\pagestyle{fancyplain}

\renewcommand{\chaptermark}[1]{\markboth{#1}{#1}}           
\renewcommand{\sectionmark}[1]{\markright{\thesection\ #1}} 

\lhead[ \fancyplain{}{\bf\thepage}  ]{ \fancyplain{}{\it\rightmark} }
\rhead[ \fancyplain{}{\it\leftmark} ]{ \fancyplain{}{\bf\thepage}   }
\chead[ \fancyplain{}{}]{\fancyplain{}{} }
\cfoot{\fancyplain{\thepage}{}}

%% file: introduccio.tex
\chapter{Introduction}
\label{cha:introduction}
\section{Summary}
\label{sec:sumary}
In this Thesis we have investigated some effects appearing in top quark and
Higgs boson decays with flavor changing neutral currents (FCNC) in the
framework of generic Two Higgs Doblet Models (2HDM) and the Minimal
Supersymmetric Standard Model (MSSM).

The Standard Model (SM) of Particle Physics interactions has
had great success in describing the strong, weak and electromagnetic interactions and its
validity has been tested up to the quantum level in past and present
accelerators, such as the LEP at CERN or the Tevatron at Fermilab. The last
great success of the SM was the discovery in 1994 of its last matter building
block, namely the top quark, with a mass of
$m_t=178.0\pm2.7\pm3.3\GeV$. However, the mechanism by which all the SM
particles get their masses is still unconfirmed, since no Higgs scalar has been
found yet. The fermions couple to the Higgs bosons with a coupling proportional
to their mass, so one expects that the large interactions between top quark and Higgs boson
particles would give rise to large quantum effects. But the FCNC processes are very
suppressed in the SM. So in some cases (specially the ones we are going to
deal with in this Ph.D. Thesis) the sole observation of these FCNC processes would
be instant evidence of new physics and could greatly help to unravel the type
of underlying Higgs model.

We have focused our work on two models: the generic Two Higgs Doblet Models
(2HDM) and the Minimal Supersymmetric Standard Model (MSSM). The 2HDM is like
the SM with an extra Higgs doblet, so it has more Higgs bosons and in
particular some charged. There are two types, I and II, and they differ in the
way they give mass to the fermions. The MSSM is an extension of the SM that
incorporates Supersymmetry (SUSY). Supersymmetry is an additional
transformation that can be added in the action of Quantum Field Theory,
leaving this action unchanged. The main phenomenological consequence is
that for any SM particle ($p$) there should exist a partner for it, which we
call {\em sparticle} ($\tilde p$) with different spin but with the same gauge
quantum numbers. This extension of the SM provides elegant
solutions to some theoretical problems of the SM, such as the {\em hierarchy
  problem}.

We have applied these two extensions to the SM to see whether they can produce
new FCNC effects.
We have computed the following FCNC decays: (1) the branching ratios ($B$) of
the top quark to Higgs bosons and charm quark in the 2HDM; (2) $B$ and number of
events at the LHC of the Higgs bosons to top and charm quarks in the 2HDM; (3)
$B$ of the Higgs bosons to bottom and strange quarks in the MSSM; (4) cross section and
number of events at the
LHC of the Higgs bosons to FCNC final states involving the
heavy quarks like the top and bottom quark in the MSSM. We also
studied the experimental signatures that would allow discover of  the nature of
these Higgs bosons in the LHC. In this study we have applied the severe
restrictions from observed low-energy FCNC processes like $b\to s\gamma$.

Decays of the top quark induced by FCNC are
known to be extremely rare events within the SM. This is so not
only for the decay modes into gauge bosons, but most notably in the case of
the Higgs channels, e.g. $t\rightarrow H^{SM}c$, with a branching fraction of
$10^{-13}$ at most. We have found that in the 2HDM the decays of the top
quark to Higgs bosons, $t\rightarrow(h^{0},H^{0},A^{0})c$, can be the most
favored FCNC modes -- comparable or even more efficient than the gluon channel
$t\rightarrow gc$.  In both cases the optimal results are obtained for
Type~II models. However, only the Higgs channels can have rates reaching the
detectable level ($10^{-5}$), with a maximum of order $10^{-4}$. Compared
with previous results obtained in the Higgs sector of the MSSM, the maximum
branching ratios are similar but have different signatures. While in the
2HDM~II there is only one Higgs boson that can reach the visible level ($h^0$
or $H^0$, but not both), in the MSSM all the channels can be competitive in
some region.

Similarly, Higgs boson decays mediated by FCNC are very much suppressed in the Standard
Model, at the level of $10^{-15}$ for Higgs boson masses of a few hundred
$\GeV$. We have computed the FCNC decays of Higgs bosons into a top quark,
$h\to t\bar{c}$ ($h=h^0,H^0,A^0$), in a
general 2HDM. The isolated top quark signature, unbalanced by any other heavy
particle, is very clean (without background noise), and should help to identify
the potential FCNC events much better than any other final state. We have computed the
maximum branching ratios and the number of FCNC Higgs boson decay events {at}
the LHC. The most favorable mode for production and subsequent FCNC decay is
the lightest CP-even state ($h^0$) in the Type II 2HDM, followed by the other
CP-even state ($H^0$) if it is not very heavy, whereas the CP-odd ($A^0$) mode can never be
sufficiently enhanced. Our calculation shows that the branching ratios of the
CP-even states may reach $10^{-5}$, and that several hundreds of events could be
collected in the highest luminosity runs of the LHC. We also point out some
strategies in which to use these FCNC decays as a handle to discriminate between 2HDM
and supersymmetric Higgs bosons.

Furthermore, we analyzed the maximum branching ratios for the FCNC decays of the neutral
Higgs bosons of the MSSM into bottom quarks, $h\to b\bar{s}$
($h=h^0,H^0,A^0$), giving the maxima in the $B(h\rightarrow
{b}\,\bar{s})\sim 10^{-4}-10^{-3}$ range. But this maximum could reach up to
$\sim10^{-2}$ depending on whether or not it is allowed a fine-tunning in the \Bbsg\
restriction, which for naturalness reasons we do not allow. We consider that the
bulk of the MSSM contribution to $B(h\rightarrow {b}\,\bar{s})$ should
originate from the strong supersymmetric sector, electroweak calculations are
in progress. These calculations show that
the FCNC modes $h\to b\bar{s}$ can be competitive with other Higgs boson
signatures and could play a helpful complementary role in identifying the
supersymmetric Higgs bosons, particularly the lightest CP-even state in the
critical LHC mass region $m_{h^0}\simeq 90-130\GeV$.

Finally, we have also analyzed the production and subsequent FCNC decay of the neutral
MSSM Higgs bosons to $tc$ and $bs$ in the LHC
collider, $pp\to h \to {t\bar{c},b\bar{s}}$ and $h\to
{t\bar{c},b\bar{s}}$ ($h=h^0,H^0,A^0$).
Only the strongly-interacting FCNC sector has been computed
because it expected to be the most important. We determined the maximum production rates
for each of these modes and identified the relevant regions of the MSSM
parameter space. The latter are
different from those obtained by maximizing only the branching ratio, due to
non-trivial correlations between the parameters that maximize/minimize each
isolated factor. The production rates for the $bs$ channel can be huge for a
FCNC process ($0.1-1 \pb$), but its detection can be problematic. The
production rates for the $tc$ channel are more modest ($10^{-3}-10^{-2}\pb$),
but its detection should be easier due to the clear-cut top quark signature. A
few thousand $tc$ events could be collected in the highest luminosity phase of
the LHC, with no counterpart in the SM.

Our general conclusion is that the physics of the processes with flavor
changing neutral currents can be very important in seeing the physics beyond the
Standard Model and to disentangle the nature of the most adequate model.
Experiments at the LHC can be crucial to unravel signs of FCNC physics beyond
the SM.

\section{Motivation}
\label{sec:motivation}
The accepted model for the interactions between elementary particles is the
Standard Model (SM)\cite{Weinberg:1967tq,Glashow:1961tr,Salam:1968nobel,Georgi:1974sy,Pati:1974yy}.
This model is composed of fundamental particles with spin $1/2$
(fermions), spin $1$ (vector bosons) and one fundamental particle with spin $0$, that is
the Higgs boson, the only undiscovered particle of the SM. But there are good
reasons to think that it is not the final model for the high energy physics.
There are different models which try to
explain the physics beyond the SM. The simplest model that extends the
SM is the 2HDM~\cite{Gunion:1989we}. The extra pieces of this model are some neutral and charged
Higgs bosons. These neutral Higgs boson are those in which we are interested in order to
compare with the SM Higgs boson. In this way we can distinguish which
model describes better the future experimental results. Another not so
simple model is the MSSM\cite{\SUSY}\footnote{For a review see
  \cite{Carena:2002es,Martin:1997ns}},
and which has some characteristics of the 2HDM and, as we have said, has
more or less twice the number of particles than the SM, as can be seen in
chapter \ref{cha:2hdm}.

The top quark is the latest-discovered elementary particle of the SM. It
almost completes all of the building blocks of the SM. But for theoretical and
aesthetic reasons we need to introduce a mechanism to give masses to the
particles. One mechanism that allows this is the Higgs mechanism. It is based
on introducing a new (undiscovered yet) particle of spin 0. So recently the Higgs boson
has become the most wanted particle. The top quark physics is very important in the
investigation of the high energy physics. It interacts with Higgs bosons with the
highest possible strength (because it is proportional to the mass of the
quark) and therefore this property may help to discover the
Higgs boson. Moreover, it provides a
big phase space, so it can decay in particles that the others cannot.

The flavor changing neutral currents are a kind of processes, especially
important to test the SM.  They are characterised by one quark changing
its flavor in the interaction (effective or not) with neutral particles
(currents). Experimentally these processes are very suppressed, especially for
the physics of the top quark, as we have seen.
In the SM the branching ratios in the top quark case are so tiny that we cannot think
of measuring them experimentally. But we could perhaps find models that, even being
depressed, can give values nearer to our experimental possibilities.

In the near and middle future, with the upgrades of the Tevatron (Run II,
TeV33), the advent of the LHC, and the construction of an $e^{+}e^{-}$ Linear
Collider (LC, nowadays called International Linear Collider ILC)~\cite{\LHC,\TESLA},
new results on top quark physics\,\cite{\TopPhys}, and possibly
on Higgs physics, will be obtained that may be extremely helpful
complementing the precious information already collected at LEP I and II from
Z and W physics. Both types of machines, the hadron colliders and the LC will
work at high luminosities and produce large amounts of top quarks. In the
LHC, for example, the production of top quark pairs will be $\sigma(t%
\overline{t})=800\;pb$ -- roughly two orders of magnitude larger than in the
Tevatron Run~II. In the so-called low-luminosity phase ($%
10^{33}\,cm^{-2}s^{-1}$) of the LHC, one expects about three $t\,\bar{t}$%
-pairs per second (ten million $t\,\bar{t}$-pairs per year!)~\cite{\Gianotti}.
And this number will be augmented by one order of magnitude in the
high-luminosity phase ($10^{34}\,cm^{-2}s^{-1}$). As for a future LC running
at e.g. $\sqrt{s}=500\;GeV$, one has a smaller cross-section $\sigma(t\bar{t}%
)=650\;fb$ but a higher luminosity factor ranging from $5\times10^{33}%
\,cm^{-2}s^{-1}$ to $5\times10^{34}\,cm^{-2}s^{-1}$ and of course a much
cleaner environment\thinspace\cite{\Miller}. With datasets from LHC and LC
increasing to several $100\,fb^{-1}/$year in the high-luminosity phase, one
should be able to pile up an enormous wealth of statistics on top quark
decays. Therefore, not surprisingly, these machines should be very useful to
analyze rare decays of the top quark and of the Higgs boson(s), viz. decays whose branching fractions
are so small ($\lesssim10^{-5}$) that they could not be seen unless the number
of collected decays is very large.

The reason for the interest in these decays is at least twofold. First, the
typical branching ratios for the rare top quark decays predicted within the
Standard Model (SM) are so small that the observation of a single event of
this kind should be ``instant evidence'', so to speak, of new physics; and
second, due to its large mass
($m_t=178.0\pm2.7\pm3.3\GeV$~\cite{Eidelman:2004wy}), the top quark could play
a momentous role in the search for Higgs physics beyond the SM. While this has
been shown to be the case for the top quark decay modes into charged Higgs
bosons, both in the Minimal Supersymmetric Standard Model (MSSM) and in a
general two-Higgs-doublet model (2HDM)~\cite {\CGGJS,\CGHS}\footnote{%
  For a review of the main features of loop-induced supersymmetric
  effects on top quark production an decay, see e.g. Ref.~\cite{\DESY}.}, we
expect that a similar situation would apply for top quark FCNC decays into non-SM
neutral Higgs bosons and for Higgs boson FCNC decays.

\section{Today's situation}
\label{sec:situation}

The search for physics beyond the Standard Model (SM) is a very relevant, if
not the most important, endeavor within the big experimental program scheduled
in the forthcoming Large Hadron Collider (LHC) experiment at
CERN~\cite{\LHC}. There are several favorite searching lines on which to
concentrate, but undoubtedly the most relevant one (due to its central role in
most extensions of the SM) is the physics of the Higgs boson(s) with all its
potential physical manifestations.

As we mention above, experimentally, processes involving Flavor Changing Neutral Current (FCNC)
have been shown to have rather low rates~\cite{\PDG}. Letting aside the
meson-meson oscillations, such as $K^0-\bar{K}^0$ and $B^0-\bar{B}^0$, the
decay processes mediated by FCNC are also of high interest and are strongly
suppressed too. For instance, we have the radiative B-meson decays, with a
typical branching ratio $B(b\rightarrow s\,\gamma)\sim10^{-4}$. But we also
have the FCNC decays with the participation of the top quark as a physical
field, which are by far the most suppressed decay
modes~\cite{\LorenzoD,Eilam:1991zc}.  Indeed, the top quark decays into gauge
bosons ($t\rightarrow c\,V$;$\;V\equiv\gamma,Z,g$) are well known to be
extremely rare events in the SM. The branching ratios are, according to
Ref.\,\cite{Eilam:1991zc}: $\sim 5\times 10^{-13}$ for the photon, slightly
above $1\times10^{-13}$ for the $Z$-boson, and $\sim4\times10^{-11}$ for the
gluon channel, or even smaller according to other estimates
\,\cite{\JASaavedra}. Similarly, the top quark decay into the SM Higgs boson,
$H^{SM}$, is a very unusual decay, typically $B(t\rightarrow c\,H^{SM})\sim
10^{-14}$\,\cite{\Mele}.

\begin{table}[h]
    \centering
    \begin{tabular}{|l||c|c|c|}\hline
        & SM & 2HDM & MSSM\\\hline\hline
        $B(t\to c \gamma)$&$ \sim 5\times10^{-13}$&$\lesssim1\times10^{-7}$&$<1\times10^{-6}$\\\hline
        $B(t\to c Z)$&$ \gtrsim 1\times10^{-13}$&$<1\times10^{-6}$&$<1\times10^{-7}$\\\hline
        $B(t\to c g)$&$ \sim
        4\times10^{-11}$&$\lesssim1\times10^{-5}$&$\lesssim1\times10^{-5}$\\\hline
        $B(t\to c H)$&$\sim10^{-13}-10^{-15}$&$\mathrm{\lesssim10^{-4}}$&$\lesssim10^{-4}$\\\hline
        $B(H\to t \bar{c})$&$\sim10^{-13}(m_H<2 m_W)$&$\mathrm{\lesssim10^{-4}}$&$
        \mathrm{\lesssim10^{-4}}$\\\hline
        &$\lesssim10^{-15}(m_H>2 m_W)$&&\\\hline
        $B(H\to b \bar{s})$&$\lesssim10^{-7} (m_H<2M_W)$&$\lesssim10^{-5}$&$\lesssim10^{-4}$\\\hline
        &$\lesssim10^{-10} (m_H>m_t)$&&$$\\\hline
    \end{tabular}
    \caption{Rare FCNC branching ratios of the top quark and the Higgs boson
      decays.}
    \label{eq:decays}
\end{table}

The reason for this rareness is simple: for FCNC top quark decays in the SM,
the loop amplitudes are controlled by down-type quarks, mainly by the bottom
quark. Therefore, the scale of the loop amplitudes is set by $m_{b}^{2}$ and
the partial widths are of order

\begin{align}
    \Gamma(t\rightarrow V\,c)&\sim
    \left(\frac{|V_{tb}^{\ast}V_{bc}|}{16\pi^2}\right)^{2}\alpha\,G_{F}^{2}\,m_{t}\,m_{b}^{4}\,F
    \sim\left(\frac{|V_{bc}|}{16\pi^2}\right)^{2}\alpha_{em}^2\alpha\,m_{t}
    \left(\frac{m_{b}}{M_{W}}\right)^{4}\,F,
    \label{GammaFCNC}\\
    \intertext{and similarly for the FCNC Higgs boson decays}
    \Gamma(H^{SM}\rightarrow
    t\,\bar{c})&\sim\left(\frac{|V_{tb}^{\ast}V_{bc}|}{16\pi^2}\right)^2\,
    \, \alpha_{W}^3\, \,m_{H}\left(
        \lambda_b^{SM}\right) ^{4}
    \sim\left(\frac{|V_{bc}|}{16\pi^2}\right)^2\,\alpha_{W}\ G_F^2
    \,m_{H}\,m_b^4\,,\\
    B(H^{SM}\rightarrow t\,\bar{c})
    &\sim\left(\frac{|V_{bc}|}{16\pi^2}\right)^2\,\alpha_{W}\
    G_F\,m_b^2\sim 10^{-13}\ \ \
    (\,{\rm for}\ m_H<2 m_W)\,,\\
    B(H^{SM}\rightarrow t\,\bar{c})
    &\sim\left(\frac{|V_{bc}|}{16\pi^2}\right)^2\,\alpha_{W}\
    G_F\,\frac{m_b^4}{m_H^2}\lesssim 10^{-15}\ \ \
    (\,{\rm for}\ m_H>2 m_W)\,,\\
    B(H^{SM}\rightarrow b\,\bar{s})
    &\sim\left(\frac{|V_{ts}|}{16\pi^2}\right)^2\,\alpha_{W}\
    G_F\,\left(\frac{m_H^4}{m_b^2}\right)\lesssim 10^{-7}\ \ \
    (\,{\rm for}\ m_H<2\,M_W)\,,
    \label{estimateBRSMhbs}\\
    B(H^{SM}\rightarrow b\,\bar{s})
    &\sim\left(\frac{|V_{ts}|}{16\pi^2}\right)^2\,\alpha_{W}\
    G_F\,\left(\frac{m_t^4}{m_H^2}\right)\lesssim 10^{-10}\ \ \
    (\,{\rm for}\ m_H>m_t)\,,\label{estimateBRSMhbs2}
\end{align}
where $\alpha$ is $\alpha_{em}$ for $V=\gamma,Z$ and $\alpha_{s}$ for $V=g$,
$G_F$ is Fermi's constant, $\alpha_{W}=g^2/4\pi$ and $g$ being the $SU(2)_L$
weak gauge coupling. Notice the presence of $\lambda_b^{SM}\sim m_b/M_W$,
which is the SM Yukawa coupling of the bottom quark in units of $g$.  The
factor $F\sim(1-m_{V}^{2}/m_{t}^{2})^{2}$ results, upon neglecting $m_{c}$,
from phase space and polarization sums. Notice that the dimensionless fourth
power mass ratio, in parenthesis in eq.~(\ref{GammaFCNC}), and the fourth
power of $\lambda_b^{SM}$ stems from the GIM mechanism and is responsible for
the ultra-large suppression beyond naive expectations based on pure dimensional
analysis, power counting and CKM matrix elements.

The GIM
mechanism~\cite{Glashow:1970gm} is related to the unitarity of the
mixing matrices between quarks. The Minimal Standard Model (SM) embeds
the GIM mechanism naturally, due to the presence of only one Higgs doublet
giving mass simultaneously to the down-type and the up-type quarks, and as a
result no tree-level FCNCs interactions appear. FCNCs are radiatively induced,
and are therefore automatically small.  However, when considering physics
beyond the SM, new horizons of possibilities open up which may radically
change the pessimistic prospects for FCNC decays involving a Higgs boson and
the top quark.  Because of the loop-induced FCNCs effects, the SM and non-SM
loops enter the FCNC observables at the same order of perturbation theory, and
new physics competes on the same footing with SM physics to generate a
non-vanishing value for these rare processes. It may well be that the non-SM
effects are dominant and become manifest. Conversely, it may happen that they
become highly constrained.

The addition of further Higgs doublets to the SM in the most general way
introduces potentially large tree-level FCNC interactions, which would predict significant FCNC
rates in contradiction with observation.
{However}, by introducing an \textit{ad-hoc} discrete symmetry these
interactions are forbidden. This gives rise to two classes of
Two-Higgs-Doublet Models (2HDM) which avoid FCNCs at the tree-level, known
conventionally as type~I and type~II 2HDMs~\cite{\Hunter}.  In Type~I 2HDM
(also denoted 2HDM~I) one Higgs doublet, $\Phi_{1}$, does not couple to
fermions at all and the other Higgs doublet, $\Phi_{2}$, couples to fermions
in the same manner as in the SM. For more details see chapter~\ref{cha:2hdm}.
In contrast, in Type~II 2HDM (also denoted
2HDM~II) one Higgs doublet, $\Phi_{1}$, couples to down quarks (but not to up
quarks) while $\Phi_{2}$ does the other way around. Such a coupling pattern is
automatically realized in the framework of supersymmetry (SUSY), in particular
in the MSSM, but it can also be arranged in non-supersymmetric extensions if
we impose a discrete symmetry, e.g. $\Phi _{1}\rightarrow-\Phi_{1}$ and
$\Phi_{2}\rightarrow+\Phi_{2}$ (or vice versa) plus a suitable transformation
for the right-handed quark fields, this symmetry is only violated by soft
terms of dimension two.

Supersymmetry (SUSY) is certainly related to Higgs boson physics,
and at the same time it may convey plenty of additional phenomenology. Ever
since its inception, SUSY has been one of the most cherished candidates for
physics beyond the SM, and as such it will be scrutinized in great detail at
the LHC. It is no exaggeration to affirm that the LHC will either prove or
disprove the existence of SUSY, at least in its most beloved low-energy
realization, namely the one which is needed to solve the longstanding
naturalness problem in the Higgs sector of the SM~\cite{Haber:1985rc}. On
the other hand, SUSY provides an appealing extension {of} the SM, which unifies the
fermionic and bosonic degrees of freedom of the fundamental particles and
provides a natural solution to the hierarchy problem. The search for SUSY
particles has been one of the main programs of the past experiments in high
energy physics (LEP, SLD, Tevatron) and continues {to play a central role} in
the present accelerator experiments (Tevatron II) and in the planning of
future experimental facilities like the LHC and the
LC. The Minimal Supersymmetric Standard Model (MSSM) is the
simplest extension of the SM which includes SUSY, and for this reason its
testing will be one of the most prominent aims of these powerful experiments.
If SUSY is realized around the TeV scale, the LHC experiments shall be able to
directly produce the SUSY particles for masses smaller than a few
TeV\,\cite{Weiglein:2004hn,Degrassi:2004ed}. On the other hand, the presence
of SUSY may also be tested indirectly through the quantum effects of the
supersymmetric particles. For one thing, it has been known since long ago that
SUSY particles may produce large virtual effects on Higgs boson
observables\,\footnote{See
  e.g.\,\cite{Guasch:1995rn,Coarasa:1996yg,Coarasa:1996qa,Guasch:1997jc,Coarasa:1997ky,
    Coarasa:1999da,Coarasa:1999db,Coarasa:1999wy,Belyaev:2001qm,Belyaev:2002eq,
    Belyaev:2002sa,Guasch:2001wv,Guasch:2003cv} and references therein. For a
  review see e.g. \cite{Carena:2002es}.}.

In Ref.\cite{\Divitiis} it was shown that the vector boson modes can be highly
enhanced within the context of the MSSM. This fact was also dealt with in great detail in
Ref. \cite{\GuaschNPo} where a dedicated study was presented of
the FCNC top quark decays into the various Higgs bosons of the MSSM (see also
\cite{\Yuan}) showing that these can be the most favored FCNC top quark
decays -- above the expectations on the gluon mode $t\rightarrow c\,g$.  For
the 2HDM it was proven that while the maximum rates for $t\rightarrow c\,g$
were one order of magnitude more favorable in the MSSM than in the 2HDM, the
corresponding rates for $t\rightarrow c\,h^0$ were comparable both for the
MSSM and the general 2HDM, namely up to the $10^{-4}$ level and should
therefore be visible both at the LHC and the
LC\,\cite{\Saavedra}.

As in the 2HDM, in MSSM one has to impose some restrictions to avoid tree-level FCNCs among
the extra predicted particles, which would induce one-loop FCNC interactions
among the SM particles. But, in fact, MSSM \textit{requires} their existence,
because of the $SU(2)_L$ gauge symmetry. Of course, low energy measurements
constrain the FCNC couplings (the most stringent being the \Bbsg). A
potentially relevant FCNC interaction is the gluino with the squarks, not very
much constrained experimentally.

Concerning the FCNC interactions of Higgs bosons with third generation quarks,
it was demonstrated long ago~\cite{Guasch:1999jp} that the leading term
corresponds to a \textit{single particle insertion approximation}. This
produces a flavor {change} in the internal squark loop propagator, since in
this case the chirality change can already take place at the
squark-squark-Higgs boson interaction vertex. Adding this to the fact that the
Higgs bosons ({in contrast to gauge bosons}) have a privileged coupling to
third generation quarks, one might expect that the FCNC interactions of the
type quark-quark-Higgs bosons in the MSSM become highly strengthened with
respect to the SM prediction.  This was already proven in the rare decay
channels $\Gamma(t\to ch)$~\cite{Guasch:1999jp} ($h$ being any of the neutral
Higgs bosons of the MSSM $h\equiv h^0,H^0,A^0$) where the maximum rate of the
SUSY-QCD induced branching ratio was found to be $B(t\to ch)\simeq 10^{-5}$,
eight orders of magnitude above the SM expectations $B(t\to cH^{SM})\simeq
10^{-13}$. Similar enhancement factors have been found in the (top,bottom)-quark-Higgs
boson interactions in other extensions of the SM,
both in the MSSM (see chapters \ref{cha:hbsSUSY_br}-\ref{cha:hbsSUSY_prod} and
Refs.~\cite{Guasch:1999jp,Guasch:1997kc,Guasch:1999ve,
  Curiel:2002pf,Demir:2003bv,Curiel:2003uk,Heinemeyer:2004by}) and in the
general two-Higgs-doublet model (2HDM)\ (see chapters \ref{cha:tch2HDM}-
\ref{cha:htc2HDM} and Ref.~\cite{Arhrib:2004xu}), and also in
other extensions of the SM-- see \,\cite{Aguilar-Saavedra:2004wm} for a
review.

The power of FCNC observables can be gauged {e.g.} by the implications of the
bottom-quark rare decay $\bsg$: the experimentally measured allowed range
$\Bbsg=(3.3\pm0.4)\times 10^{-4}$~\cite{\bsgexp} may impose
tight constraints on extensions of the SM. For example, it implies a lower
bound on the charged Higgs boson mass $\mHp\gtrsim 350\GeV$ in general type II
2HDMs~\cite{\bsgthdm}.

After this panoramic view of the FCNC processes in the SM and beyond, the work
presented in this Thesis is as follows: in chapter \ref{cha:2hdm}
(resp.~\ref{cap:MSSM}) we give the basic notations of the 2HDM (resp.~MSSM);
in chapter \ref{cha:tch2HDM} we compute the FCNC top quark decay in the 2HDM
$B(t\to c h)$; in chapter~\ref{cha:htc2HDM} we compute the Higgs boson production and FCNC
decay in the 2HDM $\sigma(pp\to h \to t\bar{c})$ at the LHC and $B(h\to t
\bar{c})$; in chapter~\ref{cha:hbsSUSY_br} we compute the FCNC Higgs boson decay in
the MSSM $B(h\to b\bar{s})$; and in chapter~\ref{cha:hbsSUSY_prod} the Higgs boson
production and decay in
the MSSM $\sigma(pp\to h \to {t\bar{c},b\bar{s}})$ and $B(h\to
{t\bar{c},b\bar{s}})$. This PhD. work is based on the following articles
\cite{Bejar:2000ub,Bejar:2001sj,Bejar:2003em,Bejar:2004rz,Glover:2004cy,Bejar:2005kv,RADCOR05}.

In chapter \ref{cha:tch2HDM} we show that within the simplest
extension of the SM, namely the general two-Higgs-doublet model, the
FCNC top quark decays into Higgs bosons,
$t\rightarrow(h^{0},H^{0},A^{0})c$, can be the most favored FCNC modes
-- comparable or even more efficient than the gluon channel
$t\rightarrow gc$.  In both cases the optimal results are obtained for
Type~II models. However, only the Higgs channels can have rates reaching
the detectable level $10^{-5}$, with a maximum of order $10^{-4}$
which is compatible with the charged Higgs bounds from radiative B-meson
decays. We compare with the previous results obtained in the Higgs
sector of the MSSM.

In chapter \ref{cha:htc2HDM} we consider the FCNC decays of Higgs bosons into
a top quark in a general two-Higgs-doublet model (2HDM). The isolated top
quark signature, unbalanced by any other heavy particle, should help to
identify the potential FCNC events much more than any other final state. We
compute the maximum branching ratios and the number of FCNC Higgs boson decay
events at the LHC collider at CERN. The most favorable mode for production
and subsequent FCNC decay is the lightest CP-even state in the Type II 2HDM,
followed by the other CP-even state, if it is not very heavy, whereas the
CP-odd mode can never be sufficiently enhanced. Our calculation shows that the
branching ratios of the CP-even states may reach $10^{-5}$, and that several
hundred events could be collected in the highest luminosity runs of the
LHC. We also point out some strategies to use these FCNC decays as a handle to
discriminate between 2HDM and supersymmetric Higgs bosons.

In chapter \ref{cha:hbsSUSY_br} we analyze the maximum branching ratios for
the FCNC decays of the neutral Higgs bosons of the MSSM into bottom and charm quarks,
$h\to b\bar{s}$ ($h=h^0,H^0,A^0$). We consistently correlate these decays with
the radiative B-meson decays $(b\to s\gamma)$. A full-fledged combined
numerical analysis is performed of these high-energy and low-energy FCNC decay
modes in the MSSM parameter space. Our calculation shows that the available
data on $\Bbsg$ severely restricts the allowed values of $B(h\rightarrow
{b}\,\bar{s})$.  While the latter could reach a few percent level in
fine-tuned scenarios, the requirement of naturalness reduces these FCNC rates
into the modest range $B(h\rightarrow {b}\,\bar{s})\sim 10^{-4}-10^{-3}$. We
expect that the bulk of the MSSM contribution to $B(h\rightarrow {b}\,\bar{s})$
should originate from the strong supersymmetric sector. Our results are
encouraging because they show that the FCNC modes $h\to b\bar{s}$ can be
competitive with other Higgs boson signatures and could play a helpful
complementary role to identify the supersymmetric Higgs bosons, particularly
the lightest CP-even state in the critical LHC mass region $m_{h^0}\simeq
90-130\GeV$.

In chapter \ref{cha:hbsSUSY_prod} we analyze the production and subsequent
decay of the neutral MSSM Higgs bosons ($h\equiv h^0,\ H^0,\ A^0$) mediated by
FCNC in the LHC collider. We have computed the $h$-production cross-section
times the FCNC branching ratio, $\sigma(pp\to h\to q{q}')\equiv\sigma(pp\to h)
\times B(h\to q{q}')$, in the LHC focusing on the strongly-interacting FCNC
sector. Here $qq'$ is an electrically neutral pair of quarks of different
flavors, the dominant modes being those containing a heavy quark: $tc$ or
$bs$. We determine the maximum production rates for each of these modes and
identify the relevant regions of the MSSM parameter space, after taking into
account the severe restrictions imposed by low energy FCNC processes. The
analyses of $\sigma(pp\to h\to q{q}')$ singles out regions of the MSSM
parameter space different from those obtained by maximizing only the branching
ratio, due to non-trivial correlations between the parameters that
maximize/minimize each isolated factor. The production rates for the $bs$
channel can be huge for a FCNC process ($0.1-1 \pb$), but its detection can be
problematic. The production rates for the $tc$ channel are more modest
($10^{-3}-10^{-2}\pb$), but its detection should be easier due to the
clear-cut top quark signature. A few thousand $tc$ events could be collected
in the highest luminosity phase of the LHC, with no counterpart in the SM.


%% file: 2hdm.tex
\chapter{Two Higgs Doublet Models (2HDM)}
\label{cha:2hdm}
\section{Introduction}
\label{sec:2hdm:introduction}

The 2HDM are models that extend minimally the Higgs sector of the SM. They
introduce one more doublet of complex scalar fields with hypercharge $Y = +1$. The
most general Lagrangian with the SM gauge symmetry $SU(2)_L\times U(1)_Y$ that
contains these Higgs bosons can be divided in three terms: a kinetic term
$\Lcin$, the Yukawa couplings term (Higgs-fermions interactions) $\LY$ and the
potential for the two Higgs doublets $\Vab$:
\begin{align}
    \label{eq:2hdm:lagrangian}
    \LHiggs &= \Lcin+\LY-\Vab\,,\\
    \Lcin &= \sum_{i=1,2} \bigl(D_{\mu}\phi_i\bigr)^\dagger \bigl(D^{\mu}\phi_i\bigr)\,,\\
    D_{\mu} &=
    \partial_{\mu}-ig\frac{\overrightarrow{\sigma}}{2}\overrightarrow{W}_{\mu}-ig'\frac{Y}{2}B_\mu\,,
\end{align}
where $D_{\mu}$ is the covariant derivative of $SU(2)_L\otimes U(1)_Y$, and
$\sigma_{i}$ are the Pauli matrices\footnote{$tr(\sigma_i
  \sigma_j)=2\delta_{ij}$}.

The Higgs potential that spontaneously breaks the symmetry $SU(2)_L\otimes
U(1)_Y$ to $U(1)_{EM}$ is\cite{Gunion:1989we,Gunion:1992hs}:
\begin{align}
    \label{eq:potencial}
    \begin{split}
    \Vab=&\lambda_1(\phi_1^\dagger\phi_1-v_1^2)^2
         +\lambda_2(\phi_2^\dagger\phi_2-v_2^2)^2+\\
    &+\lambda_3\left[(\phi_1^\dagger\phi_1-v_1^2)+(\phi_2^\dagger\phi_2-v_2^2)\right]^2+\\
    &+\lambda_4\left[(\phi_1^\dagger\phi_1)(\phi_2^\dagger\phi_2)
        -(\phi_1^\dagger\phi_2)(\phi_2^\dagger\phi_1)\right]+\\
    &+\lambda_5\left[\re(\phi_1^\dagger\phi_2)-v_1v_2\cos\xi\right]^2+\\
    &+\lambda_6\left[\im(\phi_1^\dagger\phi_2)-v_1v_2\sin\xi\right]^2\,,
    \end{split}
\end{align}
where all the $\lambda_i$ are real parameters, because the Lagrangian must be
hermitic. This is the most general Lagrangian compatible with the gauge
symmetry and the discrete symmetry $\phi_1 \rightarrow
-\phi_1$\cite{Gunion:1989we,Gunion:1992hs}, this symmetry is only violated by
soft terms of dimension two. We impose this last symmetry to
forbid the FCNC at tree level. Moreover, this potential must be bounded from
below, so the $\lambda_i$ must be non-negative. But in fact, the allowed range for
the parameters $\lambda_i$ corresponding to this minimum is a range in the
parameter space such that the square Higgs boson masses are positive and that
$V(0,0) > 0$.

In this context the minimum of the potential is:
\begin{align}
    \langle\phi_1\rangle&\equiv\begin{pmatrix}0\\v_1\end{pmatrix}\,,\\
    \langle\phi_2\rangle&\equiv\begin{pmatrix}0\\v_2e^{i\xi}\end{pmatrix},
\end{align}
which breaks the gauge symmetry giving $U(1)_{EM}$.

We need two physical parameters in order to know their value, which are
usually taken to be:
\begin{equation}
    M_W^2=\frac{1}{2} g^2 (v_1^2+v_2^2)\equiv
    g^2\frac{v^2}{2}~~,~~\tan\beta=\frac{v_2}{v_1}\,\,,
    \;\;\;0<\beta<\frac{\pi}{2}.
    \label{eq:2hdm:VEV}
\end{equation}

If we impose
$\lambda_5=\lambda_6$ (like in sypersymmetry) we can write the last two terms
of eq.~(\ref{eq:potencial}) as:
\begin{equation}
    \label{eq:lambda56}
    \left|\phi_1^\dagger\phi_2-v_1v_2e^{i\xi}\right|^2.
\end{equation}
The phase $\xi$ can disappear with a redefinition of the fields without
affecting the others terms of the potential (this phase will appear in other
terms of the total Lagrangian). Then, the Higgs potential is CP conserving.

At last, the final Higgs potential is:
\begin{align}
    \label{eq:potencialHiggs}
    \begin{split}
    \Vab=&\lambda_1(\phi_1^\dagger\phi_1-v_1^2)^2
         +\lambda_2(\phi_2^\dagger\phi_2-v_2^2)^2+\\
    &+\lambda_3\left[(\phi_1^\dagger\phi_1-v_1^2)+(\phi_2^\dagger\phi_2-v_2^2)\right]^2+\\
    &+\lambda_4\left[(\phi_1^\dagger\phi_1)(\phi_2^\dagger\phi_2)
        -(\phi_1^\dagger\phi_2)(\phi_2^\dagger\phi_1)\right]+\\
    &+\lambda_5|\phi_1^\dagger\phi_2-v_1v_2|^2\,.
    \end{split}
\end{align}

There are different forms for the Yukawa terms of the Lagrangian to satisfy
the Glashow and Weinberg \cite{Glashow:1977nt} theorem. The Glashow and
Weinberg theorem says that for a general $SU(2)\times U(1)$ gauge theory where
we demand that the neutral-current interactions conserve all quark flavor
naturally the necessary and sufficient conditions are: All quarks of fixed
charge and helicity must (1) transform according to the same irreducible
representations of weak $SU(2)$, (2) correspond to the same eigenvalue of weak
$T_3$, and (3) receive their contributions in the quark mass matrix form a
single source (either from the vacuum expectations value of a single neutral
Higgs boson or from a unique gauge-invariant bare mass term). In practice this
implies that all fermions of a given electric charge couple to no more than
one Higgs doublet.

From the Lagrangian~\eqref{eq:potencial} with the
potential~\eqref{eq:potencialHiggs} and the Yukawa terms we can obtain the
full 2HDM spectrum, as well as the interactions, which contain the usual SM
gauge interactions, the fermion-Higgs interactions, and the pure 2HDM
interactions. A detailed treatment of this Lagrangian, and the process of
derivation of the forthcoming results can be found in\,\cite{TesinaSanti}.

\subsection{{\thdm} I}
\label{sec:2hdmI}
In this model one of the Higgs doublets ($\phi_2$) couple to all the
fermions. The couplings with the quarks is of the form:
\begin{equation}
    \label{eq:lagrangia2hdmI}
    \LY^{(I)}=
    -\sum_{i,j=1}^3\left[D_{ij}^q\left(\bar{q}_L^{(i)}\phi_2\right)q_{dR}^{(j)}
        +U_{ij}^q\left(\bar{q}_L^{(i)}\tilde{\phi}_2\right)q_{uR}^{(j)}
        +\text{h.c.}\right]+\text{leptons}\,,
\end{equation}
where
\begin{align}
    \tilde{\phi}&=i\sigma_2\phi^*\,,\\
    q^{(i)}&=\begin{pmatrix}q_u^{(i)}\\q_d^{(i)}\end{pmatrix}\,,\\
    q^{(1)}=\begin{pmatrix}u\\d\end{pmatrix}\,,\quad
    q^{(2)}&=\begin{pmatrix}c\\s\end{pmatrix}\,,\quad
    q^{(3)}=\begin{pmatrix}t\\b\end{pmatrix}\,,
\end{align}

and similarly for the leptonic doublets $l^{(i)}$ that contain the neutrinos and
the leptons.

This model is very related with the minimal model (SM), being the only
difference a smaller \vev\ $v_2 < v_{\ms}$ ($v \sim
174\, GeV$) and bigger Yukawa couplings.

\subsection{{\thdm} II}
\label{sec:2hdmII}
Now one doublet ($\phi_1$) couples to the right-handed (RH) down fermions
($q_{dR},l_{dR}$) and is responsible of the down masses; the other doublet
($\phi_2$) couples to the RH up fermions ($q_{uR},l_{uR}$) and is responsible
of their masses. Taking any flavor base, i.e. one in which $f_L^{(i)}$
are isospin doublets the Lagrangian is:
\begin{equation}
    \label{eq:lagrangia2hdmII}
    \LY^{(II)}=-\sum_{i,j=1}^3
    \left[D_{ij}^q\left(\bar{q}_L^{(i)}\phi_1\right)q_{dR}^{(j)}
        +U_{ij}^q\left(\bar{q}_L^{(i)}\tilde{\phi}_2\right)q_{uR}^{(j)}
        +\text{h.c.}\right]+\text{leptons}\,.
\end{equation}
The mass matrix will be proportional to the \vev\ of the Higgs as:
\begin{align}
    M_u^{(q,l)} &= v_2U^{(q,l)}\,,\\
    M_d^{(q,l)} &= v_1D^{(q,l)}\,,
\end{align}

This is basically the Higgs sector required in the MSSM.

\subsection{{\thdm} III}
\label{sec:2hdmIII}
This is the most general \thdm\ without FCNC at tree level, being the other
two important particular cases. The Yukawa interactions in this case, using
any flavor base, is:
\begin{align}
    \LY^{(III)}=-\sum_{i,j=1}^3
    &\Bigl[D_{1,ij}^q\bigl(\bar{q}_L^{(i)}\phi_1\bigr)q_{dR}^{(j)}+
    D_{2,ij}^q\bigl(\bar{q}_L^{(i)}\phi_2\bigr)q_{dR}^{(j)}+\\
    &+U_{1,ij}^q\bigl(\bar{q}_L^{(i)}\tilde{\phi}_1\bigr)q_{uR}^{(j)}+
    U_{2,ij}^q\bigl(\bar{q}_L^{(i)}\tilde{\phi}_2\bigr)q_{uR}^{(j)}
    +\text{h.c.}\Bigr]+\\
    &+[\bar{l}Hl\quad\text{ terms}]\,,
\end{align}

where the $3\times3$ matrices $D_1,D_2,U_1,U_2$ are such that diagonalize
simultaneously with the quark mass matrix.

\section{\thdm\  spectrum}
\label{sec:2hdm:spectrum}
\subsection{Higgs sector}
\label{sec:2hdm:hmas}

We will use this structure for the doublets:
\begin{equation}
    \label{eq:phi_espinor}
    \phi_i=
    \begin{pmatrix}\phi_i^+\\\re\phi_i^0+i\im\phi_i^0\end{pmatrix}
    \hspace{1cm}i=1,2\,.
\end{equation}

These fields are not physical fields, they do not have a well defined mass, as
there are bilinear terms within the scalar fields with different fields. The
next thing is to diagonalize the mass matrix. It can be seen that the mass
matrix is a diagonal matrix in boxes for the fields a) $\phi_1^+,\phi_2^+$, b)
$\re\phi_1^0,\re\phi_2^0$ and c) $\im\phi_1^0,\im\phi_2^0$ (the real and
imaginary part can be treated separately by CP invariance). So we have to
separately diagonalize the different boxes. If we define the rotation angle as:

\begin{equation}
    R(\omega)=
    \begin{pmatrix}
        \cos\omega & \sin\omega\\ -\sin\omega & \cos\omega
    \end{pmatrix}\,,
\end{equation}
the rotations (transformations) of the fields are:

\begin{align}
    \label{eq:definiciocamps}
    \begin{pmatrix}G^{\pm}\\ H^{\pm}\end{pmatrix}
    &=R(\beta)\begin{pmatrix}\phi_1^{\pm}\\ \phi_2^{\pm}\end{pmatrix}\,,\\
    \begin{pmatrix}H^{0}\\ h^{0}\end{pmatrix}
    &=\sqrt{2}R(\alpha)\begin{pmatrix}\re\phi_1^{0}-v_1\\ \re\phi_2^{0}-v_2\end{pmatrix}\,,\\
    \begin{pmatrix}G^{0}\\ A^{0}\end{pmatrix}
    &=\sqrt{2}R(\beta)
    \begin{pmatrix}\im\phi_1^{0}\\ \im\phi_2^{0}\end{pmatrix}\,,
\end{align}
with their masses:
\begin{align}
    \label{eq:massesHiggs}
    m_{H^\pm}^2&=\lambda_4(v_1^2+v_2^2)\,,\\
    m_{A^0}^2&=\lambda_5(v_1^2+v_2^2)\,,\\
    m_{H^0,h^0}^2&=\frac{1}{2}\left[M_{11}+M_{22}\pm
        \sqrt{(M_{11}-M_{22})^2+4M_{12}^2}\right]\,,\\
    \intertext{where $M_{ij}$ are defined from the CP-even mass matrix}
    M&=\begin{pmatrix} 4v_1^2(\lambda_1+\lambda_3)+v_2^2\lambda_5 
        &(4\lambda_3+\lambda_5)v_1v_2\\
        (4\lambda_3+\lambda_5)v_1v_2
        &4v_2^2(\lambda_2+\lambda_3)+v_1^2\lambda_5
    \end{pmatrix}\,.
\end{align}
The mixing angles $\beta$ and $\alpha$ are:
\begin{align}
    \label{eq:definicio_beta}
    \tan\beta&=\frac{v_2}{v_1}\,,\\
    \sin 2\alpha&=\frac{2M_{12}}{\sqrt{(M_{11}-M_{22})^2+4M_{12}^2}}\,,\notag\\
    \cos 2\alpha&=\frac{M_{11}-M_{22}}{\sqrt{(M_{11}-M_{22})^2+4M_{12}^2}}\,.
    \label{eq:definicio_alfa}
\end{align}
Now we can redefine the parameters of the theory as:
\begin{align}
    \label{eq:freeparam}
    4\text{ masses} &: m_{h^{0}},m_{H^{0}},m_{A^{0}},m_{H^{\pm}}\\
    2\text{ mixing angles} &:\alpha,\beta
\end{align}
At tree level we get:
\begin{align}
    \frac{G_F}{\sqrt{2}}&=\frac{g^2}{8M_W^2}\,,\\
    \intertext{where we find the value of $v$}
    v=\sqrt{v_1^2+v_2^2}&=2^{-3/4}G_F^{-1/2}\sim 174\, GeV.
\end{align}

To obtain the magnitudes as functions of the physical magnitudes one has to
invert the mass equations~(\ref{eq:massesHiggs})~and~(\ref{eq:definicio_alfa}):
\begin{align}
    \lambda_1&=\frac{\cos^2\alpha\, m_{H^0}^2+\sin^2\alpha\, m_{h^0}^2}{4\, v_1^2}-
    \frac{v_2^2}{4\, v_1^2}\, \lambda_5 - \lambda_3\,,\\
    \lambda_2&=\frac{\sin^2\alpha\, m_{H^0}^2+\cos^2\alpha\, m_{h^0}^2}{4\, v_2^2}-
    \frac{v_1^2}{4\, v_2^2} \lambda_5- \lambda_3\,,\\
    \lambda_3&=\cos\alpha\,\sin\alpha\, \frac{m_{H^0}^2-m_{h^0}^2}{4\, v_1 v_2}-\frac{\lambda_5}{4}\,,\\
    \lambda_4&=\frac{m_{H^\pm}^2}{v^2}\,,\\
    \lambda_5&=\frac{m_{A^0}^2}{v^2}\,.
\end{align}

\section{Interactions in the mass-eigenstate basis}
\label{sec:2hdm:interactions}

We need to convert the interaction Lagrangian to a Lagrangian in the
mass-eigenstate basis, which is the one used in the computation of the
physical quantities. We quote only the interactions that we will need in our
studies.

\begin{itemize}
\item $W$--Higgs: this interaction is obtained from the kinetic term of
    the Lagrangian:
\begin{align}
    \begin{split}
        \mathcal{L}_{WHH}&=\frac{ig}{2}W_\mu^+
        \begin{pmatrix}G_1^+\\H_2^+\end{pmatrix}^{\dagger}
        \overleftrightarrow{\partial_\mu}\left[R(\beta-\alpha)
            \begin{pmatrix}H^0\\h^0\end{pmatrix}+i
            \begin{pmatrix}G^0\\A^0\end{pmatrix}\right]+\text{h. c.}\\
        \mathcal{L}_{WWH}&=gM_WW^2
        \begin{pmatrix}\cos(\beta-\alpha)&\sin(\beta-\alpha)\end{pmatrix}
        \begin{pmatrix}H^0\\h^0\end{pmatrix}\,.
    \end{split}
\end{align}

\item quarks--Higgs: they follow after replacing in \eqref{eq:lagrangia2hdmI}
    and \eqref{eq:lagrangia2hdmII} the mass-eigenstates Higgs fields
    \eqref{eq:definiciocamps}:

\begin{align}
    \left\{\begin{array}{c}\mathcal{L}_{Htb}^{I}\\\mathcal{L}_{Htb}^{II}
        \end{array}\right\}
    &=\frac{gV_{tb}}{\sqrt{2}\,M_{W}}\,H^{-}\overline{b}\,
    \left[ m_{t}\cot \beta \,P_{R}+m_{b}\,
        \left\{\begin{array}{c}-\cot\beta\\\tan \beta\end{array}\right\}\,
        \,P_{L}\right]\,t+\text{h.c.}\\
    \begin{split}
        \left\{\begin{array}{c}\mathcal{L}_{hqq}^{I}\\\mathcal{L}_{hqq}^{II}
            \end{array}\right\}
        &=\frac{-g\,m_{b}}{2\,M_{W}\,\left\{  
              \begin{array}{c} 
                  \sin \beta \\  
                  \cos \beta 
              \end{array} 
          \right\} }\,\overline{b}\left[ h^{0}\,\left\{  
                \begin{array}{c} 
                    \cos \alpha \\  
                    -\sin \alpha 
                \end{array} 
            \right\} +H^{0}\,\left\{  
                \begin{array}{c} 
                    \sin \alpha \\  
                    \cos \alpha 
                \end{array} 
            \right\} \right] \,b\\
        &+\frac{i\,g\,m_{b}}{2\,M_{W}}
        \left\{\begin{array}{c}-\cot\beta\\\tan \beta\end{array}\right\}\,\overline{b}%
        \,\gamma _{5}\,b\,A^{0}+\frac{i\,g\,m_{t}}{2\,M_{W}\tb}\,%
        \overline{t}\,\gamma _{5}\,t\,A^{0}\\
        &+\frac{-g\,m_{t}}{2\,M_{W}\,\sin \beta }\,\overline{t}\left[ h^{0}\,\cos 
            \alpha +H^{0}\,\sin \alpha \right] \,t\,\,.
    \end{split}
    \label{Hqq}
\end{align}

where we have used the third quark family, the
$V_{tb}$ is the corresponding element of the CKM matrix and $P_{L,R}=(1/2)(1\mp \gamma _{5})$ 
are the chiral projectors.

\item Trilinear Higgs couplings: they are summarized in \ref{tab:trilineals}, and are
    valid for Type I and Type II models. Had we not imposed the restriction
    $\lambda _{5}=\lambda _{6}$, then the trilinear rules would be explicitly
    dependent on the $\lambda _{5}$ parameter.
    \begin{table}[tb]
        \centering
        \begin{tabular}{|c|c|}
            \hline
            $H^{\pm}H^{\mp}H^0$&$-\frac{g}{\mw\stbt}
            \left[(\mHps-\mAs+\frac{1}{2}\mHs)\stbt\cbma+\right.$\\
            &$\phantom{\frac{g}{\mw\stbt}}\left.+(\mAs-\mHs)\ctbt\sbma\right]$\\\hline
            $H^{\pm}H^{\mp}h^0$&$-\frac{g}{\mw\stbt}
            \left[(\mHp^2-\mA^2+\frac12\mh^2)\sin{2\beta}\sin(\beta-\alpha)+\right.$\\
            &$\phantom{\frac{g}{\mw\stbt}}
            \left.+(\mh^2-\mA^2)\,\cos{2\beta}\,\cos(\beta-\alpha)\right]$\\\hline
            $h^0h^0H^0$&$-\frac{g\,\cos(\beta-\alpha)}{2\,M_W\,\sin{2\beta}}\,
            \left[(2\,\mh^2+\mH^2)\,\sin{2\alpha}-\right.$\\
            &$\phantom{-\frac{g}{\mw\stbt}}
            \left.-\mA^2\,(3\sin{2\alpha}-\sin{2\beta})\right]$\\\hline
            $A^0A^0H^0$&$-\frac{g}{2\,M_{W}\sin{2\beta}}\,
            \left[\mH^2\,\sin{2\beta}\cos(\beta-\alpha)+\right.$\\
            &$\phantom{-\frac{g}{\mw\stbt}}
            \left.+2(\mH^2-\mA^2)\,\cos{2\beta}\,\sin(\beta-\alpha)\right]$\\\hline
            $A^0A^0h^0$&$-\frac{g}{2\,\mw\stbt}
            \left[\mh^2\,\sin{2\beta}\sin(\beta-\alpha)+\right.$\\
            &$\phantom{-\frac{g}{\mw\stbt}}
            \left.+2(\mh^2-\mA^2)\,\cos{2\beta}\,\cos(\beta-\alpha)\right]$\\\hline
            $H^{\pm}-H^{\mp}-A^0$&$0$\\\hline
            $H^{\pm}-G^{\mp}-H^0$&$(-ig)(\mHps-\mHs)\dfrac{\sbma}{2\mw}$\\\hline
            $H^{\pm}-G^{\mp}-h^0$&$ig(\mHps-\mhs)\dfrac{\cbma}{2\mw}$\\\hline
            $H^{\pm}-G^{\mp}-A^0$&$\pm g\,\dfrac{(\mHps-\mAs)}{2\mw}$\\\hline
            $G^{\pm}-G^{\mp}-H^0$&$(-ig)\dfrac{\mHs\cbma}{2\mw}$\\\hline
            $G^{\pm}-G^{\mp}-h^0$&$(-ig)\dfrac{\mhs\sbma}{2\mw}$\\\hline
            $G^{\pm}-G^{\mp}-A^0$&$0$\\\hline
        \end{tabular}
        \caption{Feynman rules for the trilinear couplings involving the Higgs 
          self-interactions and the Higgs and Goldstone boson vertices in the Feynman 
          gauge, with all momenta pointing inward. These rules are common to both Type~I 
          and Type~II 2HDM under the conditions explained in the text. We have singled out 
          some null entries associated to CP violation.}
        \label{tab:trilineals}
    \end{table}

\end{itemize}

\section{Constraints}
\label{sec:constraints}
There are multiple constraints that must be imposed, obviously one of such
constraints is that they must reproduce the behaviour of the SM up to energy
scales probed so far.

Analysing the perturbativity of the theory one finds that the allowed range for
$\tb$ is:
\begin{equation}
    0.1<\tan\beta\lesssim60\,.  \label{eq:tbrange} 
\end{equation}

The custodial symmetry\cite{Einhorn:1981cy,Veltman:1977kh} ($SU(2)$) is a good
symmetry at tree level, so the quadratic violations of this symmetry must be
experimentally fixed. So, the one-loop corrections of the parameter $\rho$
from the 2HDM sector can not be bigger than one per mil of the SM
\thinspace\cite{Groom:2000in}:
\begin{equation}
    |\delta\rho^{\text{2HDM}}|\leqslant0.001\,.  \label{eq:drho}
\end{equation}
To be precise, the latter is the extra effect
that $\delta\rho$ can accommodate at one standard deviation
($1\,\sigma$) from the 2HDM fields beyond the SM contribution
\cite{\GuaschNPo}. This is a stringent restriction that affects the
possible mass splittings among the Higgs fields of the 2HDM, and
its implementation in our codes does severely prevent the
possibility from playing with the Higgs boson masses to
artificially enhance the FCNC contributions.

Moreover, the charged Higgs bosons have an important indirect
restriction from the radiative decays of the $B$ meson, specially the ratio
$B(B\rightarrow X_{s}\,\gamma) $ -- or $\Bbsg$ at the quark
level~\cite{\bsgexp}:
\begin{equation}
    \Bbsg=(3.3\pm0.4)\times 10^{-4}\,.
    \label{eq:CLEO}
\end{equation}
The Higgs contribution to $\Bbsg$ (that have been computed at the NLO in
QCD\,\cite{\Gambino}) is positive: bigger experimental ratio means
that the charged Higgs mass can be smaller. From the different analysis of the
literature \cite{\bsgthdm} we get:
\begin{equation}
    \label{eq:restbsg}
    m_{H^{\pm}}>350\GeV
\end{equation}
for virtually any $\tan\beta\gtrsim1$.
This bound does not apply to Type~I models because 
at large $\tan\beta$ the charged Higgs couplings are severely suppressed, 
whereas at low $\tan\beta$ we recover the previous unrestricted situation of 
Type~II models.

We can derive lower bounds for the neutral Higgs masses in these
models~\cite{Krawczyk:1997be,Krawczyk:1998wg}. For example, using the Bjorken
process $e^{+}e^{-}\rightarrow Z+h^{0}$ and the production of pairs of Higgs
boson $e^{+}e^{-}\rightarrow h^{0}(H^{0})+A^{0}$ we can get the following
restrictions in almost all the parameter
space~\cite{Abbiendi:2000ug,Abbiendi:1998rd}:
\begin{equation}
    \label{eq:restbjorken}
    m_{h^{0}}+m_{A^{0}}\begin{cases}\gtrsim100\GeV &\forall\tan\beta\\
      \gtrsim150\GeV&\tan\beta>1\end{cases}\,.
\end{equation}
In each of these cases there is a small region in the parameter space in the
ranges of the CP-even Higgs masses and CP-odd masses
around~\cite{Abbiendi:2000ug,Abbiendi:1998rd}:
\begin{equation}
    \label{eq:raco}
    m_{h,^{0}A^{0}}=20-30\GeV\,.
\end{equation}
Although, as can be seen in the electroweak precision fits in
Ref.~\cite{Chankowski:1999ta}, in the high $\tan\beta$ range a light Higgs
boson $h^{0}$ is statistically correlated with a light $H^{\pm}$, so
this situation is not favoured by the $\bsg$ restriction.
Moreover, since our interest in Type~II 
models is mainly focused in the large $\tan \beta$ regime, the corner in the 
light CP-even mass range is a bit contrived. At the end of the day  
one finds that, even in the worst situation, the strict experimental limits still allow generic 2HDM neutral scalar bosons as light as $70\,GeV$ or so. As we said, most of 
these limits apply to Type~II 2HDM's, but we will conservatively apply them 
to Type~I models as well.

Finally, the unitarity bound can be approximately formulated by imposing
that the absolute value of the trilinear coupling
of the 2HDM Higgs can not be bigger than the Trilinear coupling of the SM
Higgs:
\begin{equation} 
    \ \left| \lambda_{HHH}\right| \leqslant\left| 
        \lambda_{HHH}^{(SM)}(m_{H}=1\,TeV)\right| =\frac{3\,g\,(1\,TeV)^{2}}{2\,M_{W}}\,\,. 
    \label{eq:unitbound} 
\end{equation} 


%% file: mssm.tex
\chapter{The Minimal Supersymmetric Standard Model (MSSM)}
\label{cap:MSSM}

\section{Introduction}
\label{sec:MSSMintro}

It goes beyond the scope of this Thesis to study the formal theory of
Supersymmetry\,\cite{Wess:1992cp,Gates:1983nr}, however we would like, at
least, to give a feeling on what is it. Supersymmetry (SUSY) can be
introduced in many manners.

Let us consider the symmetries of the scattering matrix $S$, that is, those
transformations that can be reduced to an interchange of asymptotic states.
Before the discovery of Supersymmetry the only symmetries know were: the
following: (1) the ones corresponding to the Poincare group; (2) the so
called internal global symmetries, both of them ruled by a Lie algebra;
and (3) discrete
symmetries such as parity (P), charge conjugations (C) and the time reversal
(T). A 1967 theorem due to Coleman and Mandula establishes rigorously that,
under very general conditions, these are the only symmetries allowed for $S$
if we do not want to induce trivial scattering (fixed angles and speeds) in
$2\to 2$ processes.

This theorem may be eluded relaxing some of its hypotheses. The Supersymmetry
appears precisely when assuming that the generators of the new symmetry we
want to add have a spinorial character instead of a scalar one, therefore transforming
under $(\frac{1}{2},0)$ and $(0,\frac{1}{2})$ representations of the Lorentz
group. Fermionic spinorial generators necessarily have an anti-commutative
algebra, generically known as a graded Lie algebra. The algebra is not closed
with just the SUSY generators, thus it can not be understood as an internal symmetry, but it
rather forms an extension of the space-time symmetries of the Poincare
group.

Following this line of thought, one could relax some other hypotheses of the
Coleman-Mandula theorem in order to introduce new theories. SUSY is the
only known extension for the $S$ matrix symmetries. Accepting as the only valid
extensions of the Coleman-Mandula theorem conditions the presence of a graded
Lie algebra, one can show (Haag, Lopusza{\'n}ki and Sohnius theorem) that
spinorial generators different from those of SUSY are forbidden.

Next, we have to define the ``superspace'', the supersymmetric space of the
``superfields''. We add to the space-time coordinates $x$ other sets of
spinorial coordinates $(\theta,\bar\theta)$ (as many sets as the dimension of the
space-time) that are Grassmann variables, i.e.\ they anti-commute. In the case
of adding just one set it is said that we have a $N=1$ Supersymmetry and a
$N=1$ superspace:
\begin{eqnarray}
    \mathrm{space-time}&\to&N=1\;\mathrm{Superspace}\notag\\
    x^\mu&\to&(x^\mu,\theta^\alpha,\bar\theta^{\dot\alpha})
\end{eqnarray}

where $\alpha=1,2$. The supersymmetric transformations have the parameters
$(\Lambda,a,\xi,\bar\xi)$, where $\Lambda$ is the Lorentz matrix, $a$ is the
translation 4-vector, and the Weyl spinors $\xi,\bar\xi$. The generators of
SUSY transformations consist of are
the ones of the Poincare group and the new spinorial generators
$Q_\alpha$ and $\bar Q_{\dot\alpha}$, satisfying the graded Lie algebra. The
infinitesimal purely SUSY transformation of a superfield
$\Phi(x,\theta,\bar\theta)$ is:
\begin{align}
    \Phi &\to \Phi + \delta_S \Phi\\
    \delta_S&=-i(\xi^\alpha Q_\alpha+\bar\xi_{\dot\alpha}\bar Q^{\dot\alpha})
\end{align}%
The functions defined in the Superspace are polynomial functions of the
$(\theta,\bar\theta)$ variables (since
$\theta_\alpha^2=\bar\theta_{\dot\alpha}^2=0$). Thus we can decompose the
functions (superfields) of this Superspace in components of $\theta^0$,
$\theta_\alpha$, $\bar\theta_{\dot\alpha}$, $\theta_\alpha \theta_\beta$,
\ldots each of these components will be a function of the space-time
coordinates. Analogously to the space-time, we can define in the Superspace
scalar superfields, vector superfields, \ldots For example in a 4-dimensional
space time with $N=1$ supersymmetry a scalar superfield has 10 components.

We can define fields with specific properties with respect to the
$\theta$ variables\cite{Wess:1992cp,Gates:1983nr}.
A scalar \emph{chiral} field in a $4D$ $N=1$ Superspace
has 4 components:
\begin{align}
    \Phi_L&=A+\sqrt{2}\theta\psi+\theta\theta F \equiv (A,\psi,F)\\
    \Phi_R&=A^*+\sqrt{2}\bar\theta\bar\psi+\bar\theta\bar\theta F^* \equiv
    (A^*,\bar\psi,F^*)
\end{align}

where $A$ is a scalar field, $\psi$ and $\bar\psi$ are Weyl spinors
(left-handed and right handed Dirac fermions) and $F$ is an auxiliary scalar
field. This auxiliary field is not a dynamical field since its equations of
motion do not involve time derivatives. To this end we are left with a
superfield, whose components represent an ordinary scalar field and an
ordinary chiral spinor. So if nature is described by the dynamics of this
field we would find a chiral fermion and a scalar with identical quantum
numbers. That is {\em Supersymmetry relates particles which differ by spin
  1/2}. When a SUSY transformation ($Q$) acts
on a superfield it transform spin $s$ particles into spin $s\pm1/2$ particles.

Thus, for a $N=1$ SUSY, we find that to any chiral fermion there should be a
scalar particle with exactly the same properties. This fact is on the basis of
the absence of quadratic divergences in boson mass renormalization, since for
any loop diagram involving a scalar particle there should be a fermionic loop
diagram, which will cancel quadratic divergences between each other, though
logarithmic divergences remain.

Supersymmetric interactions can be introduced by means of generalized
gauge transformations, and by means of a generalized potential function, the
Superpotential, which give rise to masses, Yukawa-type interactions, and a scalar
potential. 

As no scalar particles have been found at the electroweak scale we may infer
that, if SUSY exists, it is broken. We can allow SUSY to be broken
maintaining the property that no quadratic divergences are allowed: this is
the so called Soft-SUSY-Breaking mechanism\,\cite{Girardello:1981wz}. We can
achieve this by only introducing a small set of SUSY-Breaking terms in the
Lagrangian, to wit: masses for the components of lowest spin of a
supermultiplet and triple scalar interactions. However, other terms like
explicit fermion masses for the matter fields would violate the
Soft-SUSY-Breaking condition.

The MSSM is the minimal Supersymmetric extension of the Standard Model. It
is introduced by means of a $N=1$ SUSY, with the minimum number of new
particles. Thus for each fermion $f$ of the SM there are two scalars related
to its chiral components called ``sfermions'' ($\tilde{f}_{L,R}$), for each
gauge vector $V$ there is also a chiral fermion: ``gaugino'' ($\tilde{v}$),
and for each Higgs scalar $H$ another chiral fermion: ``higgsino''
($\tilde{h}$).  In the MSSM it turns out that, in order to be able of giving
masses to up-type and down-type fermions, we must introduce two Higgs doublets
with opposite hypercharge, and so the MSSM Higgs sector is of the so called
Type II 2HDM (see chapter~\ref{cha:2hdm}, section~\ref{sec:hmas} and Ref.\,\cite{Gunion:1989we}).

To build the MSSM Lagrangian we must build a Lagrangian invariant under the
gauge group $\mathrm{SU}(3)_C\times\mathrm{SU}(2)_L\times\mathrm{U}(1)_Y$,
it must also include the superfields with the particle content of the
Table~\ref{tab:MSSMparticles} and in addition it must contain the terms that
breake supersymmetry
softly. But this Lagrangian violates the baryonic and leptonic number, so we have
to introduce an additional symmetry. In the case of the MSSM this symmetry is
the so-called $R$-symmetry. In its discrete form it relates the spin ($S$),
the baryonic number
($B$) and the leptonic number ($L$) in the so-called $R$-parity:
\begin{equation}
    R=(-1)^{2S+L+3B}
\end{equation}
so that is $1$ for the SM fields and $-1$ for its supersymmetric partners. In
the way the MSSM is implemented $R$-parity is conserved, this means that
$R$-odd particles (the superpartners of SM particles) can only be created in
pairs, also that in the final product decay of an $R$-odd particle at least
one SUSY particle exists, and that the Lightest Supersymmetric Particle
(LSP) is stable.

\section{Field content}
\label{sec:MSSMfieldcontent}

The field content of the MSSM consist of the fields of the SM plus all their
supersymmetric partners, and an additional Higgs doublet. The
Table~\ref{tab:MSSMparticles} shows all the correspondences and all the
fields. All these fields suffer some mixing, so the physical (mass
eigenstates) fields look much different from these ones, as shown in
Table~\ref{tab:MSSMmasseigen}. The gauge fields mix up to give the well known
gauge bosons of the \SM, $W^\pm_\mu$, $Z^0_\mu$, $A_\mu$, the gauginos and
higgsinos mix up to give the chargino and neutralino fields, and finally the
Left- and Right-chiral sfermions mix among themselves in sfermions of
indefinite chirality. Letting aside the intergenerational mixing between fermions
and sfermions that give rise to the well known Cabibbo-Kobayashi-Maskawa
(CKM and superCKM) matrix.

\begin{table}[ht]
    \centering
    \begin{tabular}{|c|c|c|c|c|c|}\hline
        Superfield&SM particle&Sparticle& $\mathrm{SU}(3)_C$ &
        $\mathrm{SU}(2)_L$ & $\mathrm{U}(1)_Y$ \\\hline\hline
        Matter&&&&&\\

        \begin{tabular}{c}$\hat{L}$\\$\hat{R}$\end{tabular}&
        leptons$\begin{cases}L=(\nu_l,l)_L\\R=l^-_L\end{cases}$&
        \emph{sleptons}$\begin{cases}\tilde{L}=(\tilde{\nu}_{lL},\tilde{l}_L)\\
            \tilde{R}=\tilde{l}^+_R\end{cases}$&
        \begin{tabular}{c}$1$\\$1$\end{tabular}&
        \begin{tabular}{c}$2$\\$1$\end{tabular}&
        \begin{tabular}{c}$-1$\\$2$\end{tabular}\\

        \begin{tabular}{c}$\hat{Q}$\\$\hat{U}$\\$\hat{D}$\end{tabular}&
        quarks$\begin{cases}Q=(u,d)_L\\U=u^c_L\\D=d^c_L\end{cases}$&
        \emph{squarks}$\begin{cases}\tilde{Q}=(\tilde{u}_L,\tilde{d}_L)\\
            \tilde{U}=\tilde{u}^*_R\\\tilde{D}=\tilde{d}^*_R\end{cases}$&
        \begin{tabular}{c}$3$\\$3^*$\\$3^*$\end{tabular}&
        \begin{tabular}{c}$2$\\$1$\\$1$\end{tabular}&
        \begin{tabular}{c}$1/3$\\$-4/3$\\$2/3$\end{tabular}\\

        \begin{tabular}{c}$\hat{H}_1$\\$\hat{H}_2$\end{tabular}&
        Higgs$\begin{cases}H_1=(H^0_1,H^-_1)
            \\H_2=(H^+_2,H^0_2)\end{cases}$&
        \emph{Higgsinos}$\begin{cases}\tilde{H}_1=(\tilde{H}^0_1,\tilde{H}^-_1)\\
            \tilde{H}_2=(\tilde{H}^+_2,\tilde{H}^0_2)\end{cases}$&
        \begin{tabular}{c}$1$\\$1$\end{tabular}&
        \begin{tabular}{c}$2$\\$2$\end{tabular}&
        \begin{tabular}{c}$-1$\\$1$\end{tabular}\\\hline

        Gauge&&&&&\\
        $\hat{G}$&gluon\;\;$g_\mu$&\emph{gluino}\;\;$\tilde{g}$&$8$&$0$&$0$\\
        $\hat{V}$&w\;\;$(W_1,W_2,W_3)$&
        \emph{wino}\,$(\tilde{W}_1,\tilde{W}_2,\tilde{W}_3)$&$1$&$3$&$0$\\
        $\hat{V^\prime}$&b\;\;$B^0$&
        \emph{bino}\,$(\tilde{B}^0$&$1$&$1$&$0$\\\hline
    \end{tabular}
    \caption{Particle contents of the MSSM superfields}
    \label{tab:MSSMparticles}
\end{table}

\begin{table}[ht]
    \centering
    \begin{tabular}{|c|c|c|}\hline
        Name&Mass eigenstates&Gauge eigenstates\\\hline\hline
        Higgs bosons&$h^0\;H^0\;A^0\;H^\pm$&$H^0_1\;H^0_2\;H^-_1\;H^+_2$\\\hline
        squarks&$\tilde{t}_1\;\tilde{t}_2\;\tilde{b}_1\;\tilde{b}_2\;$&
        $\tilde{t}_L\;\tilde{t}_R\;\tilde{b}_L\;\tilde{b}_R\;$\\\hline
        sleptons&$\tilde{\tau}_1\;\tilde{\tau}_2\;\tilde{\nu}_\tau$&
        $\tilde{\tau}_L\;\tilde{\tau}_R\;\tilde{\nu}_\tau$\\\hline
        neutralinos&$\tilde{N}_1\;\tilde{N}_2\;\tilde{N}_3\;\tilde{N}_4\;$&
        $\tilde{B}^0\;\tilde{W}^0\;\tilde{H}^0_1\;\tilde{H}^0_2$\\\hline
        charginos&$\tilde{C}^\pm_1\;\tilde{C}^\pm_2$&
        $\tilde{W}^\pm\;\tilde{H}^-_1\;\tilde{H}^+_2$\\\hline
    \end{tabular}
    \caption{Mass eigenstates of the MSSM particles. For notational simplicity
      only the third sfermion generation is presented.}
    \label{tab:MSSMmasseigen}
\end{table}

\section{Lagrangian}
\label{sec:MSSMlagrangian}
The \MSSM\  interactions come from three different kinds of sources:

\begin{itemize}
\item Superpotential: 
    \begin{equation}
        W=\epsilon_{ij}\left[ f \hat H_1^i \hat L^j \hat R
            +h_d \hat H_1^i \hat Q^j \hat D
            +h_u \hat H_2^j \hat Q^i \hat U-\mu \hat H_1^i \hat H_2^j
        \right]\,\,.\\
        \label{eq:superp}\end{equation}
    The superpotential contributes to the interaction
    Lagrangian~(\ref{eq:ltotal}) with two different kind of interactions. The
    first one is the Yukawa interaction, which is obtained
    from~(\ref{eq:superp}) just replacing two of the superfields by its
    fermionic field content, whereas the third superfield is replaced by its
    scalar field content:
    \begin{equation}
        \begin{array}{lcl}
            V_Y&=&\epsilon_{ij}\left[ f  H_1^i L^j R
                +h_d H_1^i Q^j D
                +h_u H_2^j Q^i U-\mu \tilde H_1^i \tilde H_2^j
            \right]\\
            ~&~&+\epsilon_{ij}\left[ f \tilde H_1^i L^j \tilde  R
                +h_d \tilde  H_1^i Q^j \tilde D
                +h_u \tilde H_2^j Q^i \tilde U
            \right]\\
            &~&+\epsilon_{ij}\left[ f \tilde H_1^i \tilde L^j R
                +h_d \tilde  H_1^i \tilde Q^j D
                +h_u \tilde H_2^j \tilde Q^i U
            \right]\\
            ~&~&+\mbox{ h.c.}\,\,.
        \end{array}
        \label{eq:vyukawa}
    \end{equation}
    The second kind of interactions are obtained by means of taking the
    derivative of the superpotential:
    \begin{equation}
        V_W=\sum_i \left|\frac{\partial W\left(\varphi\right)}{\partial
              \varphi_i}\right|^2\,\,,
        \label{eq:superder}\end{equation}
    $\varphi_i$ being the scalar components of superfields.
\item Interactions related to the gauge symmetry, which contain:
    \begin{itemize}
    \item the usual gauge interactions 
    \item the gaugino interactions:
        \begin{equation}
            V_{\tilde G \psi \tilde \psi}=
            i \sqrt{2} g_a \varphi_k \bar \lambda ^a \left( T^a \right)_{kl}
            \bar\psi_l+\mbox{ h.c.}
            \label{eq:vgfsf}\end{equation}
        where $(\varphi,\psi)$ are the spin $0$ and spin $1/2$ components of a chiral
        superfield respectively, $T^a$ is a
        generator of the gauge symmetry, $\lambda_a$ is the gaugino field and $g^a$ 
        its coupling constant.
    \item and the $D$-terms, related to the gauge structure of the theory, but that
        do not contain neither gauge bosons nor gauginos:
        \begin{equation}
            V_D=\frac{1}{2}\sum D^a D^a\,\,,
            \label{eq:vd}\end{equation}
        with
        \begin{equation}
            D^a= g^a \varphi_i^* \left(T^a\right)_{ij} \varphi_j \,\,,
        \end{equation}
        $\varphi_i$ being the scalar components of the superfields.
    \end{itemize}
\item Soft-\susy-Breaking interaction terms:
    \begin{equation}
        V_{\rm soft}^{\rm I}=\frac{g}{\sqrt{2} M_W \cos{\beta}} \epsilon_{ij}\left[
            m_l A_l H_1^i \tilde L^j \tilde R+
            m_d A_d H_1^i \tilde Q^j \tilde D-m_u A_u H_2^i \tilde Q \tilde U
        \right]+\mbox{ h.c. } \,\,.
        \label{eq:softsusy}\end{equation}
    The trilinear Soft-\susy-Breaking couplings $A_f$ can play an important role,
    specially for the third generation interactions and masses, and they are in the
    source of the large value of the bottom quark mass renormalization effects.
\end{itemize}

The full \MSSM\  Lagrangian is then:
\begin{eqnarray}
    {\cal L}_{\rm MSSM}&=&
    {\cal L}_{\rm Kinetic}+
    {\cal L}_{\rm Gauge}
    -V_{\tilde G \psi \tilde \psi}-V_D-
    V_Y-\sum_i \left|
        \frac{\partial W\left(\varphi\right)}{\partial \varphi_i}\right|^2\nonumber\\
    ~&~&
    -V_{\rm soft}^{\rm I}
    -m_1^2\, H_1^\dagger H_1
    -m_2^2\, H_2^\dagger H_2
    -m_{12}^2\, \left(H_1 H_2+H_1^\dagger H_2^\dagger\right)\nonumber\\
    ~&~&
    -\frac{1}{2}m_{\sg}\, \psi^a_{\sg} \psi^a_{\sg}
    -\frac{1}{2}M \,\tilde w_i \tilde w_i
    -\frac{1}{2}M^\prime\, \tilde B^0 \tilde B^0\nonumber\\
    ~&~&-m_{\tilde L}^2\, \tilde L^* \tilde L
    -m_{\tilde R}^2\, \tilde R^* \tilde R
    -m_{\tilde Q}^2\, \tilde Q^* \tilde Q
    -m_{\tilde U}^2\, \tilde U^* \tilde U
    -m_{\tilde D}^2\, \tilde D^* \tilde D\,\,,
    \label{eq:ltotal}\end{eqnarray}
where we have also included the Soft-\susy-breaking masses.

From the Lagrangian~(\ref{eq:ltotal}) we can obtain the full \MSSM\  spectrum,
as well as the interactions, which contain the usual \SM\  gauge interactions,
the fermion-Higgs interactions that correspond to a Type II Two-Higgs-Doublet
Model~\cite{\Hunter}, and the pure \susy\  interactions. A very detailed
treatment of this Lagrangian, and the process of derivation of the forthcoming
results can be found in\,\cite{TesiJefe}.

\section{\MSSM\  spectrum}
\label{sec:MSSMspectrum}
\subsection{Higgs boson sector}
\label{sec:hmas}
As seen in \ref{sec:2hdm:introduction},
when a Higgs doublet is added to the \SM\  there exist two possibilities for
incorporating it, avoiding Flavour Changing Neutral Currents (\FCNC) at tree
level\,\cite{\Hunter}. The first possibility is not to allow a coupling between
the second doublet and the fermion fields, this is the so called Type I
\thdm. The second possibility is to allow both Higgs doublets to couple with
fermions, the first doublet only coupling to the Right-handed down-type
fermions, and the second one to Right-handed up-type fermions, this is the so
called Type II \thdm. 

The Higgs sector of the \MSSM\  is that of a Type II \thdm\,\cite{\Hunter}, with some
\susy\  restrictions. After expanding~(\ref{eq:ltotal})
the Higgs potential reads 
\begin{eqnarray}
    V &=& m_1^2\,|H_1|^2+m_2^2\,|H_2|^2-m_{12}^2\,\left(
        \epsilon_{ij}\,H_1^i\,H_2^j+{\rm h.c.}\nonumber\right)\\
    &+&\frac{1}{8}(g^2+g'^2)\,\left(|H_1|^2-|H_2|^2\right)^2
    +\frac{1}{2}\,g^2\,|H_1^{\dagger}\,H_2|^2\,.
    \label{eq:potential}
\end{eqnarray}
This is equivalent to the 2HDM potential \eqref{eq:potencialHiggs} with the following restrictions:
\begin{align}
    \lambda_1&=\lambda_2\\
    \lambda_3&=\frac{1}{8}(g^2+g'^2)-\lambda_1\\
    \lambda_4&=2\lambda_1-\frac{1}{2}g^2\\
    \lambda_5&=\lambda_6=2\lambda_1-\frac{1}{2}(g^2+g'^2).
\end{align}
The neutral Higgs bosons fields acquire a vacuum expectation
value (\vev),
\begin{equation}
    \langle H_1 \rangle_0=\begin{pmatrix}v_1\\0\end{pmatrix}
    ~~ \langle H_2 \rangle_0 =\begin{pmatrix}0\\v_2\end{pmatrix}\,\,.
    \label{eq:VeVfields}
\end{equation}
We need two physical parameters in order to know their value, which are usually
taken to be $M_W$ and $\tb$:
\begin{align}
    M_W^2&=\frac{1}{2} g^2 (v_1^2+v_2^2)\equiv
    g^2\frac{v^2}{2}\\
    M_Z^2&=\frac{1}{2}(g^2+g'^2) v^2\equiv M_W^2\cos^2\theta_W\\
    \tan\beta&=\frac{v_2}{v_1}\,\,,\;\;\;0<\beta<\frac{\pi}{2}\\
    \tan\theta_W&=\frac{g'}{g}
    \label{eq:VEV}
\end{align}

These \vev's make the Higgs fields to mix up. There are five physical Higgs
fields: 
a couple of charged Higgs bosons ($H^\pm$); a ``pseudoscalar'' Higgs ($CP=-1$) 
$A^0$; and two scalar Higgs bosons ($CP=1$) $H^0$ (the heaviest) and  $h^0$ (the
lightest). There are also the Goldstone bosons $G^0$ and $G^\pm$. The relation
between the physical Higgs fields and that fields of~(\ref{tab:MSSMmasseigen}) is
\begin{align}
    \label{eq:HiggsDefin}
    \begin{pmatrix}-H^{\pm}_1\\ H^{\pm}_2\end{pmatrix}
    &=\begin{pmatrix}\cos\beta&-\sin\beta\\\sin\beta&\cos\beta\end{pmatrix}
    \begin{pmatrix}G^{\pm}\\ H^{\pm}\end{pmatrix}\,,\\
    \begin{pmatrix}H^{0}_1\\ H^{0}_2\end{pmatrix}
    &=\begin{pmatrix}v_1\\ v_2\end{pmatrix}
    +\frac{1}{\sqrt{2}}\begin{pmatrix}\cos\beta&-\sin\beta\\\sin\beta&\cos\beta\end{pmatrix}
    \begin{pmatrix}H^{0}\\ h^{0}\end{pmatrix}\\\notag
    &\qquad\qquad+\frac{i}{\sqrt{2}}\begin{pmatrix}-(\cos\beta&-\sin\beta)\\\sin\beta&\cos\beta\end{pmatrix}
    \begin{pmatrix}G^{0}\\ A^{0}\end{pmatrix}
\end{align}
where $\alpha$ is given in~(\ref{eq:alfa})\cite{\Hunter}.

All the masses of the Higgs sector of the \MSSM\ can be obtained with only two
parameters, the first one is $\tb$~(\ref{eq:VEV}), and the second one is a
mass; usually this second parameter is taken to be either the charged Higgs
mass $\mHp$ or the pseudoscalar Higgs mass $\mA$. We will take the last
option. From~(\ref{eq:potential}) one can obtain the tree-level mass
relations between the different Higgs particles,
\begin{align}
    \mHps&=\mAs+\mws \nonumber\,\,,\\ 
    m_{H^0,h^0}^2&=\frac{1}{2} \left(
        \mAs+\mzs\pm\sqrt{\left(\mAs+\mzs\right)^2-4\,\mAs\,\mzs
          \cos^2 2\beta}
    \right)\,\,,
    \label{eq:mh}
\end{align}
and the mixing angle between the two scalar Higgs is obtained by means of:
\begin{equation}
    \cos 2\alpha= - \cos 2 \beta
    \left(\frac{\mAs-\mzs}{\mHs-\mhs}\right)\,\,,\,\,
    \sin 2 \alpha=- \sin 2 \beta \left(\frac{\mHs+\mhs}{\mHs-\mhs}\right)\,\,.
    \label{eq:alfa}
\end{equation}
The immediate consequence of such a constrained Higgs sector, is the existence
of absolute bounds (at tree level) for the Higgs masses:
\begin{equation}
    0<\mh<\mz<\mH,~~~\mw<\mHp.
    \label{eq:MSSMmassbounds}
\end{equation}
It must be taken into account, though, that the radiative corrections, mainly
due to the top-stop supermultiplet, and also the bottom-sbottom one, are
susceptible of relaxing the limits \eqref{eq:MSSMmassbounds} in a
significant manner.

A good approximation for effective mixing angle $\alpha_\mathrm{eff}$
including only the leading one-loop contributions of top, stop, bottom
and sbottom follows from the diagonalization of the one-loop Higgs mass
matrix\cite{Dabelstein:1995hb,Kunszt:1991qe,Carena:1995wu,Carena:1995bx}:
\begin{equation}
    \mathcal{M}^2_\textrm{Higgs} = \frac{ \sin 2 \beta }{2}\left(
        \begin{array}{ll}
            \cot\beta\ \mzs + \tan\beta\ \mAs + \sigma_t + \omega_b &
            - \mzs - \mAs + \lambda_t +\lambda_b \\
            - \mzs - \mAs + \lambda_t +\lambda_b&
            \tan\beta\ \mzs + \cot\beta\ \mAs + \omega_t + \lambda_b
        \end{array} \right) \ .
    \label{glalphaap}
\end{equation}
where
\begin{eqnarray}
    \omega_t = \frac{N_C G_F m_{t}^4 }{\sqrt{2} \pi^2 \sin^2 \beta} \
    & & \hspace{-0.5cm}
    \left(  \log \ ( \frac{m_{\tilde{t}_1} m_{\tilde{t}_2} }{m_{t}^2} )
        + \frac{A_t ( A_t - \mu \cot \beta)}{ m_{\tilde{t}_1}^2 -
          m_{\tilde{t}_2}^2 } \log \frac{ m_{\tilde{t}_1}^2 }{ m_{\tilde{t}_2}^2 }
    \right. \nonumber \\ & &  \hspace*{-0.5cm}    \left. +
        \frac{A_t^2 ( A_t - \mu \cot \beta)^2 }{ ( m_{\tilde{t}_1}^2 -
          m_{\tilde{t}_2}^2 )^2} \left( 1 - \frac{m_{\tilde{t}_1}^2 + m_{\tilde{t}_2}^2}
            {m_{\tilde{t}_1}^2 - m_{\tilde{t}_2}^2} \log \frac{m_{\tilde{t}_1}}
            {m_{\tilde{t}_2}} \right) \ \right)   \nonumber \\
    \lambda_t = - \frac{N_C G_F m_{t}^4 }{\sqrt{2} \pi^2 \sin^2 \beta} \
    & & \hspace{-0.5cm}
    \left( \frac{ \mu ( A_t - \mu \cot \beta)}{ m_{\tilde{t}_1}^2 -
          m_{\tilde{t}_2}^2 } \log \frac{ m_{\tilde{t}_1}^2 }{ m_{\tilde{t}_2}^2 }
    \right. \nonumber \\ & &  \hspace*{-0.5cm}    \left. +
        \frac{2 \mu A_t ( A_t - \mu \cot \beta)^2 }{ ( m_{\tilde{t}_1}^2 -
          m_{\tilde{t}_2}^2 )^2} \left( 1 - \frac{m_{\tilde{t}_1}^2 + m_{\tilde{t}_2}^2}
            {m_{\tilde{t}_1}^2 - m_{\tilde{t}_2}^2} \log \frac{m_{\tilde{t}_1}}
            {m_{\tilde{t}_2}} \right) \ \right)   \nonumber \\
    \sigma_t = \frac{N_C G_F m_{t}^4 }{\sqrt{2} \pi^2 \sin^2 \beta} \
    & & \hspace{-0.5cm}
    \frac{ \mu^2 ( A_t - \mu \cot \beta)^2 }{ ( m_{\tilde{t}_1}^2 -
      m_{\tilde{t}_2}^2 )^2} \left( 1 - \frac{m_{\tilde{t}_1}^2 + m_{\tilde{t}_2}^2}
        {m_{\tilde{t}_1}^2 - m_{\tilde{t}_2}^2} \log \frac{m_{\tilde{t}_1}}
        {m_{\tilde{t}_2}} \right)  \ 
    \label{leadom2}
\end{eqnarray}
with the following substitutions for the bottom/sbottom factors:
\begin{eqnarray}
    (\omega_b,\lambda_b,\sigma_b) \leftrightarrow (\omega_t,\lambda_t,\sigma_t)
    &&\text{with}
    \begin{cases}t&\leftrightarrow b\\\sin\beta&\leftrightarrow\cos\beta\\
        A_t-\mu\ctb&\leftrightarrow A_b-\mu\tb\end{cases}
\end{eqnarray}
This approximate effective mixing angle $\alpha_{eff}$ is determined by
\begin{equation}
    \tan \alpha_{eff} = \frac{ - (\mAs + \mzs)\
      \tb+(\lambda_t+\lambda_b)(1+\tbs)/2}
    {\mzs + \mAs \tan^2 \beta +(\sigma_t +\omega_b -
      M_{h^0,eff}^2)(1+\tbs)}\ ,
    \label{glalpheffx}
\end{equation}
where $M_{h^0,eff}$ is the solution for the light Higgs mass:
\begin{eqnarray}
    M^2_{H^0,h^0,\, eff} & = &
    \frac{\mAs + \mzs + \omega_t + \sigma_t + \omega_b + \sigma_b}{2}
    \pm \, \left( \ \frac{ (\mAs + \mzs)^2 + ( \omega_t - \sigma_t + \sigma_b
          - \omega_b)^2}{4}\right.\nonumber\\
    & &  \left. - \mAs \mzs \cos^2 2\beta 
        + \ \frac{(\omega_t - \sigma_t + \sigma_b - \omega_b) 
          \cos 2\beta}{2}  (\mAs - \mzs)\right.\nonumber\\
    & & \left.  - \frac{(\lambda_t+\lambda_b \sin 2\beta)}{2} (\mAs + \mzs) +
        \frac{(\lambda_t+\lambda_b)^2}{4} \ \right)^{1/2} \ .
    \label{glmapp2}
\end{eqnarray}

\begin{figure}[t!]
    \begin{center}
        \includegraphics*[width=0.5\textwidth]{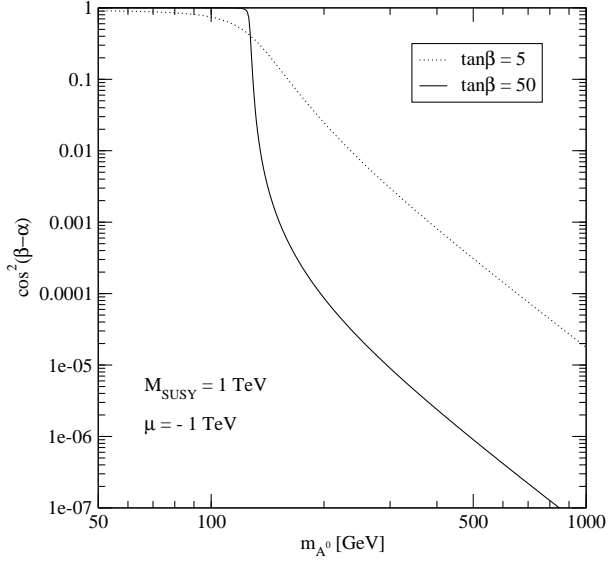}
    \end{center}
\caption{The value of $\cos^2(\beta-\alpha)$
is shown as a function of
$\ma$ for two choices of $\tan\beta = 5$ and $\tan\beta = 50$.
When radiative-corrections are included, one can define an approximate
loop-corrected angle $\alpha$ as a function of $\ma$, $\tan\beta$ and
the MSSM parameters.  In the figures above, we
have incorporated radiative corrections, assuming that
$\msusy\equiv \msq=\msd=\msup=1\TeV$.
 The decoupling effect, in which $\cos^2(\beta-\alpha)\propto \mz^4/{\ma}^4$
for $\ma\gg m_Z$,
continues to hold even when radiative corrections are included.}
    \label{cosgraph}
\end{figure}

In the limit where $\mA\gg\mz$, $\cos(\beta-\alpha)={\cal O}(\mzs/\mAs)$,
which means that the $h^0$ couplings to Standard Model particles approach
values corresponding precisely to the couplings of the SM Higgs boson.  There
is a significant region of MSSM Higgs sector parameter space in which the
decoupling limit applies, because $\cos(\beta-\alpha)$ approaches zero quite
rapidly once $\mA$ is larger than about 200~GeV, as shown in Fig.~\ref{cosgraph}.
As a result, over a significant region of the MSSM parameter space, the search
for the lightest CP-even Higgs boson of the MSSM is equivalent to the search
for the \SM\ Higgs boson.  This result is more general; in many theories of
non-minimal Higgs sectors, there is a significant portion of the parameter
space that approximates the decoupling limit.  Consequently, simulations of
the \SM\ Higgs signal are also relevant for exploring the more general Higgs
sector.

\subsection{The \SM\  sector}
\label{sub:massasm}
In this section we give some expressions to obtain some \MSSM\  parameters as a
function of the \sm\  parametrization.

As stated above~(sec. \ref{sec:hmas}) the \vev's can be obtained by means
of~(\ref{eq:VEV}), and the $Z$ mass can be obtained at tree-level by the relation:
$$
\sin^2 \theta_W=1-\frac{M_W^2}{\mzs}\,\,.
$$

Fermion masses are obtained from the Yukawa potential~(\ref{eq:vyukawa}) letting
the neutral Higgs fields acquire their \vev~(\ref{eq:VeVfields}). The up-type
fermions get their masses from the $H_2^0$ whereas $H_1^0$ gives masses to
down-type fermions, so
$$
m_u=h_u v_2 = \frac{h_u \sqrt{2} M_W \sin\beta}{g}\,\,,\,\, 
m_d=h_d v_1 = \frac{h_d \sqrt{2} M_W \cos\beta}{g}\,\,,
$$
and the Yukawa coupling can be obtained as
\begin{equation}
    \lambda_u=\displaystyle\frac{h_u}{g}=  \frac{ m_u}{\sqrt{2} M_W \sin\beta}\,\,,\,\, 
    \lambda_d=\displaystyle\frac{h_d}{g}=\frac{ m_d}{\sqrt{2} M_W \cos\beta}\,\,.
    \label{eq:Yukawasgeneric}
\end{equation}
\subsection{Sfermion sector (Flavor-diagonal case)}
\label{sec:sfmas}
The sfermion mass terms are obtained from the derivative of the
superpotential~(\ref{eq:superder}), the $D$-terms~(\ref{eq:vd}) and the 
Soft-\susy-Breaking terms~(\ref{eq:ltotal}) letting the
neutral Higgs 
fields get their \vev~(\ref{eq:VeVfields}), and one obtain the following mass
matrices:
\begin{align}
    {\cal M}_{\tilde{q}}^2 =\begin{pmatrix}
        M_{\tilde{q}_L}^2+m_q^2+\cos{2\beta}(\TqL-\Qq s_W^2) \mzs & 
        m_q M_{LR}^q\cr 
        m_q M_{LR}^q &
        M_{\tilde{q}_R}^2+m_q^2+ \cos{2\beta}\,\Qq\,s_W^2\, \mzs
    \end{pmatrix}\,,
    \label{eq:mmsq2}
\end{align}
being $Q$ the corresponding fermion electric charge, $\TqL$ the third component
of weak isospin,  $M_{{\tilde{q}}_{L,R}}$ the Soft-\susy-Breaking squark
masses~\cite{Nilles:1984ex,Haber:1985rc,Lahanas:1987uc,Ferrara87} (by
$SU(2)_L$-gauge invariance, we must have
$M_{\tilde{t}_L}=M_{\tilde{b}_L}$, whereas $M_{{\tilde{t}}_R}$,
$M_{{\tilde{b}}_R}$ are in general independent parameters),
$s_\theta=\sin\theta_W$, and 
\begin{align}
    M_{LR}^u=A_u-\mu \cot{\beta}\,\,,\,\,\nonumber\\
    M_{LR}^d=A_d-\mu \tan{\beta}\,\,.
    \label{eq:MLRdefinition}
\end{align}
We define the sfermion mixing matrix as 
($\tilde{q'}_a=\{\tilde{q'}_1\equiv \tilde{q}_L,\,\, \tilde{q'}_2\equiv
\tilde{q}_R\}$ are the weak-eigenstate squarks, and $\sq_a=\{\sq_1,\sq_2\}$
are the mass-eigenstate squark fields)
\begin{eqnarray}
    \tilde{q'}_a&=&\sum_{b}
    R_{ab}^{(q)}\tilde{q}_b,\nonumber\\ R^{(q)}&
    =&\begin{pmatrix} \cos{\theta_q} & -\sin{\theta_q} \cr 
        \sin{\theta_q} & \cos{\theta_q}\end{pmatrix}\,.
    \label{eq:rotation}
\end{eqnarray}
\begin{equation}
    R^{(q)\dagger} {\cal M}_{\tilde{q}}^2 R^{(q)}=
    {\rm diag}\{m_{\tilde{q}_2}^2,m_{\tilde{q}_1}^2\}
    \ \ \ \ \ (m_{\tilde{q}_2}\geq m_{\tilde{q}_1})\,, 
\end{equation}    
\begin{equation}
    \tan2{\theta_q} = \frac{2\,m_q\,M_{LR}^q}{
      M_{\tilde{q}_L}^2-M_{\tilde{q}_R}^2+\cos{2\beta}(\TqL-2\Qq s_W^2) \mzs}\,.
    \label{eq:thetarot}
\end{equation}

From eq.~(\ref{eq:mmsq2})  we can see that the sfermion
mass is dominated by the Soft-\susy-Breaking parameters  ($M_{\tilde f}
\gg m_f \mbox{ for } f \not= \mbox{top}$), and that the non-diagonal terms could
be neglected, except in the case of the top squark (and bottom squark at large $\tb$), however we will maintain
those terms, the reason is that, although the $A$ parameters do not play any
role when computing the sfermion masses, they do play a role in the
Higgs-sfermion-sfermion coupling, and thus it
has an effect on the Higgs self-energies. Moreover these $A$ parameters are
constrained by the approximate (necessary) condition of absence of
colour-breaking minima,
\begin{equation}
    A_q^2<3\,(m_{\tilde{t}}^2+m_{\tilde{b}}^2+M_H^2+\mu^2)\,,
    \label{eq:necessary}
\end{equation}
where $m_{\tilde{q}}$
is of the order of the average squark masses for
$\tilde{q}=\tilde{t},\tilde{b}$~\cite{Frere:1983ag,Claudson:1983et,Kounnas:1983td,Gunion:1987qv}.

All the Soft-\susy-Breaking parameters are free in the strict \MSSM, however some
simplifications must be done to be able of making a feasible numerical
analysis. As the main subject of study are the third generation squarks we 
make a separation between them and the rest of sfermions. This separation is
justified by the evolution of the squark masses from the (supposed) unification
scale down to the electroweak scale~\cite{Martin:1997ns} (see also
section~\ref{sec:parameters} for a more detailed discussion).

So we will use the following approximations:
\begin{itemize}
\item equality of the diagonal elements of eq.~(\ref{eq:mmsq2})  
    \begin{equation}
        \label{eq:msqDiag}
        {\cal M}_{\tilde{q}}^2{}_D\equiv{\cal M}_{\tilde{q}}^2{}_{11}={\cal M}_{\tilde{q}}^2{}_{22}\,\,,
    \end{equation}
    for each charged slepton and each squark of the
    the first and second generation.
\item the first and second generation squarks share the same value of the $A$
    parameter~(\ref{eq:MLRdefinition}) and the mass parameter~(\ref{eq:msqDiag})
\item sleptons also share the same value for~(\ref{eq:msqDiag}) and $A$
    parameters~(\ref{eq:MLRdefinition}).
\end{itemize}

\subsection{Sfermion sector (Non-flavor-diagonal case)}
\label{sec:sfmas_nondiag}

Very important for our FCNC studies is
when the squark mass matrix does not diagonalize with the same matrix as the
one for the quarks. We introduce then intergenerational mass terms for the
squarks, but in order to prevent the number of parameters from being too
large, we have allowed (symmetric) mixing mass terms only for the left-handed
squarks. This simplification is often used in the MSSM, and is justified by
RGE analysis\cite{Duncan:1983iq}.

The flavor mixing terms are introduced through the parameters $\delta_{ij}$ defined as
\begin{equation}
  \label{eq:defdelta}
(M^2_{LL})_{ij}=m_{ij}^2\equiv\delta_{ij}\,m_i\,m_j\,\,,
\end{equation}
where $m_i$ is the mass of the left-handed $i$ squark, and $m^2_{ij}$ is the mixing
mass matrix element between the generations $i$ and $j$. Thus we must
diagonalize two $6\times6$ mass matrices in order to obtain the mass-eigenstates
squark fields. Generalizing the notation in Sec.~\ref{sec:sfmas} we introduce the
$6\times6$ mixing matrices as follows:
\begin{eqnarray}
  \label{eq:definicioR6gen}
  \sq^{\prime}_{\alpha}&=&\sum_\beta R^{(q)}_{\alpha\beta}\sq_{\beta}^{}\\
  R^{(q)\dagger}{\cal M}^{2}_{\sq} R&=&{\cal M}_{\sq D}^{2}
  ={\rm diag}\{m^2_{\sq_1},\ldots,m^2_{\sq_6}\}\,\,,\,\,q\equiv u,\,d  \,\,,
\end{eqnarray}
where ${\cal M}^{2}_{(\su,\sd)}$ is the $6\times6$ square mass matrix for
up-type (or down-type) squarks in the \EW\ basis, with indices
$\alpha=1,2,3,\ldots,6\equiv\su_L,\su_R,\ldots,\stopp_R$ for up-type
squarks, and similarly for down-type squarks. In this study we are
only interested in the up-type quarks-squarks system, so we will drop out the
$(q)$ super-index in the forthcoming expressions. The rotation matrix $R$
introduces gluino mediated tree-level \FCNC\ between quarks and squarks.

\begin{figure}
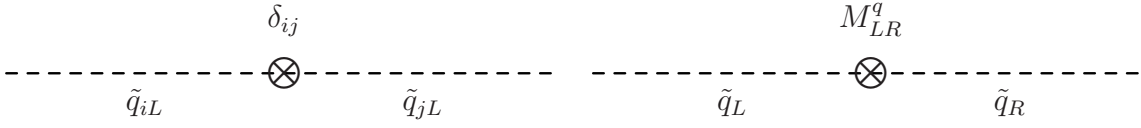

    \centering
    \begin{tabular}{cc}
        \propmix{\neutralscalar{\sq_{iL}}}{\neutralscalar{\sq_{jL}}}{\delta_{ij}}&
        \propmix{\neutralscalar{\sq_{L}}}{\neutralscalar{\sq_{R}}}{M_{LR}^q}
    \end{tabular}
    \caption{Diagrams in the mass insertion approximation}
    \label{fig:massinsertion}
\end{figure}

To analyze the contributions from these flavor and chiral mixed squarks we can
use the so-called mass
insertion approximation. This is based on the fact that the $\delta_{ij}$
parameters are small, so instead of diagonalizing the $6\times6$ squared mass
matrix we can treat them as an interaction (see Fig.~\ref{fig:massinsertion}):

\begin{align}
    \begin{array}{rcccc}
        {\cal L} &\ni& \begin{pmatrix} \sq_{iL}^* & \sq_{jL}^* \end{pmatrix}
        \begin{pmatrix} p^2-m_i^2 & m^2_{ij} \\
            m^2_{ji} & p^2 - m_j^2 \end{pmatrix}
        \begin{pmatrix}\sq_{iL}\\ \sq_{jL}\end{pmatrix} \\
        &=&
        \begin{pmatrix} \sq_{iL}^* & \sq_{jL}^* \end{pmatrix}
        \begin{pmatrix} p^2-m_i^2 & 0 \\
            0 & p^2 - m_j^2 \end{pmatrix}
        \begin{pmatrix}\sq_{iL}\\\sq_{jL}\end{pmatrix}
        &+&
        \begin{pmatrix} \sq_{iL}^* & \sq_{jL}^*\end{pmatrix}
        \begin{pmatrix} 0 & m^2_{ij} \\
            m^2_{ji} & 0 \end{pmatrix}
        \begin{pmatrix}\sq_{iL}\\ \sq_{jL}\end{pmatrix}
    \end{array}
\end{align}
and similarly for the left-right squark matrix.

\section{Interactions in the mass-eigenstate basis}
\label{sec:MSSMinteractions}

We need to convert the MSSM interaction Lagrangian \eqref{eq:ltotal}
to a Lagrangian in the mass-eigenstate basis,
which is the one used in the computation of the physical quantities. As the
expression for the full interaction Lagrangian in the \MSSM\ is rather lengthy
we quote only the interactions that we will need in our studies.

\begin{itemize}
\item quark--quark--neutral Higgs: this is the usual Yukawa interaction from Type II
    2HDM, in the \MSSM\  it follows after replacing in~(\ref{eq:vyukawa}) the
    mass-eigenstate Higgs fields~(\ref{eq:HiggsDefin}):
    \begin{eqnarray}
        {\cal L}_{Hqq}&=&
        -\frac{gm_d}{2\mw\cos\beta}\left[\left(\cos\alpha\,H^0-\sin\alpha\,h^0\right)\bar
            dd - i \sin\beta\, \bar d \gamma_5 d\,A^0 \right]\nonumber \\
        ~&~& -\frac{g m_u}{2\mw \sin\beta}
        \left[\left(\sin\alpha\, H^0 +\cos\alpha\, h^0\right) \bar u u
            - i \cos\beta\, \bar u \gamma_5 u\, A^0 \right]\,\,,
        \label{LqqH}
    \end{eqnarray}
    where we have replaced the Yukawa couplings $h_i$ in favour of masses and \tb.

\item quark--gluon interactions: this is the usual \QCD\  Lagrangian
    \begin{equation}
        \label{eq:QCDlagrangian}
        {\cal L}_{QCD}=\frac{g_s}{2}\,
        G_{\mu}^c\,\lambda^c_{ij}\,\bar{q}_{i}\,\gamma^\mu\,q_{j}  \,\,.
    \end{equation}

\item quark--squark--gluino: the supersymmetric version of the strong
    interaction is obtained from~(\ref{eq:vgfsf}):
    \begin{eqnarray}
        \label{eq:SUSYQCDFCNClagrangian}
        {\cal L}_{\sg q\tilde{q}}&=& -\frac{g_s}{\sqrt{2}}\, \bar{\psi}_c^{\sg} \left[
            R_{5\alpha}^*\,\pl-R_{6\alpha}^*\,\pr \right] \sq^*_{\alpha,i}\, \lambda_{ij}^c\, t_j\nonumber\\
        &-& \frac{g_s}{\sqrt{2}}\, \bar{\psi}_c^{\sg} \left[
            R_{3\alpha}^*\,\pl-R_{4\alpha}^*\,\pr \right] \sq^*_{\alpha,i}\, \lambda_{ij}^c\, c_j\nonumber\\
        &-& \frac{g_s}{\sqrt{2}}\, \bar{\psi}_c^{\sg} \left[
            R_{1\alpha}^*\,\pl-R_{2\alpha}^*\,\pr \right] \sq^*_{\alpha,i}\, \lambda_{ij}^c\, u_j\,\,.
    \end{eqnarray}
    \begin{equation}
        {\cal L}_{\sg q\tilde{q}}=- \frac{g_s}{\sqrt{2}}
        \tilde q_{a,i}^{*}\, \bar \psi^{\sg}_c 
        \left(\lambda^c\right)_{ij} \left(R_{1a}^{(q)*} \pl - R_{2a}^{(q)*} \pr \right) q_j
        +\mbox{ h.c.}\,\,,
        \label{eq:Lqsqglui}\end{equation}
    where $\lambda^c$ are the Gell-Mann matrices.

\item squark--squark--neutral Higgs: the origin of this interaction is twofold, on one
  side the superpotential derivative~(\ref{eq:superder}), and on the other the
  Soft-\susy-Breaking trilinear interactions. It is convenient to express the
  results in the $\sq_L-\sq_R$ basis, and use the transformation
  \ref{eq:definicioR6gen} or \ref{eq:rotation}.
  \begin{eqnarray}
      {\cal L}_{h\sq\sq}&=&-\frac{g\mz}{\ct}
      \sum_{i=u,d}\left[(T_3^{iL}-Q_i\sts)\sq_{iL}^{*}\sq_{iL}+
          Q_i\sts\sq_{iR}^{*}\sq_{iR}\right]\notag\\
       &&\hspace{1cm}\times\left(H^0\cos(\beta+\alpha)-h^0\sin(\beta+\alpha)\right)\notag\\
       &-&\frac{g m_d^2}{\mw \cbt}(\sd_{L}^{*}\sd_{L}+\sd_{R}^{*}\sd_{R})
       (H^0\ca-h^0\sa)\notag\\
       &-&\frac{g m_u^2}{\mw \sbt}(\su_{L}^{*}\su_{L}+\su_{R}^{*}\su_{R})
       (H^0\ca-h^0\sa)\\
       &-&\frac{g m_d}{2\mw \cbt}(\sd_{R}^{*}\sd_{L}+\sd_{L}^{*}\sd_{R})\notag\\
       &&\hspace{1cm}\times\left[(-\mu\sa+A_d\ca)H^0+(-\mu\ca-A_d\sa)h^0\right]\notag\\
       &-&\frac{g m_u}{2\mw \sbt}(\su_{R}^{*}\su_{L}+\su_{L}^{*}\su_{R})\notag\\
       &&\hspace{1cm}\times\left[(-\mu\ca+A_u\sa)H^0+(+\mu\sa+A_u\ca)h^0\right]\notag\\
       &-&\frac{ig m_u}{2\mw}(\mu+A_d\tb)(\sd_{R}^{*}\sd_{L}-\sd_{L}^{*}\sd_{R})A^0\notag\\
       &-&\frac{ig m_u}{2\mw}(\mu+A_u\ctb)(\su_{R}^{*}\su_{L}-\su_{L}^{*}\su_{R})A^0\notag
      \label{eq:Lhsqsq}
  \end{eqnarray}

\end{itemize}

\section{Flavor changing neutral currents}
\label{sec:mssm:fcnc}

\begin{figure}
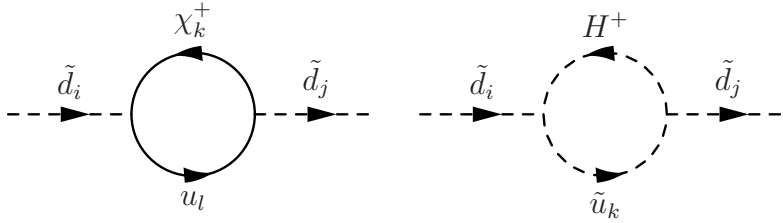

    \centering
    \begin{tabular}{cc}
        \autoenermix
        {\scalar{\sd_i}}{\scalar{\sd_j}}
        {\fermion{u_l}}{\fermion{\chi^+_k}}
        &
        \autoenermix
        {\scalar{\sd_i}}{\scalar{\sd_j}}
        {\scalar{\su_k}}{\scalar{H^+}}
    \end{tabular}
    \caption{Feynman graphs contributing to the squark mixing.}
    \label{fig:squarkmixing}
\end{figure}

\begin{figure}
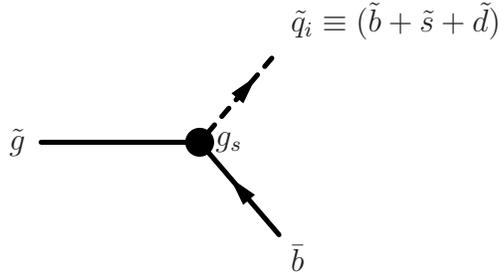

    \centering
    \ginosqsqver
    \caption{Tree level FCNC strong interactions.}
    \label{fig:diag_gqsq}
\end{figure}

The most general MSSM includes tree-level FCNCs among the extra
predicted particles, which induce one-loop FCNC interactions among
the SM particles. Given the observed smallness of these
interactions, tree-level SUSY FCNCs are usually avoided by
including one of the two following assumptions: either the SUSY
particle masses are very large, and their radiative effects are
suppressed by the large SUSY mass scale; or the soft SUSY-breaking
squark mass matrices are aligned with the SM quark mass matrix, so
that both mass matrices are simultaneously diagonal. However, if
one looks closely, one soon realizes that the MSSM does not only
include the possibility of tree-level FCNCs, but it actually
\textit{requires} their existence~\cite{Duncan:1983iq}. Indeed,
the requirement of $SU(2)_L$ gauge invariance means that the
up-left-squark mass matrix can not be simultaneously diagonal to
the down-left-squark mass matrix, and therefore these two matrices
can not be simultaneously diagonal with the up-quark and the
down-quark mass matrices, that is, unless both of them are
proportional to the identity matrix. {But even then we
could not take such a possibility too seriously, for the radiative
corrections would produce non-zero elements in the non-diagonal
part of the mass matrix (i.e. induced by $H^\pm$ and $\chi^\pm$, see
Fig.~\ref{fig:squarkmixing}). All in all, we naturally expect
tree-level FCNC interactions mediated by the SUSY partners of the
SM particles. As an example in the MSSM that one can not set the FCNC Higgs
bosons interactions to zero without being inconsistent notice that
$\Gamma(\stopp \to c \chi^0)$ is UV divergent in the absent of these
couplings~\cite{Hikasa:1987db}.
The potentially largest FCNC interactions are those
originating from the strong supersymmetric (SUSY-QCD) sector of
the model (viz. those interactions involving the
squark-quark-gluino couplings, see Fig.~\ref{fig:diag_gqsq}), and in chapters \ref{cha:hbsSUSY_br} and
\ref{cha:hbsSUSY_prod} we mainly
concentrate on those. These couplings induce FCNC loop effects on
more conventional fermion-fermion interactions, like the
gauge boson-quark vertices $Vqq'$.}

\begin{figure}
    \centering
    \begin{tabular}{cc}
        \resizebox{0.5\textwidth}{!}{
  \input{feyn_def.tex}
  \fmfframe(30,30)(30,30){\begin{fmfgraph*}(200,70)
        \fmfpen{thick} \fmfstraight
        \fmfleft{i1,li,i2} \fmfright{o1,lo,o2}
        \fermion{\bar{d}}{}{o1,p1} \fermion{\bar{s}}{}{p1,i1}
        \fermion{d}{,label.side=left}{i2,p2}
        \fermion{s}{,label.side=left}{p2,o2} \noline{}{}{li,pl,lo}
        \fmflabel{$K^0$}{li} \fmflabel{$\bar{K^0}$}{lo}
        \fmfv{decor.shape=circle,decor.filled=shaded,
          foreground=red,decor.size=.4w}{pl}
        \fmfipath{k[]}
        \fmfiset{k1}{fullcircle scaled .5w xscaled .18
          shifted (0w,.5h)}
        \fmfiset{k2}{fullcircle scaled .5w xscaled .18
          shifted (1w,.5h)}
        \fmfcmd{fill k1 withcolor 0.8white;}
        \fmfcmd{fill k2 withcolor 0.8white;}
    \end{fmfgraph*}}
}&
        \resizebox{0.5\textwidth}{!}{
  \input{feyn_def.tex}
  \fmfframe(30,30)(30,30){\begin{fmfgraph*}(200,70)
        \fmfpen{thick}
        \fmfleft{i1} \fmfright{o1,o2}
        \fermion{b}{,label.side=left}{i1,v1}
        \fermion{s}{,label.side=left,tension=0.5}{v1,o2}
        \boson{\gamma}{,label.side=right,tension=0.5}{v1,o1}
        \fmfv{decor.shape=circle,decor.filled=shaded,
          foreground=red,decor.size=.2w}{v1}
    \end{fmfgraph*}}
}
    \end{tabular}
    \caption{Diagram of the processes $K^0-\bar{K}^0$ and $B\to X_{s}\,\gamma$ at
      quark level}
    \label{fig:k0k0b_bsg}
\end{figure}
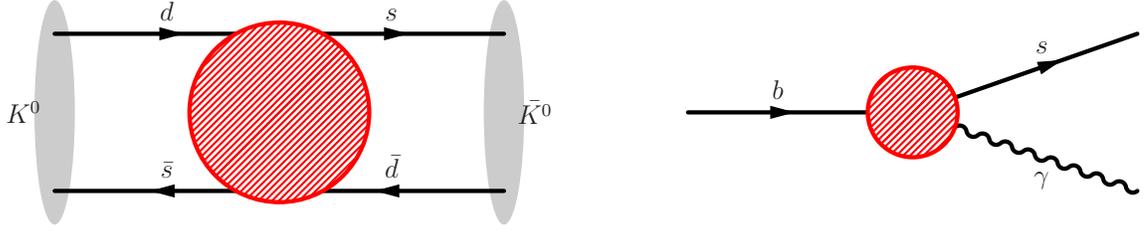

\begin{figure}
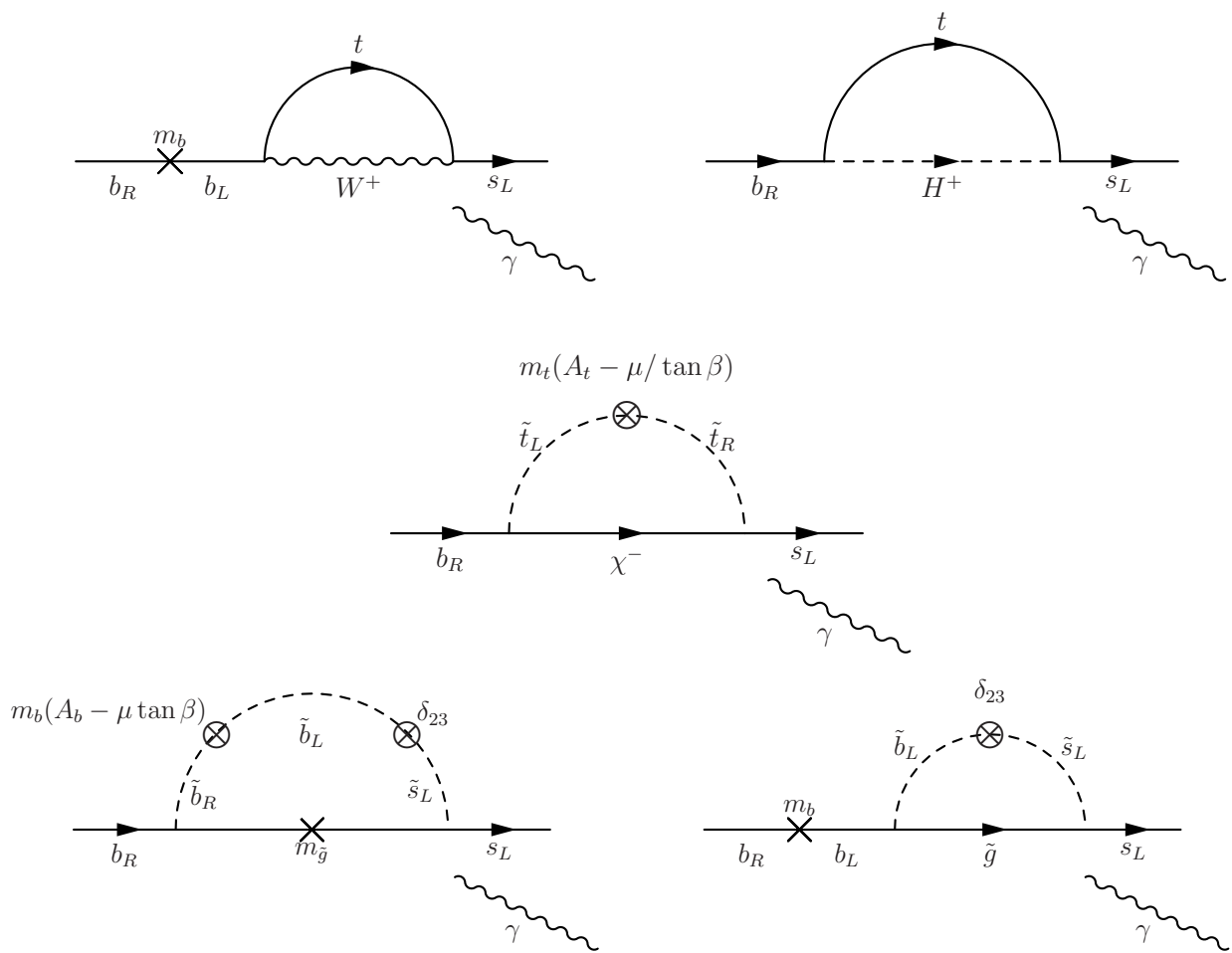

    \centering
    \begin{tabular}{cc}
        \resizebox{0.5\textwidth}{!}{\bsgSM}&
        \resizebox{0.5\textwidth}{!}{\bsgHiggs}\\[1cm]
        \multicolumn{2}{c}{\resizebox{0.5\textwidth}{!}{\bsgcharginos}}\\
        \resizebox{0.5\textwidth}{!}{\bsggluino}&
        \resizebox{0.5\textwidth}{!}{\bsggluinosub}\\
    \end{tabular}
    \caption{Leading contributions to the $b\to s\gamma$ Wilson operator $O_7$.}
    \label{fig:bsgwilson}
\end{figure}

{Of course, low energy meson physics puts tight
constraints on the possible value of the FCNC couplings,
especially for the first and second generation squarks which are
sensitive to the data on} $K^0-\bar{K}^0$ (see Fig.~\ref{fig:k0k0b_bsg}) and
$D^0-\bar{D}^0$~\cite{\BurasGabbiani}. The third generation system is,
in principle, very loosely constrained since present data on
$B^0-\bar{B}^0$ mixing still leaves much room for FCNCs and
the most stringent constraints are given by the $\Bbsg$
measurement~\cite{\bsgexp}. Therefore, the relevant FCNC gluino
coupling $\delta_{23}$ \cite{\BurasGabbiani}
is not severely bound at present. The lack
of tight FCNC constraints in the top-bottom quark doublet enables
the aforementioned lower bound on the charged Higgs boson mass in
the MSSM to be easily avoided, to wit: by arranging that the
SUSY-electroweak ({SUSY-EW}) contribution to $\Bbsg$ from
the top-squark/chargino loops screens partially the charged Higgs
boson contribution. This situation can be naturally fulfilled if
the higgsino mass parameter ($\mu$) and the soft SUSY-breaking
top-squark trilinear coupling ($A_t$) satisfy the relation
$\mu\,A_t < 0$~\cite{\bsgold}. The relevant Wilson operator in the effective
theory involves a chirality flip:
\begin{equation}
    O_7 = \frac{e}{16 \pi^2}\ \mb\ (\bar{s}_L \  \sigma^{\mu\nu} \ b_{R} )\
F_{\mu\nu}.
\end{equation}
In Fig.~\ref{fig:bsgwilson} one can see the leading contributions to this
operator. The most important contribution is expected to be SUSY-QCD
one. Roughly speaking its amplitude goes like:
\begin{equation}
A^{SUSY-QCD}(b\to s\gamma) \sim \deltatt \frac{\mb (A_b -\mu\tb)
}{\msusy^2}\times \frac{1}{\mg}.
\end{equation}
This amplitude can be as large as minus twice the SM contribution, so the
total MSSM amplitude can
respect the experimental $b\to s\gamma$ restriction. We would regard this choice
as a fine-tuning of the parameters, hence unnatural, but in fact it is excluded
experimentally\cite{Gambino:2004mv}.

On the other hand, if we assume that the squark mass matrices are diagonal at a certain energy
scale ($\Lambda$) and write the Renormalization Group
Equations\cite{Duncan:1983iq}, then non-diagonal terms are generated in the
left-left sector, because $\lambda_u$ and $\lambda_d$ are not simultaneously diagonal.
\begin{eqnarray*}
    (M^d_{RR})^2&=&\mu_0(\sd_R) {\mathbbm 1} +\mu_1(\sd_R) \lambda_{d}^\dagger\lambda_d \\
    (M^u_{RR})^2&=&\mu_0(\su_R) {\mathbbm 1} +\mu_1(\su_R) \lambda_{u}^\dagger\lambda_u \\
    (M^d_{LL})^2&=&\mu_0(\sd_L) {\mathbbm 1} +\mu_1(\sd_L) \lambda_d\lambda_{d}^\dagger
    + \mu_2(\sd_L) \lambda_u\lambda_{u}^\dagger\\
    (M^u_{LL})^2&=&\mu_0(\su_L) {\mathbbm 1} +\mu_1(\su_L) \lambda_u\lambda_{u}^\dagger
    + \mu_2(\su_L) \lambda_d\lambda_{d}^\dagger
\end{eqnarray*}
And then the FCNC terms are communicated from the up to the down sector by the
CKM matrix, due to the $SU(2)$ gauge invariant:
\begin{equation}
  (M^d_{LL})^2  = V_{CKM}^\dagger \ \times  (M^u_{LL})^2 \times \ V_{CKM} \\
\end{equation}

The FCNC gluino interactions also induce large contributions to
$\Bbsg$, see Fig.~\ref{fig:bsgwilson}. {It should be noted} however that the leading
contributions to the {$Vqq'$ FCNC interactions from the third
quark generation} correspond to a \textit{double insertion} term,
in which the squarks propagating in the loop suffer a double
mutation: a flavor conversion and a chirality transition. This
fact has been demonstrated in the $\Bbsg$ observable
itself~\cite{Borzumati:1999qt}, as well as {in} the FCNC rare
decay width $\Gamma(t\to cg)$~\cite{Guasch:1999jp}. As a
consequence, the loose limits on the third generation FCNC
interactions derived under the assumption that the leading terms
contributing to $\bsg$ correspond to the \textit{single particle
insertion approximation}~\cite{\BurasGabbiani} are not valid and
more complex expressions must be taken into
account~\cite{Besmer:2001cj}.

The FCNC effects in the $K^0-\bar{K^0}$ system can be described with an
effective Lagrangian with a mass matrix as:
\begin{equation}
    \begin{pmatrix} m_{K}^2 & A\\ A & m_{K}^2\end{pmatrix}
\Rightarrow \mbox{diagonalize} \Rightarrow 
m_{K_{1,2}}^2= m_{K}^2 \pm A
\end{equation}
so the signal of FCNC can determined by the mass difference of the
eigenstates:
\begin{equation}
    \Delta m_K \equiv | m_{K_1} - m_{K_2} |
\end{equation}
In contrast the experimental signal for $\bsg$ is directly its branching
ratio. The experimental measurements
are\cite{\CLEO,Barate:1998vz,Abe:2001hk,Aubert:2002pd,Eidelman:2004wy}:
\begin{align}
    B(b\to s \gamma) &= (3.3 \pm 0.4) \times 10^{-4}\\
    \Delta m_K &= (3.483\pm 0.006)\times 10^{-12}\MeV&
      (d\bar{s}\leftrightarrow\bar{d}s) \\ m_K&=497.648\pm0.022\MeV\notag\\
    \Delta m_D &< 4.6 \times 10^{-11} \MeV&
      (c\bar{u}\leftrightarrow\bar{c}u)  \\ m_D&=1864.6\pm0.5 \MeV\notag\\
    \Delta m_B&=(3.304\pm0.046)\times 10^{-10}\MeV&
      (b\bar{d}\leftrightarrow\bar{b}d)\\
      m_B&=5279.4\pm 0.5 \MeV\notag
\end{align}
that translated to $\delta_{ij}$ restrictions they read:
\begin{equation}
    \begin{array}{rcl}
        \delta_{12}&\lesssim&.1 \,\sqrt{m_{\tilde u}\,m_{\tilde c}}/{500 \GeV} \\
        \delta_{13}&\lesssim&.098\,\sqrt{m_{\tilde u}\,m_{\tilde t}}/{500 \GeV} \\
        \delta_{23}&\lesssim&8.2 \,m_{\tilde c}\,m_{\tilde t}/{(500 \GeV)^2}
    \end{array}\label{eq:4}
\end{equation}

\section{\MSSM\  parametrization}
\label{sec:limits}
\subsection{\MSSM\  parameters}
\label{sec:parameters}
Since SUSY is a broken symmetry we have to deal with a plethora of
Soft-\susy-Breaking parameters, namely
\begin{itemize}
\item $\tb$,
\item masses for Left- and Right-chiral sfermions,
\item a mass for the Higgs sector,
\item gaugino masses,
\item triple scalar couplings for squarks and Higgs.
\end{itemize}

This set of parameters is often simplified to allow a comprehensive
study. Most of these simplifications are based on some universality assumption
at the unification scale. In minimal supergravity (mSUGRA) all the parameters
of the \MSSM\ are computed from a restricted set of parameters at the
Unification scale, to wit: \tb; a common scalar mass $m_0$; a common fermion
mass for gauginos $m_{1/2}$; a common trilinear coupling for all sfermions
$A_0$; and the higgsino mass parameter $\mu$. Then one computes the running of
each one of these parameters down to the \EW\ scale, using the Renormalization
Group Equations (RGE), and the full spectrum of the \MSSM\ is found.

We will not restrict ourselves to a such simplified model. As stated in the
introduction we treat the \MSSM\ as an effective Lagrangian, to be embedded in
a more general framework that we don't know about. This means that essentially
all the parameters quoted above are free. However for the kind of studies we
have performed there is an implicit asymmetry of the different particle
generations. We are mostly interested in the phenomenology of the third
generation, thus we will treat top and bottom supermultiplet as distinguished
from the rest. This approach is well justified by the great difference of the
Yukawa couplings of top and bottom with respect to the rest of fermions. We
are mainly interested on effects on the Higgs sector, so the smallness of the
Yukawa couplings of the first two generations will result on small effects in
our final result. We include them, though, in the numerical analysis and the
numerical dependence is tested. On the other hand, if we suppose that there is
unification at some large scale, at which all sfermions have the same mass,
and then evolve these masses to the \EW\ scale, then the RGE have great
differences\cite{Martin:1997ns}. Slepton RGE are dominated by \EW\ gauge
interactions, 1st and 2nd generation squarks RGE are dominated by \QCD, and
for the 3rd generation squarks there is an interplay between \QCD\ and Yukawa
couplings. Also, as a general rule, the gauge contribution to the RGE
equations of left- and right-handed squark masses are similar, so when Yukawa
couplings are not important they should be similar at the \EW\ scale.

With the statement above in mind we can simplify the \MSSM\  spectrum by taking
an unified parametrization for 1st and 2nd generation squarks (same for
sleptons). As we have seen in Sect.~\ref{sec:sfmas} we will use: a common
mass\footnote{Note that after diagonalization of the squark mass matrix the
  physical masses will differ slightly.} for $\su_L$ and $\sq_R$
($\msup$); an unified
trilinear coupling $A_u$ for 1st and 2nd generation; a common mass for all 
$\tilde \nu_L$ and $\tilde l_R$ ($\mstau$); and a common trilinear coupling
$A_\tau$.

For the third generation we will use different trilinear couplings $A_t$ and
$A_b$, as these can play an important role in the kind of processes we are
studying (see chapter~\ref{cha:hbsSUSY_br} and \ref{cha:hbsSUSY_prod}). Stop
masses can present a large gap (due to its Yukawa couplings), being the
right-handed stop the lightest one. We will use a common mass for both chiral
sbottoms, which we parametrize with the lightest sbottom mass ($\msbo$), and
the lightest stop quark mass ($\msto$), as the rest of mass inputs in this
sector. This parametrization is useful in processes where squarks only appear
as internal particles in the loops (such as the ones studied in
chapters~\ref{cha:hbsSUSY_br} and \ref{cha:hbsSUSY_prod}), as one-loop
corrections to these parameters would appear as two-loop effects in the
process subject of study. Gluino mass (\mg), on the other hand, is let free.

For the Higgs sector two choices are available, we can use the
pseudoscalar mass $\mA$, or the charged Higgs mass $\mHp$. Both choices are on
equal footing. As the neutral Higgs particles are the main element for most of our
studies we shall use its mass as input parameter in most of our work.

Standard model parameters are well known, we will use present determinations
of \EW\  observables\cite{Eidelman:2004wy}
\begin{eqnarray}
    \mz &=& 91.1899\pm 0.0021 \GeV\nonumber\\
    \mw&=&80.418\pm 0.054 \GeV\nonumber\\
    G_F&=& (1.16639 \pm 0.00001)\times 10^{-5} \GeV^{-2} \nonumber\\
    \alpha_{em}^{-1}(\mz)&=&128.896 \pm 0.090 \nonumber\\
    \label{eq:EWprecisiondata}
\end{eqnarray}
\QCD\  related observables are not so precise. On the other hand as the main
results are not affected by specific value of these observables we will use the
following ones
\begin{eqnarray}
    \label{eq:parametersQCD}
    \mt&=&175 \GeV \nonumber\\
    \mb&=&5 \GeV \nonumber\\
    \alpha_s(\mz)&=&0.11\,\,
\end{eqnarray}

\subsection{Constraints}
\label{sec:MSSMconst}
The \MSSM\  reproduces the behaviour of the \SM\  up to energy scales probed so
far\,\cite{Hollik:2004Zuoz}. Obviously  this is not for every point of the full
parameter space!

There exists direct limits on sparticle masses based on direct searches
at the high energy colliders (LEP II, SLC, Tevatron). Although hadron colliders
can achieve larger center of mass energies than $e^+e^-$ ones, their samples
contain large backgrounds that make the analysis more difficult. This drawback
can be avoided if the ratio signal-to-background is improved, in fact they can
be used for precision measurements of ``known'' observables (see
e.g.\,\cite{Gianotti:1998yf}). $e^+e^-$ colliders samples are more clean, and
they allow to put absolute limits on particle masses in a model independent
way.

\begin{figure}[t!]
\begin{center}
\resizebox{\textwidth}{!}{
\includegraphics[width=5cm]{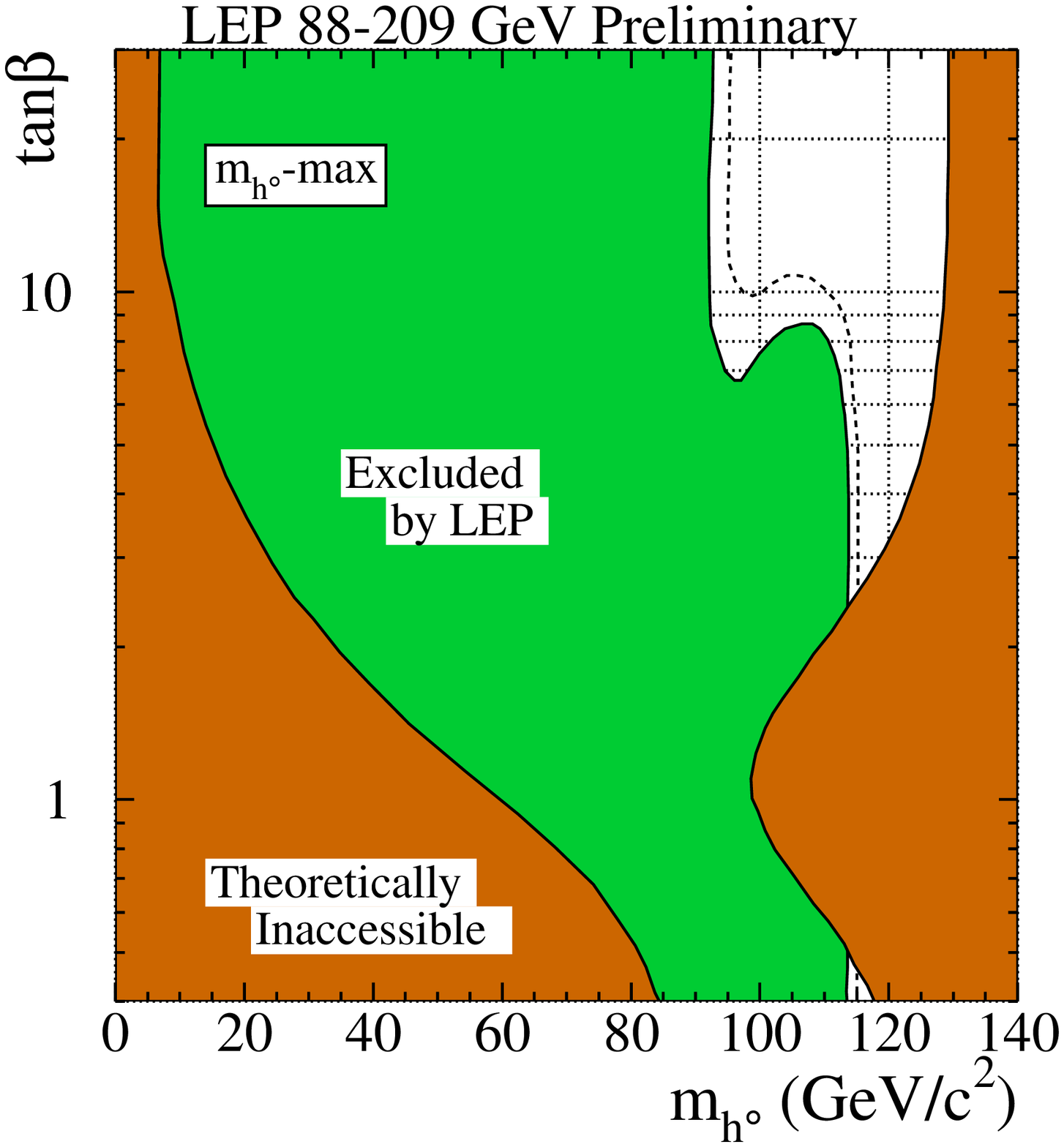}
\hfill
\includegraphics[width=5cm]{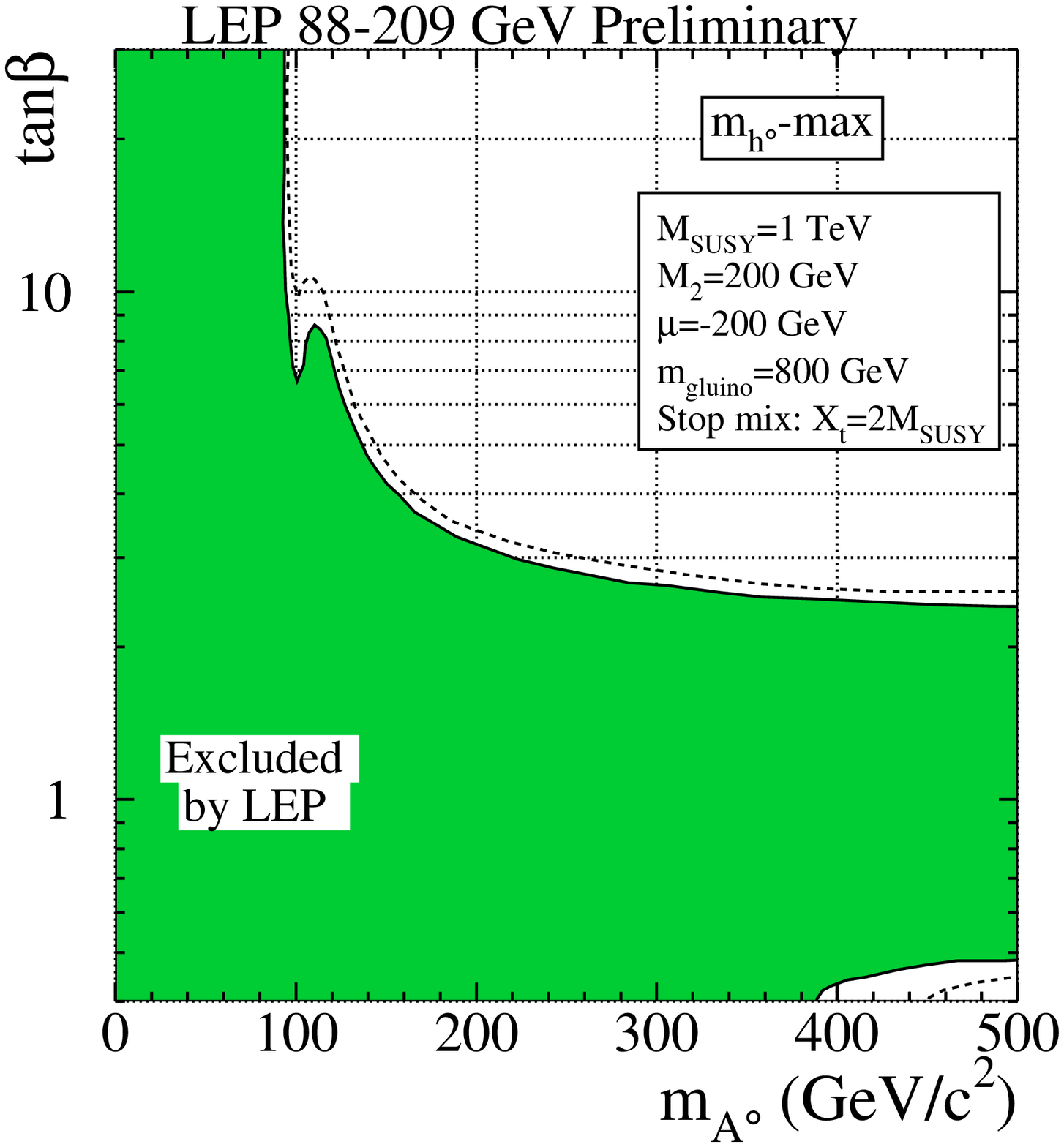}
}
\end{center}
\caption{\label{mh200} LEP2 contours of the 95\% CL exclusion
limits for MSSM Higgs sector parameters as a function of $\tb$ and
(a) $\mh$ and (b) $\ma$ (in GeV), taken from Ref.~\cite{:2001xx}.
The contours shown have been obtained for MSSM Higgs parameters chosen
according to the maximal mixing benchmark of Ref.~\cite{Carena:1999xa}.}
\end{figure}

\begin{figure}[ht!]
    \begin{center}
        \includegraphics[width=0.5\textwidth]{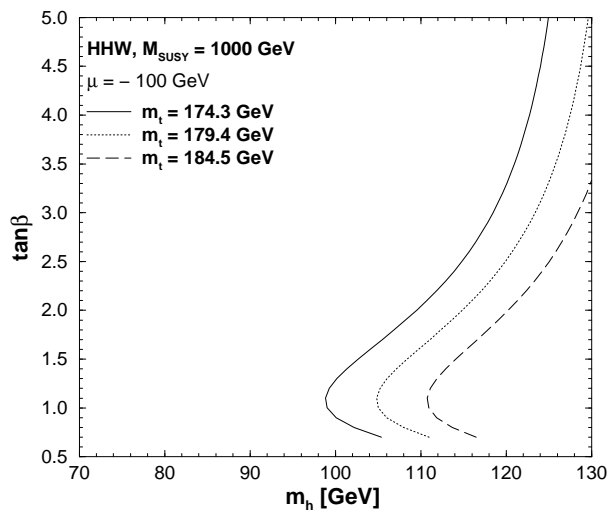}
    \end{center}
    \caption{Theoretically exclusion limits as in Fig.~\ref{mh200} for three
      different values of the top-quark mass, $\mt = 174.3, 179.4, 184.5
      \GeV$. Taken from Ref.~\cite{Heinemeyer:1999zf}}
    \label{fig:mhmt}
\end{figure}

No significant evidence for a Higgs signal has been detected at
LEP~\cite{:2001xx}.
As a result, one can obtain bounds on the possible MSSM Higgs
parameters.  These limits are often displayed in the $\ma$--$\tb$
plane, see Fig.~\ref{mh200}, although there is additional dependence on various MSSM
parameters that effect the radiative corrections to the Higgs masses
as discussed above.
In representative scans of the MSSM parameters, the LEP Higgs Working
Group~\cite{:2001xx}
finds that $\mh>91.0$~GeV and $\ma>91.9$~GeV at 95\% CL.  These
limits actually correspond to the large $\tb$ region in which $Zh^0$
production is suppressed, as shown in Fig.~\ref{mh200}.
In this case, the quoted Higgs limits arise
as a result of the non-observation of $h^0A^0$ and $H^0A^0$ production.
As $\tb$ is lowered, the limits on $\mh$ and $\mA$ become more
stringent.  In this regime, the $h^0A^0$ production is suppressed
while the $Zh^0$ production rate approaches its SM value.  Thus, in this
case, the SM Higgs limit applies ($\mh\gtrsim 114\GeV$\cite{Eidelman:2004wy}) as shown in
Fig.~\ref{mh200}(a).
The precise region of MSSM Higgs parameter space that is excluded
depends on the values of the
MSSM parameters that control the Higgs mass radiative corrections.
For example, a conservative exclusion
limit is obtained in the maximal mixing scenario, since
in this case the predicted value of $\mh$ as a function of $\mA$ and
$\tb$ is maximal (with respect to changes in the other MSSM parameters).
The excluded regions of the MSSM Higgs parameter space based on the
maximal mixing benchmark scenario of Ref.~\cite{Carena:1999xa}, are shown in
Fig.~\ref{mh200}, and correspond to the exclusion of the range
$0.5<\tb<2.4$ at the 95\%~CL.  However, the $\tb$
exclusion region can still be significantly reduced (even to the point
of allowing all $\tb$ values) by, {\it e.g.}, taking $\msusy=2$~TeV
and $\mt=180$~GeV, see Fig~\ref{fig:mhmt} (which still lies within the error bars of the 
experimentally measured value), and allowing for the theoretical
uncertainty in the prediction of ${\mh}^{\mathrm{max}}$~\cite{Heinemeyer:1999zf}.

Similarly, no evidence for the charged Higgs boson has yet been found.
The LEP Higgs Working Group quotes a limit of $\mHp>78.6$~GeV at 
95\% CL~\cite{unknown:2001xy}, which holds for a more
general non-supersymmetric two-Higgs doublet model and assumes
only that the $H^+$ decays dominantly into $\tau^+\nu_\tau$ and/or
$c \bar s$.  Although the MSSM tree-level bound $\mHp\geq\mw$ can be
relaxed somewhat by radiative corrections, the LEP bound quoted above
provides no useful additional constraints on the MSSM Higgs sector.
Actual fits to the \MSSM\
parameter space project a preferred value for the charged Higgs mass of
$\mHp\simeq 120\GeV$\,\cite{Hollik:1997qp}.

Hadron colliders bounds are not so restrictive as those from $e^+e^-$
machines. Most bounds on squark and gluino masses are obtained by supposing
squark mass unification in simple models, such as mSUGRA. At present
approximately the limits, with certain conditions,
on squarks (1st and 2nd generation) and gluino masses are\,\cite{Eidelman:2004wy}
\begin{equation}
    \label{eq:limitsCDFD0squarks}
    m_{\sq} > 250 \GeV \,\,,\,\, \mg > 195 \GeV \,\,.
\end{equation}

From the top quark events at the Tevatron a limit on the branching ratio
$B(t \rightarrow H^+\,b)$ can be extracted, and thus a limit on the $\tb-\mHp$
relation\cite{Coarasa:1996qa,Guasch:1997jc}.

Finally indirect limits on sparticle masses are obtained from the \EW\
precision data. We apply these limits through all our computations by computing
the contribution of sparticles to these observables and requiring that they
satisfy the bounds from \EW\  measurements. We require new contributions to 
the $\rho$ parameter to be smaller than present experimental error on
it, namely
\begin{equation}
    \label{eq:deltarhoNEW}
    |\delta\rho_{\rm new}| < 0.003\,\,\,(3\sigma)\,.
\end{equation}
We notice that as $\delta \rho_{\rm new}$ is also the main contribution from
sparticle contributions to $\Delta r$\,\cite{TesiDavid}, new contributions to
this parameter are also below experimental constrains. Also the corrections in
the $\alpha$- and $G_F$-on-shell renormalization schemes will not differ
significantly~\cite{Guasch:1999tk}.

We recall that at the moment the MSSM may escape the indirect bound
from $b\rightarrow s\,\gamma$ because the positive charged Higgs virtual
contributions can be compensated for by negative stop
and chargino loops, if they are not too heavy. Therefore, in the
MSSM the charged Higgs can stay relatively light, $\mHp\gtrsim
120\,GeV$, just to comply with the aforementioned LEP 200 bounds
on $\mA$\,\cite{\LEPHWG}.

There exist also theoretical constrains to the parameters of the \MSSM. As a
matter of fact the \MSSM\  has a definite prediction: there should exist a
light neutral scalar Higgs boson $h^0$. Tree-level analysis put this bound to
the $Z$ mass, however the existence of large radiative corrections to the
Higgs bosons mass relations grow this limit up to $\sim 130 \GeV$\cite{Carena:1995wu,Carena:1995bx}. More recently the
two-loop radiative corrections to Higgs mass relations in the \MSSM\  have been
performed\cite{Carena:2000dp,Espinosa:2000df,Heinemeyer:1998jw,Hempfling:1993qq,RADCORhaber}, and the present
upper limit on $\mh$ is
\begin{equation}
    \label{eq:UPh0limit}
    \mh \leq 130-135 \GeV\,\,.
\end{equation}
Of course if the \MSSM\ is
extended in some way this limit can be evaded, though not to values larger of
$\sim 200\GeV$\,\cite{Masip:1998jc,Masip:1998za}.

Another theoretical constraint is the necessary condition~(\ref{eq:necessary}) on
squark trilinear coupling ($A$) to avoid colour-breaking minima. This constraint
is easily implemented when the $A$ parameters are taken as inputs, but if we
choose a different set of inputs (such as the mixing angle $\theta_{\sq}$)
then it constrains the parameter space in a
non-trivial way.

Whatever the spectrum of the \MSSM\ is, it should comply with the benefits
that \susy\ introduces into the \SM\ which apply if the following condition is
fulfilled:
\begin{equation}
    \label{eq:MSUSYUPLIMIT}
    M_{\rm SUSY} \lesssim {\cal O} (1 \mbox{ TeV})\,\,.
\end{equation}
If supersymmetric particles have masses heavier than the TeV scale then
problems with GUT's appear. This statement does not mean that \susy\ would
not exist, but that then the \SM\ would not gain practical benefit from the
inclusion of \susy.

A similar upper bound is obtained when making cosmological analyses, in these
type of analyses one supposes the neutralino to be part of the cold dark
matter of the universe, and requires its annihilation rate to be sufficiently
small to account for the maximum of cold dark matter allowed for cosmological
models, while at the same time sufficiently large so that its presence does
not becomes overwhelming. Astronomical observations also restrict the
parameters of \susy\ models, usually in the lower range of the mass parameters
(see e.g.\,\cite{Grifols:1996hi,Peacock:2004Zuoz,Pauss:2004Zuoz}).

For the various RGE analyses to hold the couplings of the \MSSM\ should be
perturbative all the way from the unification scale down to the \EW\ scale. This
implies, among other restrictions, that top and bottom quark Yukawa couplings should
be below certain limits ($h_{t,b}^2/4\pi<1$).  In terms of \tb\ this amounts it to be confined in
the approximate interval
\begin{equation}
    \label{eq:tanbetalim}
    1 < \tb \lesssim 70\,\,.
\end{equation}

There are
additional phenomenological restrictions that bring the lower
bound on $\tb$ to roughly $2.4$ for the so-called maximal $m_h^0$
scenario, and $10.5$ for the so-called no mixing
scenario\,\cite{\LEPHWG}, the upper bound being the same\cite{Carena:2002es}.

In the MSSM case we use the more restrictive limits \eqref{eq:tanbetalim}, see
Fig.~\ref{mh200}, whereas in the 2HDM model the lower limit can be smaller as shown
in \eqref{eq:tbrange}. Any deviation from this framework of restrictions will
only be for demonstrational purpouses, and will be explicitly quoted in the
text.


%% file: feyn_def.tex
\fmfcmd{%
  style_def mass_boson expr p=
  cdraw (wiggly p);
  shrink (1);
  cfill (arrow p);
  endshrink;
  enddef;
}

\fmfcmd{%
  style_def gluino expr p=
    cdraw (curly p);
  shrink (2);
 draw_plain p;
  cfill (arrow p);
  endshrink;
  enddef;
}

\fmfcmd{%
  vardef crossedcircle expr p =
  polycross 4;
  fullcircle;
  enddef;
}



%% file: tch2HDM/tch2HDM.tex
\chapter{Loop Induced FCNC Decays of the Top Quark in a
 General 2HDM}
\label{cha:tch2HDM}
\section{Introduction} 

As we have said in the introduction, in the 
LHC the production of top quark pairs will be $\sigma(t%
\overline{t})=860\;pb$ -- {roughly two orders of magnitude larger than in 
the Tevatron Run II}. In the so-called low-luminosity phase ($%
10^{33}\,cm^{-2}s^{-1}$) of the LHC one expects about three $t\,\bar{t}$%
-pair per second (ten million $t\,\bar{t}$-pairs per year!)~\cite{\Gianotti}. 
And this number will be augmented by one order of magnitude in the 
high-luminosity phase ($10^{34}\,cm^{-2}s^{-1}$). As for a future LC running 
at e.g. $\sqrt{s}=500\;GeV$, one has a smaller cross-section $\sigma(t\bar{t}%
)=650\;fb$ but a higher luminosity factor ranging from $5\times10^{33}%
\,cm^{-2}s^{-1}$ to $5\times10^{34}\,cm^{-2}s^{-1}$ and of course a much 
cleaner environment\thinspace\cite{\Miller}. With datasets from LHC and LC 
increasing to several $100\,fb^{-1}/$year in the high-luminosity phase, one 
should be able to pile up an enormous wealth of statistics on top quark 
decays. Therefore, not surprisingly, these machines should be very useful to 
analyze rare decays of the top quark, viz. decays whose branching fractions 
are so small ($\lesssim10^{-5}$) that they could not be seen unless the 
number of collected top decays is very large. 
 
The situation is dramatic with the top quark decay into the SM  Higgs boson,
$B(t\rightarrow c\,H_{SM})\sim10^{-13}-10^{-15}$ $(m_{t}=175\,GeV;\;M_{Z}%
\leq M_{H}\leq2\;M_{W})$\thinspace\cite{\Mele}. This extremely tiny rate is 
far out of the range to be covered by any presently conceivable high 
luminosity machine. On the other hand, the highest FCNC top quark rate in 
the SM, namely that of the gluon channel $t\rightarrow c\,g$, is still $6$ 
orders of magnitude below the feasible experimental possibilities at the 
LHC. All in all the detection of FCNC decays of the top quark at visible 
levels (viz. $\,B(t\rightarrow c\,X)>10^{-5}$) by the high luminosity 
colliders round the corner (especially LHC and LC) seems doomed to failure 
in the absence of new physics. Thus the possibility of large enhancements of 
some FCNC channels up to the visible threshold, particularly within the 
context of the general 2HDM and the 
MSSM, should be very welcome. Unfortunately, although the FCNC decay modes 
into electroweak gauge bosons $V_{ew}=W,Z$ may be enhanced a few orders of 
magnitude, it proves to be insufficient to raise the meager SM rates 
mentioned before up to detectable limits, and this is true both in the 2HDM 
-- where $B(t\rightarrow V_{ew}\,c)<10^{-6}$~\cite{\GEilam} -- and in the 
MSSM -- where $B(t\rightarrow V_{ew}\,c)<10^{-7}$ except in highly unlikely 
regions of the MSSM parameter space~\cite{\Lopez}\footnote{%
Namely, regions in which there are wave-function renormalization thresholds 
due to (extremely fortuitous!) sharp coincidences between the sum of the 
sparticle masses involved in the self-energy loops and the top quark mass. 
See e.g. Ref.~\cite{\GJHS} for similar situations already in the conventional  
$t\rightarrow W\,b$ decay within the MSSM. In our opinion these narrow 
regions should not be taken too seriously.}. In this respect it is a lucky 
fact that these bad news need not to apply to the gluon channel, which could 
be barely visible ($B(t\rightarrow g\,c)\lesssim10^{-5}$) both in the
MSSM~\cite{\Divitiis,\NP} and in the general 2HDM~\cite{\GEilam}. But, most  
significant of all, they may not apply to the non-SM Higgs boson channels $%
t\rightarrow(h^{0},H^{0},A^{0})+c$ either. As we shall show in the sequel, 
these Higgs decay channels of the top quark could lie above the visible 
threshold for a parameter choice made in perfectly sound regions of 
parameter space! 
 
While a systematic discussion of these ``gifted'' Higgs channels
was made in Ref.~\cite{\NP} for the MSSM 
case and in other models\footnote{Preliminary SUSY analysis of the Higgs channels
are given in \cite 
{\Hewett}, but they assume the MSSM Higgs mass relations at the 
tree-level. Therefore these are particular cases of the general MSSM 
approach given in~\cite{\NP}. Studies beyond the MSSM (e.g. including 
R-parity violation) and also in quite different contexts from the present 
one are available in the 
literature, see e.g.\thinspace \cite{\FCNCSM}.}, 
to the best of our knowledge there is no similar study in the general 
2HDM. And we believe that this study is 
necessary, not only to assess what are the chances to see traces of new 
(renormalizable) physics in the new colliders round the corner but also to 
clear up the nature of the virtual effects; in particular to disentangle 
whether the origin of the hypothetically detected FCNC decays of the top 
quark is ultimately triggered by SUSY or by some alternative, more generic, 
renormalizable extension of the SM such as the 2HDM or generalizations 
thereof. Of course the alleged signs of new physics could be searched for 
directly through particle tagging, if the new particles were not too heavy. 
However, even if accessible, the corresponding signatures could be far from 
transparent. In contrast, the indirect approach based on the FCNC processes 
has the advantage that one deals all the time with the dynamics of the top 
quark. Thus by studying potentially new features beyond the well-known SM 
properties of this quark one can hopefully uncover the existence of the 
underlying new interactions. 
 
\section{Relevant fields and interactions in the 2HDM} 
 
%
%
%
%
%
%

We will mainly focus our interest on the loop induced FCNC decays  
\begin{equation} 
t\rightarrow c\;h\;\;\;(h=h^{0},H^{0},A^{0})\,,  \label{Higgschannels} 
\end{equation} 
in which any of the three possible neutral Higgs bosons from a general 2HDM 
can be in the final state. However, as a reference we shall compare 
throughout our analysis the Higgs channels with the more conventional gluon 
channel  
\begin{equation} 
t\rightarrow c\;g\,.  \label{gchannel} 
\end{equation} 
\begin{figure}[t] 
\begin{center} 
\includegraphics{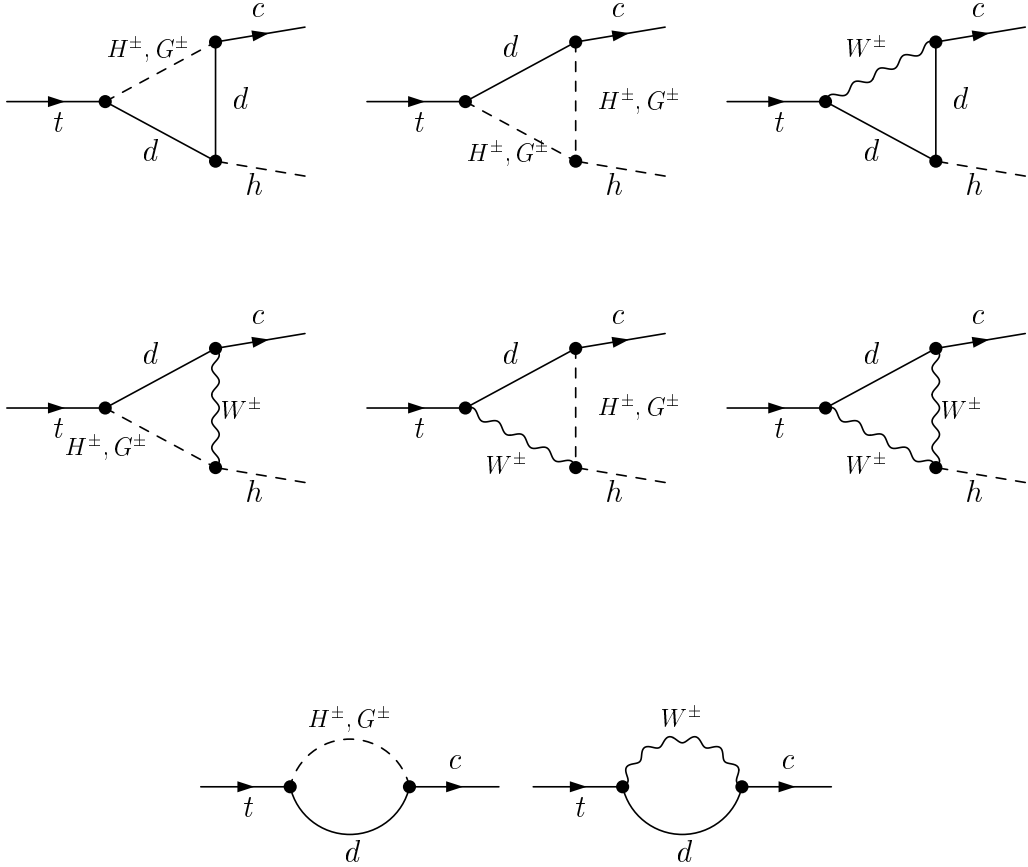}
\end{center} 
\caption{One-loop vertex diagrams contributing to the FCNC top quark decays (%
\ref{Higgschannels}). Shown are the vertices and mixed self-energies with 
all possible contributions from the SM fields and the Higgs bosons from the 
general 2HDM. The Goldstone boson contributions are computed in the Feynman 
gauge.} 
\label{fig:1} 
\end{figure}%
Although other quarks could participate in the final state of these 
processes, their contribution is obviously negligible -- because it is 
further CKM-suppressed. The lowest order diagrams entering these decays are 
one-loop diagrams in which Higgs, quarks, gauge and Goldstone bosons -- in 
the Feynman gauge -- circulate around. While the diagrams for the decays (%
\ref{Higgschannels}) are depicted in Fig.~\ref{fig:1}, the ones for the 
decay (\ref{gchannel}) are not shown~\cite{\GEilam}. 

We wish to concentrate here on Type~I and Type~II models of a sufficiently
generic nature, to wit, those which are 
characterized by the following set of free parameters:  
\begin{equation} 
(m_{h^{0}},m_{H^{0}},m_{A^{0}},m_{H^{\pm}},\tan\alpha,\tan\beta )\,, 
\label{freeparam} 
\end{equation} 
$\tb$ is a key parameter in our analysis. The numerical analysis that we 
perform in the next section does not depend in any essential way on the
simplification $\lambda _{5}=\lambda _{6}$,\eqref{eq:lambda56}.
In essence we have just traded $\lambda _{5}$ for $%
m_{A^{0}}^{2}$ in these rules and so by varying with respect to $m_{A^{0}}$ 
we do explore most of the quantitative potential of the general 2HDM. 

Since we shall perform our calculation in the on-shell scheme, we understand 
that the physical inputs are given by the electromagnetic coupling and the 
physical masses of all the particles:  
\begin{equation} 
(e,M_{W},M_{Z},m_{h^{0}},m_{H^{0}},m_{A^{0}},m_{H^{\pm}},m_{f})\,. 
\label{onshell} 
\end{equation} 
The remaining parameters, except the Higgs mixing angles, are understood to 
be given in terms of the latter, e.g. the $SU(2)$ gauge coupling appearing 
in the previous formulae and in Table~\ref{tab:trilineals} is given by $g=e/s_{w}$ , 
where the sinus of the weak mixing angle is defined through $%
s_{w}^{2}=1-M_{W}^{2}/M_{Z}^{2}$. It should be clear that, as there are no 
tree-level FCNC decays of the top quark, there is no need to introduce 
counterterms for the physical inputs in this calculation. In fact, the 
calculation is carried out in lowest order (``tree level'') with respect to 
the effective $tch$ and $tcg$ couplings and so the sum of all the one-loop 
diagrams (as well as of certain subsets of them) should be finite in a 
renormalizable theory, and indeed it is. 
 
\section{Numerical analysis} 
 
From the previous interaction Lagrangians and Feynman rules it is now 
straightforward to compute the loop induced FCNC rates for the decays (\ref 
{Higgschannels}) and (\ref{gchannel}). We shall refrain from listing the 
lengthy analytical formulae as the computation is similar to the one 
reported in great detail in Ref.~\cite{\NP}. Therefore, we will limit 
ourselves to exhibit the final numerical results. The fiducial ratio on 
which we will apply our numerical computation is the following:  
\begin{equation} 
B^{j}(t\rightarrow h+c)=\frac{\Gamma^{j}(t\rightarrow h+c)}{\Gamma 
(t\rightarrow W^{+}+b)+\Gamma^{j}(t\rightarrow H^{+}+b)}\,\,, 
\label{fiducialH} 
\end{equation} 
for each Type $j=I,II$ of 2HDM and for each neutral Higgs boson $h=h^{0}$, $%
H^{0}$, $A^{0}$. While this ratio is not the total branching fraction, it is 
enough for most practical purposes and it is useful in order to compare with 
previous results in the literature. Notice that for $m_{H^{\pm}}>m_{t}$ (the 
most probable situation for Type~II 2HDM's, see below) the ratio (\ref 
{fiducialH}) reduces to $B^{j}(t\rightarrow h+c)=\Gamma^{j}(t\rightarrow 
h+c)/\Gamma(t\rightarrow W^{+}+b)$, which is the one that we used in Ref.  
\cite{\NP}. It is understood that $\Gamma^{j}(t\rightarrow h+c)$ above is 
computed from the one-loop diagrams in Fig.~\ref{fig:1}, with all quark 
families summed up in the loop. Therefore, consistency in 
perturbation theory requires to compute $\Gamma(t\rightarrow W^{+}+b)$ and $%
\Gamma (t\rightarrow H^{+}+b)$ in the denominator of~(\ref{fiducialH}) only 
at the tree-level (for explicit expressions see e.g.~\cite{\CGGJS}). As 
mentioned in Sec. 2, we wish to compare our results for the Higgs channels (%
\ref{Higgschannels}) with those for the gluon channel~(\ref{gchannel}), so 
that we similarly define  
\begin{equation} 
B^{j}(t\rightarrow g+c)=\frac{\Gamma^{j}(t\rightarrow g+c)}{\Gamma 
(t\rightarrow W^{+}+b)+\Gamma^{j}(t\rightarrow H^{+}+b)}\,\,. 
\label{fiducialG} 
\end{equation} 
We have performed a fully-fledged independent analytical and numerical 
calculation of $\Gamma^{j}(t\rightarrow g+c)$ at one-loop in the context of 
2HDM~I and II. Where there is overlapping, we have checked the numerical 
results of Ref.~\cite{\GEilam}, but we point out that they agree with us only 
if $\Gamma(t\rightarrow H^{+}+b)$ is included in the denominator of eq.~(\ref 
{fiducialG}), in contrast to what is asserted in that reference in which $%
B(t\rightarrow g+c)$ is defined without the charged Higgs channel 
contribution. 
 
We have performed part of the analytical calculation of the diagrams for 
both processes~(\ref{Higgschannels}) and~(\ref{gchannel}) by hand and we 
have cross-checked our results with the help of the numeric and algebraic 
programs FeynArts, FormCalc and LoopTools~\cite{\FeynArts}, with which we 
have completed the rest of the calculation. In particular, the cancellation 
of UV divergences in the total amplitudes was also verified by hand. In 
addition we have checked explicitly the gauge invariance of the total 
analytical amplitude for the process~(\ref{gchannel}), which is a powerful 
test. And we have confirmed that our code reproduces the SUSY Higgs contribution 
of Ref.\cite{\NP} when we turn on the MSSM Higgs mass relations.
 
As mentioned above, a highly relevant parameter is $\tan\beta$, which must 
be restricted to the approximate range \eqref{eq:tbrange}
in perturbation theory\footnote{%
Some authors~\cite{\Berger} claim that perturbativity allows $\tan\beta$ to 
reach values of order $100$ and beyond, and these are still used in the 
literature. We consider it unrealistic and we shall not choose $\tan\beta$ 
outside the interval~(\ref{eq:tbrange}). Plots versus $\tan\beta$, however, 
will indulge larger values just to exhibit the dramatic enhancements of our 
FCNC top quark rates.}. It is to be expected from the various couplings 
involved in the processes under consideration that the low $\tan\beta$ 
region could be relevant for both the Type~I and Type~II 2HDM's. In 
contrast, the high $\tan\beta$ region is only potentially important for the 
Type~II. However, the eventually relevant regions of parameter space are 
also determined by the value of the mixing angle $\alpha$, as we shall see 
below. 
 
Of course there are several restrictions that must be respected by our 
numerical analysis, see \ref{sec:constraints}, as the $\rho$-parameter
restriction, \eqref{eq:drho}, and the $\bsg$ restriction, \eqref{eq:CLEO}.
With the charge Higgs boson mass restrictions in section \ref{sec:constraints}
\footnote{At the time of this work this restriction was
$m_{H^{\pm}}>(165-200)\,\,GeV$ for virtually any $\tan\beta\gtrsim1$ \cite
{\Borzumati,\Chankowski}.}
in principle the top quark decay $t\rightarrow 
H^{+}+b$ is still possible in 2HDM~I; but also in 2HDM~II, if $m_{H^{\pm}} $ 
lies near the lowest end of the previous bound, and in this case that decay 
can contribute to the denominator of eqs.~(\ref{fiducialH})-(\ref{fiducialG}%
). 
 
\begin{figure}[t] 
\begin{center} 
\resizebox{\textwidth}{!}{\includegraphics{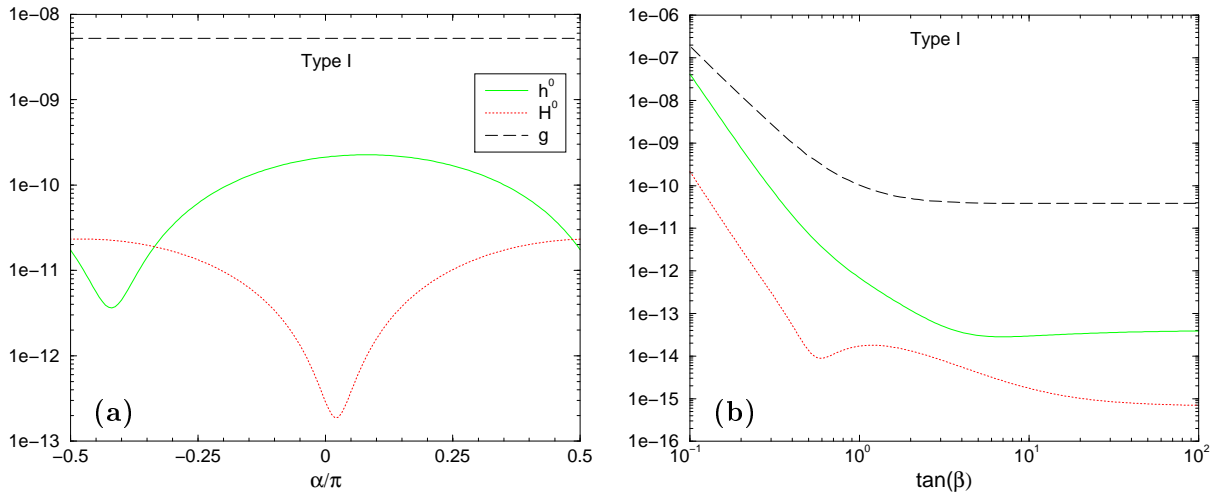}} 
\end{center} 
\caption{Evolution of the FCNC top quark fiducial ratios~(\ref{fiducialH})-(%
\ref{fiducialG}) in Type~I 2HDM versus: \textbf{(a)} the mixing angle $%
\protect\alpha$ in the CP-even Higgs sector, in units of $\protect\pi$;  
\textbf{(b)} $\tan\protect\beta$. The values of the fixed parameters are as 
in eqs.~(\ref{inputsmixing}) and (\ref{inputsmasses}).} 
\label{fig:2} 
\end{figure}

\begin{figure}[t] 
\begin{center} 
\resizebox{\textwidth}{!}{\includegraphics{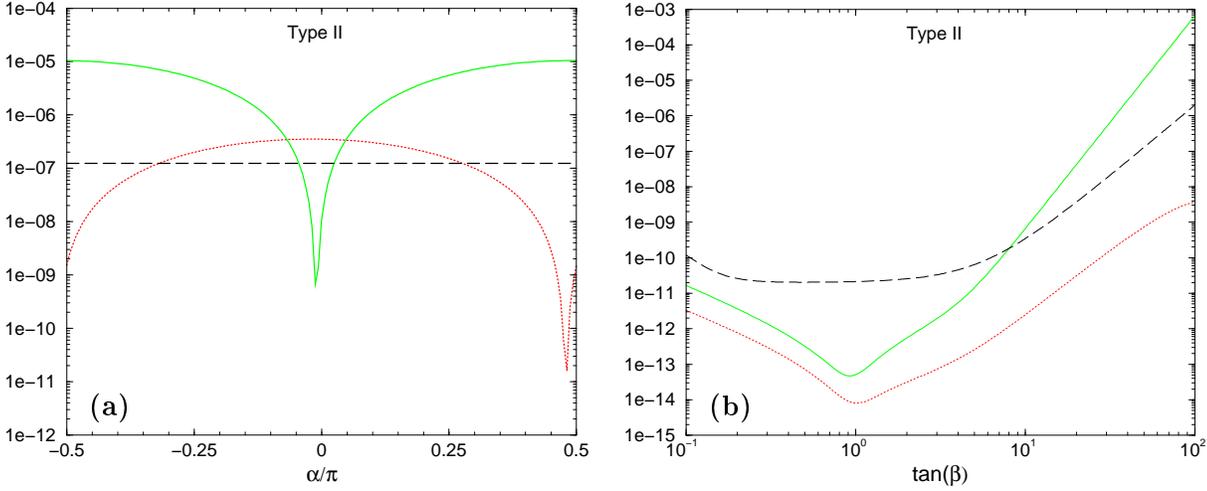}} 
\end{center} 
\caption{As in Fig.~\ref{fig:2}, but for the 2HDM~II. The plot in (b) 
continues above the bound in eq.~(\ref{eq:tbrange}) just to better show the 
general trend.} 
\label{fig:3} 
\end{figure}

\begin{figure}[t] 
\begin{center} 
\resizebox{\textwidth}{!}{\includegraphics{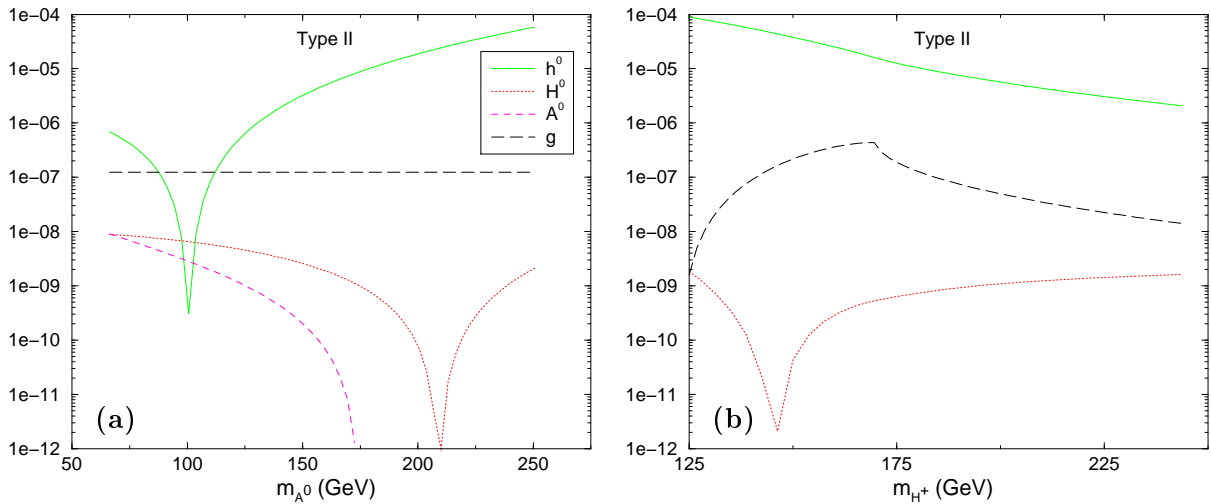}} 
\end{center} 
\caption{Evolution of the FCNC top quark fiducial ratios~(\ref{fiducialH})-(%
\ref{fiducialG}) in Type~II 2HDM versus: \textbf{(a)} the CP-odd Higgs boson 
mass $m_{A^{0}}$; \textbf{(b) }the charged Higgs boson mass $m_{H^{\pm}}$. 
The values of the fixed parameters are as in eqs.~(\ref{inputsmixing}) and (%
\ref{inputsmasses}). The plot in (b) starts below the bound $%
m_{H^{\pm}}>165\,GeV$ mentioned in the text to better show the general 
trend. } 
\label{fig:4} 
\end{figure}

\begin{figure}[t] 
\begin{center} 
\resizebox{\textwidth}{!}{\includegraphics{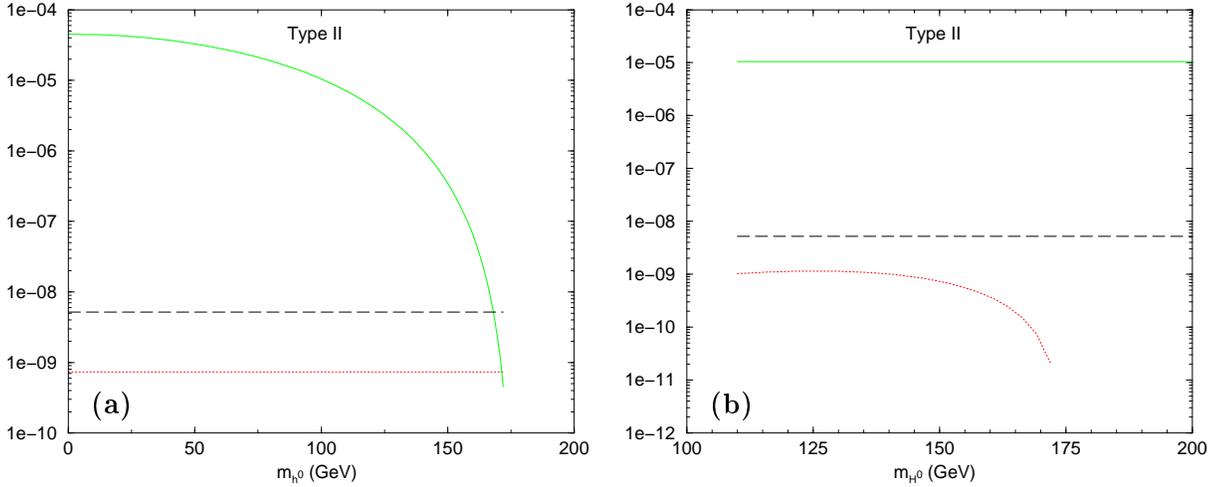}} 
\end{center} 
\caption{As in Fig.~\ref{fig:4}, but plotting versus: \textbf{(a) }the 
lightest CP-even Higgs boson mass $m_{h^{0}}$; \textbf{(b) }the heaviest 
CP-even Higgs boson mass $m_{H^{0}}$.} 
\label{fig:5} 
\end{figure}

The combined set of independent conditions turns out to be quite effective 
in narrowing down the permitted region in the parameter space, as can be 
seen in Figs.~\ref{fig:2}-\ref{fig:5} where we plot the fiducial FCNC rates~(%
\ref{fiducialH})-(\ref{fiducialG}) versus the parameters~(\ref{freeparam}). 
The cuts in some of these curves just reflect the fact that at least one of 
these conditions is not fulfilled. 
 
After scanning the parameter space, we see in Figs.~\ref{fig:2}-\ref{fig:3} 
that the 2HDM~I (resp. 2HDM~II) prefers low values (resp. high values) of $%
\tan\alpha$ and $\tan \beta$ for a given channel, e.g. $t\rightarrow 
h^{0}\,c $. Therefore, the following choice of mixing angles will be made to 
optimize the presentation of our numerical results:  
\begin{align} 
\text{2HDM I} & :\ \tan\alpha=\tan\beta=1/4\,\,;  \notag \\ 
\text{2HDM II} & :\ \tan\alpha=\tan\beta=50\,.  \label{inputsmixing} 
\end{align} 
We point out that, for the same values of the masses, one obtains the same 
maximal FCNC rates for the alternative channel $t\rightarrow H^{0}\,c$ 
provided one just substitutes $\alpha\rightarrow\pi/2-\alpha$. Equations~(%
\ref{inputsmixing}) define the eventually relevant regions of parameter 
space and, as mentioned in section~\ref{sec:constraints}, depend on the values of the mixing angles $%
\alpha$ and $\beta$, namely $\beta \simeq\alpha\simeq0$ for Type~I and $%
\beta\simeq\alpha\simeq\pi/2$ for Type~II.

Due to the $\alpha \rightarrow \pi /2-\alpha $ symmetry of the maximal rates 
for the CP-even Higgs channels, it is enough to concentrate the numerical 
analysis on one of them, but one has to keep in mind that the other channel 
yields the same rate in another region of parameter space. Whenever a mass 
has to be fixed, we choose conservatively the following values for both 
models:  
\begin{equation} 
m_{h^{0}}=100\,GeV\,,\;m_{H^{0}}=150\,GeV\,,\;\;m_{A^{0}}=m_{H^{\pm 
}}=180\,GeV\,.  \label{inputsmasses} 
\end{equation} 
Also for definiteness, we take the following values for some relevant SM\ 
parameters in our calculation:  
\begin{equation} 
m_{t}=175\,GeV\,,\;m_{b}=5\,GeV\,,\,\,\alpha 
_{s}(m_{t})=0.11\,,\;\, V_{cb}=0.040\,,  \label{SMparam} 
\end{equation} 
and the remaining ones are as in \cite{\PDBMM}. Notice that our choice of $%
m_{A^{0}}$ prevents the decay $t\rightarrow \;A^{0}\,c$ from occurring, and 
this is the reason why it does not appear in Figs.~\ref{fig:2}-~\ref{fig:3}. The variation 
of the results with respect to the masses is studied in Figs.~\ref{fig:4}-%
\ref{fig:5}. In particular, in Fig.~\ref{fig:4} we can see the (scanty) rate 
of the channel $t\rightarrow A^{0}\,c$ when it is kinematically allowed. In 
general the pseudoscalar channel is the one giving the skimpiest FCNC rate. 
This is easily understood as it is the only one that does not have trilinear 
couplings with the other Higgs particles (Cf. Table~\ref{tab:trilineals}). While it 
does have trilinear couplings involving Goldstone bosons, these are not 
enhanced. The crucial role played by the trilinear Higgs self-couplings in 
our analysis cannot be underestimated as they can be enhanced by playing 
around with both (large or small) $\tan \beta $ \emph{and} also with the 
mass splittings among Higgses. This feature is particularly clear in Fig.~%
\ref{fig:4}a where the rate of the channel $t\rightarrow h^{0}\,c$ is 
dramatically increased at large $m_{A^{0}}$, for fixed values of the other 
parameters and preserving our list of constraints. Similarly would happen 
for $t\rightarrow H^{0}\,c$ in the corresponding region $\alpha \rightarrow 
\pi /2-\alpha $. 
 
From Figs.~\ref{fig:2}a and \ref{fig:2}b it is pretty clear that the 
possibility to see FCNC decays of the top quark into Type~I Higgs bosons is 
plainly hopeless even in the most favorable regions of parameter space -- 
the lowest (allowed) $\tan\beta$ end. In fact, the highest rates remain 
neatly down $10^{-6}$, and therefore they are (at least) one order of 
magnitude below the threshold sensibility of the best high luminosity top 
quark factory in the foreseeable future (see Section 4). We remark, in Fig.~%
\ref{fig:2}, that the rate for the reference decay $t\rightarrow g\,c$ in the 
2HDM~I is also too small but remains always above the Higgs boson rates. 
Moreover, for large $\tan\beta$ one has, as expected, $B^{I}\,($ $%
t\rightarrow g\,c)\rightarrow B^{SM}\,($ $t\rightarrow 
g\,c)\simeq4\times10^{-11}$ because in this limit all of the charged Higgs 
couplings in the 2HDM~I (the only Higgs couplings involved in this decay) 
drop off. Due to the petty numerical yield from Type~I models we refrain 
from showing the dependence of the FCNC rates on the remaining parameters.

Fortunately, the meager situation just described does not replicate for Type 
II Higgs bosons. For, as shown in Figs.~\ref{fig:3}a and \ref{fig:3}b, the 
highest potential rates are of order $10^{-4}$, and so there is hope for 
being visible. In this case the most favorable region of parameter space is 
the high $\tan\beta$ end in eq.~(\ref{eq:tbrange}). Remarkably, there is no 
need of risking values over and around $100$ (which, as mentioned above, are 
sometimes still claimed as perturbative!) to obtain the desired rates. But 
it certainly requires to resort to models whose hallmark is a large value of  
$\tan\beta$ of order or above $m_{t}/m_{b}\gtrsim35$. As for the dependence 
of the FCNC rates on the various Higgs boson masses (Cf. Figs.~\ref{fig:4}-%
\ref{fig:5}) we see that for large $m_{A^{0}}$ the decay $t\rightarrow 
h^{0}\,c$ can be greatly enhanced as compared to $t\rightarrow g\,c$; and of 
course, once again, the same happens with $t\rightarrow H^{0}\,c$ in the 
alternative region $\alpha\rightarrow \pi/2-\alpha$. We also note (from the 
combined use of Figs.~\ref{fig:3}b, \ref{fig:4}a and \ref{fig:4}b) that in 
the narrow range where $t\rightarrow H^{+}\,b$ could still be open in the 
2HDM~II, the rate of $t\rightarrow h^{0}\,c$ becomes the more visible the 
larger and larger is $\tan\beta$ and $m_{A^{0}}$. Indeed, in this region one 
may even overshoot the $10^{-4}$ level without exceeding the upper bound~(%
\ref{eq:tbrange}) while also keeping under control the remaining constraints, 
in particular eq.~(\ref{eq:drho}). Finally, the evolution of the rates~(\ref 
{fiducialH})-(\ref{fiducialG}) with respect to the two CP-even Higgs boson 
masses is shown in Figs.~\ref{fig:5}a and \ref{fig:5}b. The neutral Higgs 
bosons themselves do not circulate in the loops (Cf. Fig.~\ref{fig:1}) but 
do participate in the trilinear couplings (Cf. Table~\ref{tab:trilineals}) and so the 
evolution shown in some of the curves in Fig.~\ref{fig:5} is due to both the 
trilinear couplings and to the phase space exhaustion. 
 
Turning now to the light scalar and pseudoscalar corners in parameter space 
mentioned above, it so happens that, after all, they prove to be of little 
practical interest in our case. Ultimately this is due to the quadratic 
Higgs boson mass differences entering $\delta\rho$ which make very 
difficult to satisfy the bound~(\ref{eq:drho}). The reason being that for Type 
II models the limit $m_{H^{\pm}}\gtrsim165\,GeV$ from $b\rightarrow 
s\,\gamma $ implies that the constraint~(\ref{eq:drho}) cannot be preserved in 
the presence of light neutral Higgses. In actual fact the analysis shows 
that if e.g. one fixes $m_{h^{0}}=20-30\;GeV$, then the minimum $m_{A^{0}}$ 
allowed by $\delta\rho$ is $100\,GeV$ and the maximum rate~(\ref{fiducialH}) 
is of order $10^{-6}$. Conversely, if one chooses $m_{A^{0}}=20-30\;GeV$, 
then the minimum $m_{h^{0}}$ allowed by $\delta\rho$ is $120\,GeV$ and the 
maximum rate (\ref{fiducialH}) is near $10^{-4}$. Although in the last case 
the maximum rate is higher than in the first case, it is just of the order 
of the maximum rate already obtained outside the light mass corners of 
parameter space. On the other hand, these light mass regions do not help us 
in Type~I models either. Even though for these models we do not have the $%
b\rightarrow s\,\gamma$ bound on the charged Higgs, we still have the direct 
LEP 200 bound $m_{H^{\pm}}\gtrsim78.7\,GeV$~\cite{\TJunk} which is of course 
weaker than the CLEO bound. As a consequence the $\delta\rho$ constraint can 
be satisfied in the 2HDM~I for neutral Higgs bosons lighter than in the 
corresponding 2HDM~II case, and one does get some enhancement of the FCNC 
rates. Specifically, one may reach up to $10^{-6}$. However, the maximal 
rates~(\ref{fiducialH}) for the 2HDM~I Higgs bosons are so small (see Figs.~%
\ref{fig:2}a-\ref{fig:2}b) that this order of magnitude enhancement is 
rendered immaterial. The upshot is that the top quark FCNC processes are not 
especially sensitive to the potential existence of a very light Higgs boson 
in either type of 2HDM. 
 
\section{Discussion and conclusions} 
 
The sensitivities to FCNC top quark decays for $100\,fb^{-1}$ of integrated 
luminosity in the relevant colliders are estimated to be~\cite{\Frey}:  
\begin{align} 
\mathrm{\mathbf{LHC:}}B(t\rightarrow c\,X) & \gtrsim5\times10^{-5}\,,  \notag \\  
\mathrm{\mathbf{LC:}}B(t\rightarrow c\,X) & \gtrsim5\times10^{-4}\,,  
\label{sensitiv} \\ 
\mathrm{\mathbf{TEV33:}}B(t\rightarrow c\,X) & \gtrsim5\times10^{-3}\,\,\,.  
\notag 
\end{align} 
This estimation has been confirmed by a full signal-background analysis 
for the hadron colliders in Ref.\cite{\Saavedra}. 
From these experimental expectations and our numerical results it becomes 
patent that whilst the Tevatron will remain essentially blind to this kind 
of physics, the LHC and the LC will have a significant potential to observe 
FCNC decays of the top quark beyond the SM. Above all there is a possibility 
to pin down top quark decays into neutral Higgs particles, eq.~(\ref 
{Higgschannels}), within the framework of the general 2HDM~II provided $%
\tan\beta\gtrsim\,m_{t}/m_{b}\sim35$. The maximum rates are of order $%
10^{-4} $ and correspond to the two CP-even scalars. This conclusion is 
remarkable from the practical (quantitative) point of view, and also 
qualitatively because the top quark decay into the SM Higgs particle is, in 
notorious contradistinction to the 2HDM~II case, the less favorable top 
quark FCNC rate in the SM. On the other hand, we deem practically hopeless 
to see FCNC decays of the top quark in a general 2HDM~I for which the 
maximum rates are of order $10^{-7}$. This order of magnitude cannot be 
enhanced unless one allows $\tan\beta\ll0.1$, but the latter possibility is 
unrealistic because perturbation theory breaks down and therefore one cannot 
make any prediction within our approach. 
 
We have made a parallel numerical analysis of the gluon channel $%
t\rightarrow c\;g$ in both types of 2HDM's. We confirm that this is another 
potentially important FCNC mode of the top quark in 2HDM extensions of the 
SM~\cite{\GEilam} but, unfortunately, it still falls a bit too short to be 
detectable. The maximum rates for this channel lie below $10^{-6}$ in the 
2HDM~I (for $\tan\beta>0.1)$ and in the 2HDM~II (for $\tan\beta<60$), and so 
it will be hard to deal with it even at the LHC. 
 
We are thus led to the conclusion that the Higgs channels~(\ref{Higgschannels}), 
more specifically the CP-even ones, give the highest potential rates for top 
quark FCNC decays in a general 2HDM~II. Most significant of all: they are  
\emph{the only} FCNC decay modes of the top quark, within the simplest 
renormalizable extensions of the SM, that have a real chance to be seen in 
the next generation of high energy, high luminosity, colliders. 
 
The former conclusions are similar to the ones derived in Ref.~\cite{\NP} for 
the MSSM case, but there are some conspicuous differences on which we wish 
to elaborate a bit in what follows~\cite{\Carmel}. First, in the general 
2HDM~II the two channels $t\rightarrow(h^{0},H^{0})\;c$ give the same 
maximum rates, provided we look at different (disjoint) regions of the 
parameter space. The $t\rightarrow A^{0}\;c$ channel is, as mentioned, 
negligible with respect to the CP-even modes. Hereafter we will discard this 
FCNC top quark decay mode from our discussions within the 2HDM context. On 
the other hand, in the MSSM there is a most distinguished channel, viz. $%
t\rightarrow h^{0}\;c$, which can be high-powered by the SUSY stuff all over 
the parameter space. In this framework the mixing angle $\alpha$ becomes 
stuck once $\tan\beta$ and the rest of the independent parameters are given, 
and so there is no possibility to reconvert the couplings between $h^{0}$ 
and $H^{0}$ as in the 2HDM. Still, we must emphasize that in the MSSM the 
other two decays $t\rightarrow H^{0}\;c$ and $t\rightarrow A^{0}\;c$ can be 
competitive with $t\rightarrow h^{0}\;c$ in certain portions of parameter 
space. For example, $t\rightarrow H^{0}\;c$ becomes competitive when the 
pseudoscalar mass is in the range $110\,GeV<m_{A^{0}}\lesssim170\,GeV$~\cite 
{\NP}. The possibility of having more than one FCNC decay~(\ref{Higgschannels}%
) near the visible level is a feature which is virtually impossible in the 
2HDM~II. Second, the reason why $t\rightarrow h^{0}\;c$ in the MSSM is so 
especial is that it is the only FCNC top quark decay~(\ref{Higgschannels}) 
which is always kinematically open throughout the whole MSSM parameter 
space, while in the 2HDM all of the decays (\ref{Higgschannels}) could be, 
in the worse possible situation, dead closed. Nevertheless, this is not the 
most likely situation in view of the fact that all hints from high precision 
electroweak data seem to recommend the existence of (at least) one 
relatively light Higgs boson~\cite{\TJunk,\Hagiwara}. This is certainly an 
additional motivation for our work, as it leads us to believe that in all 
possible (renormalizable) frameworks beyond the SM, and not only in SUSY, we 
should expect that at least one FCNC decay channel (\ref{Higgschannels}) 
could be accessible. Third, the main origin of the maximum FCNC rates in the 
MSSM traces back to the tree-level FCNC couplings of the gluino~\cite{\NP}. 
These are strong couplings, and moreover they are very weakly restrained by 
experiment. In the absence of such gluino couplings, or perhaps by further 
experimental constraining of them in the future, the FCNC rates in the MSSM 
would boil down to just the electroweak (EW) contributions, to wit, those 
induced by charginos, squarks and also from SUSY Higgses. The associated 
SUSY-EW rate is of order $10^{-6}$ at most, and therefore it is barely 
visible, most likely hopeless even for the LHC. In contrast, in the general 
2HDM the origin of the contributions is purely EW and the maximum rates are 
two orders of magnitude higher than the full SUSY-EW effects in the MSSM. It 
means that we could find ourselves in the following situation. Suppose that 
the FCNC couplings of the gluino get severely restrained in the future and 
that we come to observe a few FCNC decays of the top quark into 
Higgs bosons, perhaps at the LHC and/or the LC. Then we would immediately 
conclude that these Higgs bosons could not be SUSY-MSSM, whilst 
they could perhaps be CP-even members of a 2HDM~II. Fourth, the gluino 
effects are basically insensitive to $\tan\beta$, implying that the maximum 
MSSM rates are achieved equally well for low, intermediate or high values of  
$\tan\beta,$ whereas the maximum 2HDM~II rates (comparable to the MSSM ones) 
are attained only for high $\tan\beta$. 
 
The last point brings about the following question: what could we possibly 
conclude if the gluino FCNC couplings were not further restricted by 
experiment and the tagging of certain FCNC decays of the top quark into 
Higgs bosons would come into effect? Would still be possibly to discern 
whether the Higgs bosons are supersymmetric or not? The answer is, most 
likely yes, provided certain additional conditions would be met. 
 
There are many possibilities and corresponding strategies, but we will limit 
ourselves to point out some of them. For example, let us consider the type 
of signatures involved in the tagging of the Higgs channels. In the favorite 
FCNC region~(\ref{inputsmixing}) of the 2HDM~II, the combined decay  $t\rightarrow 
h\;\,c\rightarrow cb\overline{b}$ is possible only for $h^{0}$ or for $H^{0}$%
, but not for both -- Cf. Fig.~\ref{fig:3}a -- whereas in the MSSM, $h^{0}$ 
together with $H^{0}$, are highlighted for $110\,GeV<m_{A^{0}}<m_{t}$, with 
no preferred $\tan \beta $ value. And similarly, $t\rightarrow A^{0}\;c$ is 
also non-negligible for $m_{A^{0}}\lesssim120\,GeV$\thinspace \cite{\NP}. Then 
the process $t\rightarrow h\;\,c\rightarrow cb\overline{b}$ gives 
rise to high $p_{T}$ charm-quark jets and a recoiling $b\overline{b}$ pair 
with large invariant mass. It follows that if more than one distinctive 
signature of this kind would be observed, the origin of the hypothetical 
Higgs particles could not probably be traced back to a 2HDM~II.  
 
One might worry that in the case of $h^{0}$ and $H^{0}$ they could also (in 
principle) decay into electroweak gauge boson pairs $h^{0},H^{0}\rightarrow 
V_{ew}\overline{V}_{ew}$, which in some cases could be kinematically 
possible. But this is not so in practice for the 2HDM~II if we stick to our 
favorite scenario, eq.~(\ref{inputsmixing}). In fact, we recall that the 
decay $h^{0}\rightarrow V_{ew}\overline{V}_{ew}$ is not depressed with 
respect to the SM Higgs boson case provided $\beta -\alpha =\pi /2$, and 
similarly for $H^{0}\rightarrow V_{ew}\overline{V}_{ew}$ if $\beta -\alpha =0 
$. However, neither of these situations is really pinpointed by FCNC physics 
because we have found $\beta \simeq \pi /2$ in the most favorable region of 
our numerical analysis, and moreover $\alpha $ was also seen there to be 
either $\alpha \simeq \pi /2$ (for $h^{0}$) or $0$ (for $H^{0}$), so both 
decays $h^{0},H^{0}\rightarrow V_{ew}\overline{V}_{ew}$ are suppressed in 
the regions where the FCNC rates of the parent decays $t\rightarrow 
(h^{0},H^{0})\;\,c$ are maximized. Again, at variance with this situation, 
in the MSSM case $H^{0}\rightarrow V_{ew}\overline{V}_{ew}$ is perfectly 
possible -- not so $h^{0}\rightarrow V_{ew}\overline{V}_{ew}$ due to the 
aforementioned upper bound on $m_{h^{0}}$ -- because $\tan \beta $ has no 
preferred value in the most favorable MSSM decay region of $t\rightarrow 
H^{0}\;\,c$. Therefore, detection of a high $p_{T}$ charm-quark jet against 
a $V_{ew}\overline{V}_{ew}$ pair of large invariant mass could only be 
advantageous in the MSSM, not in the 2HDM. Similarly, for $\tan \beta \gtrsim%
1$ the decay $H^{0}\rightarrow h^{0}\;h^{0}$ (with real or virtual $h^{0}$) 
is competitive in the MSSM~\cite{\Djouadi} in a region where the parent FCNC 
top quark decay is also sizeable. Again this is impossible in the 2HDM~II 
and therefore it can be used to distinguish the two (SUSY and non-SUSY) 
Higgs frames. 
 
Finally, even if we place ourselves in the high $\tan\beta$ region both for 
the MSSM and the 2HDM~II, then the two frameworks could still possibly be 
separated provided that two Higgs masses were known, perhaps one or both of 
them being determined from the tagged Higgs decays themselves, eq.~(\ref 
{Higgschannels}). Suppose that $\tan\beta$ is numerically known (from other 
processes or from some favorable fit to precision data), then the full 
spectrum of MSSM Higgs bosons would be approximately determined (at the tree 
level) by only knowing one Higgs mass, a fact that could be used to check 
whether the other measured Higgs mass becomes correctly predicted. Of 
course, the radiative corrections to the MSSM Higgs mass relations can be 
important at high $\tan\beta$~\cite{\Hollik}, but these could be taken into 
account from the approximate knowledge of the relevant sparticle masses 
obtained from the best fits available to the precision measurements within 
the MSSM. If there were significant departures between the predicted mass 
for the other Higgs and the measured one, we would probably suspect that the 
tagged FCNC decays into Higgs bosons should correspond to a 
non-supersymmetric 2HDM~II. 
 
At the end of the day we see that even though the maximum FCNC rates for the 
MSSM and the 2HDM~II are both of order $10^{-4}$ -- and therefore 
potentially visible -- at some point on the road it should be possible to 
disentangle the nature of the Higgs model behind the FCNC decays of the top 
quark. Needless to say, if all the fuss at CERN~\cite{\TJunk} about 
the possible detection of a Higgs boson would eventually be confirmed, this 
could still be interpreted as the discovery of one neutral member of an 
extended Higgs model. Obviously the combined Higgs data from LEP 200 and the 
possible discovery of FCNC top quark decays into Higgs bosons at the LHC/LC 
would be an invaluable cross-check of the purportedly new phenomenology. 
 
We emphasize our most essential conclusions in a nutshell: i) Detection of 
FCNC top quark decay channels into a neutral Higgs boson would be a blazing 
signal of physics beyond the SM; ii) There is a real chance for seeing rare 
events of that sort both in generic Type~II 2HDM's and in the MSSM. The 
maximum rates for the leading FCNC processes~(\ref{Higgschannels}) and (\ref 
{gchannel}) in the 2HDM~II (resp. in the MSSM) satisfy the relations  
\begin{equation} 
B(t\rightarrow g\,c)<10^{-6}(10^{-5})<B(t\rightarrow h\,c)\sim10^{-4}\,, 
\end{equation} 
where it is understood that $h$ is $h^{0}$ or $H^{0}$, but not both, in the 
2HDM~II; whereas $h$ is most likely $h^{0}$, but it could also be $H^{0}$ 
and $A^{0}$, in the MSSM ; iii) Detection of more than one Higgs channel 
would greatly help to unravel the type of underlying Higgs model. 
 
The pathway to seeing new physics through FCNC decays of the top quark is 
thus potentially open. It is now an experimental challenge to accomplish 
this program using the high luminosity super-colliders round the corner. 
 
\bigskip 
 

%% file: htc2HDM/htc2HDM.tex
\chapter{Higgs Boson FCNC Decays into Top Quark in a
  General 2HDM}
\label{cha:htc2HDM}
\section{Introduction}

When considering physics beyond
the SM, new horizons of possibilities open up which may radically
change the pessimistic prospects for FCNC decays involving a Higgs
boson and the top quark. For example, in Ref.\cite{\Divitiis} it
was shown that the vector boson modes can be highly enhanced
within the context of the Minimal Supersymmetric Standard Model
(MSSM) \,\cite{\SUSY}. This fact was also dealt with in great
detail in Ref. \cite{\GuaschNPo} where in addition a dedicated
study was presented of the FCNC top quark decays into the various
Higgs bosons of the MSSM (see also \cite{\Yuan}), showing that
these can be the most favored FCNC top quark decays -- above the
expectations on the gluon mode $t\rightarrow c\,g$. A similar
study is performed in the chapter \ref{cha:tch2HDM}
for the FCNC top quark decays into Higgs
bosons in a general two-Higgs-doublet model (2HDM).

In the previous
chapter analysing
the FCNC top quark decays in 2HDM extensions of the SM
it was proven that while the
maximum rates for $t\rightarrow c\,g$ were one order of magnitude
more favorable in the MSSM\cite{Guasch:1999jp} than in the 2HDM, the corresponding
rates for $t\rightarrow c\,h^0$ were comparable both for the MSSM
and the general 2HDM, namely up to the $10^{-4}$ level and should
therefore be visible both at the LHC and the
LC\,\cite{\Saavedra}.

Similarly, one may wonder whether the FCNC decays of the Higgs
bosons themselves can be of some relevance. Obviously the
situation with the SM Higgs is essentially hopeless, so again we
have to move to physics beyond the SM. Some work on these decays,
performed in various contexts including the MSSM, shows that
these effects can be important \,\cite{\Hou,\Curiel,\Demir,
\Brignole}, as seen in chapter \ref{cha:hbsSUSY_br}. This could be expected, at least for
heavy quarks in the MSSM, from the general SUSY study (including
both strong and electroweak supersymmetric effects) of the FCNC
vertices $h\,t\,c$ ($h=h^0\,,H^0\,,A^0$) made in  Ref.
\cite{\GuaschNPo}. However, other frameworks could perhaps be
equally advantageous. Here we are particularly interested in the
FCNC Higgs decay modes into top quark within a general 2HDM,
which have not been studied anywhere in the literature to our
knowledge. It means we restrict to Higgs bosons heavier than
$m_t$. From the above considerations, and most particularly on
the basis of the detailed results obtained in
the previous chapter \ref{cha:tch2HDM} one may expect that some of the
decays of the Higgs bosons
\begin{equation}\label{htcFCNC}
    h\rightarrow t\,\bar{c}\,,\ \  \ \ h\rightarrow \bar{t}\,c\,\ \
    \ \ \ (h=h^0\,,H^0\,,A^0)
\end{equation}
in a general 2HDM can be substantially enhanced and perhaps can be
pushed up to the visible level, particularly for $h^0$ which is
the lightest CP-even spinless state in these
models\,\cite{\Hunter}. This possibility can be of great relevance
on several grounds. On the one hand the severe degree of
suppression of the FCNC Higgs decay in the SM obviously implies
that any experimental sign of Higgs-like FCNC decay (\ref{htcFCNC})
would be instant evidence of physics beyond the SM. On the other
hand, the presence of an isolated top quark in the final state,
unbalanced by any other heavy particle, is an unmistakable
carrier of the FCNC\,  signature. Finally, the study of the
maximum FCNC rates for the top quark modes (\ref{htcFCNC}) within
the 2HDM, which is the simplest non-trivial extension of the SM,
should serve as a fiducial result from which more complicated
extensions of the SM can be referred to. Therefore, we believe
there are founded reasons to perform a thorough study of the FCNC
Higgs decays in minimal extensions of the Higgs sector of the SM
and see whether they can be of any help to discover new physics.

The study is organized as follows. In Section \ref{sec:htc2HDM:expected_branching} we summarize the
2HDM interactions most relevant for our study and estimate the
expected FCNC rates of the Higgs decays in the SM and the general
2HDM. In Section \ref{sec:htc2HDM:numerial_analysis}
 a detailed numerical analysis of the one-loop
calculations of the FCNC decay widths and production rates of
FCNC Higgs events is presented. Finally, in Section
\ref{sec:htc2HDM:conclusions} we discuss
the reach of our results and its phenomenological implications,
and deliver our conclusions.

\section{Expected branching ratios in the SM and the 2HDM}
\label{sec:htc2HDM:expected_branching}

Before presenting the detailed numerical results of our
calculation, it may be instructive to estimate the typical
expected widths and branching ratios ($B$) both for the SM decay
$H^{SM}\rightarrow t\,\bar{c}\ $ and the non-standard decays
(\ref{htcFCNC}) in a general 2HDM. This should be especially useful
in this kind of rare processes, which in the strict context of
the SM are many orders of magnitude out of the accessible range.
Therefore, one expects to be able to grossly reproduce the order
of magnitude from simple physical considerations based on
dimensional analysis, power counting, CKM matrix elements and
dynamical features. By the same token it should be possible to
{guess at the} potential enhancement factors in the 2HDM extension
of the SM. In fact, guided by the previous criteria the FCNC decay
width of the SM Higgs of mass $m_H$ into top quark is expected to
be of order
\begin{equation}
    \Gamma(H^{SM}\rightarrow
    t\,\bar{c})\sim\left(\frac{1}{16\pi^2}\right)^2\,
    |V_{tb}^{\ast}V_{bc}|^{2}\, \alpha_{W}^3\, \,m_{H}\left(
        \lambda_b^{SM}\right) ^{4}
    \sim\left(\frac{|V_{bc}|}{16\pi^2}\right)^2\,\alpha_{W}\ G_F^2
    \,m_{H}\,m_b^4\,,\label{estimateSM1}
\end{equation}
where $G_F$ is Fermi's constant and $\alpha_{W}=g^2/4\pi$, $g$
being the $SU(2)_L$ weak gauge coupling. We have approximated the
loop form factor by just a constant prefactor. Notice the
presence of $\lambda_b^{SM}\sim m_b/M_W$, which is the SM Yukawa
coupling of the bottom quark in units of $g$. The fourth power of
$\lambda_b^{SM}$ in (\ref{estimateSM1}) gives the non-trivial
suppression factor reminiscent of  the GIM mechanism after
summing over flavors. Since we are maximizing our estimation, a
missing function related to kinematics and polarization sums,
$F(m_{t}/m_{H})\sim(1-m_{t}^{2}/m_{H}^{2})^{2}$, has been
approximated to one. To obtain the (maximized!) branching ratio it
suffices to divide the previous result by
$\Gamma(H^{SM}\rightarrow b\,\bar{b})\sim\alpha_W\,\left(
    \lambda_b^{SM}\right) ^{2}\,m_H\sim G_F\,m_H\,m_b^2\ $  to obtain
\begin{equation}
    B(H^{SM}\rightarrow t\,\bar{c})
    \sim\left(\frac{|V_{bc}|}{16\pi^2}\right)^2\,\alpha_{W}\
    G_F\,m_b^2\sim 10^{-13}\,,\label{estimateBRSMhtc}
\end{equation}
with $V_{bc}=0.04$, $m_b=5\,GeV$. In general this $B$ will be
even smaller, specially for higher Higgs boson masses ($m_H>2\,M_W
$) for which the vector boson Higgs decay modes $H^{SM}\rightarrow
W^+\,W^-(Z\,Z)$ can be kinematically available and become
dominant. In this case it is easy to see that
$B(H^{SM}\rightarrow t\,\bar{c})$ will be suppressed by an
additional factor of $m_b^2/m_H^2$, which amounts at the very
least to two additional orders of magnitude suppression, bringing
it to a level of less than $10^{-15}$. Already the optimized
branching ratio (\ref{estimateBRSMhtc}) will remain invisible to all
foreseeable accelerators in the future! To obtain the
corresponding maximized estimation for the 2HDM we use the couplings from
\eqref{Hqq} and the trilinear couplings \ref{tab:trilineals}.

Let us now first assume large $\tan\beta$ and restrict to Type
II models. From the interaction Lagrangians above it is clear
that we may replace $\lambda_b^{SM}\rightarrow
\lambda_b^{SM}\,\tan\beta$ in the previous formulae for the
partial width. Moreover, the leading diagrams in the 2HDM
contain the trilinear Higgs couplings $\lambda_{H^+\,H^-\,h}$.
Therefore, the maximum $B$ associated to the FCNC decays
(\ref{htcFCNC}) in a general 2HDM II should be of
order\footnote{Here we have normalized the $B$ with respect to
the $h\rightarrow b\bar{b}$ channel only, because the gauge boson
modes will be suppressed in the relevant FCNC region, Cf. Section
3.}
\begin{equation}
    B^{II}(h\rightarrow t\,\bar{c})
    \sim\left(\frac{|V_{bc}|}{16\pi^2}\right)^2\,\alpha_{W}\
    G_F\,m_b^2\,\tan^2\beta\ \lambda^2_{H^+\,H^-\,h}\,,
    \label{estimateBR2HDM1}
\end{equation}
where $\lambda_{H^+\,H^-\,h}$ is defined here in units of $g$ and
dimensionless as compared to Table~\ref{tab:trilineals}. Clearly
a big enhancement factor $\tan^2\beta$ appears, but this does not
suffice. Fortunately, the trilinear couplings
$\lambda_{H^+\,H^-\,h}$ for $h=h^0,H^0$ (but not for $h=A^0$)
carry two additional sources of potential enhancement (Cf.
Table~\ref{tab:trilineals}) which are absent in the MSSM case.
Take e.g. $h^0$, then we see that under appropriate conditions
(for example, large $\ta$ and large $\tb$) the trilinear coupling
behaves as $\lambda_{H^+\,H^-\,h^0}\sim
(m_{h^0}^2-m_{A^0}^2)\,\tan\beta/(M_W\,\mHp)$, and in this case
\begin{equation}
    B^{II}(h^0\rightarrow t\,\bar{c})
    \sim\left(\frac{|V_{bc}|}{16\pi^2}\right)^2\,\alpha_{W}\
    G_F\,m_b^2\,\tan^4\beta\,
    \left(\frac{m_{A^0}^2-m_{h^0}^2}{M_W\,\mHp}\right)^2 \,.
    \label{estimateBR2HDM2}
\end{equation}
So finally $B^{II}(h^0\rightarrow t\,\bar{c})$, and of course
$B^{II}(h^0\rightarrow \bar{t}\,{c})$, can be augmented by a huge
factor $\tan^4\beta$ times the square of the relative splitting
among the CP-even Higgs decaying boson mass and the CP-odd Higgs
mass. Since the neutral Higgs bosons do not participate in the
loop form factors (see \ref{fig:1}), it is clear
that various scenarios can be envisaged where these mass
splittings can be relevant. In the next section this behaviour
will be borne out by explicit calculations showing that
$h^0\rightarrow t\,\bar{c}$ can be raised to the visible level in
the case of the Type II model. As for the Type I model the Higgs
trilinear coupling enhancement is the same, but in the charged
Higgs Yukawa coupling all quarks go with a factor $\cot\beta$;
hence when considering the leading terms in the loops that
contribute one sees that in the corresponding expression
(\ref{estimateBR2HDM1}) the term $m_b^2\,\tan^2\beta$ is traded
for $m_t\,m_c\,\cot^2\beta$, which is negligible at high
$\tan\beta$. Both sources of enhancement are needed, and this
feature is only tenable in the 2HDM II.  Of course one could
resort to the range $\tan\beta\ll 1$ for the Type I models, but
this is not theoretically appealing. For example, for
$\tb\lesssim 0.1$  the top quark Yukawa coupling
$g_t=g\,m_t/(\sqrt{2}\,M_W\,\sin\beta)$, which is present in the
interaction Lagrangians above, is pushed into the non-perturbative
region $g_t^2/16\pi^2\gtrsim1$ and then our calculation would not
be justified. And what is worse: for the 2HDM I we would actually
need $\tb\leq {\cal O}(10^{-2})$ to get significant FCNC rates!
In short, we consider that $B^{I}(h\rightarrow
t\,\bar{c}+\,\bar{t}\,c)$ is essentially small (for all $h$), and
that these decays remain always invisible to speak of. Hereafter
we abandon the study of the decays (\ref{htcFCNC}) for the 2HDM I
and restrict ourselves to the general 2HDM II.

\section{Numerical analysis}
\label{sec:htc2HDM:numerial_analysis}

\begin{figure}
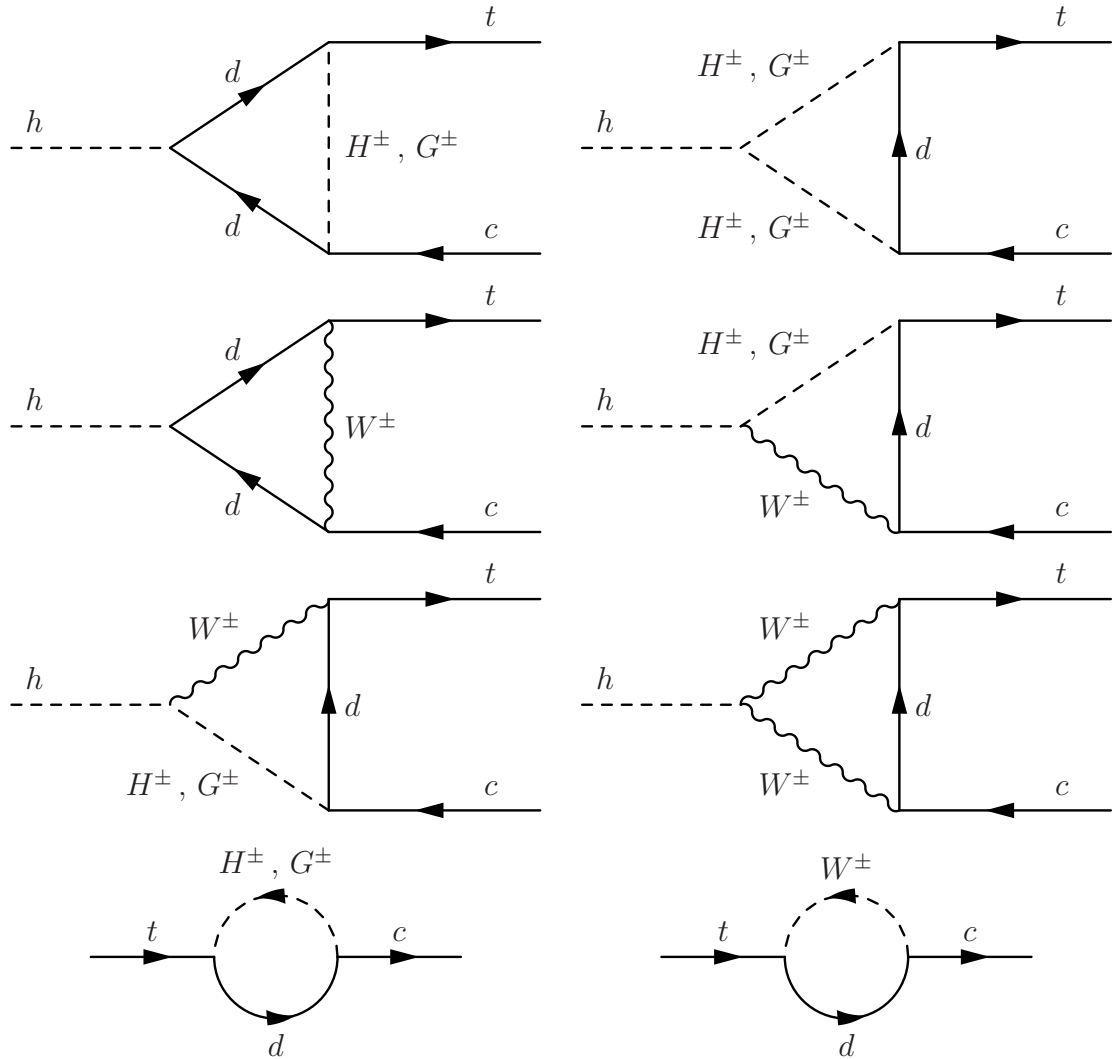

    \begin{center}
        \begin{tabular}{cc}
            \htcvhu & \htcvhdos \\
            \htcvhtres & \htcvhquatre \\
            \htcvhcinc & \htcvhsis \\
            \tcmixHG & \tcmixW \\
        \end{tabular}
    \end{center}
    \caption{One-loop vertex diagrams contributing to the FCNC Higgs decay
      (\ref{htcFCNC}). Shown are the vertices and mixed
      self-energies with all possible contributions from the SM fields and the
      Higgs bosons from the general 2HDM. The Goldstone boson contributions
      are computed in the Feynman gauge.}
    \label{fig:htc_diag}
\end{figure}

\begin{figure}
    \begin{center}
        \begin{tabular}{cc}
            
  \input{feyn_def.tex}
  \fmfframe(0,15)(0,5){\begin{fmfgraph*}(200,100)
        \fmfpen{\mypen}
        \fmfleft{i1,i2}
        \fmfright{o1}
        \gluon{g}{,label.side=left}{i1,v1}
        \gluon{g}{,label.side=left}{i2,v2}
        \higgsz{h}{,label.side=right}{o1,v3}
        \fermion{t\coma b}{}{v1,v2}
        \fermion{t\coma b}{,label.side=left}{v2,v3}
        \fermion{t\coma b}{}{v3,v1}
        \fmfforce{.4w,0h}{v1}
        \fmfforce{.4w,1h}{v2}
        \fmfforce{.7w,0.5h}{v3}
    \end{fmfgraph*}}
&
  \input{feyn_def.tex}
  \fmfframe(0,10)(0,10){\begin{fmfgraph*}(140,100)
        \fmfpen{\mypen}
        \fmfleft{i1,i2}
        \fmfright{o1,o2,o3}
        \gluon{g}{}{v1,i2}
        \fermion{t\coma b}{}{v1,o3}
        \fermion{t\coma b}{,label.side=left,tension=.5}{v2,v1}
        \higgsz{h}{,tension=.5}{v2,o2}
        \fermion{t\coma b}{,label.side=right,tension=.5}{v3,v2}
        \fermion{t\coma b}{,label.side=left}{o1,v3}
        \gluon{g}{}{i1,v3}
    \end{fmfgraph*}}

        \end{tabular}
        \caption{Leading order Feynman diagrams for the Higgs boson
          production at the LHC}
        \label{fig:ggh}
  \end{center}
\end{figure}
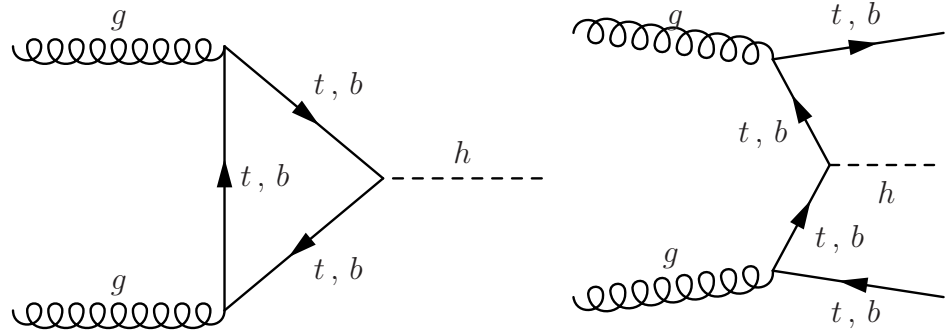

Let us now substantiate the previous claims and provide the
precise numerical results of the full one-loop calculation of
$B^{II}(h\rightarrow t\,\bar{c}+\bar{t}\,{c})$\,\footnote{Here
and throughout we use the notation $B^{II}(h^0\rightarrow
t\,\bar{c}+\bar{t}\,{c})\equiv B^{II}(h^0\rightarrow t\,\bar{c})+
B^{II}(h^0\rightarrow \bar{t}\,{c})$.} as well as of the LHC
production rates of these FCNC events. We refer the reader to previous
chapters for more details. The diagrams for the decays
(\ref{htcFCNC}) are shown in Fig.~\ref{fig:htc_diag}, and the
diagrams for the productions can be seen in Fig.~\ref{fig:ggh}.
In what follows we
present the final results of our
numerical analysis together with a detailed discussion,
interpretation and phenomenological application. We have
performed the calculations with the help of the numeric and
algebraic programs FeynArts, FormCalc and
LoopTools~\cite{\FeynArts}. The calculation must obviously be
finite without renormalization, and indeed the cancellation of UV
divergences in the total amplitudes was verified explicitly.

The input set for our numerical analysis is given by the data row
\begin{equation}\label{row}
    (\mh,\mH,\mA,\mHp,\ta,\tb)
\end{equation}
made out of six independent parameters in the general 2HDM.
Remaining inputs as in \cite{\PDBMM}. In practice there are some
phenomenological restrictions on the data (\ref{row}) which were
already described in the chapter \ref{cha:2hdm},
particularly\,\cite{\OPAL}. Again, a key parameter is $\tb$, with the
restriction \eqref{eq:tbrange}.
In practice, since Type I models are not considered, the effective
range for our calculation will be the high $\tb$ end of
(\ref{eq:tbrange}).
\begin{figure}
    \centering
    \includegraphics[width=0.8\textwidth]{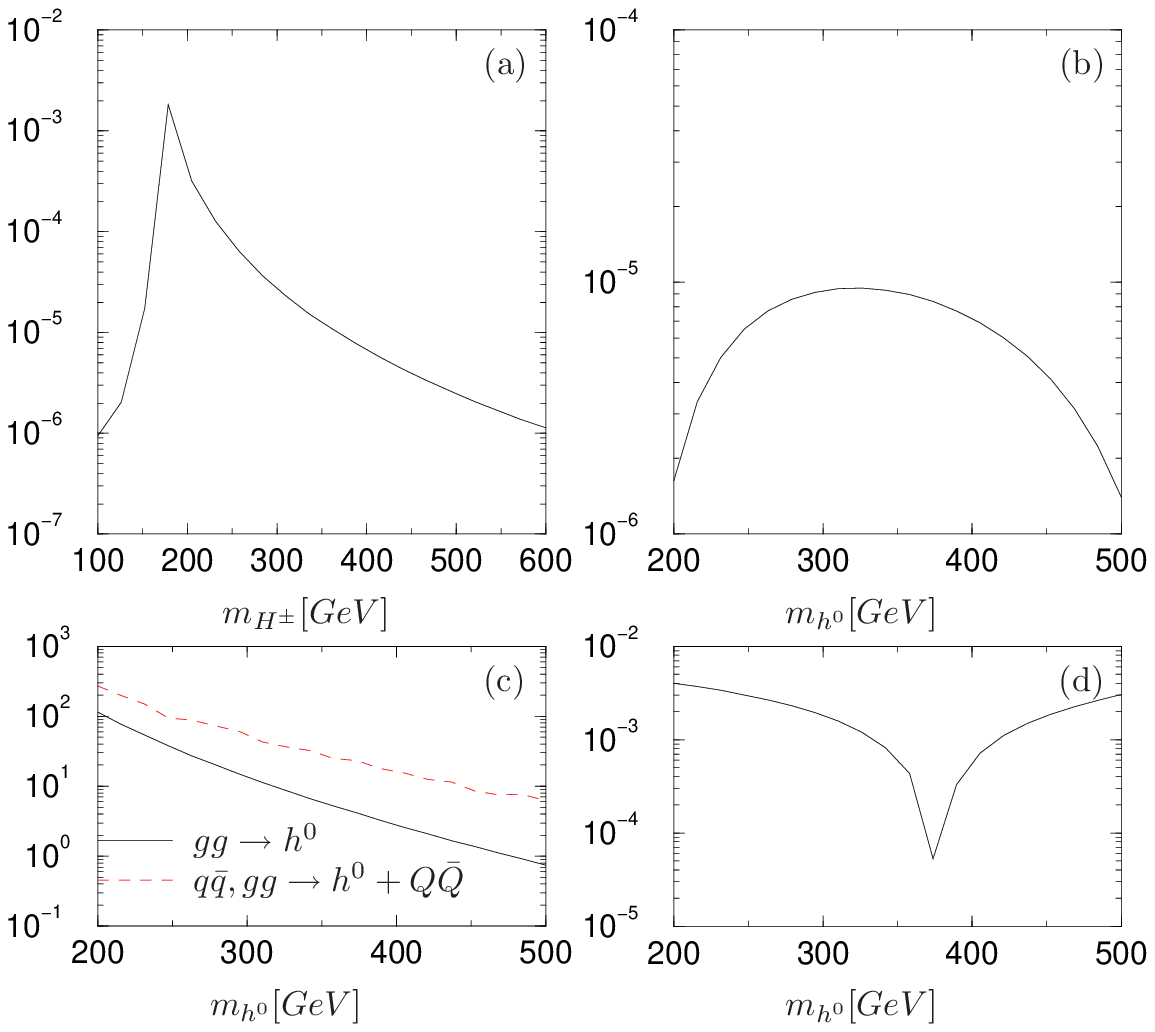}
    \caption{\textbf{(a)}$\, B^{II}(h^0\rightarrow
    t\,\bar{c}+\bar{t}\,c)$
      versus $\mHp$; \textbf{(b)} Idem, versus $\mh$; \textbf{(c)} The
      production cross-section (in $pb$) of $h^0$ at the LHC versus its mass;
      \textbf{(d)} $\delta\rho^{\rm 2HDM}$ versus $\mh$, see the text. In these figures, when a parameter is not
      varied  it is  fixed as in eq.(\ref{inputparam}).}
      \label{fig:4figslh}
\end{figure}
\begin{figure}
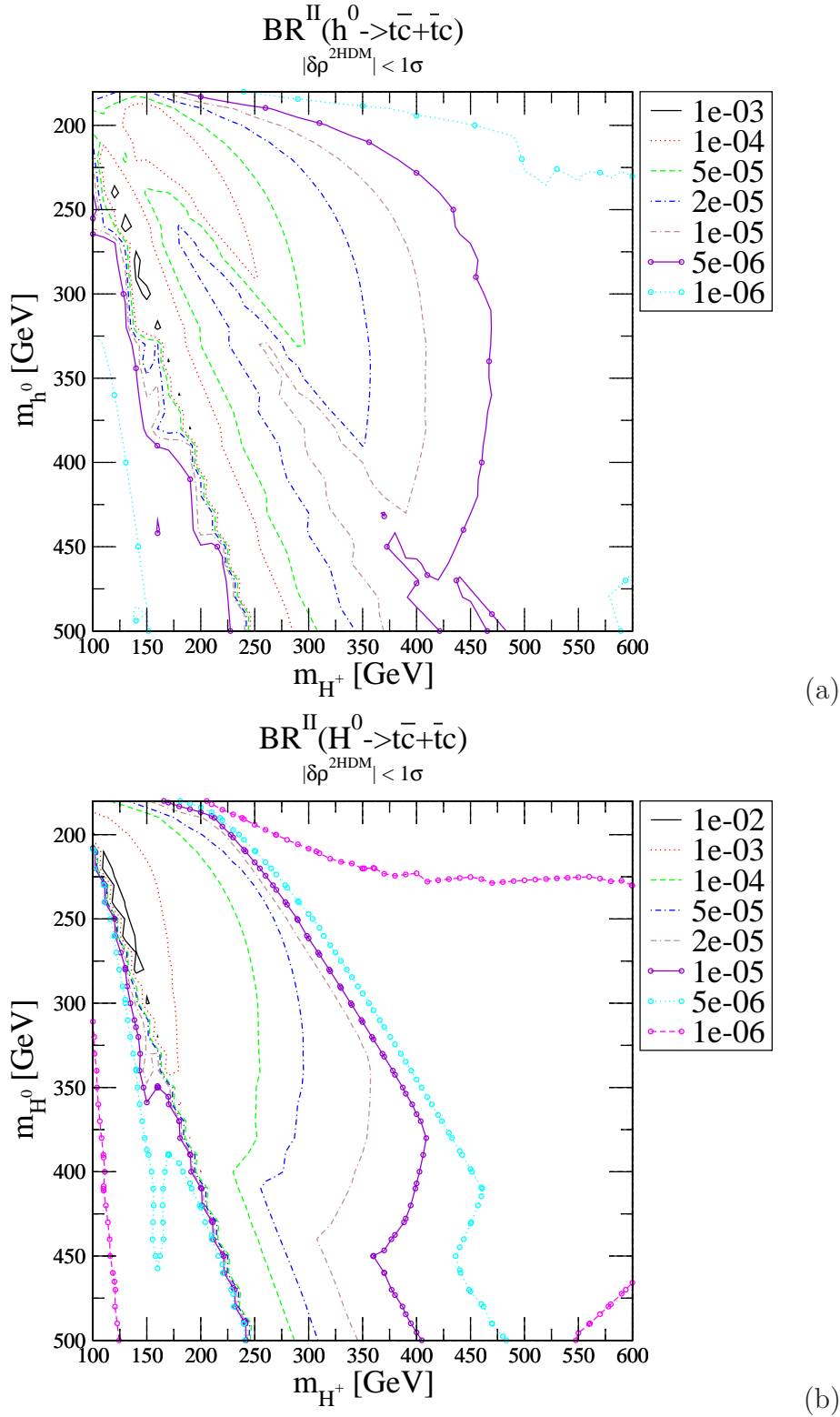

    \begin{tabular}{cc}
        \includegraphics*[width=0.70\textwidth]{br1} & (a) \\
        \includegraphics*[width=0.70\textwidth]{brhh1} & (b) \\
    \end{tabular}
    \caption{Contour lines
    in the $(\mHp,\mh)$-plane for the branching ratios (2HDM II case)  \textbf{(a)}
     $B^{II}(h^0\rightarrow t\,\bar{c}+\bar{t}\,c$) and \textbf{(b)} $B^{II}(H^0\rightarrow
      t\,\bar{c}+\bar{t}\,c$) assuming  $\delta\rho^{\rm 2HDM}$
      at $1\,\sigma$.}
    \label{fig:br1}
\end{figure}
\begin{figure}
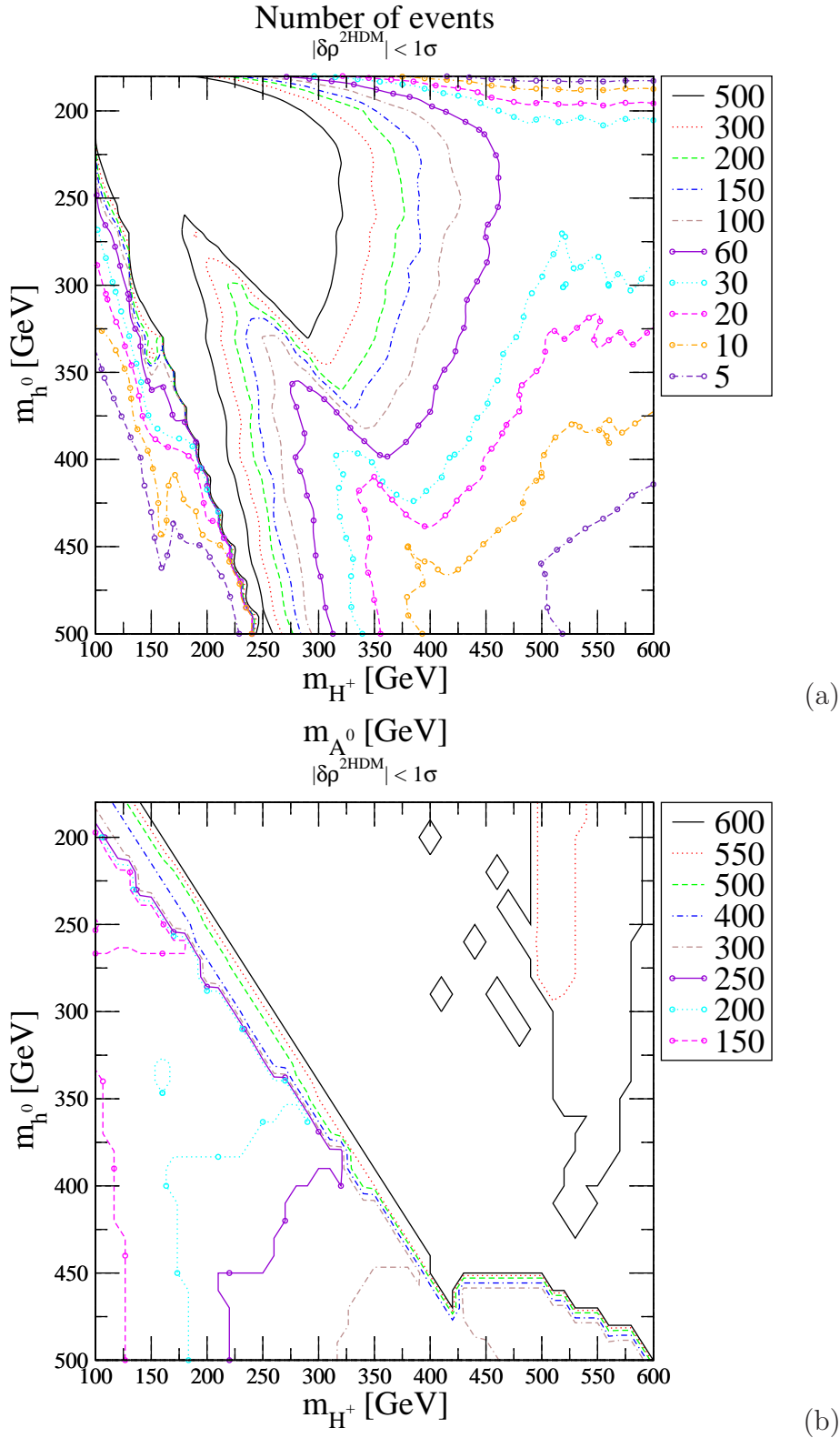

    \begin{tabular}{cc}
        \includegraphics*[width=0.70\textwidth]{number1} & (a) \\
        \includegraphics*[width=0.70\textwidth]{ma1} & (b) \\
    \end{tabular}
    \caption{\textbf{(a)} Contour lines in the $(\mHp,\mh)$-plane for the
      maximum number of light CP-even Higgs FCNC events $h^0\rightarrow
      t\,\bar{c}+\bar{t}\,c$ produced at the LHC for
      $100\,fb^{-1}$ of integrated luminosity within $\delta\rho^{\rm
        2HDM}$ at $1\,\sigma$; \textbf{(b)}
      Contour lines showing the value of $\mA$ that maximizes the number of
      events.}
    \label{fig:numberma1s}
\end{figure}
\begin{figure}
    \begin{tabular}{cc}
        \includegraphics*[width=0.70\textwidth]{number3} & (a) \\
        \includegraphics*[width=0.70\textwidth]{ma3} & (b)
    \end{tabular}
    \caption{As in Fig.~\ref{fig:numberma1s} but within $\delta\rho^{\rm
    2HDM}$  at $3\,\sigma$.}
    \label{fig:numberma3s}
\end{figure}

\begin{figure}
    \begin{tabular}{cc}
        \includegraphics*[width=0.70\textwidth]{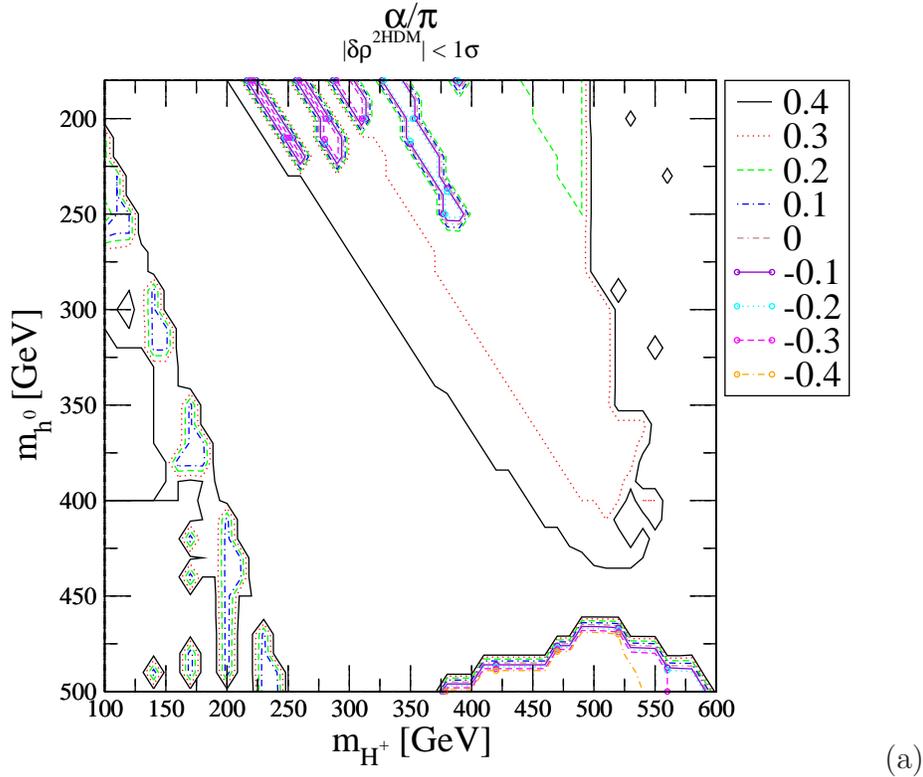} & (a) \\
        \includegraphics*[width=0.70\textwidth]{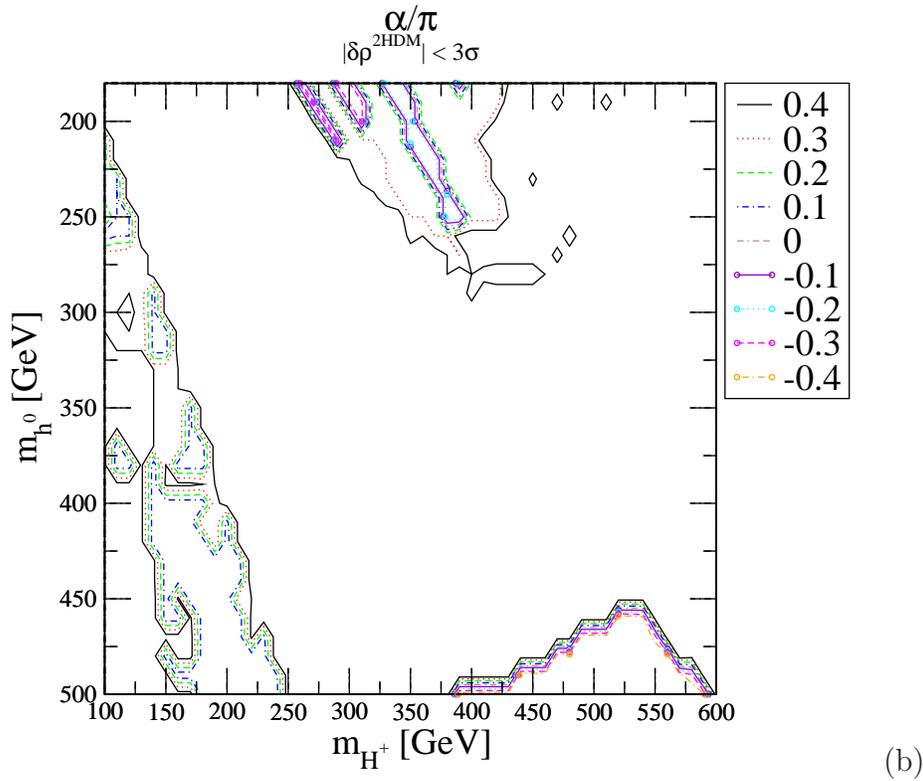} & (b)
    \end{tabular}
    \caption{Contour lines $\alpha/\pi=const.$ ($\alpha$ is the mixing angle in the CP-even sector)
    corresponding to Figs.~\ref{fig:numberma1s}-\ref{fig:numberma3s}
    for $\delta\rho^{\rm 2HDM}$  at
      \textbf{(a)} $1\,\sigma$ and \textbf{(b)} $3\,\sigma$ .}
    \label{fig:a13}
\end{figure}
\begin{figure}
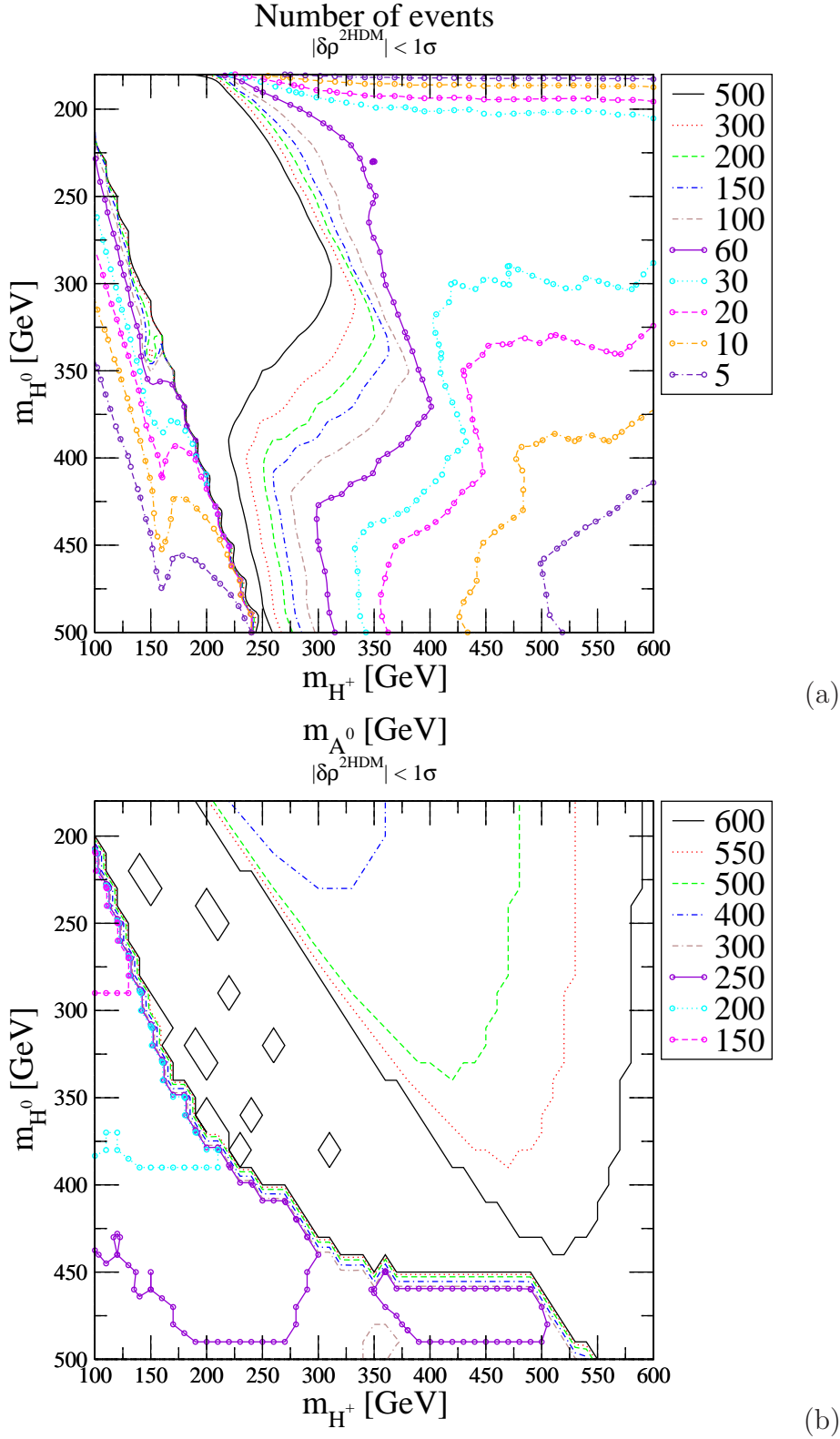

    \begin{tabular}{cc}
        \includegraphics*[width=0.70\textwidth]{numberhh1} & (a) \\
        \includegraphics*[width=0.70\textwidth]{mahh1} & (b)
    \end{tabular}
    \caption{\textbf{(a)} Contour lines in the $(\mHp,\mH)$-plane for the
      maximum number of heavy CP-even Higgs FCNC events $H^0\rightarrow
      t\,\bar{c}+\bar{t}\,c$ (2HDM II case) produced at the LHC for
      $100\,fb^{-1}$ of integrated luminosity within $\delta\rho^{\rm 2HDM}$
        at $1\,\sigma$.}
    \label{fig:numbermahh1s}
\end{figure}
\begin{figure}
    \begin{tabular}{cc}
        \includegraphics*[width=0.70\textwidth]{numberhh3} & (a) \\
        \includegraphics*[width=0.70\textwidth]{mahh3} & (b)
    \end{tabular}
    \caption{As in Fig.~\ref{fig:numbermahh1s} but within $\delta\rho^{\rm 2HDM}$
        at $3\,\sigma$.}
    \label{fig:numbermahh3s}
\end{figure}

\begin{figure}
    \begin{tabular}{cc}
        \includegraphics*[width=0.70\textwidth]{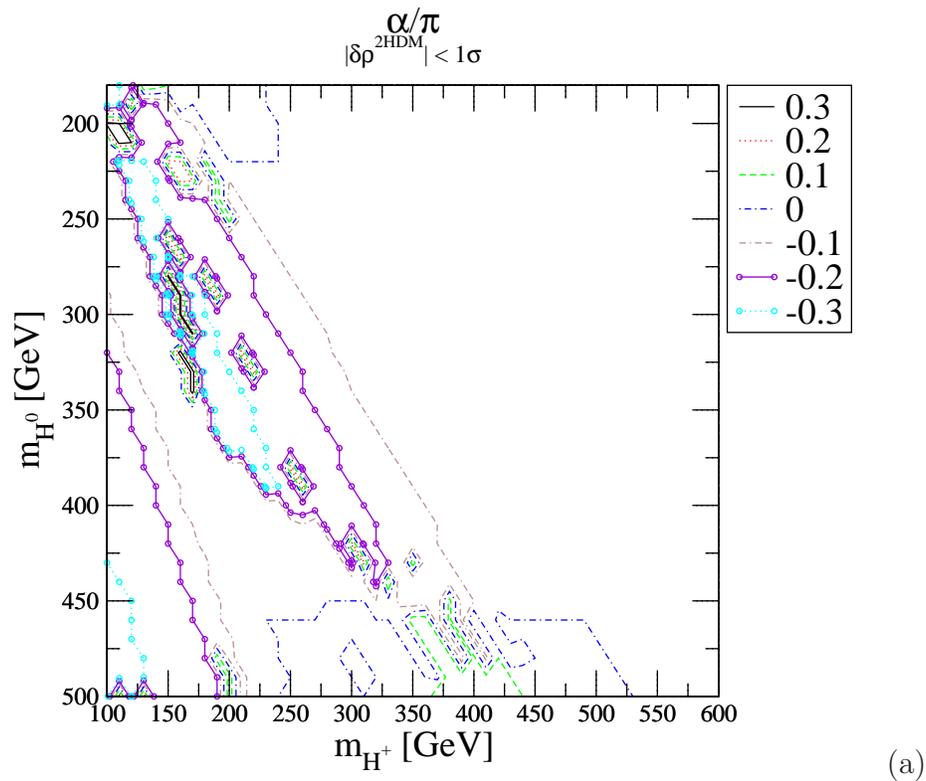} & (a) \\
        \includegraphics*[width=0.70\textwidth]{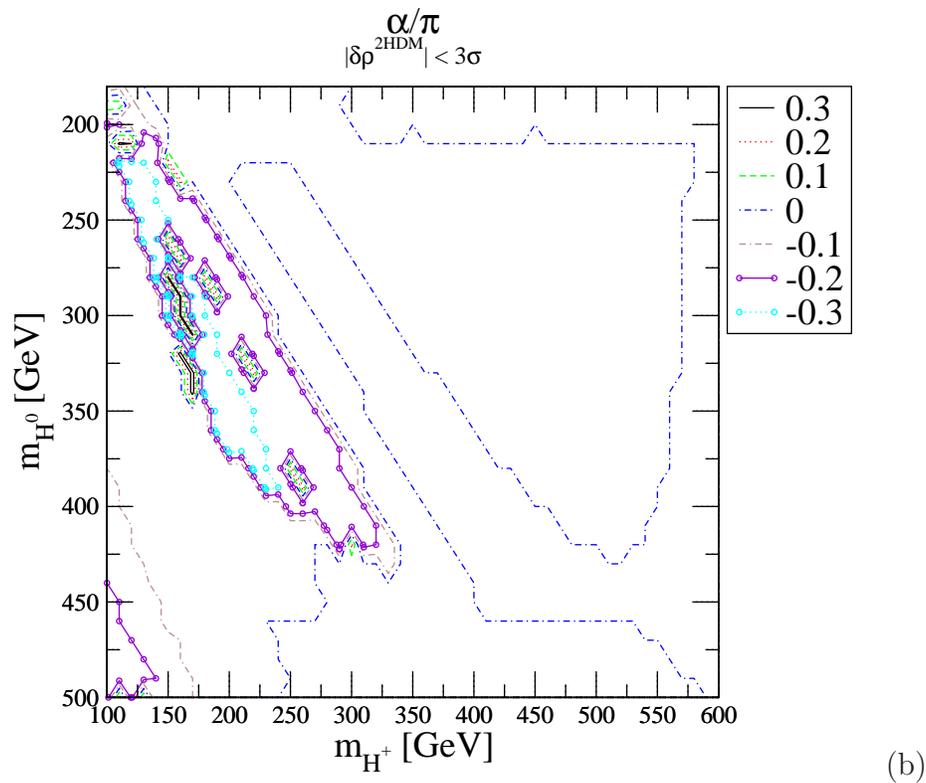} & (b)
    \end{tabular}
    \caption{Contour lines $\alpha/\pi=const.$ as in Fig.\ref{fig:a13}, but for the heavy CP-even Higgs.}
    \label{fig:ahh13}
\end{figure}

With these restrictions in mind we have computed the number of
FCNC Higgs decay events into top quark at the LHC:
\begin{equation}\label{CS}
pp\to\ h +X\rightarrow t\,\bar{c}\,(\bar{t}\,c) +X \ \ \ \
(h=h^0,H^0,A^0)\,.
\end{equation}
The necessary cross-sections to compute the production of neutral
Higgs bosons at this collider, including all known QCD
corrections, have been computed by adapting the codes HIGLU 1.0
and HQQ 1.0\,\cite{Spira} -- originally written for the MSSM
case\,\cite{\Spirat}-- to the general case of the
2HDM\footnote{{We have used the default parton distribution
functions and  renormalization/factorization scales used in these
programs, namely GRV94 with $\mu_R=\mu_F=m_h$  for HIGLU, and
CTEQ4L with $\mu_R=\mu_F=\sqrt{\hat{s}}\equiv\sqrt{(p_h+p_Q
+p_{\bar Q})^2}$ for  HQQ.}}. Folding the cross-sections with
the one-loop branching ratios of the processes (\ref{htcFCNC}) we
have obtained the number of FCNC Higgs decay events at the LHC.
Let us first consider the branching ratios themselves. In
Fig.~\ref{fig:4figslh}a,b we show $B^{II}(h^0\rightarrow
t\,\bar{c}+\bar{t}\,{c})$ for the lightest CP-even state. In
particular, Fig.~\ref{fig:4figslh}a shows $B^{II}(h^0\rightarrow
t\,\bar{c}+\bar{t}\,c)$  versus the charged Higgs mass $\mHp$. In
this figure we fix the values of the parameters in (\ref{row})
which are not varying as follows:
\begin{equation}
\label{inputparam}
\begin{array}{l}
(\mh=350\,GeV,\mH=600\,GeV,\mA=550\,GeV,\mHp=375\,GeV,\\
\ta=30,\tb=60)
\end{array}
\end{equation}
After crossing a local maximum (associated to a pseudo-threshold
of the one-loop vertex function involving the $h^0\,H^+\,H^-$
coupling) the subsequently falling behavior of the $B$ with
$\mHp$ clearly shows that the previously discussed bounds on
$\mHp$ are quite relevant. The branching ratio, however, stays
within $10^{-6}-10^{-5}$ for a wide range of heavy charged Higgs
masses extending up to $\mHp\leq 600\,GeV$ in
Fig.~\ref{fig:4figslh}a. Hence, for $\mHp$ heavy enough to
satisfy the indirect bounds from radiative $B$-meson
decays\,\cite{\Gambino}, the maximum $B$ is still sizeable. In
Fig.~\ref{fig:4figslh}c the production rate of $h^0$ bosons at the
LHC is shown as a function of $\mh$, for fixed parameters
(\ref{row}). The production cross-sections for the subprocesses
\begin{equation}\label{CS2}
gg\to\ h^0 +X\,, \ \ \ \ \ gg,qq\to\ h^0 +Q\bar{Q}\,,
\end{equation}
contributing to (\ref{CS}) in the case of the light CP-even Higgs
$h^0$ are explicitly separated in Fig.~\ref{fig:4figslh}c. The
gluon-gluon fusion process proceeds at one-loop and the
$h^0 Q\bar{Q}$ associated production proceeds at tree-level\,\cite{\Spiratt}.
Similar subprocesses and results apply for $H^0$ and $A^0$
production. At large $\tb$ and the larger the Higgs
boson masses the particular associated production mechanism with
the bottom quark, $Q=b$, i.e. $h^0\ b\bar{b}$, becomes dominant by
far. All other mechanisms for Higgs boson production in Type II
models\,\cite{\Spirat,\Spiratt,\CoarasaJS}, like vector-boson fusion
(which contributes also to $h^0 Q\bar{Q}$ when $Q$ are light quarks),
vector-boson bremsstrahlung ($q\bar{q}\to h\,V$) and associated $t\,\bar{t}$
production, are subdominant at large $\tb$ and can be neglected
for our purposes. Admittedly, some of these mechanisms can be
relevant for Higgs boson production in the case of the Type I
2HDM at low $\tb$, but we have already warned that the
corresponding FCNC branching ratios are never sufficiently high.

The control over $\delta\rho^{\rm 2HDM}$ is displayed in
Fig.~\ref{fig:4figslh}d. Recall that $\delta\rho$ is not sensitive
to the mass splitting between $\mh$ and $\mH$, because of
$CP$-conservation in the gauge boson sector, but it does feel all
the other mass splittings among Higgs bosons, charged and neutral.
A more systematic search of $B$ values in the parameter space is
presented in Figs. \ref{fig:br1}a,b corresponding to
$B^{II}(h^0\rightarrow t\,\bar{c}+\bar{t}\,c)$ and
$B^{II}(H^0\rightarrow t\,\bar{c}+\bar{t}\,c)$ respectively. Here
we have scanned independently on the parameters (\ref{row}) while
holding the $\delta\rho^{\rm 2HDM}$ bound at $1\,\sigma$. The
contour lines in these figures represent the locus of points in
the $(\mHp,\mh)$-plane giving maximized values of the $B$ in the
2HDM II. Let us remark that the highest value of $\tb$ is always
preferred, and therefore all these contour lines correspond to
$\tb=60$.

In practice, to better assess the possibility of detection at the
LHC, one has to study the production rates of the FCNC events.
These are determined by combining the production cross-sections of
neutral 2HDM II Higgs bosons at the LHC and the FCNC branching
ratios. If we just adopt the mild LEP bound $\mHp\gtrsim 80\,GeV$
and let $\mHp$ approach the maximum in Fig. \ref{fig:4figslh}a
then the $B$ can be as large as $10^{-3}$ and the number of FCNC
events can be huge, at the level of ten thousand per
$100\,fb^{-1}$ of integrated luminosity. But of course the region
near the maximum is too special. Moreover, if we switch on the
above mentioned indirect bound from $b\rightarrow
s\,\gamma$\,\cite{\Gambino}, then the typical $B$ is much smaller
(of order $10^{-5}$) and the number of events is reduced
dramatically, at a level of hundred or less for the same
integrated luminosity. On the other hand it may well happen that
there are regions of parameter space where $B\sim 10^{-5}$ (see
Fig. \ref{fig:br1}) but the production cross-section is too small
because the decaying Higgs boson is too heavy. Therefore, it is
the product of the two quantities that matters.

The systematic search of the regions of parameter space with the
maximum number of FCNC events for the light CP-even Higgs is
presented in the form of contour lines in the multiple
Figs.~\ref{fig:numberma1s} and \ref{fig:numberma3s}. For instance, each isoline in
Figs.~\ref{fig:numberma1s}a and~\ref{fig:numberma3s}a corresponds to a fixed number of
produced FCNC events at the LHC while keeping the value of
$\delta\rho^{\rm 2HDM}$ within $1\,\sigma$ or $3\,\sigma$
respectively of its central experimental value. When scanning over
the parameter space (\ref{row}) we have found again that
$\tan\beta$ is preferred at the highest allowed value
($\tan\beta=60$) -- for Type II models. We have also determined
(see Figs.~\ref{fig:numberma1s}\,b and \ref{fig:numberma3s}\,d) the corresponding contour
lines for $\mA$ associated to these events. The $\mA$-lines are
important because the FCNC processes under consideration are
sensitive to the mass splittings between $\mA$ and the
corresponding decaying Higgs boson, see e.g.
eq.(\ref{estimateBR2HDM2}) and Table~\ref{tab:trilineals}. The
combined figures \ref{fig:numberma1s} and \ref{fig:numberma3s} are very useful because
they give a panoramic view of the origin of our results in the
parameter space. To complete the map of the numerical analysis we
provide Fig. \ref{fig:a13} in which we have projected the contour
lines of the CP-even mixing angle $\alpha$ associated to the
previous plots. For a given contour line $\alpha/\pi=const.$, the
set of inner points have a value of $\alpha/\pi$ smaller than the
one defined by the line itself. In particular, the large domains
in Figs. \ref{fig:a13}a,b without contour lines correspond to
$\alpha/\pi>0.4$ and so to relatively large (and positive)
$\tan\alpha$.  There are a few and small neighborhoods where the
FCNC rates for $h^0$ can be sizeable also for small $\ta$.

Knowing that high $\tan\alpha$ is generally preferred by
$h^0\rightarrow t\,\bar{c}+\bar{t}\,c$, and noting from
Figs.~\ref{fig:numberma1s} and \ref{fig:numberma3s} that large mass splittings between $\mh$
and $\mA$ are allowed, we find that the trilinear coupling
$\lambda_{H^+\,H^-\,h^0}$ can take the form
$\lambda_{H^+\,H^-\,h^0}\sim
(m_{h^0}^2-m_{A^0}^2)\,\tan\beta/(M_W\,\mHp)$. Hence it provides a
substantial additional enhancement beyond the $\tan\beta$ factor.
One can check from the approximate formula (\ref{estimateBR2HDM2})
that the maximum FCNC branching ratios $B^{II}(h^0\rightarrow
t\,\bar{c}+\bar{t}\,c)$ can eventually reach the $10^{-5}$ level
even in regions where the charged Higgs boson mass preserve the
stringent indirect bounds from radiative $B$-meson
decays\,\cite{\Gambino}. These expectations are well in agreement
with the exact numerical analysis presented in
Fig.\,\ref{fig:br1}, thus showing that eq.(\ref{estimateBR2HDM2})
provides a reasonable estimate, and therefore a plausible
explanation for the origin of the maximum contributions. As a
matter of fact, we have checked that the single (finite)
  Feynman diagram giving rise to the estimation~(\ref{estimateBR2HDM2})
  -- the one-loop vertex Feynman diagram with a couple of charged Higgs bosons
  and a bottom quark in the loop -- reproduces the full result with an
  accuracy better than 10\% for $\tb\gtrsim 10-20$. At lower $\tb$ values
  large deviations are possible but, as warned before,
  eq.~(\ref{estimateBR2HDM2}) is expected to be valid only at large
  $\tb$. Furthermore, for low values of $\tb\lesssim 20$ the FCNC $B$s are
  too small to be of any phenomenological interest. The
exact numerical analysis is of course based on the full expression
for the branching ratio
\begin{equation}
    B^{II}(h^0\rightarrow t\,\bar{c}+\bar{t}\,c)=\frac{\Gamma
      (h^0\rightarrow t\,\bar{c}+\bar{t}\,c)}{\Gamma (h^0\rightarrow
      b\,\bar{b})+\Gamma (h^0\rightarrow
      t\,\bar{t})+\Gamma(h^0\rightarrow V\,V)+\Gamma(h^0\rightarrow H\,H)}\,, \label{fullBRh}
\end{equation}
where all decay widths in the denominator of this formula have
been computed at the tree-level in the 2HDM II, since this
provides a consistent description of eq.~(\ref{fullBRh})
at leading order. Here we have defined
\begin{equation}\label{totalw1}
\Gamma(h^0\rightarrow V\,V)\equiv \Gamma(h^0\rightarrow
W^+W^-)+\Gamma(H^0\rightarrow Z\,Z)\,,
\end{equation}
\begin{equation}\label{totalw2}
\Gamma(h^0\rightarrow H\,H)\equiv \Gamma(h^0\rightarrow
A^0\,A^0)+\Gamma(h^0\rightarrow H^+\,H^-)\,.
\end{equation}
We disregard the loop induced decay channels, since they have
  branching ratios below the percent level all over the parameter
  space. The $\tau$-lepton decay channel is also neglected, since it is
  suppressed by a factor of ${\cal O}(10^{-2})$ with respect the $b\bar{b}$-channel in the whole 2HDM
  parameter space.
In general the effect of the gauge boson channels $h^0\rightarrow
W^+W^-, ZZ$ in the $B$ (\ref{fullBRh}) is not so important as in
the SM, actually for $\beta=\alpha$ they vanish in the $h^0$ case
because they are proportional to $\sin^2(\beta-\alpha)$. This is
approximately the case for large $\ta$ and large $\tb$, the
dominant FCNC region for $h^0$ decay (Cf. Fig.\ref{fig:a13}a and
\ref{fig:a13}b).  In this region, the mode $h^0\rightarrow
t\,\bar{t}$ is, when kinematically allowed, suppressed:
$B(h^0\rightarrow
t\,\bar{t})\propto\cos^2\alpha/\sin^2\beta\rightarrow 0$ (Cf.
Eq.\,(\ref{Hqq})). On the other hand there are domains in our
plots where the decays $h^0\rightarrow H^+\,H^-$ and
$h^0\rightarrow A^0\,A^0$ are kinematically possible and
non-(dynamically) suppressed. Indeed, this can be checked from
the explicit structure of the trilinear couplings $h^0\,H^+\,H^-$
and $h^0\,A^0\,A^0$ in Table~\ref{tab:trilineals}; in the dominant
region for the decays $h^0\rightarrow t\,\bar{c}+\bar{t}\,{c}$
both of these couplings are $\tb$-enhanced. Nevertheless the
decay $h^0\rightarrow A^0\,A^0\,$ is only possible for
$\mA<\mh/2\,$, and since the optimal FCNC regions demand the
largest possible values of $\mA$, this decay is kinematically
blocked there. On the other hand the mode $h^0\rightarrow
H^+\,H^-$ is of course allowed if we just take the aforementioned
direct limits on the 2HDM Higgs boson masses. But it is never
available if we apply the indirect bound from $b\rightarrow
s\,\gamma$ on the charged Higgs mass mentioned above, unless
$\mh>2\mHp>700\,GeV$, in which case $h^0$ is so heavy that its
production cross-section is too small for FCNC studies to be
further pursued.

The corresponding results for the heavy CP-even Higgs boson are
displayed in Figs. \ref{fig:numbermahh1s}, \ref{fig:numbermahh3s} and \ref{fig:ahh13}. The
exact formula for the $B$ in this case reads
\begin{equation}
    B^{II}(H^0\rightarrow t\,\bar{c}+\bar{t}\,c)=\frac{\Gamma
      (H^0\rightarrow t\,\bar{c}+\bar{t}\,c)}{\Gamma (H^0\rightarrow
      b\,\bar{b})+\Gamma (H^0\rightarrow
      t\,\bar{t})+\Gamma(H^0\rightarrow V\,V)+\Gamma(H^0\rightarrow H\,H)}\,,
      \label{fullBRH}
\end{equation}
where we have defined
\begin{equation}\label{totalw3}
\Gamma(H^0\rightarrow V\,V)\equiv \Gamma(H^0\rightarrow
W^+W^-)+\Gamma(H^0\rightarrow Z\,Z)\,,
\end{equation}
\begin{equation}\label{totalw4}
\Gamma(H^0\rightarrow H\,H)\equiv \Gamma(H^0\rightarrow
h^0\,h^0)+\Gamma(H^0\rightarrow A^0\,A^0)+\Gamma(H^0\rightarrow
H^+\,H^-)\,.
\end{equation}
From the contour lines in Figs.\,\ref{fig:numbermahh1s}a and~\ref{fig:numbermahh3s}a it is
patent that the number of FCNC top quark events stemming from
$H^0$ decays is comparable to the case of the lightest Higgs
boson. However, Fig. \ref{fig:ahh13}a,b clearly reveals that
these events are localized in regions of the parameter space
generally different from the $h^0$ case, namely they prefer
$\ta\simeq 0$. Even so, there are some ``islands'' of events at
large $\ta$. This situation is complementary to the one observed
for $h^0$ in Fig.\,\ref{fig:a13}. However, in both cases
these isolated regions are mainly concentrated in the segment
$\mHp< 350\,GeV$. Therefore, if the bound on $\mHp$ from
$b\rightarrow s\,\gamma$ is strictly preserved, it is difficult to
find regions of parameter space where the two CP-even states of a
general 2HDM II may both undergo a FCNC decay of the type
(\ref{htcFCNC}).

In the dominant regions of the FCNC mode $H^0\rightarrow
t\,\bar{c}$ (where $\ta$ is small and $\tb$ is large), the decay
of $H^0$ into the $t\bar{t}$ final state is suppressed:
$B(H\rightarrow
t\,\bar{t})\propto\sin^2\alpha/\sin^2\beta\rightarrow 0$. In the
same regions the gauge boson channels in (\ref{fullBRH}) are
suppressed too because $\Gamma(H^0\rightarrow W^+W^-, ZZ)\propto
\cos^2(\beta-\alpha)$. In principle the heavy CP-even Higgs boson
$H^0$ also could (as $h^0$) decay into $A^0\,A^0$ and $H^+\,H^-$.
But there is a novelty here with respect to the $h^0$ decays, in
that there could be regions where  $H^0$ could decay into the
final state $h^0\,h^0$. This contingency has been included
explicitly in eq.(\ref{totalw4}). However, in practice, neither
one of these three last channels is relevant in the optimal FCNC
domains of parameter space. First, the decay $H^0\rightarrow
h^0\,h^0$, although it is kinematically possible, is dynamically
suppressed in the main FCNC region for $H^0$. This can be seen
from Table~\ref{tab:trilineals}, where the trilinear coupling
$H^0\,h^0\,h^0$ becomes vanishingly small at large $\tb$ and
small $\ta$. Second, the coupling ${H^0A^0A^0}$ in
Table~\ref{tab:trilineals} is non-suppressed in the present
region, but again the mode $H^0\rightarrow A^0\,A^0$ is
kinematically forbidden in the optimal FCNC domains because the
latter favor large values of the CP-odd mass (see
Figs.\,\ref{fig:numbermahh1s}b and~\ref{fig:numbermahh3s}b). Third, although in these domains
the decay $H^0\rightarrow H^+\,H^-$ is also non-dynamically
suppressed (see the corresponding trilinear coupling in
Table~\ref{tab:trilineals}), it becomes kinematically shifted to
the high mass range $\mH> 700\,GeV$ if we switch on the indirect
bound from $b\rightarrow s\,\gamma$. Obviously, in  this latter
case the $H^0$ production cross section becomes too small and the
FCNC study has no interest. All in all the contributions from
(\ref{totalw1}),(\ref{totalw2}),(\ref{totalw3}) and
(\ref{totalw4}) are irrelevant for $\mh,\mH<700\,GeV$ as their
numerical impact on $B^{II}(h^0,H^0\rightarrow
t\,\bar{c}+\bar{t}\,c)$ is negligible. Our formulae
(\ref{fullBRh}) and (\ref{fullBRH}) do contain all the decay
channels and we have verified explicitly these features.

As remarked before, in general the most favorable regions of
parameter space for the FCNC decays of $h^0$ and $H^0$  do not
overlap much. The trilinear Higgs boson self-couplings in
Table~\ref{tab:trilineals} (also the fermionic ones) are
interchanged when performing the simultaneous substitutions
$\alpha\rightarrow\pi/2-\alpha$ and $\mh\rightarrow\mH$
\, (see last chapter \ref{cha:tch2HDM}). Furthermore, the LHC production rates of the
neutral Higgs bosons fall quite fast with the masses of these
particles, as seen e.g. in Fig.~\ref{fig:4figslh}b for the $h^0$
state. As a consequence that exchange symmetry on the branching
ratios does not go over to the final event rates, so in practice
the number of FCNC events from $H^0$ decays are smaller (for the
same values of the other parameters) as compared to those for
$h^0$; thus $H^0$ requires e.g. lighter charged Higgs masses to
achieve the same number of FCNC events as $h^0$. As for the
CP-odd state $A^0$, we have seen that it plays an important
indirect dynamical role on the other decays through the trilinear
couplings in Table~\ref{tab:trilineals}, but its own FCNC decay
rates never get a sufficient degree of enhancement due to the
absence of the relevant trilinear couplings, so we may discard it
from our analysis.

We notice that this picture is consistent with the decoupling
limit in the 2HDM: for
  $\alpha\to\beta$, the heaviest  CP-even Higgs boson ($H^0$)
  behaves as the SM Higgs boson, whereas $h^0$ decouples from the electroweak gauge bosons
  and may develop enhanced
  couplings to up and down-like quarks, depending on whether $\tan\beta$ is small or
  large respectively;
  in the opposite limit   ($\alpha\to\beta-\pi/2$), it is $h^0$ that behaves as
  $H^{SM}$, while $H^0$
  decouples from gauge bosons and may develop the same enhanced couplings to quarks as
  $h^0$ did in the previous case.
  Indeed these are the situations that we
  find concerning the FCNC decay rates.
We recall that the numerical results presented in our figures
correspond to an integrated luminosity of $100\,fb^{-1}$.
However, the combined ATLAS and CMS detectors might eventually
accumulate a few hundred inverse femtobarn\,\cite{ATLAS,CMS}.
Therefore, hopefully, a few hundred FCNC events (\ref{htcFCNC})
could eventually be collected in the most optimistic scenario.
Actually, the extreme rareness of these events in the SM suggests
that if only a few of them could be clearly disentangled, it
should suffice to claim physics beyond the SM.

\section{Discussion and conclusions}
\label{sec:htc2HDM:conclusions}

Detection strategies at the CERN-LHC collider for the search of
the SM Higgs boson, and also for the three spinless fields of the
MSSM Higgs sector, have been described in great detail in many
places of the
literature\,\cite{\Hunter,ATLAS,CMS,\Gianotti,\CarenaHaber}, but not
so well for the corresponding charged and neutral Higgs bosons of
the general 2HDM. The result is that the discovery of the SM
Higgs boson is guaranteed at the LHC in the whole presumed range
$100\,GeV\lesssim m_H\lesssim 1\,TeV$. However, the discovery
channels are different in each kinematical region and sometimes
the most obvious ones are rendered useless. For example, due to
the huge irreducible QCD background from $b\,\bar{b}$ dijets, the
decay mode $H^{SM}\rightarrow b\,\bar{b}$ is difficult and one
has to complement the search with many other channels,
particularly
$H^{SM}\rightarrow\gamma\,\gamma$\,\,\cite{ATLAS,CMS}. We have
shown in this work that there are scenarios in the 2HDM parameter
space where alternative decays, like the FCNC modes
$h^0\rightarrow t\,\bar{c}+\bar{t}\,c$ and $H^0\rightarrow
t\,\bar{c}+\bar{t}\,c$, can also be useful. For instance, in the
$h^0$ case, this situation occurs when $\tan\beta$ and
$\tan\alpha$ are both large and the CP-odd state is much heavier
than the CP-even ones. The potential enhancement is then
spectacular and it may reach up to ten billion times the SM value
$B(H^{SM}\rightarrow t\,\bar{c})\sim 10^{-15}$, thereby bringing
the maximum value of the FCNC branching ratio $B(h^{0}\rightarrow
t\,\bar{c})$ to the level of $\sim 10^{-5}$. As a matter of fact,
the enhancement would be much larger were it not because we
eventually apply the severe (indirect) lower bound  on the
charged Higgs mass from $b\rightarrow s\,\gamma$\,\cite{\Gambino}.
Although these decays have maximal ratios below
$B(h\rightarrow\gamma\gamma)\sim 10^{-3}$, they should be
essentially free of QCD background \footnote{Misidentification of
$b$-quarks as
 $c$-quarks in $tb$ production might be a source of background to
 our FCNC events. However, to rate the actual impact of that
 misidentification one would need a dedicated simulation of the
 signal versus background, which is beyond the scope of this
 work.}.

While in the MSSM almost the full $(m_{A^0},\tan\beta)$-parameter
space is covered, with better efficiency at high $\tan\beta$
though, we should insist that within the general 2HDM the tagging
strategies are not so well studied and one would like to have
further information to disentangle the MSSM scenarios from the
2HDM ones. Here again the study of the FCNC Higgs decays can play
a role. Of course the statistics for the FCNC Higgs decays is poor
due to the weakness of the couplings and the large masses of the
Higgs bosons to be produced. However, in the favorable regions,
which are generally characterized by large values of $\tan\beta$
and of $\tan\alpha$, one may collect a few hundred events of the
type (\ref{htcFCNC})-- mainly from $h^0$ -- in the high luminosity
phase of the LHC.  As we have said, this is basically due to the
enormous enhancement that may undergo the FCNC decay rates, but
also because in the same regions of parameter space where the
$B$s are enhanced, also the LHC production rates of the Higgs
bosons can be significantly larger (one order of magnitude) in
the 2HDM II as compared to the SM.

Interestingly enough, in many cases one can easily distinguish
whether the enhanced FCNC events (\ref{htcFCNC}) stem from the
dynamics of a general, unrestricted, 2HDM model, or rather from
some supersymmetric mechanisms within the MSSM. This is already
obvious from the fact that the ranges of neutral and charged
Higgs boson masses in the 2HDM case can be totally incompatible
with the corresponding ones in the MSSM. But there are many other
ways to discriminate these rare events. For instance, in the 2HDM
case the CP-odd modes $A^0\rightarrow t\,\bar{c}+\bar{t}\,c$ are
completely hopeless whereas in the MSSM they can be
enhanced\, see chapter \ref{cha:hbsSUSY_prod},\cite{\GuaschNPo,\Curiel, \Demir}. Using this
information in combination with the masses of potentially
detected Higgs bosons could be extremely useful to pinpoint the
supersymmetric or non-supersymmetric nature of them. We may
describe a few specific strategies. As it was first shown in
Ref.\,\cite{\GuaschNPo}, the leading SUSY-FCNC effects associated
to the $h\,t\,c$ vertices ($h=h^0\,,H^0\,,A^0$) come from the
FCNC gluino interactions which are induced by potentially large
misalignments of the quark and squark mass
matrices\,\cite{\Duncan}. These effects are not particularly
sensitive to $\tan\beta$ and they can be very sizeable for both
high and moderately low values of this parameter. This sole fact
can be another distinguishing feature between FCNC events
(\ref{htcFCNC}) of MSSM or 2HDM origin. If, for example, a few of
these events were observed and at the same time the best MSSM
fits to the electroweak precision data would favor moderate
values of $\tan\beta$, say in the range $10-20$, then it is clear
that those events could originate in the FCNC gluino interactions
but in no way within the context of the general 2HDM. In this
respect it should be mentioned that the  FCNC gluino couplings
became more restricted from the low-energy meson data
\,\cite{\Besmer}, and will presumably become further restricted in
the near future. The reason being that the same couplings are
related, via $SU(2)$ gauge invariance and CKM rotation, to those
affecting the down-like quark sector, which will most likely
become constrained by the increasingly more precise low-energy
meson physics\,\cite{\Besmer,\Pokorsky}. In that circumstance the
only source of FCNC Higgs decays in the MSSM will stem purely
from the electroweak interactions within the super-CKM basis.
Then, in the absence of these SUSY-QCD FCNC effects, we could
judiciously conclude from the work of Ref.\cite{\GuaschNPo} -- in
which both the SUSY-QCD \textit{and} the SUSY electroweak
contributions were computed for the $h\,t\,c$ vertices -- that
the FCNC rates in the MSSM should diminish dramatically (two to
three orders of magnitude). In such case we can imagine the
following ``provocative'' scenario. Suppose that the LHC finds a
light neutral Higgs boson of mass $\lesssim 140\,GeV$
(suggestively enough, in a mass range near the MSSM upper bound
for $\mh$!) and subsequently, or about simultaneously, a charged
Higgs boson and another neutral Higgs boson both with masses
around $400\,GeV$ or more. At this point one could naively
suspect that a MSSM picture out these findings is getting somehow
confirmed. If, however, later on a few FCNC events (\ref{htcFCNC})
are reported and potentially ascribed to the previously
discovered heavy neutral Higgs boson (presumably $H^0$), then
the overall situation could not correspond at all to the MSSM,
while it could be perfectly compatible with the 2HDM II.
Alternatively, suppose that the FCNC gluino couplings were not
yet sufficiently restricted, but (still following the remaining
hypotheses of the previous example) a third neutral Higgs boson
(presumably $A^0$) is found, also accompanied with a few FCNC
events. Then this situation would be incompatible with the 2HDM
II, and in actual fact it would put forward strong (indirect)
evidence of the MSSM!!\, \

We should also mention that there are other FCNC Higgs decay
modes, as for example $h\rightarrow b\,\bar{s}+\bar{b}\,s$, which
could be, in principle, competitive with the top quark modes
(\ref{htcFCNC}). In some cases these bottom modes can be highly
enhanced in the MSSM case\,\cite{\Curiel,\Demir}.  Actually, a more
complete assessment of the FCNC bottom modes in the MSSM case is studied in
the chapter \ref{cha:hbsSUSY_br} --
namely one which takes also into account the supersymmetric
contributions to the highly restrictive radiative B-meson decays
-- shows that they are eventually rendered at a similar level of
the top modes under study in most of the parameter
space.

To summarize, the FCNC decays of the Higgs bosons into top quark
final states can be a helpful complementary strategy to search for
signals of physics beyond the SM in the LHC. Our
comprehensive numerical analysis shows that the FCNC studies are
feasible for CP-even Higgs masses up to about $500\,GeV$. While
the statistics of these FCNC decays is of course poor, the
advantage is that a few tagged and well discriminated events of
this sort could not be attributed by any means to the SM, and
therefore should call for various kinds of new physics. In this
work we have shown that a general 2HDM II is potentially
competitive to be ultimately responsible for these FCNC decays,
if they are ever found, and we have exemplified how to
discriminate this possibility from the more restricted one
associated to the MSSM.


%% file: hbsSUSY_br/hbsSUSY_br.tex
\chapter{Higgs Boson FCNC Decays into Bottom Quarks
  in the MSSM}
\label{cha:hbsSUSY_br}
\section{Introduction}
\label{sect:introduction}

The most general MSSM includes tree-level FCNCs among the extra
predicted particles, which induce one-loop FCNC interactions among
the SM particles, as discussed in section \ref{sec:mssm:fcnc}.

Concerning the FCNC interactions of Higgs bosons with third
generation quarks, it was demonstrated long ago~\cite{Guasch:1999jp}
that the leading term corresponds to a \textit{single particle
insertion approximation}, which produces a flavor {change} in the
internal squark loop propagator, since in this case the chirality
change can already take place at the squark-squark-Higgs boson
interaction vertex. Adding this to the fact that the Higgs bosons
({in contrast to gauge bosons}) have a privileged coupling
to third generation quarks, one might expect that the FCNC
interactions of the type quark-quark-Higgs bosons in the MSSM
{become highly strengthened with respect to the SM prediction}.
This was already proven in the rare decay channels $\Gamma(t\to
ch)$~\cite{Guasch:1999jp} ($h$ being any of the neutral Higgs bosons
of the MSSM $h\equiv h^0,H^0,A^0$), where the maximum rate of the
SUSY-QCD induced branching ratio was found to be $B(t\to
ch)\simeq 10^{-5}$, eight orders of magnitude above the SM
expectations $B(t\to cH^{SM})\simeq 10^{-13}$. Similar
enhancement factors have been found in the top-quark-Higgs boson
interactions in other extensions of the SM, as seen in chapters
\ref{cha:tch2HDM}-\ref{cha:hbsSUSY_prod}.

\begin{figure}[tb]
    \begin{center}
        \begin{tabular}{cc}
            
      \input{feyn_def.tex}
      \fmfframe(0,0)(15,0){\begin{fmfgraph*}(200,80)
            \fmfpen{\mypen}
            \fmfleft{i1}
            \fmfright{o2,o1}
            \fmf{dashes,label.side=left,label={\hspace*{-1.5cm}$h$}}{i1,v1}
            \fmf{fermion,label.side=left,label={\hspace*{1.5cm}$b$}}{v2,o1}
            \fmf{fermion,label.side=right,label={\hspace*{1.5cm}$s$}}{o2,v3}
            \scalar{\sd_\alpha}{}{v1,v2}
            \gluino{\sg}{,label.side=right}{v3,v2}
            \scalar{\sd_\beta}{,label.side=right}{v3,v1}
            \fmfforce{.6w,1h}{v2}
            \fmfforce{.6w,0}{v3}
            \fmfforce{.3w,.5h}{v1}
        \end{fmfgraph*}}
&
    \input{feyn_def.tex}
    \autoenermixm{\fermion{b}}{\fermion{s}}{\scalar{\sd_\alpha}}{\gluino{\sg}}

        \end{tabular}
        \caption{One-loop \susy-\QCD\ vertex diagram contributing to the decay
          $\hbs$ and diagrams contributing to mixed $b-s$ self-energy.
          $\sd_{\{\alpha,\beta\}}$ represent mass-eigenstate down type squarks
          of any generation\label{fig:hbs}.}
    \end{center}
\end{figure}
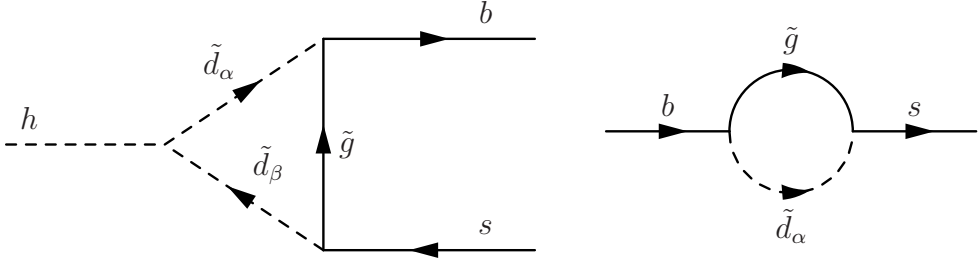

{From the experience of the previous calculations
with the top quark, we expect similar enhancements in the FCNC
interactions of the MSSM Higgs bosons with the bottom quark.
Indeed, the purpose of this work is to quantify, in a reliable
way, the MSSM expectations on the FCNC Higgs boson decay modes}
\begin{equation}\label{hbsFCNC}
h\rightarrow {b}\,\bar{s}\,, \ \ \ h\rightarrow \bar{b}\,s\ \ \ \
\ \ (h=h^0,H^0,A^0)\,.
\end{equation}
The Feynman diagrams for these decays are depicted in Fig.~\ref{fig:hbs}.
There are other FCNC decay modes involving light quarks. However,
only these bottom quark channels are relevant, as the remaining
FCNC decays into light quarks are negligible in the MSSM.
Moreover, the FCNC decays of Higgs bosons into bottom quarks are
specially interesting as they can provide an invaluable tool to
discriminate among different extended Higgs boson scenarios in
the difficult LHC range $90<m_h<130\GeV$~\cite{\LHC}.

{In this work we present what we believe is the first realistic
estimate of the SUSY-QCD  contributions to the FCNC branching
ratios of the MSSM Higgs bosons into bottom quark. Specifically,
we compute}
\begin{equation}
\Bhbs=\frac{\Gamma(h\to q\,q')}{\Gamma(h\to X)}\equiv
\frac{\Gamma(h\rightarrow {b}\,\bar{s})+\Gamma(h\rightarrow
\bar{b}\,s)}{\sum_i \Gamma(h\to X_i)} \label{eq:hbs-def}
\end{equation}
for the three Higgs bosons of the MSSM, $h=h^0,H^0,A^0$, where
$\Gamma(h\to X)$ is the -- consistently computed -- total width
in each case. The maximization process of the above branching
ratios in the MSSM parameter space is performed on the basis of a
simultaneous analysis of the relevant partial decay widths and of
the branching ratio of the low-energy FCNC process $b\to
s\,\gamma$, whose value is severely restricted by
experiment~\cite{\bsgexp}. It turns out that the maximum FCNC
rates that we find disagree quite significantly with Ref.~\cite{\Madrid}
as they do not impose the restriction of $\Bbsg$.\footnote{See also
Ref.~\cite{Demir:2003bv} for a
  combined analysis of flavor-violating and CP-violating MSSM couplings.}

{The structure of the study is as follows.
In Section \ref{sect:partialwidths} we estimate the expected
branching ratios and describe the structure of
Eq.\,(\ref{eq:hbs-def}) in the MSSM in more detail; in Section
\ref{sect:numerical} we present the numerical analysis, and in
Section \ref{sect:conclusions} we deliver our conclusions.}

\section{Partial widths and branching ratios}
\label{sect:partialwidths}

As seen in eqs.~(\ref{estimateBRSMhbs}) and (\ref{estimateBRSMhbs2}) the
branching ratios into bottom quark are much larger than the Higgs
boson FCNC branching ratio into top quark in the
SM\, as seen in chapter \ref{cha:htc2HDM}. However, even in the case
(\ref{estimateBRSMhbs}) it is still too small to have a chance for
detection in the LHC. It is clear that unless new physics comes to
play the process $H^{SM}\rightarrow b\,\bar{s}$ (and of course
$H^{SM}\rightarrow \bar{b}\,{s}$) will remain virtually invisible.
Nonetheless the result (\ref{estimateBRSMhbs}) is not too far from
being potentially detectable, and one might hope that it should
not be too difficult for the new physics to boost it up to the
observable level.

{Consider how to estimate the potentially
augmented rates for the MSSM processes (\ref{hbsFCNC}), if only
within a similarly crude approximation as above. Because of the
strong FCNC gluino couplings mentioned in Section
\ref{sect:introduction} and the $\tb$-enhancement inherent to the
MSSM Yukawa couplings (see Ref.\,\cite{Guasch:1999jp} for details),
we may expect several orders of magnitude increase of the
branching ratios (\ref{eq:hbs-def}) as compared to the previous SM
result. A naive approach might however go too far. For instance,
one could look at the general structure of the couplings and
venture an enhancement factor typically of order
$(\alpha_s/\alpha_W)^2\,\tan^2\beta\ |\delta_{23}/V_{ts}|^2$, which
for $\delta_{23}\lesssim 1$ and $\tb>30$ could easily rocket the
SM result some $5-6$ orders of magnitude higher, bringing perhaps
one of the MSSM rates (\ref{eq:hbs-def}) to the ``scandalous''
level of $10\%$ or more. But of course only a more elaborated
calculation, assisted by a judicious consideration of the various
experimental restrictions, can provide a reliable result. As we
shall see, a thorough analysis generally disproves the latter
overestimate.}

The detailed computation of the SUSY-QCD one-loop partial decay
widths $\Ghbs$ in (\ref{eq:hbs-def}) within the MSSM follows
closely that of $\Gamma(t\to ch)$ (see Ref.~\cite{Guasch:1999jp}).
The rather cumbersome analytical expressions will not be listed here
as they are an straightforward adaptation of those presented in
the aforementioned references. However, there are a few
subtleties that need to be pointed out.
{One of them is related to the calculation of the total
widths $\Gamma(h\to X)$ for the three Higgs bosons
$h=h^0,H^0,A^0$ in the MSSM}.
{As long as $\Ghbs$ in the numerator of Eq.~(\ref{eq:hbs-def})
is computed at leading order, the denominator has to be computed
also} at leading order, otherwise an artificial enhancement of
$\Bhbs$ can be generated. For example, including the
next-to-leading (NLO) order QCD corrections to $\Gamma(h\to
b\bar{b})$ reduces the decay width by a significant
amount~\cite{\hbbQCD}. {Then, to be consistent,} the NLO
(two-loop) contributions to $\Ghbs$ should also be included.
{Similarly}, the one-loop SUSY-QCD corrections to $\Gamma(h\to
b\bar{b})$ can be very large and negative~\cite{Coarasa:1996yg},
which would enhance $\Bhbs$. At the same time these corrections
also contribute to $\Ghbs$, such that contributions to the
numerator and denominator of Eq.~(\ref{eq:hbs-def}) compensate
(at least partially) each other. Therefore the same order of
perturbation theory must be used in both partial decay widths
entering the observable $\Bhbs$ to obtain a consistent result.
{By the same token, using running masses in the numerator
of (\ref{eq:hbs-def}) is mandatory, if they are used in the
denominator.}
{Last, but not least, consistency with the experimental
bounds on related observables should also be taken into account.
In this respect an essential role is played by the constraints on
the FCNC couplings from the measured value of $\Bbsg$. They must
be included in this kind of analysis, if we aim at a realistic
estimate of the maximal rates expected for the FCNC processes
(\ref{hbsFCNC}) in the MSSM. In our calculation we have used the
full one-loop MSSM contributions to $\Bbsg$ as given
in~\cite{\Urban}}\,\footnote{Ref.\cite{\Urban} contains a partial two-loop
  computation of $\Bbsg$, but only the one-loop contributions have been
used for the present work.}.

Let us now summarize the {conditions under which we have
performed the computation and the approximations and assumptions
made in the present analysis}:
\begin{itemize}
\item We include the full one-loop SUSY-QCD contributions to the partial
  decay widths $\Ghbs$ in (\ref{eq:hbs-def}).
\item We assume that FCNC mixing terms appear only in the
  {LH-chiral} sector of the squark mixing matrix. This is
  the most natural assumption, and, moreover, it was proven in
  Ref.~\cite{Guasch:1999jp} that the presence of FCNC terms in the
  {RH-chiral sector} enhances the partial widths by a factor two
  at most -- not an order of magnitude.
\item The Higgs bosons total decay widths $\Gamma(h\to X)$ {are} computed
  at leading order, including all the relevant channels: $\Gamma(h\to
  f\bar{f},ZZ,W^+W^-,gg)$. The off-shell decays
  $\Gamma(h\to ZZ^*,W^{\pm}W^{\mp*})$ have also been
  included. {The one-loop decay rate $\Gamma(h\to gg)$
  has been taken from~\cite{Spira:1995rr} {and the off-shell decay
  partial widths have been computed explicitly and found perfect agreement with the old
  literature on the subject\,\cite{Keung:1984hn}.}
  We have verified that some of the aforementioned higher order decays
   are essential to consistently compute the
  total decay width of $\Gamma(h^0\to X)$ in certain regions of the parameter
  space where the maximization procedure probes domains in which some
(usually leading) two-body processes become {greatly} diminished.
  We have checked that our
  implementation of
  the various Higgs boson decay rates is consistent with the results
  of \texttt{HDECAY}~\cite{\hdecay}. However, care must be
  exercised if using the full-fledged result from \texttt{HDECAY}.
  For example, it would be inconsistent,
  and numerically significant, to compute the
  total widths $\Gamma(h\rightarrow X)$ with this program and at the same time
  to compute the SUSY-QCD one-loop partial widths $\Ghbs$ without including the leading
  conventional QCD effects through e.g. the running quark masses.}

\item The Higgs sector parameters (masses and CP-even mixing angle
  $\alpha$) {have} been treated using the leading $\mt$ and $\mb\tb$
  approximation  to the one-loop result~\cite{\Dabels}. For comparison, we also
  perform the analysis using the tree-level approximation.
\item We include the {constraints} on the {MSSM parameter space} from
  $\Bbsg$. {We adopt $\Bbsg=(2.1-4.5)\times 10^{-4}$ as the experimentally allowed
  range within three standard deviations}~\cite{Hagiwara:2002fs}. {Only the
SUSY-QCD contributions
  induced from tree-level FCNCs are considered in the present work.}

\end{itemize}
Running quark masses ($m_q(Q)$) and strong coupling constants
($\alpha_s(Q)$) are used throughout. More details are given
below, as necessary.

\section{Full one-loop SUSY-QCD calculation: Numerical analysis}
\label{sect:numerical}

Given the setup described in Section \ref{sect:partialwidths}, we
have performed a {systematic scan} of the MSSM parameter
space with the following restrictions:
\begin{equation}
\begin{array}{rcl}
\delta_{23}&<&10^{-0.09} \simeq 0.81  \\
A_b&=&-1500 \cdots 1500 \GeV\\
\mu&=&-1000 \cdots 1000 \GeV\\
m_{\squark}&=&150 \cdots 1000 \GeV
\end{array}
\label{eq:scan-parameters-br}
\end{equation}
and the following fixed parameters:
\begin{equation}
\begin{array}{rcl}
\tan\beta&=&50 \\
m_{\sbottom_L}&=&m_{\sbottom_R}=m_{\stopp_R}=\mg=m_{\squark}\\
A_t&=&-300 \GeV\,\,.
\end{array}
\label{eq:scan-fixed}
\end{equation}
Here $m_{\sbottom_{L,R}}$ are the  left-{chiral} and
right-chiral bottom-squark soft-SUSY-breaking mass parameters, and
$m_{\squark}$ is a common mass for the strange- and down-squark
left- and right-chiral  soft SUSY-breaking mass parameters.
{Following the same notation as in~\cite{Guasch:1999jp}, the
parameter $\delta_{23}$  represents the mixing between the second
and third generation squarks. Let us recall its definition:}
\begin{equation}\label{delta23}
\delta_{23}\equiv \frac{m^2_{\sbottom_L
\sstrange_L}}{m_{\sbottom_L} m_{\sstrange_L}}\,,
\end{equation}
$m^2_{\sbottom_L \sstrange_L}$ {being} the non-diagonal
term in the squark mass matrix squared mixing the second and third
generation left-chiral squarks. {The parameter
$\delta_{23}$ is a fundamental parameter in our analysis as it
determines the strength of the tree-level FCNC interactions
induced by the supersymmetric strong interactions, which are then
transferred to the loop diagrams of the Higgs boson FCNC decays
(\ref{hbsFCNC}).}

\begin{figure}
\begin{tabular}{cc}
\resizebox{!}{6cm}{\includegraphics{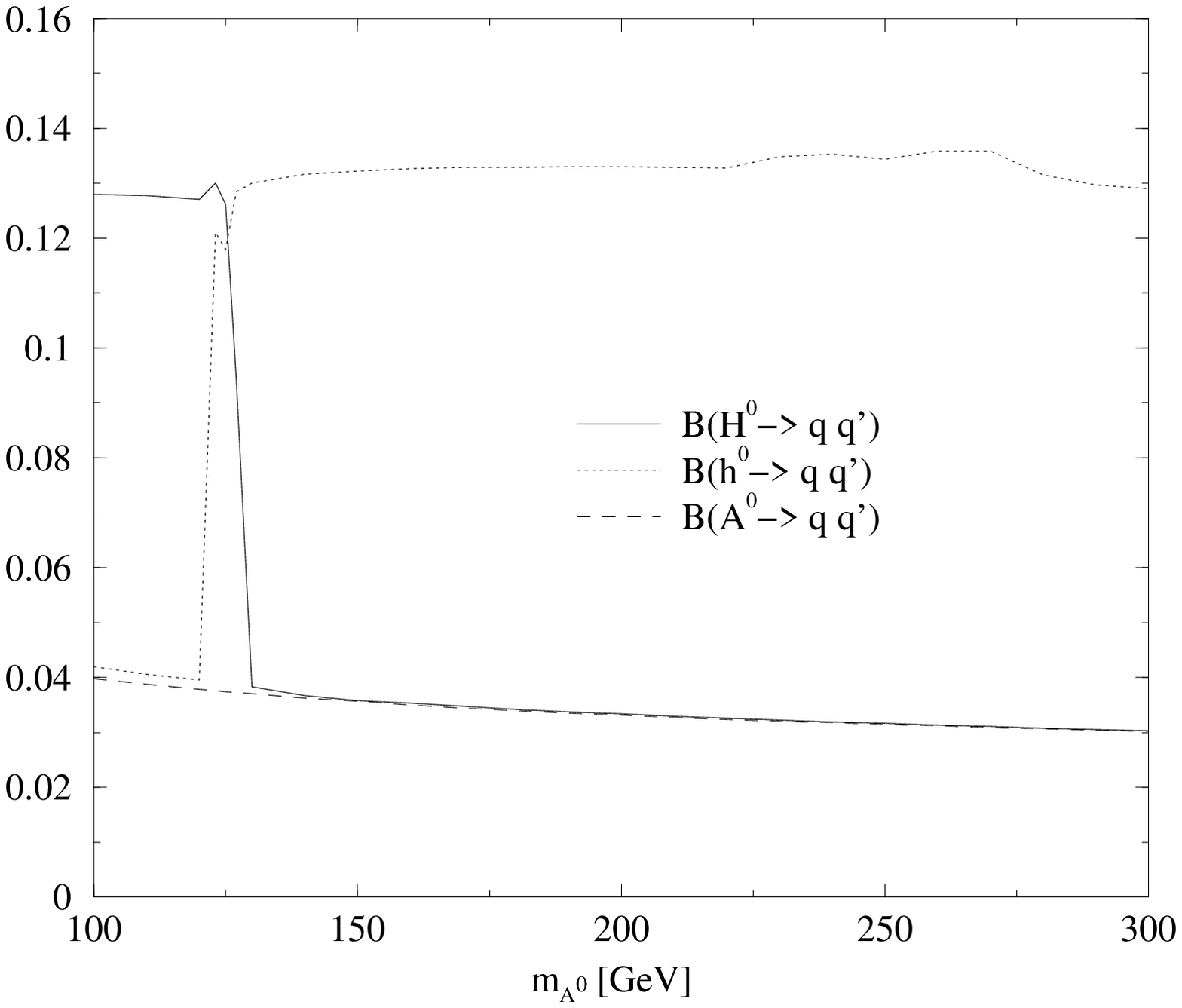}} &
\resizebox{!}{6cm}{\includegraphics{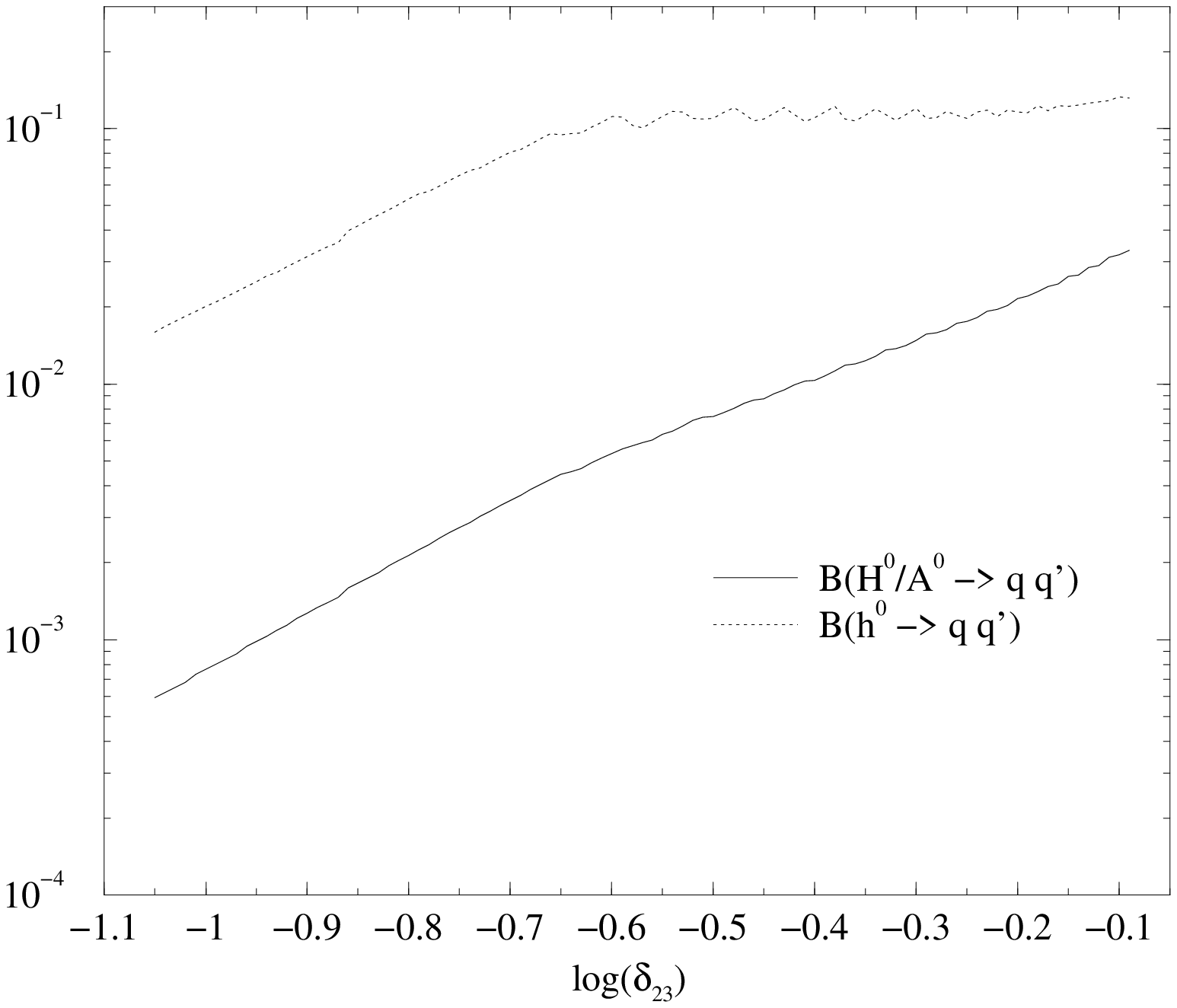}} \\
(a) & (b)
\end{tabular}
\caption{{Maximum SUSY-QCD contributions to
$B(h\rightarrow q\,q')$, Eq.\,(\ref{eq:hbs-def}), as a
  function of} \textbf{a)} $\ma$ and \textbf{b)} $\delta_{23}$ for
  $\ma=200\GeV$.}\label{fig:maximfull}
\end{figure}

\begin{table}
\centerline{\begin{tabular}{|c||c|c|c|} \hline Particle &  $H^0$
& $h^0$ & $A^0$ \\\hline\hline 
\Bhbs &  $3.3\times 10^{-2}$ & $1.3\times 10^{-1}$ & $3.3\times
10^{-2}$\\
\hline $\Gamma(h\to
X)$ & $11.0 \GeV$ & $1.6\times 10^{-3} \GeV$ & $11.3 \GeV$
\\\hline $\delta_{23}$ & $10^{-0.09}$& $10^{-0.1}$ &
$10^{-0.09}$\\\hline $m_{\squark}$ & $975 \GeV$ &  $975 \GeV$ &
$975 \GeV$\\\hline $A_b$ & $1500 \GeV$ & $730 \GeV$ &
$1290\GeV$\\\hline $\mu$ & $980 \GeV$ & $1000 \GeV$ & $980 \GeV$
\\\hline \Bbsg &  $4.42\times 10^{-4}$ &  $4.23\times 10^{-4}$
&$4.50\times 10^{-4}$ \\\hline
\end{tabular}}
\caption{Maximum values of $\Bhbs$ and corresponding SUSY
parameters for
  $\ma=200\GeV$.}\label{tab:maxim1}
\end{table}

{The result of the scan is depicted in Fig.~\ref{fig:maximfull}.
To be specific: Fig.~\ref{fig:maximfull}a shows  the maximum value
$\Bhbsmax$ of the FCNC decay rate (\ref{eq:hbs-def}) under study
as a function of $\ma$; Fig.~\ref{fig:maximfull}b displays
$\Bhbsmax$ as a function of the mixing parameter $\delta_{23}$ for
$\ma=200\GeV$}. Looking at Fig.~\ref{fig:maximfull} three facts
strike the eye immediately :
{i) the maximum is huge ($ 13\%$!) for a FCNC rate, actually it is as
big as initially guessed from the rough estimates made in Section
\ref{sect:partialwidths}}; ii) very large values of $\delta_{23}$
are allowed; iii) the maximum rate is independent of the
pseudo-scalar Higgs boson mass $\mA$.
{We will now analyze facts ii) and iii) in turn,
and will establish their incidence on fact i)}. For further
reference, in Table~\ref{tab:maxim1} we show the numerical values
of $\Bhbsmax$ together with the parameters which maximize the
rates for $\ma=200\GeV$.

\begin{figure}[tb]
\begin{tabular}{cc}
\resizebox{!}{6cm}{\includegraphics{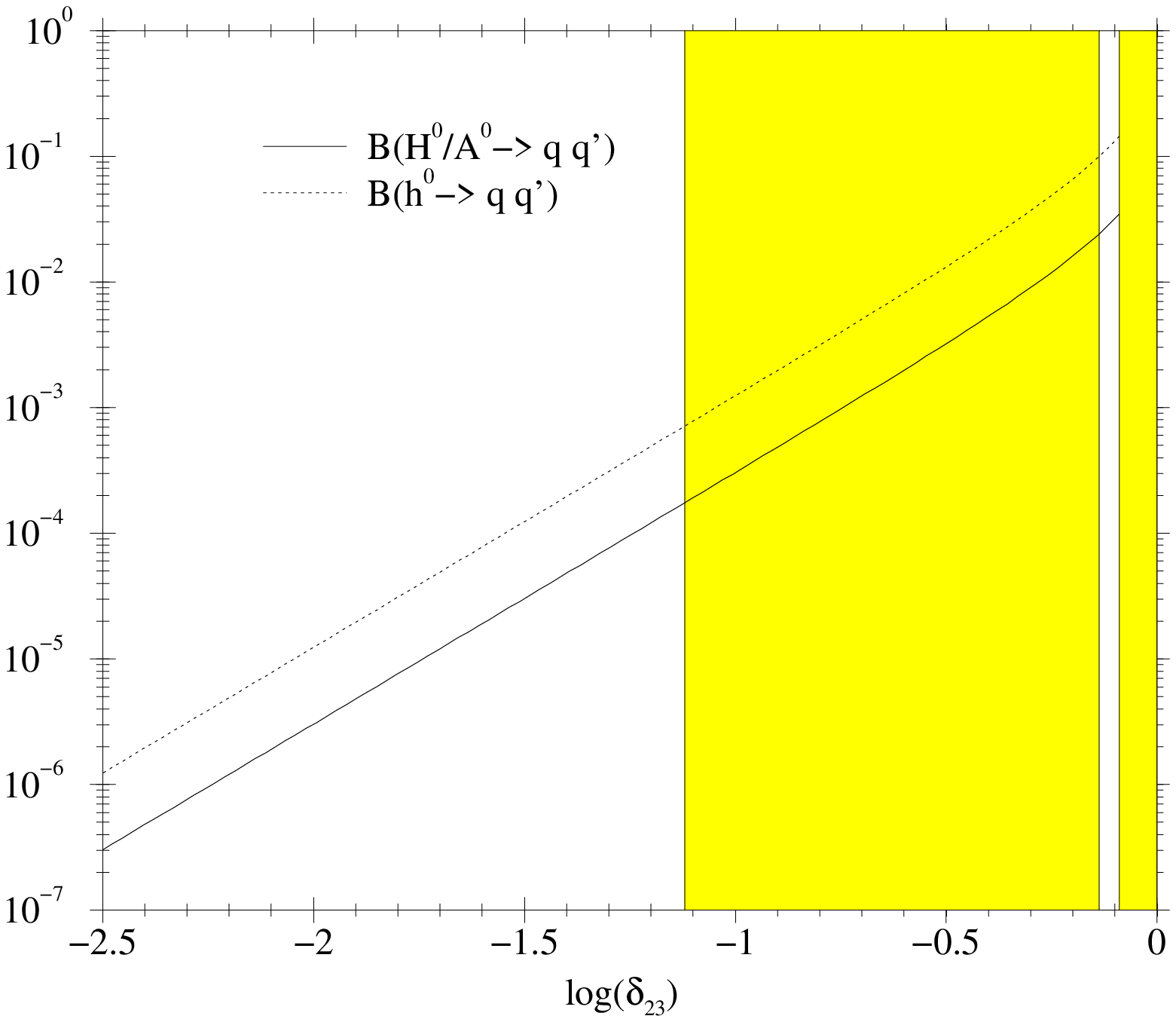}} &
\resizebox{!}{6cm}{\includegraphics{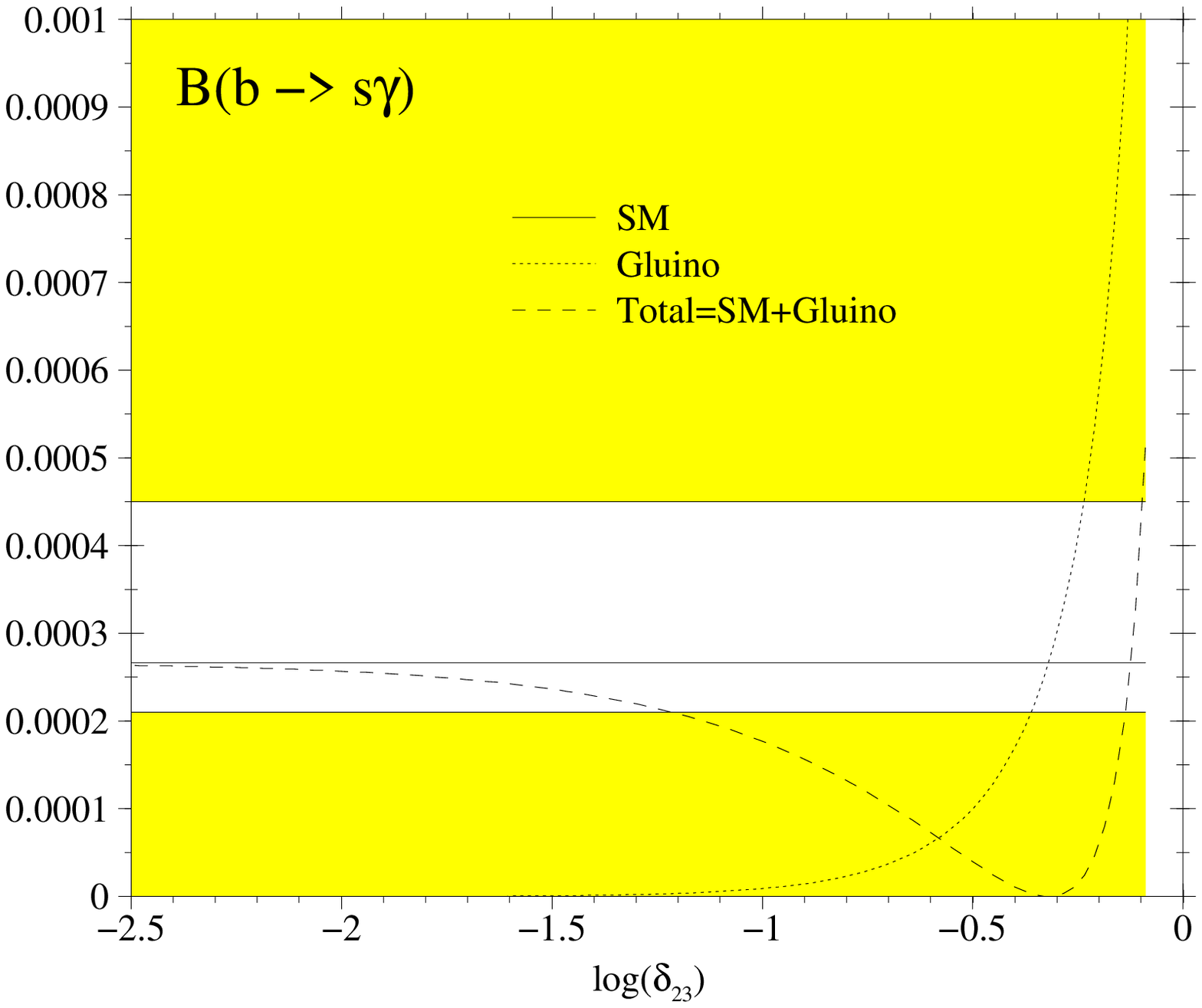}} \\
(a) & (b)
\end{tabular}
\caption{$\Bhbs$ and $\Bbsg$ as a function of $\delta_{23}$ for
the
  parameters that maximize $\Bhzbs$ in
  Table~\protect{\ref{tab:maxim1}}. The shaded region is excluded experimentally.}\label{fig:d23bsg}
\end{figure}

\begin{figure}
\begin{tabular}{cc}
\resizebox{!}{6cm}{\includegraphics{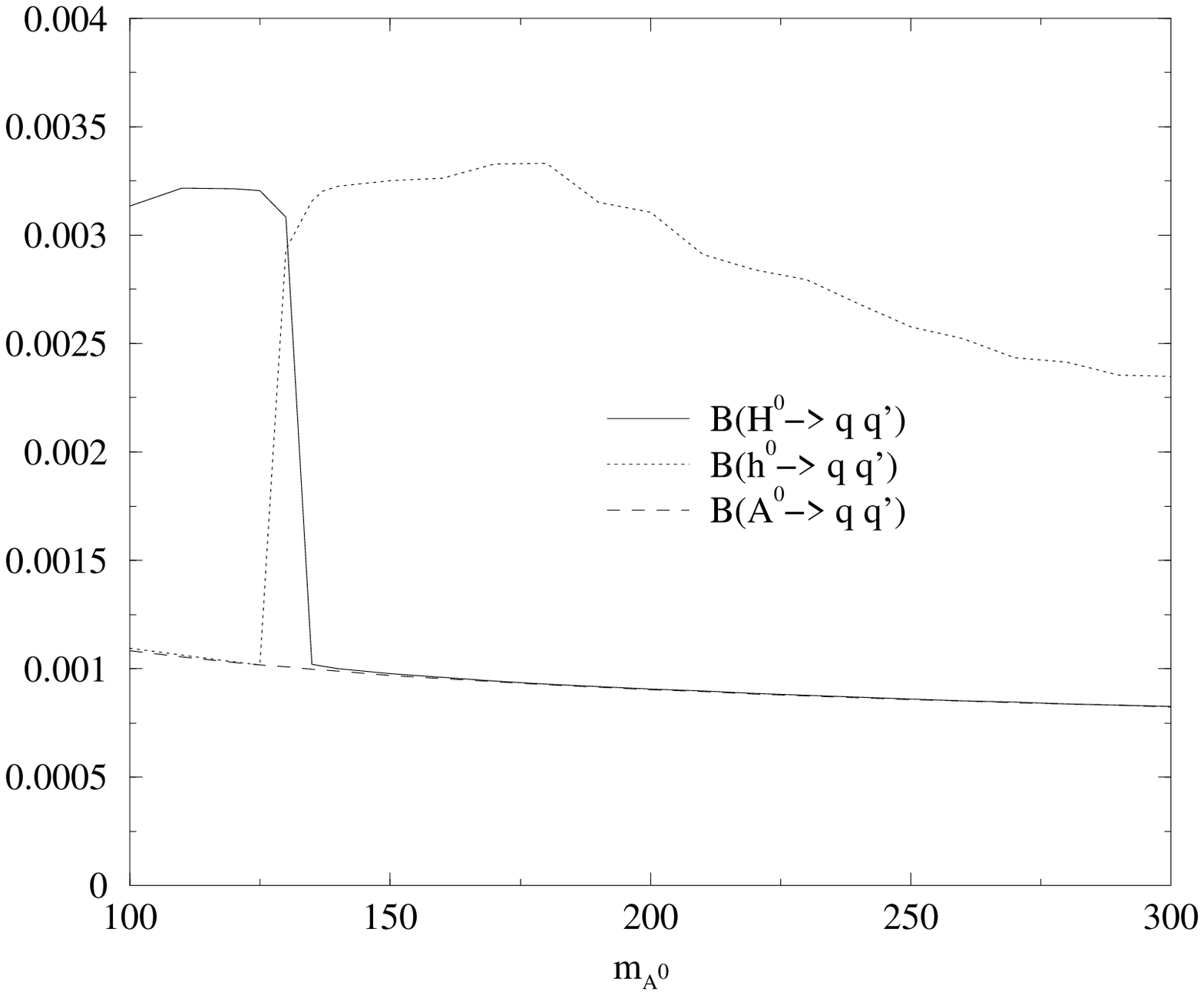}} &
\resizebox{!}{6cm}{\includegraphics{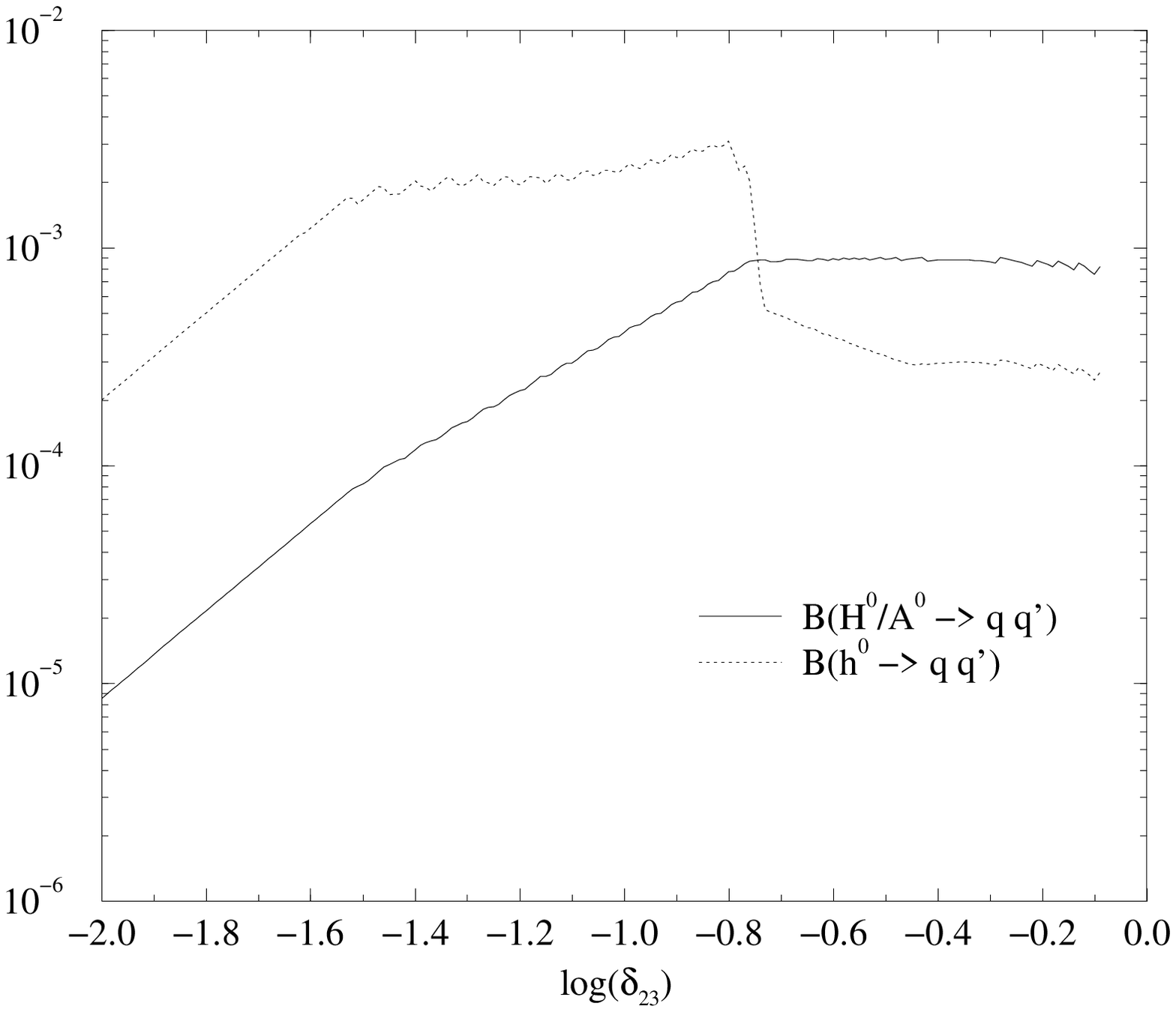}} \\
(a) & (b)
\end{tabular}
\caption{Maximum value of the SUSY-QCD contributions to $\Bhbs$
as a
  function of \textbf{a)} $\ma$ and \textbf{b)} $\delta_{23}$ for
  $\ma=200\GeV$, for the scenario excluding the \textit{window} regions.\label{fig:maximnowindow}}
\end{figure}

\begin{table}[tb]
\centerline{\begin{tabular}{|c||c|c|c|} \hline Particle &  $H^0$
& $h^0$ & $A^0$ \\
\hline\hline \Bhbs &  
$9.1\times 10^{-4}$ & $3.1\times 10^{-3}$ & $9.1\times 10^{-4}$\\
\hline $\Gamma(h\to
X)$ & $11.2 \GeV$ & $1.4\times 10^{-3} \GeV$ & $11.3 \GeV$
\\\hline $\delta_{23}$ & $10^{-0.43}$& $10^{-0.8}$ &
$10^{-0.43}$\\\hline $m_{\squark}$ & $1000 \GeV$ &  $975 \GeV$ &
$1000 \GeV$\\\hline $A_b$ & $-1500 \GeV$ & $-1500 \GeV$ & $-1500
\GeV$\\\hline $\mu$ & $-460 \GeV$ & $-1000 \GeV$ & $-460 \GeV$
\\\hline \Bbsg &  $4.49\times 10^{-4}$ &  $4.48\times 10^{-4}$
&$4.49\times 10^{-4}$ \\\hline
\end{tabular}}
\caption{Maximum values of $\Bhbs$ and corresponding SUSY
parameters for
  $\ma=200\GeV$ excluding the \textit{window} region.\label{tab:maximnowindow}}
\end{table}

One would expect that a large value of $\delta_{23}$ should
induce a large gluino contribution to $\Bbsg$. In fact it does!
However our automatic scanning process picks up the corners of
parameter space where the gluino contribution alone is much
larger than the SM contribution, but opposite in sign, such that
both contributions destroy themselves partially leaving a result
in accordance with the experimental {constraints}. We
examine this behaviour in Fig.~\ref{fig:d23bsg}, where we show
the {values} of $\Bhbs$ together with $\Bbsg$ as a
function of $\delta_{23}$ for the parameters which maximize {the
FCNC rate of the lightest CP-even state $h^0$} in
Table~\ref{tab:maxim1}. 
We see that, for small values of
$\delta_{23}$, the gluino contribution to $\Bbsg$ is small, and
the total $\Bbsg$ prediction is close to the SM
expectation. {In contrast,  as $\delta_{23}$ steadily grows, $\Bbsg$
decreases fast (meaning a dramatic cancellation between the two
contributions) until reaching a point where $\Bbsg=0$}. From
there on it starts to grow with a large slope, and in its race
eventually crosses the allowed $\Bbsg$ region.
{The crossing is very fast, and so rather ephemeral in the
$\delta_{23}$ variable, and it leads to the appearance of a narrow
allowed \textit{window}
{at} large $\delta_{23}$ values, see  Fig.~\ref{fig:d23bsg}a}.
 {We would regard the choice
of this window as a fine-tuning of parameters, hence unnatural}
\footnote{At the time of writing this Thesis this fine-tuning
is excluded experimentally\cite{Gambino:2004mv}.}.
For this reason we reexamine the $\Bhbs$ ratio by performing
{a new scan of the MSSM parameter space in which we
exclude the fine-tuned (or \textit{window}) region}. The result
for $\mA=200\GeV$ can be seen in Table~\ref{tab:maximnowindow} and
Fig.~\ref{fig:maximnowindow}. This time we see that the maximum
values of $\Bhbs$ are obtained for much lower values of
$\delta_{23}$, and the maximum rates have decreased more than one
order of magnitude with respect to Table~\ref{tab:maxim1},
{reaching the level of few per mil}. These FCNC rates can still
be regarded as fantastically large. Had we included the
{SUSY-EW}
  contributions to $\Bbsg$, further cancellations might have
  occurred between the SUSY-EW and the SUSY-QCD amplitudes. Even more: since each
    contribution depends on a separate set of parameters, one would be able to
    find a set of parameters in the SUSY-EW sector which creates an
    amplitude that compensates the SUSY-QCD contributions for almost any
    point of the SUSY-QCD parameter
    space. But of course this would be only at the price of
    performing some
    fine tuning, which is not the approach we want to follow here.

    On the other hand further
      contributions to $\Bbsg$ might exist. In the most general MSSM, flavor-changing
      interactions  for the right-chiral squarks
      ($\delta_{23RR}$), and mixing left- and right-chiral squarks
      ($\delta_{23LR}$) can be introduced. The latter can produce
      significant contributions to 
    $\Bbsg$, changing the allowed parameter space. The
    introduction of $\delta_{23LR}$ can produce two possible outcomes:
    First, in certain regions of the parameter space, the contributions
    of $\delta_{23LR}$ and $\delta_{23}$ are of the same sign, enhancing
    each other. In this situation, the maximum allowed value of
    $\delta_{23}$ is obtained for $\delta_{23LR}=0$. Second, in other
    regions of the parameter space the two contributions would
    compensate each other, producing an overall value of $\Bbsg$ in
    accordance with experimental constraints, even though each
    contribution would be much larger. Again, we would regard these
    compensations as unnatural, and would discard that region of the
    parameter space. In the following we will require that the SUSY-QCD
    contributions induced by $\delta_{23}$ do not compensate the SM
    ones to give an acceptable 
    value of $\Bbsg$; this is equivalent to 
    the condition that the SUSY-QCD amplitude  represents a
    small contribution to the total $\Bbsg$ value, and is therefore
    independent of the inclusion of the other 
    contributions (SUSY-EW,
    $\delta_{23LR}$).\footnote{The analysis of
        Ref.~\cite{Demir:2003bv} follows the opposite approach, that is:
      to find the fine-tuning conditions imposed by low energy data that
    allow for the largest possible value of the FCNC parameters.}

\begin{figure}[tb]
\begin{tabular}{cc}
\resizebox{!}{6cm}{\includegraphics{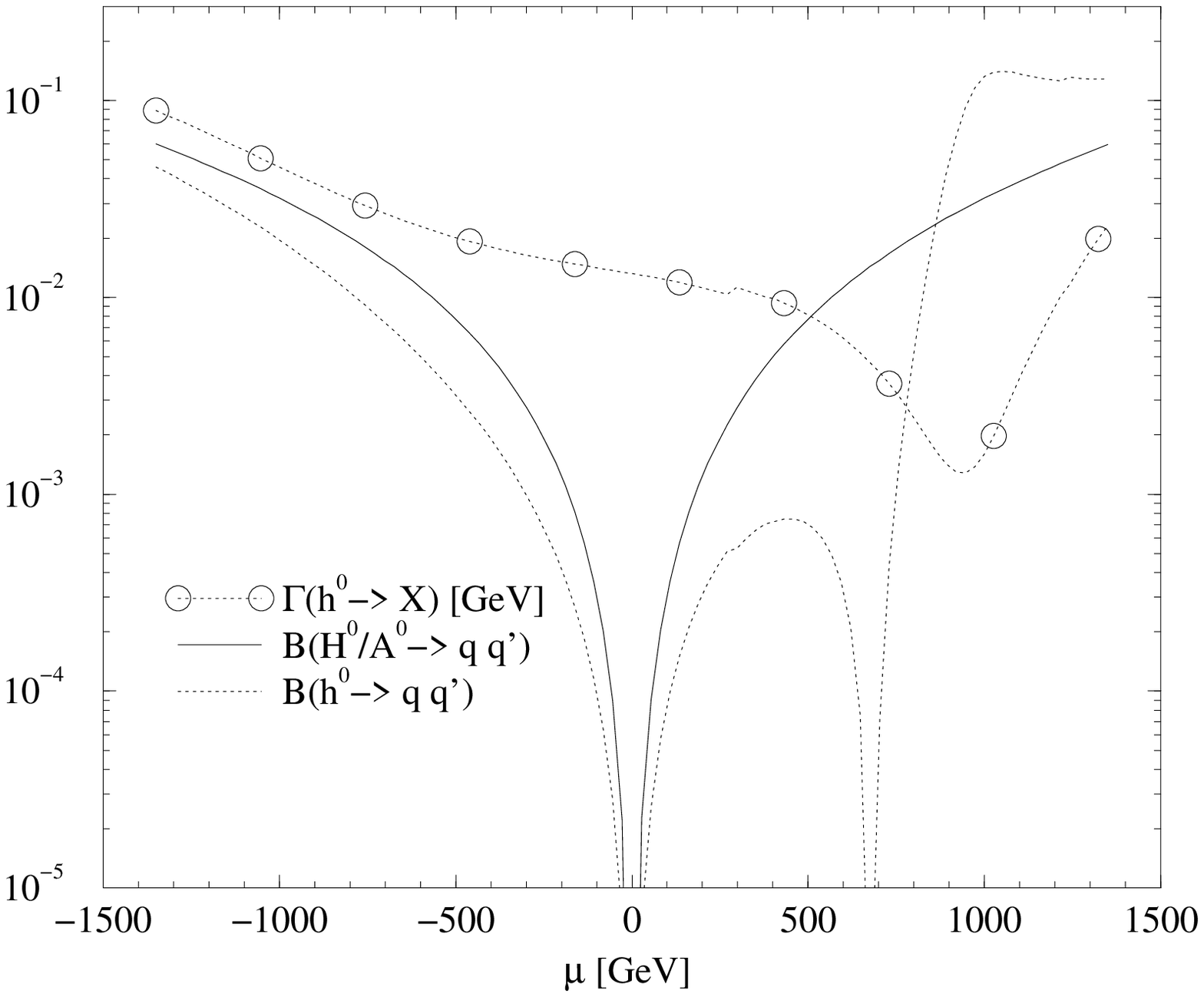}} &
\resizebox{!}{6cm}{\includegraphics{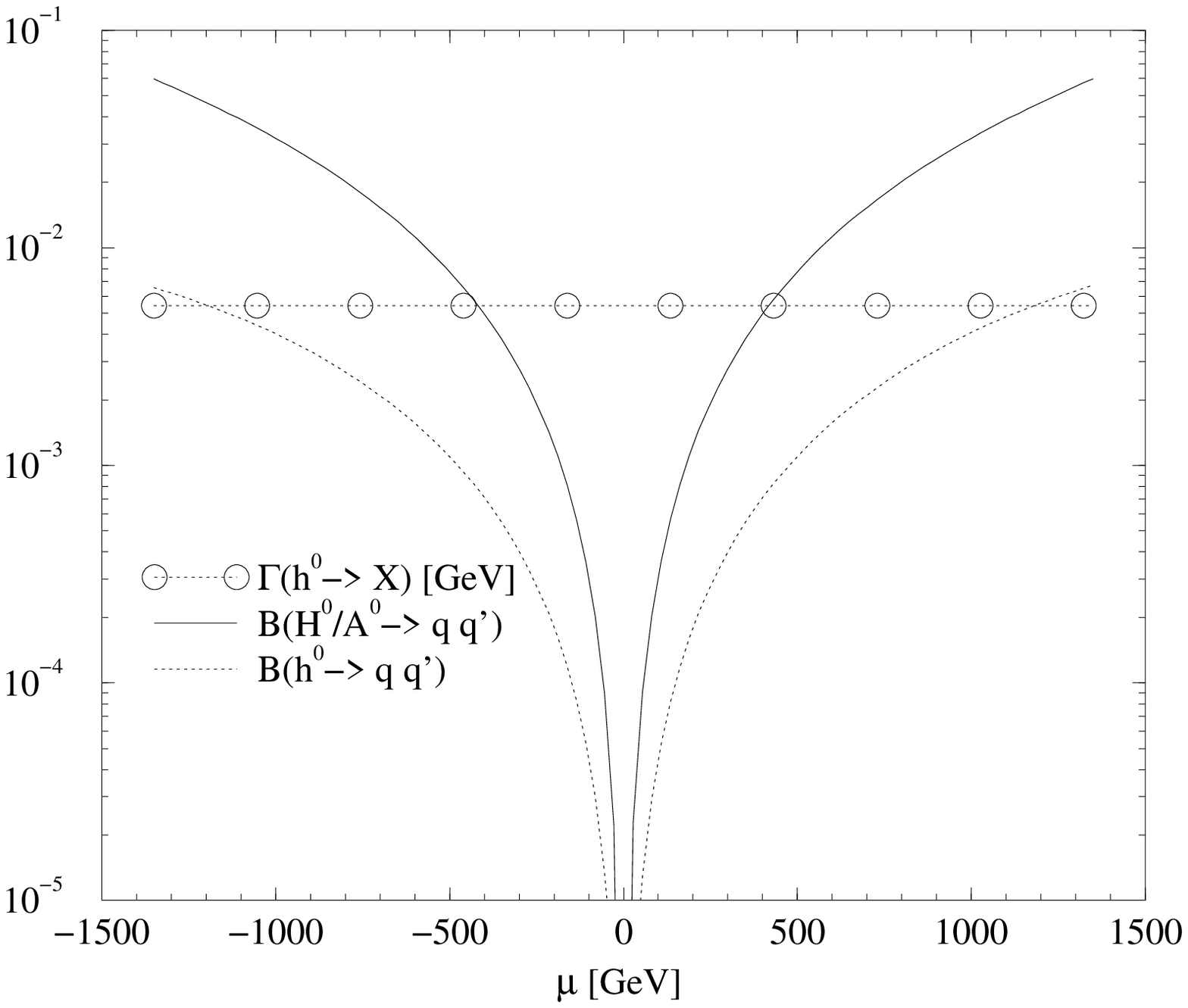}} \\
(a) & (b) 
\end{tabular}
\caption{$\Bhbs$ and $\Gamma(h^0\to X)$ (in\GeV) as a function of $\mu$ for
  \textbf{a)} one-loop $\alpha$ angle; \textbf{b)} tree-level $\alpha$
  angle, and for the
  parameters that maximize $\Bhzbs$ in
  Table~\ref{tab:maxim1}.
  The $H^0$ and $A^0$ curves coincide. The $\Bbsg$ {constraint} is
  not shown.\label{fig:alphamu}}
\end{figure}

We turn now our view to the second fact, namely the independence
of the maximum rates with respect to $\ma$. {We will show
that it also plays a central role as to the enhancement of
$\Bhbs$}. Actually, a good hint is given by the small values of
the lightest Higgs boson decay width in Tables~\ref{tab:maxim1}
and~\ref{tab:maximnowindow}, $\Gamma(h^0\to X)\sim 2\times
10^{-3}\GeV$. The maximization process of $\Bhzbs$ does not only
find the parameters for which $\Ghzbs$ is maximum, but also the
parameters for which $\Gamma(h^0\to X)$ is minimum. Specifically,
since $\Gamma(h^0 \to b\bar{b})$ is the dominant decay decay
channel of $h^0$ for large $\tb$, the maximum of $\Bhzbs$ is
produced in the parameter range of the so-called \textit{small
$\alpha_{eff}$
  scenario}~\cite{Carena:2002qg}, that is, a parameter range where the
radiative corrections make the CP-even Higgs boson mixing angle
$\alpha$ vanish (or very small), such that the leading partial
decay width $\Gamma(h^0\to b\bar{b})$ is strongly suppressed. The
consequences of this scenario have been extensively studied in
Ref.~\cite{Carena:1999bh}. {As advertised in Section
\ref{sect:partialwidths}, the possibility that the maximization
process explores these regions of the parameter space is the
reason why the leading higher order decay channels, and also the
leading three-body decay modes have to be taken into account in
the computation of the total width.}

{In Fig.~\ref{fig:alphamu} we plot the value of the various
branching ratios $\Bhbs$ and of the total width of the lightest
CP-even Higgs boson, $\Gamma(h^0\to X)$, as a function of the
higgsino mass parameter $\mu$}, the rest of the parameters being
{those of the third column of Table~\ref{tab:maxim1},
i.e. the ones that maximize the branching ratio $\Bhzbs$}.
Fig.~\ref{fig:alphamu}a shows that $\Gamma(h^0\to X)$ has a deep
minimum in the range of $\mu$ corresponding to the maximum of
$\Bhzbs$, {which reaches the level of a few percent}. If, instead
of using the radiatively corrected $\alpha$ value we use the
tree-level expression, we obtain the result shown in
Fig.~\ref{fig:alphamu}b. Here the total decay width of the Higgs
boson is independent of $\mu$, and $\Bhbs$ does not show any
peak. {Actually in this case the branching ratio for $h^0$
becomes smaller than that of $H^0$ and $A^0$ for all $\mu$.} The
maximization procedure in Fig.~\ref{fig:maximfull} selects for
each value of $\ma$ the MSSM parameters corresponding to the
small $\alpha_{eff}$ scenario for that specific value of $\ma$.
Of course, this discussion regarding the $h^0$ channels for large
values of $\ma$ has a correspondence {with} the $H^0$ channel for
low values\footnote{{Large or low values here means
$\ma>m_{h^0}^{\rm max}$ or $\ma<m_{h^0}^{\rm max}$, i.e. above or
below the maximum possible value for the mass of the lightest
Higgs boson $h^0$, respectively}.} of $\ma$.

{As indicated in Section \ref{sect:partialwidths},
we have used a one-loop approximation for the Higgs
sector}~\cite{\Dabels}, instead of the more sophisticated
complete two-loop result present in the
literature~\cite{Carena:2000dp,Espinosa:2000df}. However, we
should stress that the exact MSSM parameters at which the small
$\alpha_{eff}$ scenario is realized are not important for the
sake of the present analysis. All that matters is that some
portion of the parameter space exists, for which $\Gamma(h^0\to
b\bar{b})$ is strongly suppressed, {but $\Ghzbs$ is not}.

\begin{table}
\centerline{\begin{tabular}{|c||c|c|c|} \hline Particle &  $H^0$
& $h^0$ & $A^0$ \\
\hline\hline \Bhbs & 
 $9.0\times 10^{-4}$ & $1.3\times 10^{-4}$ & $9.0\times 10^{-4}$\\
\hline $\Gamma(h\to
X)$ & $11.3 \GeV$ & $5.4\times 10^{-3} \GeV$ & $11.3 \GeV$
\\\hline $\delta_{23}$ & $10^{-0.43}$& $10^{-0.28}$ &
$10^{-0.43}$\\\hline $m_{\squark}$ & $1000 \GeV$ &  $1000 \GeV$ &
$1000 \GeV$\\\hline $A_b$ & $-1500 \GeV$ & $-1500 \GeV$ & $-1500
\GeV$\\\hline $\mu$ & $-460 \GeV$ & $-310 \GeV$ & $-460 \GeV$
\\\hline \Bbsg &  $4.49\times 10^{-4}$ &  $4.50\times 10^{-4}$
&$4.49\times 10^{-4}$ \\\hline
\end{tabular}}
\caption{Maximum values of $\Bhbs$ and corresponding SUSY
parameters for
  $\ma=200\GeV$, using the tree-level expressions for the Higgs sector,
  and excluding the \textit{window} region.\label{tab:maximnowindowtree}}
\end{table}

\begin{figure}
\begin{tabular}{cc}
\resizebox{!}{6cm}{\includegraphics{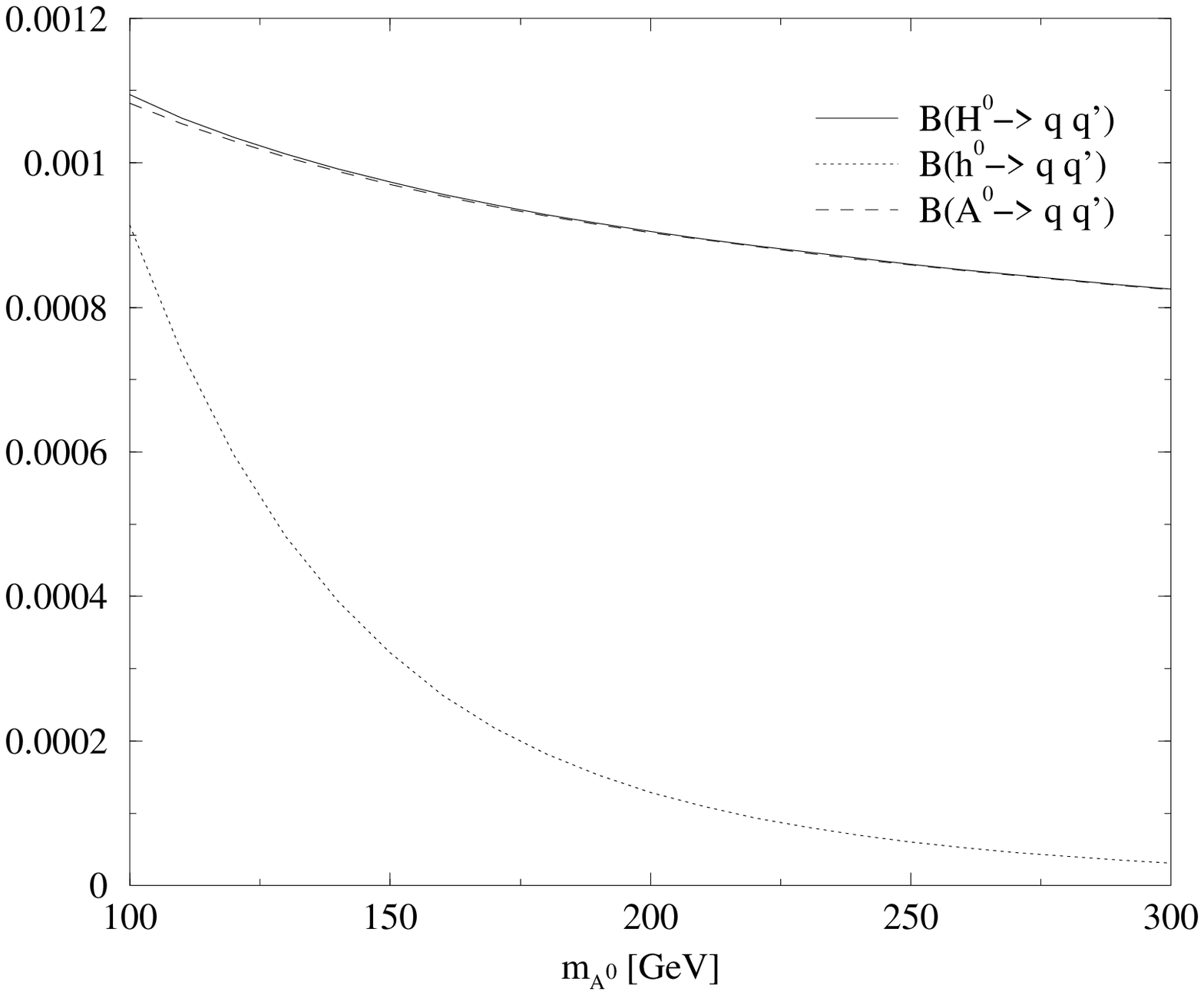}} &
\resizebox{!}{6cm}{\includegraphics{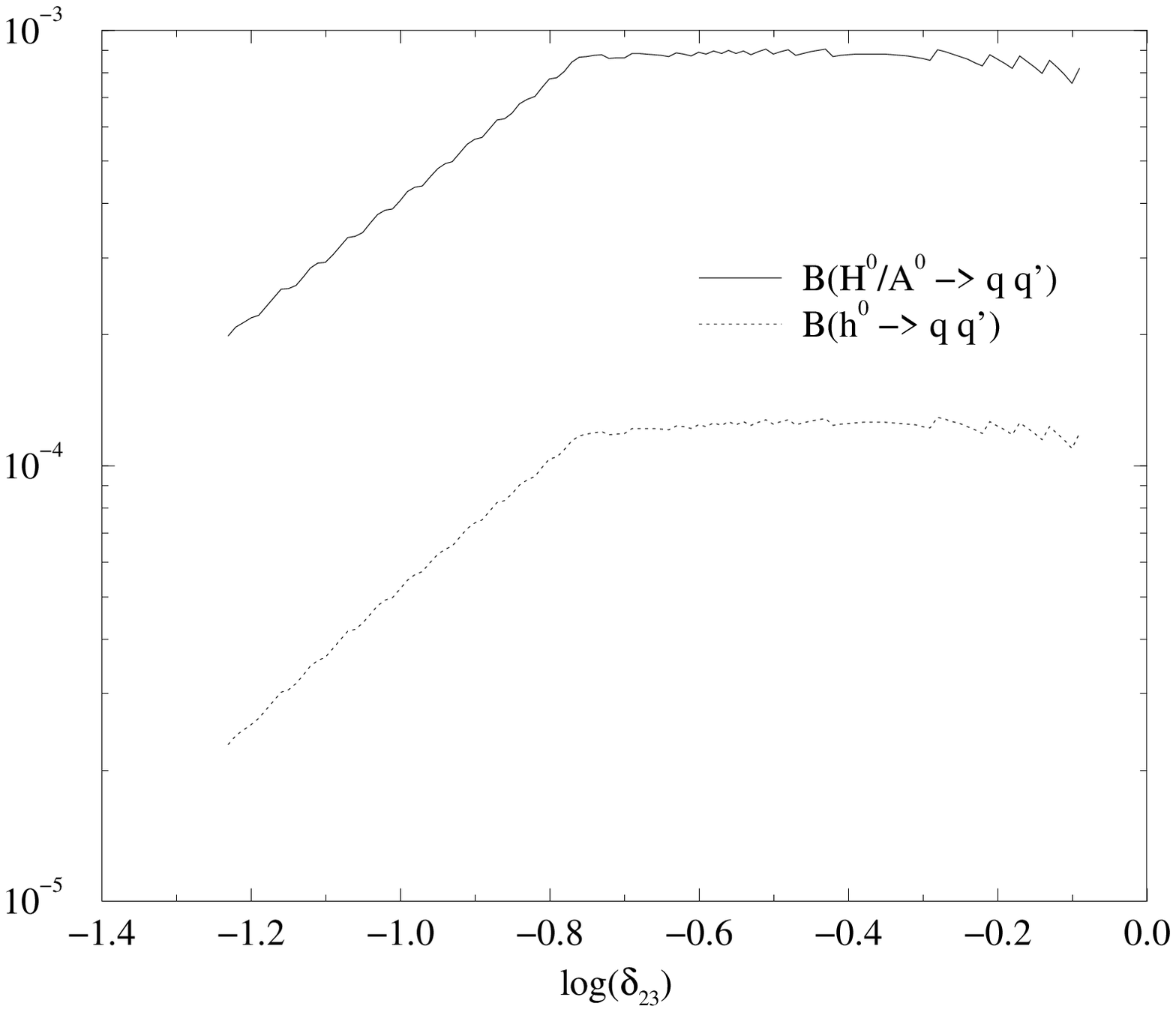}} \\
(a) & (b)
\end{tabular}
\caption{Maximum value of the SUSY-QCD contributions to $\Bhbs$
as a
  function of \textbf{a)} $\ma$ and \textbf{b)} $\delta_{23}$, for
  $\ma=200\GeV$ and for the scenario excluding the \textit{window} region
  and using the tree-level expressions for the Higgs sector
  parameters.\label{fig:maximnowindowtree}}
\end{figure}

{To compare the maximum value of $\Bhzbs$ obtained with and
without the small $\alpha_{eff}$ scenario, we have performed the
maximization procedure using the tree-level expressions for the
{Higgs sector parameters}}. The result is shown {in}
Table~\ref{tab:maximnowindowtree} and
Fig.~\ref{fig:maximnowindowtree}. {In this case $\Bhzbs$ is
reduced by a sizeable factor of $\gtrsim 20$ with respect {to}
Table~\ref{tab:maximnowindow}, whereby the $h^0$ rate descends
about an order of magnitude below that of the $H^0/A^0$ channels
which remain basically unchanged}. Notice also that
$\Gamma(h^0\to X)$ is larger than in previous tables. In spite of
the reduction, achieving a FCNC ratio  $\Bhzbs\sim 1.3 \times
10^{-4}$ is a remarkable result, {three orders of magnitude}
larger than the maximum SM rate (\ref{estimateBRSMhbs}), and only
one order of magnitude below the rare decay $B(h^0\to
\gamma\gamma)\sim 10^{-3}$.
{Also worth noticing
in Fig.~\ref{fig:maximnowindowtree}b (and
Fig.~\ref{fig:maximnowindow}b) is the fact that $\Bhbsmax$ is
essentially flat in $\delta_{23}$ in the upper range {down to}
$\delta_{23}\sim10^{-0.8}\simeq 0.16$}. The reason lies in the
correlation between $\Bhbs$ and $\Bbsg$. In order to comply with
the (non-fine-tuned) value of $\Bbsg$ for large $\delta_{23}$, the
absolute value of the $\mu$ parameter must be small. When
$\delta_{23}$ decreases, $|\mu|$ can grow to larger values,
leaving the overall {maximum rates} $\Bhbsmax$
effectively unchanged (see Eq.~(\ref{eq:approxleading}) below).

The maximization process selects a squark mass scale in the
vicinity of the maximum values used in the scanning procedure. We
should point out, however, that the same order of magnitude for
$\Bhbs$ could be obtained with a much lower squark {mass}
scale. {In this case} the lighter squark masses induce a
much larger $\Bbsg$ value, and $\delta_{23}$ is much more
constrained. For example,
if we perform a {scan} in the parameter
space~(\ref{eq:scan-parameters-br}), but fixing the squark mass scale to be
$m_{\squark}<500\GeV$, we obtain the following values for the {maximal} branching
ratios for $\ma=200\GeV$:
\begin{equation}
\Bhzbsmax=1.4\times 10^{-5}\   , \ 
\BHAbsmax=9.2\times 10^{-5}\   ,
\label{eq:brlowscale}
\end{equation}
with $\delta_{23}\sim 10^{-0.6}$, $\mu\sim-110\GeV$, and we have limited
ourselves to the scenario avoiding the 
\textit{window} regions and using the tree-level expression for the Higgs
sector parameters. These numbers have to be compared with
Table~\ref{tab:maximnowindowtree}. 

The reason behind this \textit{scale independence} admits an explanation
in terms of an effective Lagrangian approach, in which one
can estimate the leading effective coupling to behave approximately as
\ref{fig:hbsSQCDvermiapprox}\cite{Demir:2003bv}:
\begin{figure}
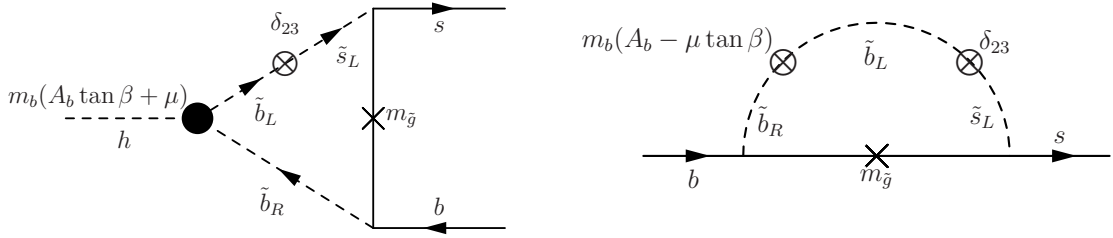

    \centering
    \begin{tabular}{cc}
        \resizebox{0.45\textwidth}{!}{\hbsSQCDvermiapprox}&
        \resizebox{0.45\textwidth}{!}{\bsSQCDmixmiapprox}
    \end{tabular}
    \caption{One-loop \susy-\QCD\ vertex diagram contributing to the decay
          $\hbs$ and diagrams contributing to mixed $b-s$ self-energy in the
          mass approximation}
    \label{fig:hbsSQCDvermiapprox}
\end{figure}
\begin{eqnarray}
g_{h b\bar{s}}   &\simeq& \frac{g\mb}{\sqrt{2}\mw \cos\beta} \frac{2
    \alpha_s}{3 \pi}  \delta_{23} \frac{-\mu \,
    \mg}{M_{SUSY}^2} \left\{ \begin{array}{cc}
 \sin(\beta-\alpha)& (H^0)\\
    \cos(\beta-\alpha) & (h^0)\\
 1& (A^0)
  \end{array}\right.\ \ .
\label{eq:approxleading}
\end{eqnarray}
Aside from {ensuring (at least) a} partial SUSY scale
independence of the leading terms, this expression also shows
that $\Bhbs$ has a weak dependence on the soft-SUSY-breaking
trilinear coupling $A_b$. 
The observed situation is
similar to the flavor-conserving $hb\bar{b}$ interactions, where
the cancellation of the $A_b$ terms at leading order has been
proven~\cite{\SpiraGuasch}. It also shows that the
leading {non-decoupling SUSY contributions to $\Ghzbs$
eventually fade out as the decoupling limit of the Higgs sector is
approached: $\cos(\beta-\alpha)\to 0$}. We have found (using the
tree-level expression for $\alpha$) that the non-leading
({SUSY-decoupling}) contributions to $\Ghzbs$ dominate for
$\mA\gtrsim 450\GeV$, inducing a value $\Ghzbsmax\sim 1.2\times
10^{-5}$, with $\delta_{23}\sim 10^{-1}$, $\mu\sim1000\GeV$.

\begin{figure}
\begin{tabular}{cc}
\resizebox{!}{6cm}{\includegraphics{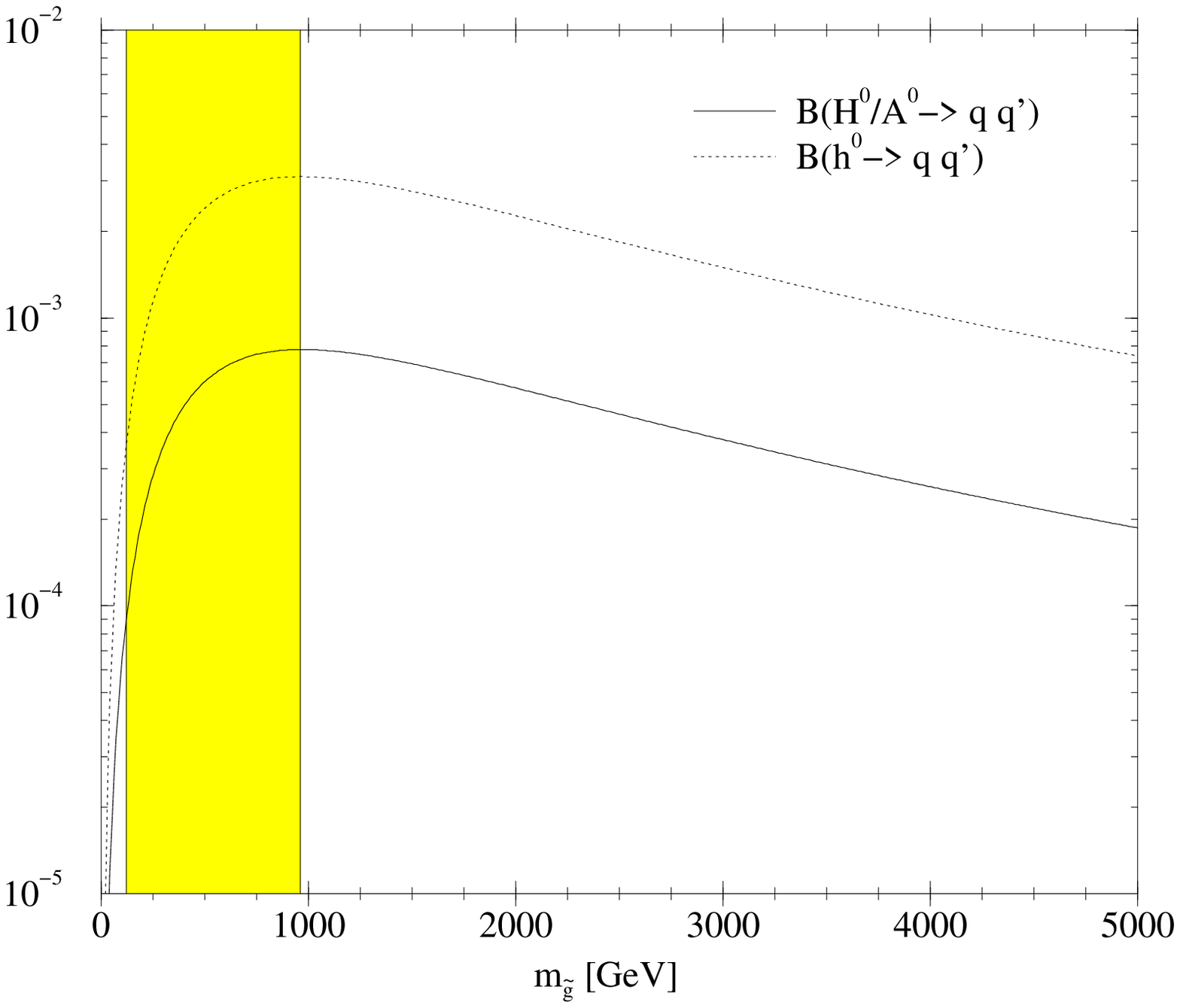}} &
\resizebox{!}{6cm}{\includegraphics{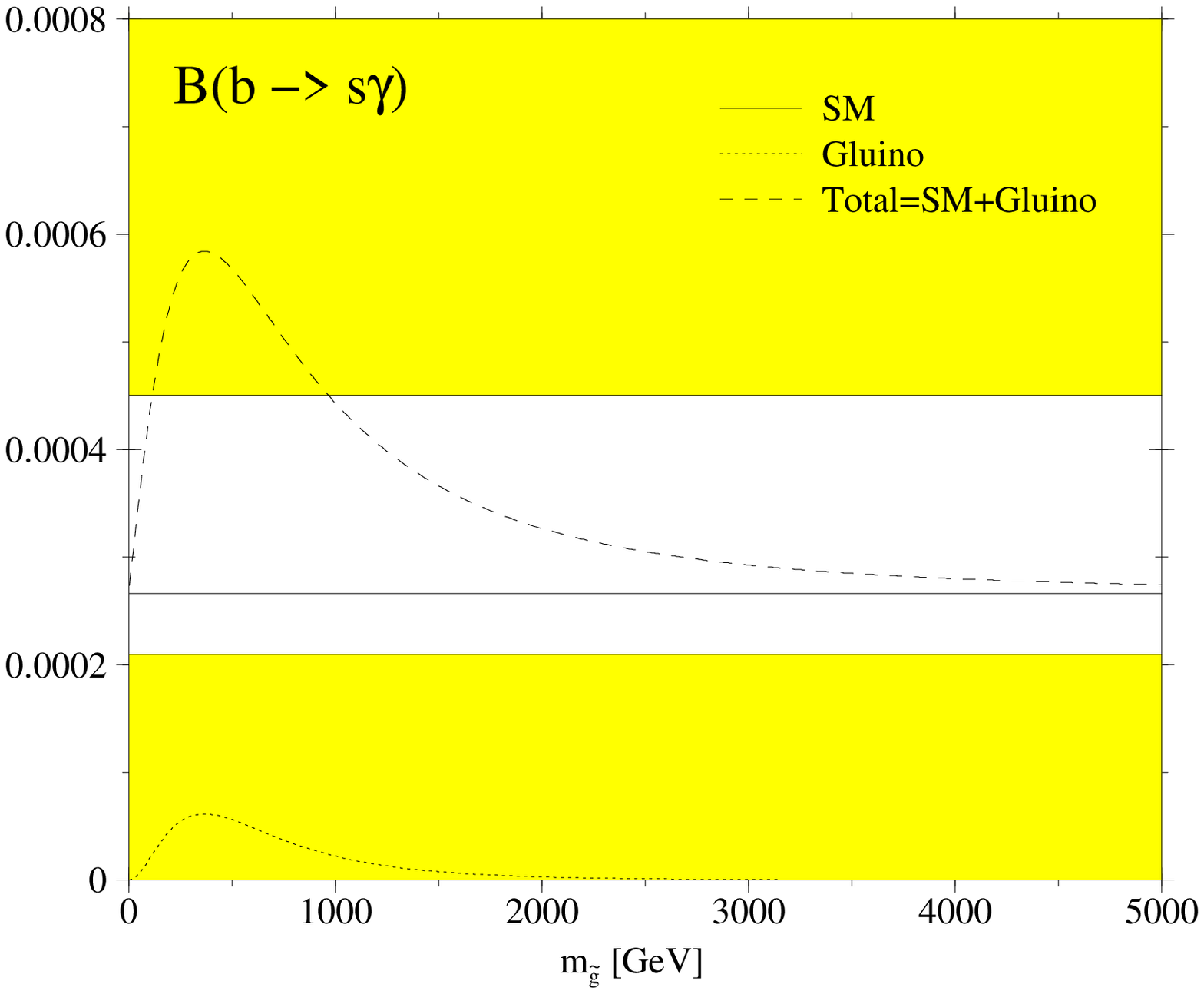}} \\
(a) & (b)
\end{tabular}
\caption{$\Bhbs$ and $\Bbsg$ as a function of $\mg$ for the
  parameters 
  that maximize $B(h^0\to b\bar{s})$ excluding the window
  region {(see third
  column of Table~\protect{\ref{tab:maximnowindow}}).} The shaded region is excluded experimentally.\label{fig:mg}}
\end{figure}

\begin{figure}
\begin{tabular}{cc}
\resizebox{!}{6cm}{\includegraphics{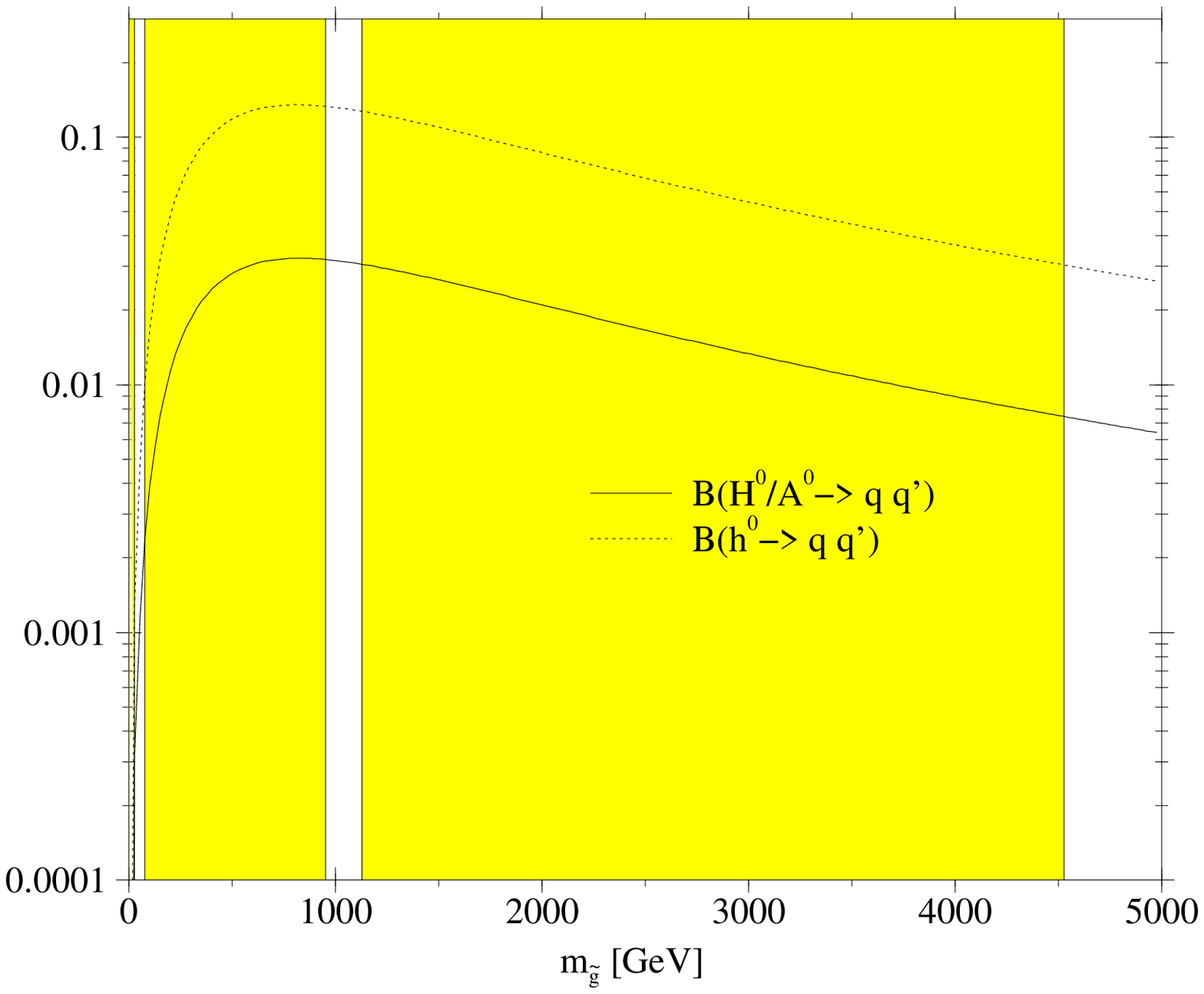}} &
\resizebox{!}{6cm}{\includegraphics{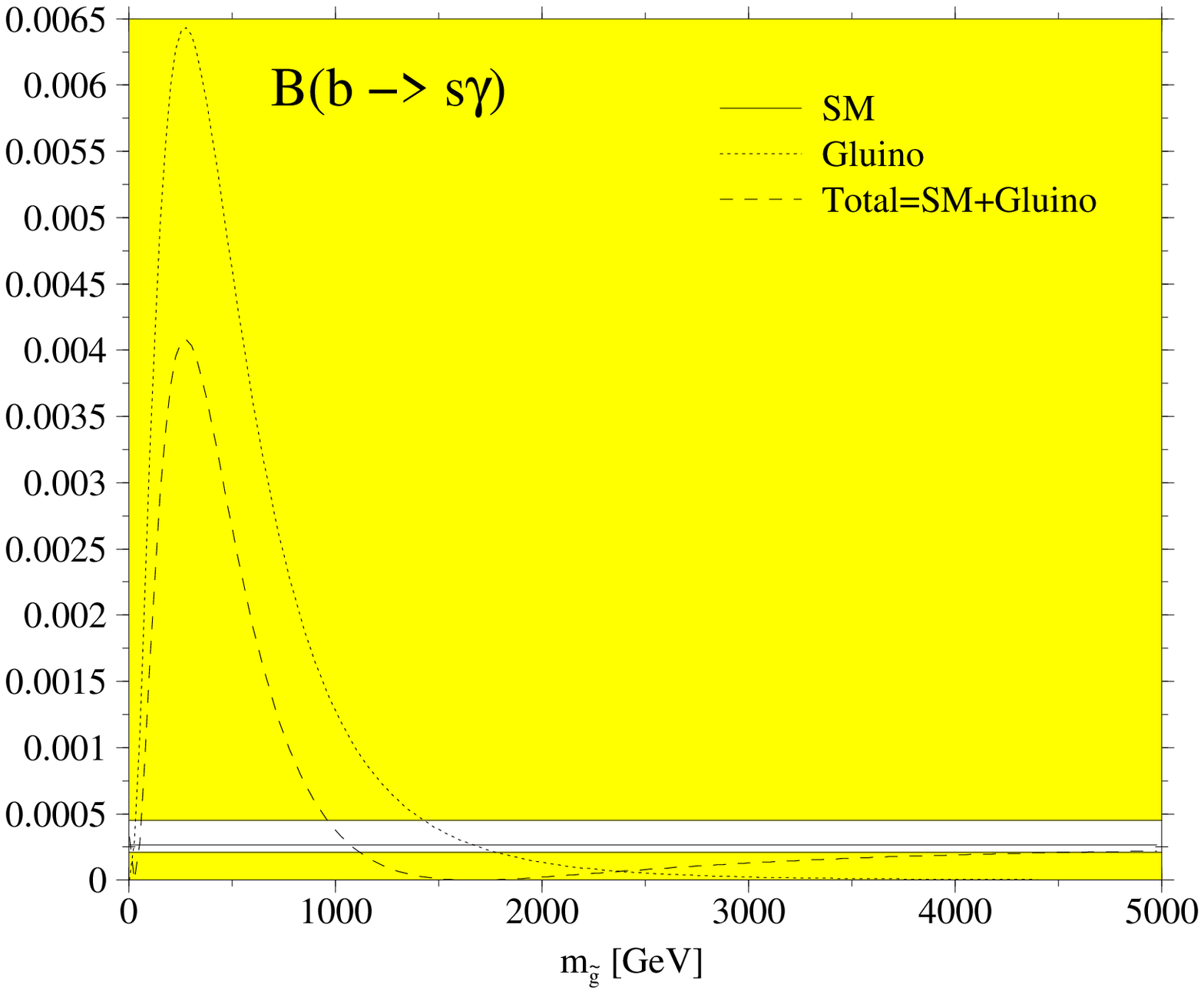}} \\
(a) & (b)
\end{tabular}
\caption{As in Fig.\,\ref{fig:mg}, but including the window region. {The remaining parameters are fixed as in the third
  column of Table~\protect{\ref{tab:maxim1}}.}}\label{fig:escletxes}
\end{figure}

{We further investigate the role of the scale of SUSY masses,
  and the fine-tuning behaviour in Figs.~\ref{fig:mg}
  and~\ref{fig:escletxes}. {In these figures we give up the equality
  $\mg=m_{\squark}$~(\ref{eq:scan-fixed}), the squark masses are fixed
  at the values stated in Tables~\ref{tab:maximnowindow}
  and~\ref{tab:maxim1} respectively.} Fig.~\ref{fig:mg} shows}
the
    values of $\Bhbs$ {for the three Higgs decays} and of $\Bbsg$
    as a function of the gluino mass for the
    parameters that maximize $\Bhzbs$ when the
    \textit{window} regions are excluded ({third column of
    Table~\ref{tab:maximnowindow}}). Here we see that, while the gluino
    contribution to $\Bbsg$ decouples \textit{fast} as a function of
    $\mg$, {its contribution} to $\Bhbs$ is fairly
    sustained. {Indeed},
    between $\mg=1\TeV$ and $\mg=5\TeV$ $\Bhzbs$ decreases
    only by a factor $\sim 1/4$, while the gluino contribution to $\Bbsg$
    becomes negligible
 at $\mg=5\TeV$ and we recover the SM
  prediction.
As a consequence, {the
    maximum rates $\Bhbs$ that we have found are robust}, in the sense that
    further theoretical refinements and experimental results that change
    the allowed range of $\Bbsg$ can easily {be compensated for by a slight
    increase} of the gluino mass ($\mg$), which would leave the
    prediction for $\Bhbs$ {essentially} unchanged.
    We note in Fig.\,\ref{fig:escletxes} the corresponding behaviour of
    $\Bhzbs$ and  $\Bbsg$  in the presence of fine-tuning, i.e. as in
    Table\,\ref{tab:maxim1}. {In contrast to} the previous
    case, here we observe the presence of  two tiny windows in the
    regions $\mg=25-75\GeV$ and $\mg=950-1125\GeV$. In the middle region
    $\mg=75-950\GeV$, $\Bbsg$ is one order of magnitude larger than the allowed
    experimental range, and in the region above $\mg=1125\GeV$ it only
    enters the allowed region for $\mg>4500\GeV$. In this region
    $\Bhzbs$ is still large, but at the price of having a gluino five
    times heavier than the rest of the SUSY spectrum. This is another
    manifestation of the large fine-tuning that governs this region of
    the parameter space.

\begin{table}
\begin{tabular}{|c||c|c||c|c||c|c||}
\hline Particle & \multicolumn{2}{c||}{$H^0$} &
\multicolumn{2}{c||}{$h^0$} & \multicolumn{2}{c||}{$A^0$} \\
\hline & $\Gamma(\GeV)$ & \Bhbs & $\Gamma(\GeV)$ & \Bhbs &
$\Gamma(\GeV)$ & \Bhbs \\\hline
 \parbox{2cm}{\vspace{0.1cm} small-$\alpha_{eff}$ \textit{window}\vspace{0.1cm}}   &
11.0  & $ 3.3\times 10^{-2}$  & $ 1.6\times 10^{-3}$ & $
1.3\times 10^{-1}$  & 11.3  & $ 3.3\times 10^{-2}$\\\hline
 \parbox{2cm}{\vspace{0.1cm} tree-Higgs \textit{window} \vspace{0.1cm}} &
11.3 & $ 3.3\times 10^{-2}$ & $ 5.4\times 10^{-3}$ & $ 4.3\times
10^{-3}$ & 11.3 & $ 3.3\times 10^{-2}$\\\hline
\parbox{2cm}{\vspace{0.1cm} small-$\alpha_{eff}$ no-\textit{window}\vspace{0.1cm}}
& 11.2 & $ 9.1\times 10^{-4}$ & $ 1.4\times 10^{-3}$ & $
3.1\times 10^{-3}$ & 11.3 & $ 9.0\times 10^{-4}$\\\hline
 \parbox{2cm}{\vspace{0.1cm} tree-Higgs no-\textit{window} \vspace{0.1cm}}  &
11.3 & $ 9.1\times 10^{-4}$ & $ 5.4\times 10^{-3}$ & $ 1.3\times 10^{-4}$ & 11.3 & $ 9.0\times 10^{-4}$\\\hline
 $\tb=5$  &
0.11 & $ 2.0\times 10^{-3}$ & $ 6.0\times 10^{-3}$ & $ 1.7\times 10^{-4}$ & 0.11 & $ 2.1\times 10^{-3}$\\\hline
 \parbox{2cm}{\vspace{0.1cm}$\tb=5$ tree Higgs\vspace{0.1cm}}  &
0.12 & $ 1.9\times 10^{-3}$ & $ 4.4\times 10^{-3}$ & $ 2.6\times 10^{-4}$ & 0.11 & $ 2.1\times 10^{-3}$\\\hline
 \parbox{2cm}{\vspace{0.1cm}$\tb=5$ no-\textit{window}\vspace{0.1cm} }  &
0.15 & $ 3.8\times 10^{-4}$ & $ 9.7\times 10^{-3}$ & $ 1.1\times 10^{-4}$ & 0.11 & $ 5.1\times 10^{-4}$\\\hline
\end{tabular}
\caption{Maximum values of $\Bhbs$ and corresponding $\Gamma(h\to
X)$ for the different
  scenarios studied in this work.\label{tab:conclu}}
\end{table}

{Up to this point we have used the high $\tb$ value
{quoted in Eq.\,(\ref{eq:scan-fixed})}. But we have also looked at
the impact of varying $\tb$ on $\Bhbsmax$}. Since the latest LEP
data restricts $\tb\gtrsim 2.5$, we have used a moderate value of
$\tb=5$. Note that, at low $\tb$, the small $\alpha_{eff}$
scenario does not arise. {As a consequence similar results
are obtained} using either the tree-level or one-loop expressions
for the Higgs sector parameters. {We find that the three
branching ratios $\Bhbsmax$ at $\tb=5$ stay in the same order of
magnitude as in the scenarios with $\tb=50$ (default case) with
the tree-level Higgs sector and no-window (Cf.
Table\,\ref{tab:maximnowindow})}.

\section{Remarks and conclusions}
\label{sect:conclusions}

The main numbers of our analysis are put in a nutshell in
Table~\ref{tab:conclu}, {where} we show the results presented
previously, together with some other scenarios and the low $\tb$
case. The computed maximum values of $\Bhbs$ must
  not be taken as exact numbers {in practice}, but order of magnitude results. 
The implications that can be derived from
Table~\ref{tab:conclu} can be synthesized as follows:
\begin{enumerate}
\item The SUSY-QCD contributions can enhance  the maximum
  expectation for the FCNC decay rates $\Bhbs$ {enormously}. {This
  is seen by comparing the results of Table~\ref{tab:conclu} with the maximum
  value of $B(H^{SM}\rightarrow b\bar{s})$
  considered in Eq.\,(\ref{estimateBRSMhbs})}. The optimized MSSM
  branching ratios are at the very least 3 orders of magnitude
  bigger than the SM result.

\item If no special circumstances apply, that is, {if} no fine-tuning occurs
  between the parameters contributing to $\Bbsg$ in the MSSM, and {if} $\Gamma(h^0\to b\bar{b})$ is not
  suppressed, the maximum
  rates are $\Bhzbsmax\simeq 1.3\times 10^{-4}$, $\BHAbsmax\simeq 9\times 10^{-4}$. This corresponds to the
  {\textit{tree-Higgs}/\textit{no-window}} scenario in
  Table~\ref{tab:conclu}.
  \label{sc:noloopnofine}
\item If, however, $\Gamma(h^0\to b\bar{b})$ is suppressed by the radiative corrections
to the CP-even mixing angle $\alpha$, then $\Bhzbs$ can be an
order of magnitude larger: $\Bhzbsmax\sim 3\times 10^{-3}$. This
corresponds to the
  small $\alpha_{eff}$ scenario, and is indicated by small-$\alpha_{eff}$/ no-\textit{window} in
  {Table}~\ref{tab:conclu}. The FCNC branching ratio that we find for $h^0$ in this case should
  be considered as the largest possible one within the conditions of naturalness (no
  fine-tuning).
  \label{sc:loopnofine}
\item On the other hand, if fine-tuning between the gluino and the SM contributions to $\Bbsg$
  is allowed, but the small-$\alpha_{eff}$ scenario is not realized,
  then $\Bhzbsmax$ grows one order of
  magnitude up to $\Bhzbsmax\sim 4\times 10^{-3}$, whereas $\BHAbsmax\sim
  3\times 10^{-2}$. {This corresponds to the case labelled
  \textit{tree-Higgs/window} in Table~\ref{tab:conclu}}.
  \label{sc:noloopfine}
\item When both special conditions take place simultaneously, viz. fine-tuning in
$\Bbsg$ (triggered by a very special choice of the $\delta_{23}$
parameter in a narrow window range) and small $\alpha_{eff}$
scenario (independent of assumptions on $\delta_{23}$), we reach
an over-optimistic situation where $\Bhzbsmax$ could reach the
$\sim 10\%$ level. This is the case referred to as
small-$\alpha_{eff}$/ \textit{window} in Table~\ref{tab:conclu}.
  \label{sc:loopfine}
  \item If $\tb$ is low/moderate, then $\Bhbsmax$ lie in the lower
  range $\sim 10^{-4}$, which can grow an order or magnitude for
  $\BHAbsmax$ in fine-tuned scenarios (last three rows in
  Table~\ref{tab:conclu}).
\end{enumerate}

Although the large FCNC rates mentioned in points
\ref{sc:noloopfine} and \ref{sc:loopfine} above seem to offer a
rather tempting perspective, we will not elaborate on them any
further since in our opinion the fine-tuning requirement inherent
in them is too contrived. 
On the other hand, points~\ref{sc:noloopnofine}
and~\ref{sc:loopnofine} offer a moderate, but certainly much more
realistic scenario, which in no way frustrates our hopes to
potentially detect the FCNC Higgs boson decays (\ref{hbsFCNC}). 
Indeed, in the case described in point \ref{sc:noloopnofine}, $\Bhbs$
can be at most of order $10^{-4}$.
But this is still a fairly
respectable FCNC branching ratio (comparable to that of
$b\rightarrow s\gamma$) and it may lead to a large number of
events at a high luminosity collider\, as seen in chapter
\ref{cha:hbsSUSY_prod} for the MSSM and the chapter \ref{cha:tch2HDM} for the
general 2HDM.
Moreover, if $\Gamma(h\to X)$ becomes suppressed (e.g. by
realizing the small $\alpha_{eff}$ scenario, point~\ref{sc:loopnofine})
then $\Bhzbs$ can be 
enhanced {by} an additional order of magnitude. 

{Our analysis correlates the values of $\Bhbs$ with that of
  $\Bbsg$, taking into account only the SUSY-QCD contributions due to
  flavor mixing parameters among the left-chiral squarks. The presence
  of several other competing contributions to $\Bbsg$ alters the borders
  of the allowed parameter space:}
\begin{itemize}
\item {For
  the fine-tuned scenarios, the presence and
  position of the allowed \textit{window} regions in the parameter space
  depends significantly on all the contributions, and therefore also
  does the maximum value of $\Bhbs$. Outside the \textit{window} regions,
  the computed value of $\Bbsg$ can only be made consistent with the
  experimental range, by means of a large splitting between the squark
  and gluino  masses.}
\item {For the non-fine-tuned scenarios, the
  inclusion of further contributions to $\Bbsg$ also alters the allowed
  parameter space, but the condition of non-fine-tuning ensures
  precisely that the change in the allowed range of $\delta_{23}$ is
  smooth, and the corresponding change in $\Bhbsmax$ is not dramatic. }
\end{itemize}

Of course, the question immediately arises on what will happen if
the data from present $B$-meson factories further constrains the
$\delta_{23}$ parameter. In that case, we should take into
account the (charged-current induced) SUSY-EW contributions to
$\Bhbs$, see \cite{\Madridt}). However, we can advance that
the SUSY-EW effects on $\Bhbsmax$ that we find are in
the ballpark of $\Bhzbsmax\sim 3 \times 10^{-5}$ and
$\BHAbsmax\sim 1\times10^{-5}$ for a non-fine-tuned scenario,
{while} $\Bhzbsmax \sim 2\times 10^{-4}$ and $\BHAbsmax\sim
8\times 10^{-5}$ for a fine-tuned scenario. {From the analysis of
the chapter \ref{cha:htc2HDM} we expect that even with these
impoverished MSSM rates the number of FCNC events of that sort
should be non-negligible at the LHC.}

Even though we have detected the existence of corners of
the MSSM parameter space where a Higgs boson FCNC branching ratio
can barely reach the $10\%$ level (cf. the narrow windows in
Fig.\,\ref{fig:d23bsg} and Fig.\,\ref{fig:escletxes}), we insist
once more that they should be considered rather unlikely as they
are associated to fine tuning of the parameters and excluded
experimentally\cite{Gambino:2004mv}. Moreover, in
contrast to Ref~\cite{\Madrid}, we find that it is the lightest
CP-even state, $h^0$, the one that could have the largest FCNC
branching ratio.

To conclude, we have presented a first realistic estimate of
the branching ratios of the Higgs boson FCNC decays (\ref{hbsFCNC})
within the MSSM, assuming that the SUSY-QCD corrections can be as
large as permitted by the experimental constraints on $\Bbsg$.
{We have carried out a systematic and self-consistent
maximization of the branching ratios (\ref{eq:hbs-def}) taking
into account this crucial experimental constraint. At the end of
the day the results that we obtain, especially for the lightest
CP-even Higgs boson of the MSSM, are fairly large: $\Bhzbsmax\sim
10^{-4}-10^{-3}$}. These MSSM rates turn out to be {between three
to four orders of magnitude} larger than the maximum SM rate
(\ref{estimateBRSMhbs}), but not five or six orders as naive
expectations indicated. Whether this branching ratio is
measurable at the LHC~\cite{\LHC} or at a high energy $e^+e^-$
Linear Collider~\cite{\TESLA} can only be established by means of
specific experimental analyses. However, on the basis of related
studies in the general 2HDM (chapter \ref{cha:htc2HDM}) and from the MSSM
(chapter \ref{cha:hbsSUSY_prod}), we can foresee that an important
number of FCNC events (\ref{hbsFCNC}) can be potentially collected
at the LHC. {They could play a complementary, if not decisive,
role in the identification of low-energy Supersymmetry}. In this
work we have dealt only with the maximum rates induced by the
SUSY-QCD sector of the model.

%% file: hbsSUSY_prod/hbsSUSY_prod.tex
\chapter{Production and FCNC decay of MSSM Higgs bosons into heavy
  quarks in the LHC}
\label{cha:hbsSUSY_prod}
\section{Introduction}
The rareness
of the FCNC Higgs boson decay modes, as seen in the previous
chapter~\ref{cha:hbsSUSY_br} are an ideal laboratory to look for non-standard
interactions superimposed onto the SM ones.  Similar
considerations apply to the FCNC processes associated to the $Hbs$
vertex, but in this case it is more difficult to pin down the
phenomenological signatures. Some work along these lines has
already been done, both in the MSSM (see chapter \ref{cha:hbsSUSY_br} and
Refs.~\cite{Guasch:1999jp,Guasch:1997kc,Guasch:1999ve,
Curiel:2002pf,Demir:2003bv,Curiel:2003uk,Heinemeyer:2004by})
and in the general two-Higgs-doublet model
(2HDM)\ (see chapters \ref{cha:tch2HDM} and \ref{cha:htc2HDM} and Refs.~\cite{Bejar:2001sj,Arhrib:2004xu}),
and also in other extensions of the SM-- see
\,\cite{Aguilar-Saavedra:2004wm} for a review. Up to now, the
main effort has been concentrated in computing the FCNC decay
modes at one-loop within the new physics, and also in getting a
realistic estimate of the maximum branching ratios expected. It
is not enough to compute the FCNC branching ratios in, say the
MSSM, and then evaluate them in some favorable region of the
parameter space, for one has to preserve at the same time the
stringent bounds on other observables in which the same physics
can be applied, like the aforementioned low-energy $b\rightarrow
s\gamma$ decay. This kind of correlated study was done very
carefully in chapter~\ref{cha:hbsSUSY_br} for the specific Higgs boson
FCNC decays into bottom quarks within the MSSM, $h\rightarrow
b\,s\ (h=h^0,H^0,A^0)$. In this work we extend the latter work by
computing also the top quark Higgs boson FCNC decay modes of the
heavy MSSM Higgs bosons, $h\rightarrow t\,c\ (h=H^0,A^0)$ under
the same restrictions (recall that $h^0$ cannot participate in
this decay because $\mh<m_t$ in the
MSSM\,\cite{Carena:2002es}). Furthermore, in this work we carry
out an additional step absolutely necessary to make contact with
experiment, namely we combine the FCNC decay branching ratios of
the MSSM Higgs bosons (into both top and bottom quarks) with
their MSSM production cross-sections in order to estimate the
maximum number of FCNC events expected at LHC energies and
luminosities. Only in this way one can assess in a practical way
the probability of detecting such processes at the LHC.
This computation was done for the general 2HDM\ in chapter~\ref{cha:htc2HDM},
but to the best of our knowledge the
corresponding calculation in the MSSM case is not available. In
this work we perform this calculation and compare the FCNC
results obtained for the MSSM and 2HDM scenarios.

The relevant observable quantity on which we shall focus hereafter
is the cross-section for the production of electrically neutral
pairs of heavy quarks of different flavors at the LHC, whose
origin stems from the FCNC decays of the neutral Higgs bosons of
the MSSM, $h=h^0,\,H^0,\,A^0$\,\cite{\Hunter,Carena:2002es}. Thus,
we aim at the quantity
\begin{eqnarray}
        \sigmapphqq
        &\equiv&
        \sigma(pp\to h X)B(\hqq)\equiv \sigma(pp\to h X) \frac{\Gamma(\hqq)}{\Gamma(h\to X)}\nonumber\\
        &\equiv& \sigma(pp\to h X) \frac{\Gamma(h\to
          q\,\bar{q'}+\bar{q}\,q')}{\sum_i \Gamma(h\to X_i)}\ \ \
        (qq'\equiv bs \mbox{ or } tc )\,.
    \label{eq:hqq-def}
\end{eqnarray}
Here $\Gamma(h\to X)$ is the -- consistently computed -- total
width in each case. In order to asses the possibility to measure
these processes at the LHC, we have performed a scan of the MSSM
parameter space to find the maximum possible value of the
production rates~(\ref{eq:hqq-def}) under study. The computation
of the combined production rate is necessary, since the
correlations among the different factors are important. For
example, in chapter~\ref{cha:hbsSUSY_br} it was shown that the maximum
branching ratio for the lightest MSSM Higgs boson, $B(h^0\to
b\bar{s})$, is obtained in the regions of the parameter space
where the coupling $h^0\,b\,\bar{b}$ is strongly suppressed by
quantum effects. On the other hand, the associated production
$\sigma(pp\to h^0 b\bar{b})$ is one of the leading processes for
the production of the lightest MSSM Higgs boson. It is clear then
that, in the regions where $B(h^0\to b\bar{s})$ is largest,
$\sigma(pp\to h^0 b\bar{b})$ will be suppressed. Therefore, the
maximum FCNC production rate at the LHC can only be obtained by
the combined analysis of the two relevant factors in
(\ref{eq:hqq-def}) (viz. the branching ratio and the Higgs boson
production cross-section). We will see that the effects from each
factor are different in different regions of the parameter space.
Moreover, the realistic production FCNC rates (\ref{eq:hqq-def})
in the MSSM parameter space can be obtained only by including the
restrictions imposed by the simultaneous analysis of the
branching ratio of the low-energy process $b\to s\,\gamma$, whose
range of values is severely limited by experiment~\cite{\bsgexp}.
As in chapter~\ref{cha:hbsSUSY_br}, in this work we limit ourselves to
supersymmetric FCNC interactions mediated by the strongly
interacting sector of the MSSM, i.e. the SUSY-QCD
flavor-violating interactions induced by the gluinos. The
corresponding analysis for the electroweak supersymmetric FCNC
effects requires a lengthy separate presentation, and will be
reported elsewhere\,\cite{\FCNCEW}.

The work is organized as follows. In Section
\ref{sec:hbsSUSY_prod:numerical-analysis} we describe the general setting for
our numerical analysis. In Section \ref{sec:analysis-bs} we present the LHC
production rates of Higgs boson decaying into bottom quarks through
supersymmetric FCNC interactions. In Section~\ref{sec:analysis-tc} we present
the corresponding FCNC rates for the top quark channel. Finally, in
Section~\ref{sec:hbsSUSY_prod:conclusions} we compare the MSSM results with
the 2HDM results, and deliver our conclusions.

\section{General setting for the numerical analysis}
\label{sec:hbsSUSY_prod:numerical-analysis}

We have performed the calculations with the help of the numeric
programs \texttt{HIGLU},
\texttt{PPHTT}~\cite{Spira,Spira:1995mt,Spira:1995rr,Spira:1997dg}
and LoopTools~\cite{Hahn:1998yk,LTuser,vanOldenborgh:1989wn}. The
calculation must obviously be finite without renormalization, and
indeed the cancellation of UV divergences using either
dimensional regularization or dimensional reduction -- the two
methods giving the same results here -- in the total amplitudes
was verified explicitly. In the following we will detail the
approximations used in our computation:
\begin{itemize}
\item We include the full one-loop SUSY-QCD contributions to the FCNC partial
  decay widths $\Ghbs$ in the observable (\ref{eq:hqq-def}).
\item We assume that FCNC mixing terms appear only in the
  {LH-chiral} sector of the $6\times 6$ squark mixing matrix. Therefore, this matrix has only
  non-diagonal blocks in the LH-LH sector. This is
  the most natural assumption from the theoretical point of view\,\cite{Duncan:1983iq}, and,
  moreover, it was proven in Ref.~\cite{Guasch:1999jp} that the presence of FCNC terms in the
  {RH-chiral sector} would enhance the partial widths by a factor two
  at most -- not an order of magnitude.
\item The Higgs sector parameters (masses and CP-even mixing angle
  $\alpha$) {have} been treated using the leading $\mt$ and $\mb\tb$
  approximation  to the one-loop result~\cite{\Dabels}. For comparison, we also
  perform the analysis using the tree-level approximation.
\item The Higgs bosons total decay widths $\Gamma(h\to X)$ are computed
  at leading order, including all the relevant channels: $\Gamma(h\to
  f\bar{f},ZZ,W^+W^-,gg)$. The off-shell decays
  $\Gamma(h\to ZZ^*,W^{\pm}W^{\mp*})$ have also been
  included. This is necessary to consistently compute the
  total decay width of $\Gamma(h^0\to X)$ in regions of the parameter
  space where the maximization of the cross-section (\ref{eq:hqq-def})
  is obtained at the expense of greatly diminishing the partial decay widths of the two-body
  process $h^0\rightarrow b\bar{b}$ (due to dramatic quantum effects that may reduce
  the CP-even mixing angle  $\alpha$ to small values~\cite{Carena:2002qg}).
  The one-loop decay rate $\Gamma(h\to gg)$
  has been taken from~\cite{Spira:1995rr} and the off-shell decay
  partial widths have been recomputed explicitly and found perfect agreement with the old
  literature on the subject\,\cite{Keung:1984hn}.
\item The MSSM Higgs boson production cross-sections at the LHC have been computed
using the programs  \texttt{HIGLU 2.101} and \texttt{PPHTT
1.1}~\cite{Spira,Spira:1995mt,Spira:1995rr,Spira:1997dg}. These
programs include the following channels: gluon-gluon fusion
$\sigma(pp(gg)\to h)$, associated production with top-quarks
$\sigma(pp \to h t\bar{t})$ and associated production with
bottom-quarks $\sigma(pp\to h b\bar{b})$. In order to have a
consistent description, we have used the leading order
approximation for all channels. The corresponding Feynamn diagrams are
depicted in~\ref{fig:ggh}.  The QCD renormalization scale is
set to the default values for each program, namely $\mu_0=\Mh$ for
\texttt{HIGLU} and  $\mu_0=(\Mh+2 M_Q)/2$ for \texttt{PPHTT}. We
have used the set of CTEQ4L Parton Distribution
Functions~\cite{Lai:1996mg}.
\end{itemize}
Running quark masses ($m_q(Q)$) and strong coupling constants
($\alpha_s(Q)$) are used throughout, with the renormalization scale set
to the decaying Higgs boson mass in the decay processes.
More details are given
below, as necessary.

Using this setup, we have performed a maximization of the FCNC
cross-section, Eq.~(\ref{eq:hqq-def}), in the MSSM parameter space
with the following restrictions on the parameters:
\begin{equation}
    \begin{array}{c|c|c}
        qq'&bs&tc\\\hline\\
        \deltatt&\multicolumn{2}{c}{<10^{-0.09} \simeq 0.81}\\
        \tan\beta & 50 & 5\\
        A_t&-300 \GeV&|A_t|\leq3\msusy\\
        A_b&|A_b|\leq3\msusy&300 \GeV\\
        \mu&\multicolumn{2}{c}{(-1000 \cdots 1000) \GeV}\\
        m_{\tilde{q_i}}&\multicolumn{2}{c}{\msdl=\msdr=\msur=\mg\equiv\msusy}\\
        \msusy&\multicolumn{2}{c}{(150 \cdots 1000) \GeV}\\
        \mA& \multicolumn{2}{c}{(100\cdots 1000)\GeV}\\
         M_{\tilde{q_i}}& \multicolumn{2}{c}{2\,M_{\tilde{q_i}}>\mH+
         50\GeV}\\
         &
         \multicolumn{2}{c}{M_{\tilde{q}_i}+M_{\tilde{q}_j}>\mA+
         50\GeV \, \, (i\neq j)}\\
        \\\hline
    \end{array}
    \label{eq:scan-parameters}
\end{equation}
Here $m_{\tilde{q_i}}$ are the  LH-{chiral} and RH-chiral squark
soft-SUSY-breaking mass parameters, $m_{\tilde{q}_{L,R}}$, common
for the three generations; $M_{\tilde{q_i}}$ are the physical
masses of the squarks, and $\Mh$ is the mass of the decaying
Higgs boson $h=h^0,H^0,A^0$. These masses are fixed at the
tree-level by the values of $(\tb,\mA)$ and the SM gauge boson
masses and couplings\,\cite{\Hunter}. Due to the structure of the
Yukawa couplings in the MSSM, the value of $\tb$ is fixed at a
high (small) value for the bottom (top) quark channel as
indicated. The parameter $\mA$ (the mass of the CP-odd Higgs
boson) is assumed to vary in the range indicated in
(\ref{eq:scan-parameters}). At one loop these masses receive
corrections from the various SUSY fields, and therefore depend on
the values of the remaining parameters in
Eq.\,(\ref{eq:scan-parameters}). The characteristic SUSY mass
scale $\msusy$ defines the typical mass of the squark and gluino
masses\,\footnote{Our programs are able to deal with completely
arbitrary masses for each squark, but we are forced to make some
simplifications in order to provide a reasonable analysis within
a manageable total CPU time, see below.}. The rest of the
parameters of the squark sector are determined by this setup. For
instance, by $SU(2)$ gauge invariance we have $\msul=\msdl$.
Following the same notation as in~\cite{\GuaschNPo}, the
  parameter $\delta_{23}$  represents the mixing between the second
  and third generation of LH-chiral squarks. Let us recall its definition:
\begin{equation}
    \delta_{23}\equiv \frac{m^2_{\sbottom_L
        \sstrange_L}}{m_{\sbottom_L} m_{\sstrange_L}}\,,
\end{equation}
$m^2_{\sbottom_L \sstrange_L}$ {being} the non-diagonal term in
the squark mass matrix squared mixing the second and third
generation of LH-chiral squarks -- and an equivalent definition
for the up-type quarks.\footnote{Recall that the $\delta_{ij}$
  parameters in the up-sector are related to the corresponding
  parameters in the down-sector by the Cabibbo-Kobayashi-Maskawa matrix,
  see e.g.\cite{Gabbiani:1996hi,Misiak:1997ei}.}
The parameter $\delta_{23}$ is a fundamental quantity in our
analysis as it determines the strength of the tree-level FCNC
interactions induced by the supersymmetric strong interactions,
which are then transferred to the loop diagrams of the Higgs
boson FCNC decays in Eq.\,(\ref{eq:hqq-def}). The last two
restrictions in Eq.~(\ref{eq:scan-parameters}) ensure that the
(heavy) Higgs boson decay channels into a pair of squarks are
kinematically forbidden. We have checked explicitly for some of
the heavy Higgs boson channels ($h^0$ can never do it in practice)
that we obtain the same results if we remove these conditions and
include  the partial widths $\Gamma(h\to \tilde{q}\tilde{q}^*)$
in the denominator of (\ref{eq:hqq-def}). Strictly speaking this
condition could be implemented, in the case of the $H^0$ boson,
by just requiring $\mH<2\,M_{\tilde{q_i}}$, but we have made it
stronger by including an additive term. This term is arbitrary
(provided it is not very small) and acts as a buffer, namely it
impedes that by an appropriate choice of the squark masses we can
approach arbitrarily close the threshold from above, and
therefore avoids artificial enhancement effects in our loop
calculations (see below). Similarly, for the CP-odd Higgs boson,
the condition expressed in (\ref{eq:scan-parameters}) ensures
that we avoid a similar kind of enhancement. In this case,
however, the condition is a bit different because the
$A^0\,\tilde{q} \tilde{q}^*$ vertex can only exist with squarks
of different chirality types ($A^0\,\tilde{q_L} \tilde{q_R}^*$)
or, equivalently, with different mass eigenstates
($A^0\,\tilde{q_i} \tilde{q_j}^*$).
We have used fixed values for the soft-SUSY-breaking trilinear
couplings $A_t$ and $A_b$ for the $bs$ and $tc$ channels
respectively. Our results are essentially independent of these
values and their signs.

The task of scanning the MSSM parameter space in order to
maximize $\sigmapphqq$ for the various Higgs bosons is quite
demanding and highly CPU-time consuming, even under the
conditions imposed in Eq.~(\ref{eq:scan-parameters}). As stated
in the introduction, our code includes also the restrictions on
the MSSM parameter space due to the experimental constraint on
$\Bbsg$, and therefore contains the full one-loop SUSY-QCD
amplitude for $b\rightarrow s\gamma$ constructed from the FCNC
interactions induced by the gluinos. The scan was carried out
with the help of two entirely different methods. In the first
method we used a systematic procedure based on dividing the
parameter subspace~(\ref{eq:scan-parameters}) into a lattice
which we filled with points distributed in a completely
homogeneous way. The second is a Monte-Carlo based method, first
proposed in~\cite{Brein:2004kh}. We have adapted the well-known
Vegas integration program\cite{Lepage:1977sw} to generate a
sufficient number of ``interesting'' points in our parameter
subspace. The total number of points used in this case was far
smaller than in the first method. Obviously the lattice procedure
gives more accurate results by increasing arbitrarily the total
number of points, but the CPU time becomes prohibitively long for
the whole analysis. This is so even after factoring out in a
suitable way the phase-space integrals of the Higgs boson
production processes, so that these integrals are computed only
once for every fixed Higgs boson mass and for all the MSSM points
of our scan in the parameter subspace (\ref{eq:scan-parameters}).
The second method is comparatively much faster, but it still
involves a quite respectable amount of CPU time for the whole
analysis. We found that the partial results obtained by the two
methods are compatible at the level of $10-20\%$. For the study
of our FCNC processes we consider that this level of accuracy
should be acceptable, and for this reason all of the plots that
we present in this work have been finally computed with the
Vegas-based procedure. This also explains the wiggling appearance
observed in the profiles of the curves presented in Sections 3-4.
For any given abscissa point in each one of these curves, the
corresponding value on the vertical (ordinate) axis is somewhere
within a band whose width lies around $10-20\%$ of the central
value.

A few words on the effects of the $\Bbsg$ constraint in our
analysis are now in order. The SUSY-QCD contribution to $\Bbsg$
can be quite large, in fact as large as the SM one, and with any
sign. This raises the possibility of ``fine-tuning'' between the
two type of contributions in certain (narrow) regions of the
parameter space. As a consequence we could highly optimize our
FCNC rates in these regions without being in conflict with the
experimentally measured $\Bbsg$ band. We have checked that in
these regions the number of  FCNC events can be artificially
augmented by one or two orders of magnitude. Our scanning
procedure indeed finds automatically these fine-tuning domains.
However, we have systematically avoided them in the presentation
of our analysis (for more details see chapter~\ref{cha:hbsSUSY_br}). In all
of our plots, therefore, we show the results obtained for the
non-fine-tuned case only. We adopt $\Bbsg=(2.1-4.5)\times
10^{-4}$ as the experimentally allowed range within three standard
deviations~\cite{\PDG}.

\section{Analysis of the bottom-strange channel}
\label{sec:analysis-bs}

\begin{figure}[tp]
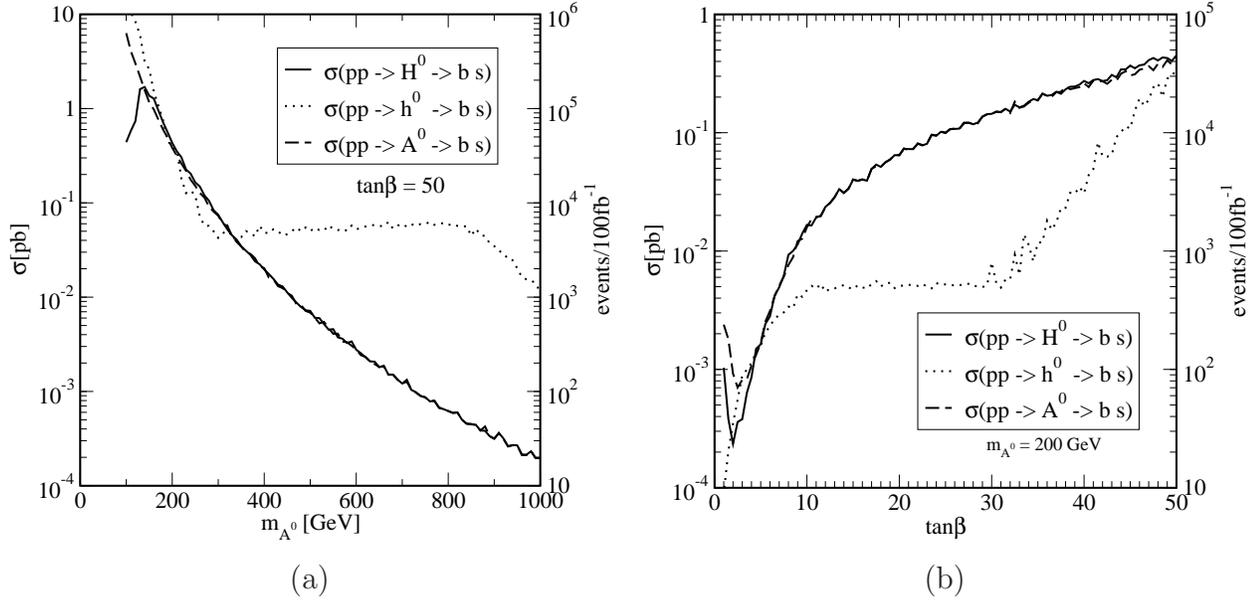

    \begin{tabular}{cc}
        \includegraphics*[height=0.45\textwidth]{hbs_prod_ma} &
        \includegraphics*[height=0.45\textwidth]{hbs_prod_tb} \\
        (a) & (b)
    \end{tabular}
    \caption{Maximum SUSY-QCD contributions to $\sigma(pp\rightarrow
      h\rightarrow b\,s)$, Eq.\,(\ref{eq:hqq-def}), as a function of
      \textbf{(a)} $\mA$ (at fixed $\tb$) and \textbf{(b)} $\tan\beta$ (at fixed $\mA$).
      In each plot the
      left-vertical axis provides the cross-section in pb and the
      right-vertical axis tracks the number of events per $100 \fb^{-1}$
      of integrated luminosity.}
    \label{fig:hbs-prod-ma-tb}
\end{figure}

The main result of the numerical scan for the bottom channel is
shown in Fig.~\ref{fig:hbs-prod-ma-tb}.  To be specific:
Fig.~\ref{fig:hbs-prod-ma-tb}a shows the maximum values of the
production cross-sections $\sigmapphbs$ for the three MSSM Higgs
bosons $h=h^0,H^0,A^0$ at the LHC, as a function of $\mA$;
Fig.~\ref{fig:hbs-prod-ma-tb}b displays the cross-section as a
function of $\tan\beta$. In this plot we indicate the value of
$\sigmapphbs$ (in pb) in the left-vertical axis, and at the same
time we track number of FCNC events (per $100 \fb^{-1}$ of
integrated luminosity) on the right vertical axis. Looking at
Fig.~\ref{fig:hbs-prod-ma-tb} one can see immediately that at
large $\tb$: i) the maximum number of events is remarkably high ($
10^6$ events!) for a FCNC process; ii) there is a sustained
region in the $h^0$ channel, comprising the interval
$300\GeV\lesssim\mA\lesssim900\GeV$, with a flat value of
$5\times 10^3$ events; iii) the chosen value of $\tan\beta=50$ is
not critical for $H^0$ and $A^0$ as long it is larger than $10$.
In Fig.~\ref{fig:hbs-prod-ma-tb}b  we see that the dependence on
$\tb$ is essentially the same for $H^0$ and $A^0$, but for $h^0$
it is quite different: in the region
$10\lesssim\tan\beta\lesssim30$ the cross-section remains below
$10^{-2}\pb$, but for $\tan\beta>30$ it starts climbing fast up
to $0.3\pb$ at $\tan\beta=50$. The number of events here reaches
a few times $10^4$ for all channels (for fixed
$\mA=200\GeV$)\,\footnote{As already advertised, in reading the plots
  and tables in this work, 
  one must keep in mind that they are the result of a Monte-Carlo sampling
  near the region of the maximal values.}.
For further reference, in Table~\ref{tab:hbs-maxims} we show the
numerical values of $\sigmapphbs$ together with the parameters
which maximize the production for $\tb=50$ and $\mA=200\GeV$.
We include the value of $B(h\to bs)$ at the maximization point of
the FCNC cross-section. We notice that at this point the
lightest Higgs boson $h^0$ is the one having the smallest branching
ratio. This is in contrast to the situation when one maximizes the
branching ratios independently of the number of events\,as in
chapter~\ref{cha:hbsSUSY_br}.

\begin{figure}[tp]
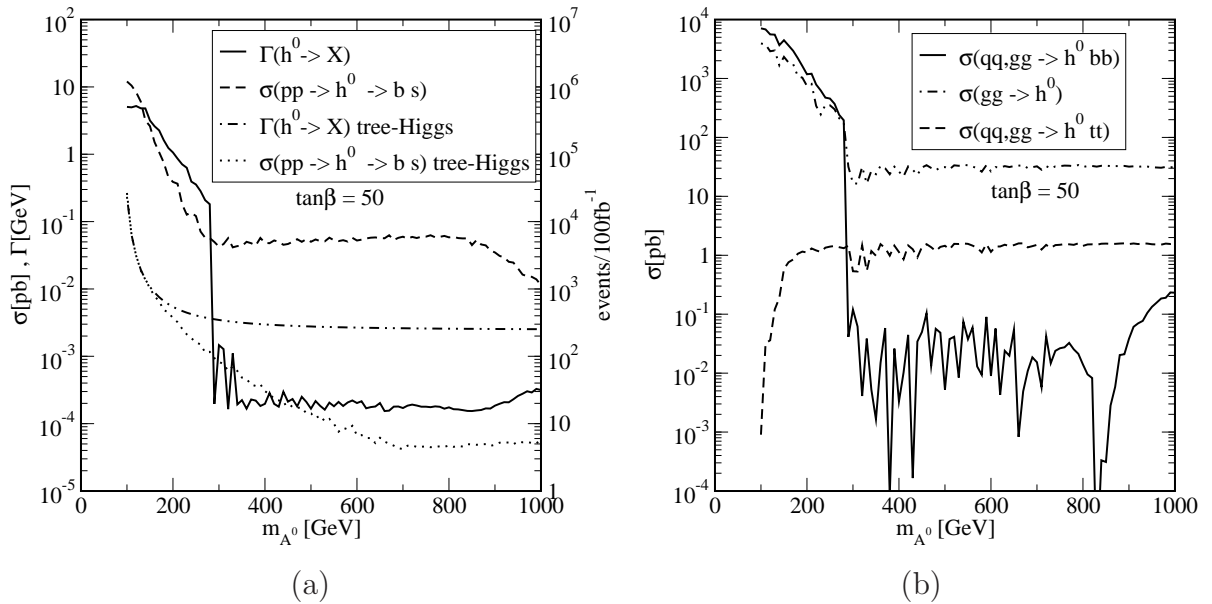

    \begin{tabular}{cc}
        \includegraphics*[height=0.45\textwidth]{hbs_prod_gamma_tree_ma} &
        \includegraphics*[height=0.45\textwidth]{hbs_prods_ma} \\
        (a) & (b)
    \end{tabular}
    \caption{\textbf{(a)} $h^0$ production cross-section and decay width as a
      function of $\mA$ with the Higgs mass relations at tree-level
      and at
      one-loop. \textbf{(b)} Different contributions to the $h^0$
      production cross-section as a function of $\mA$
      corresponding to the maximization of $\sigmapphzbs$ using
      the one-loop Higgs mass relations. }
    \label{fig:hbs-prod-gamma}
\end{figure}

In Fig.~\ref{fig:hbs-prod-gamma}a we show the effect on
$\sigmapphzbs$, and on the total decay width
$\Gamma(h^0\rightarrow X)$, of using the Higgs boson sector at the
tree-level or at one-loop in our computation. It is well-known
that the $h^0$ couplings to quarks are particularly sensitive to
this issue, and for this reason we focus on the lightest CP-even
Higgs boson for these considerations. We can appreciate the
correlations among the different factors that enter the production
rate~(\ref{eq:hqq-def}). The plotted values for $\sigmapphzbs$
and $\Gamma(h^0\rightarrow X)$ in Fig.~\ref{fig:hbs-prod-gamma}a
correspond precisely to the parameters that
maximize~(\ref{eq:hqq-def}) at the tree-level or at one loop in
each case. Fig.~\ref{fig:hbs-prod-gamma}b shows a comparison of
the various $h^0$-production mechanisms for the values that
maximize $\sigmapphzbs$ at one-loop. Remarkably, the effect of the
radiative corrections in the Higgs sector amounts to an
enhancement of our maximal FCNC rates of up to three orders of
magnitude. As we mentioned in chapter~\ref{cha:hbsSUSY_br}
the maximum of $B(h^0\to b\bar{s})$ is
attained under the conditions of the so-called ``small
$\alpha_{\rm eff}$ scenario''\,\cite{Carena:2002qg,Carena:1999bh},
where the two-body decay $h^0\to b\bar{b}$ is strongly suppressed
due to a corresponding suppression of the $h^0\,b\,\bar{b}$
coupling. Since $\Gamma(h^0\to b\bar{b})$ usually dominates the
total width $\Gamma(h^0\to X)$, the latter also diminishes
drastically (at the level of the partial width of a $h^0$
three-body decay, as mentioned above). In this scenario the
production cross-section $\sigma(pp\to h^0 b\bar{b})$ is also
suppressed, so the final result is a compromise between the
suppression of $\Gamma(h^0\to b\bar{b})$ and the possible
enhancement (or at least sustenance) of $\sigma(pp\to h^0\,X)$ by
other mechanisms other than the associated production with bottom
quarks (like the mechanism of gluon-gluon fusion, see
Fig.~\ref{fig:hbs-prod-gamma}b).
\begin{table}
    \center
    \begin{tabular}{|c||c|c|c|}
        \hline
        $h$ &  $H^0$ & $h^0$ & $A^0$ \\\hline\hline
        \sigmapphbs &  $0.45\pb$ & $0.34\pb$ & $0.37\pb$ \\\hline
        events/$100\fb^{-1}$ & $4.5\times10^4$ & $3.4\times 10^4$ & $3.7\times10^4$\\\hline
        $B(h\to bs)$ & $9.3\times 10^{-4} $& $2.1\times 10^{-4} $& $8.9\times10^{-4} $ \\\hline
        $\Gamma(h\to X)$ & $10.9\GeV$ & $1.00\GeV$ & $11.3\GeV$
        \\\hline
        $\delta_{23}$ & $10^{-0.62}$ & $10^{-1.32}$ & $10^{-0.44}$ \\\hline
        $m_{\squark}$ & $990\GeV$ &  $670\GeV$ & $990\GeV$ \\\hline
        $A_b$ & $-2750\GeV$ & $-1960\GeV$ & $-2860\GeV$ \\\hline
        $\mu$ & $-720\GeV$ & $-990\GeV$ & $-460\GeV$ \\\hline
        \Bbsg & $4.50\times 10^{-4}$ & $4.47\times 10^{-4}$ & $4.39\times
        10^{-4}$ \\\hline
    \end{tabular}
    \caption{Maximum value of $\sigmapphbs$ (and of the number of
      $bs$ events per $100\,fb^{-1}$) in the LHC,  for $\mA=200\GeV$ and
      $\tan\beta=50$. Shown are also the corresponding values of the
      relevant branching ratio $B(h\to bs)$ and of the total width of
      the Higgs bosons, together with the values of the SUSY
      parameters. The last row includes $B(\bsg)$.}
    \label{tab:hbs-maxims}
\end{table}

Indeed, for $\mA\lesssim300\GeV$ we can check in
Fig.~\ref{fig:hbs-prod-gamma}b that the most relevant factor for
maximizing the FCNC cross-section is the enhancement of the $h^0$
production channel in association with bottom quarks,
$\sigma(pp\to h^0 b\bar{b})$. This channel operates through the
$b\bar{b}$-fusion vertex $b\bar{b}\rightarrow h^0$ and is highly
enhanced at large $\tb$. In the region $\mA\lesssim300\GeV$ stays
as the dominant mechanism for $h^0$ production, although one can
see that the alternate $gg$-fusion mechanism remains all the way
non-negligible. The corresponding effect on our FCNC cross-section
(\ref{eq:hqq-def}) is nevertheless not obvious because this same
parameter choice does also maximize the total width of $h^0$,
mainly through the enhancement of $\Gamma(h^0 \to b\,\bar{b})$.
There is a delicate interplay of various factors here. In
particular, in the region $\mA<300\GeV$ the maximized partial
width of the FCNC process $h^0\rightarrow bs$ (which is a
function of all the parameters in (\ref{eq:scan-parameters})) is
larger than in the region $\mA>300\GeV$, but at the same time the
total width becomes smaller in the latter region. Overall, the
result is that $\sigmapphzbs$ is larger in the former $\mA$ range
than in the latter. Furthermore, when we cross ahead the limit
$\mA\simeq 300\GeV$ the dominant $h^0$-production channel changes
turn: the associated $h^0$-production with bottom quarks falls
abruptly down (see explanations below) and the gluon-gluon fusion
mechanism takes over, so that in this range the $h^0$-production
cross-section becomes completely dominated by $gg\rightarrow
h^0$. We note that while this mechanism is not so efficient at
its maximum as the associated production, it has a virtue: it is
non-suppressed in the entire range of $\mA$. As a result, in the
region above $\mA\gtrsim 300\GeV$ a small value of the
$h^0\,b\,\bar{b}$ coupling enhances $B(h^0\to bs)$ while it does
not dramatically suppress the total cross-section $\sigma(pp\to
hX)$. For $\mA>300\GeV$ our sampling procedure finds the maximum
by selecting the points in parameter space corresponding to the
small $\alpha_{\rm eff}$ scenario, where $B(h^0\to bs)$ takes the
highest values and $\Gamma(h^0\to X)$ and $\sigma(pp\to h^0
b\bar{b})$ are strongly suppressed -- the latter staying below
the associated Higgs boson production with top quarks!\, In this
way the net Higgs boson cross-section $\sigma(pp\to h^0)$ is not
drastically reduced thanks to the sustenance provided by the
$gg$-fusion channel.
\begin{figure}[tp]
    \begin{tabular}{cc}
        \includegraphics*[height=0.45\textwidth]{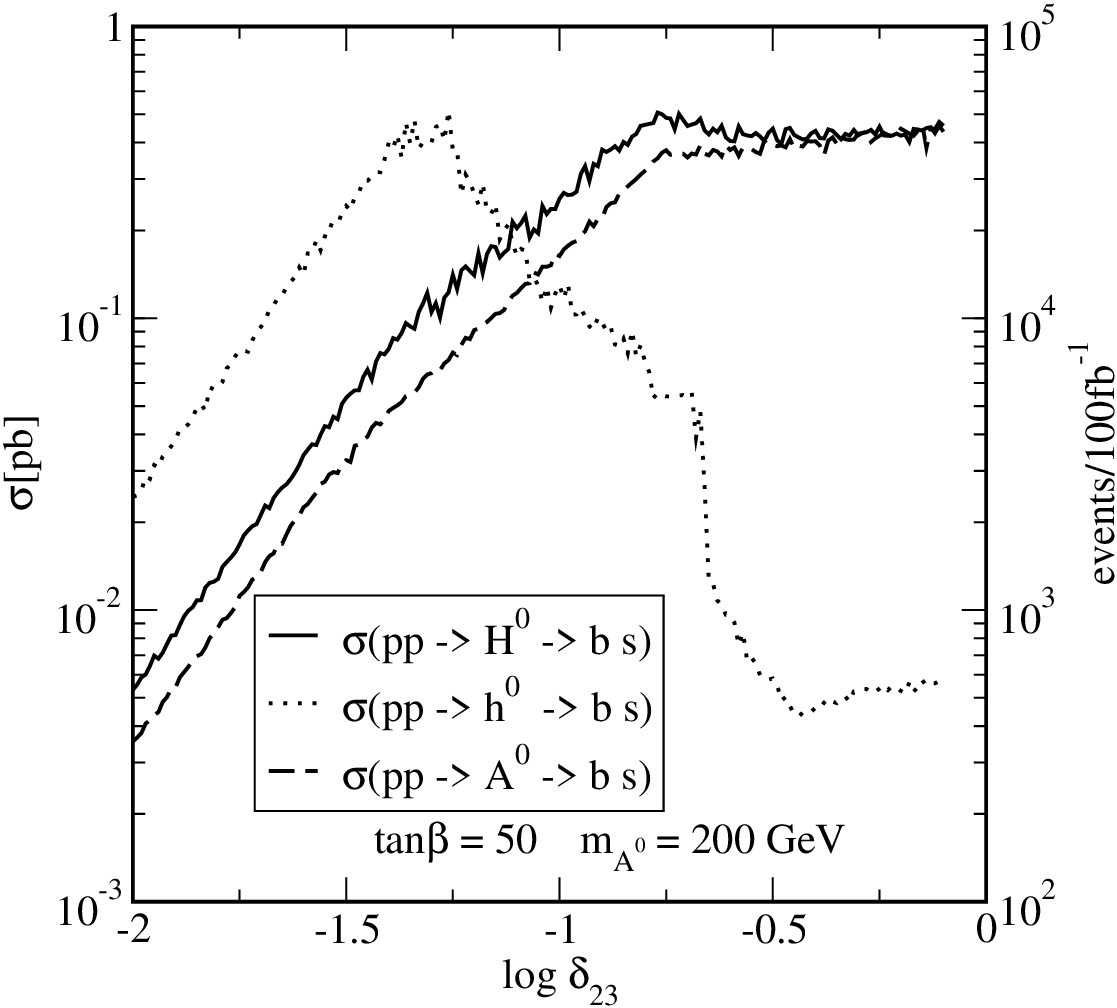} &
        \includegraphics*[height=0.45\textwidth]{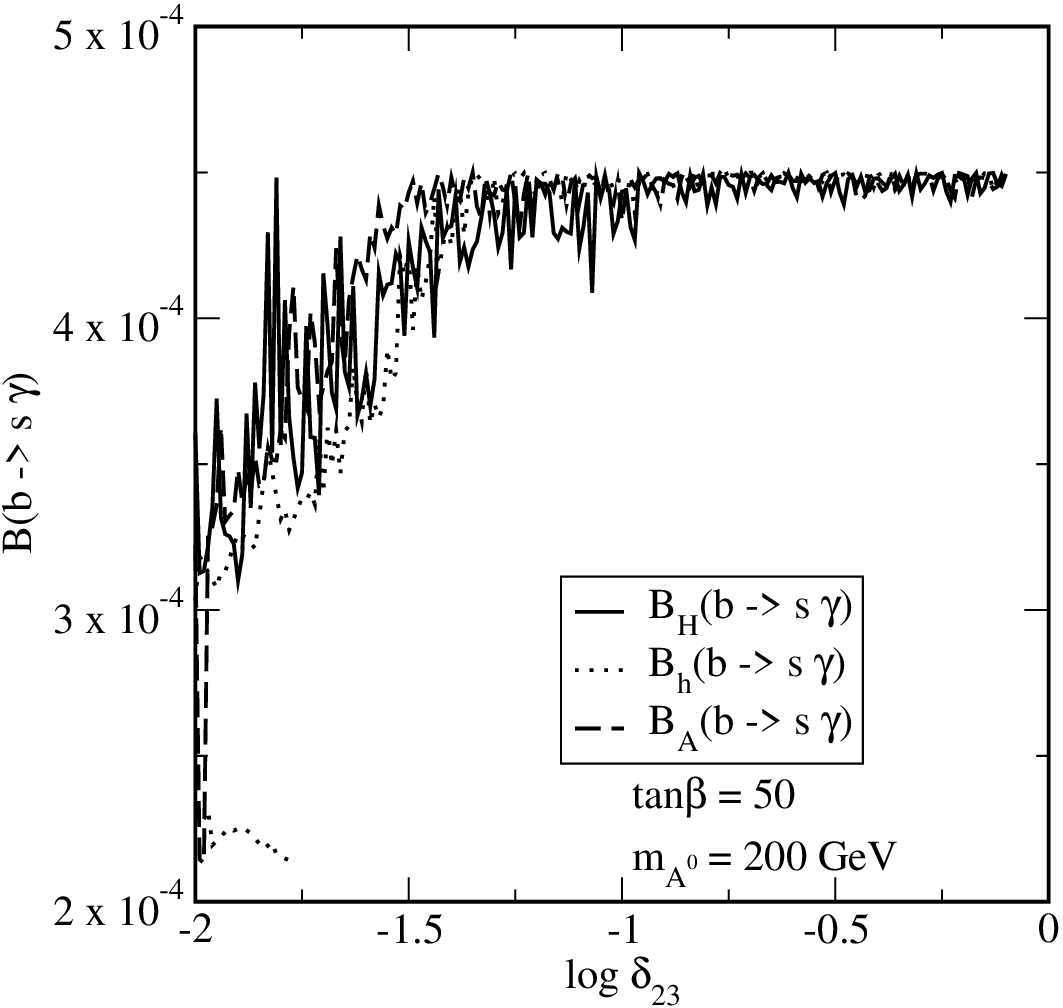} \\
        (a) & (b)
    \end{tabular}
    \caption{Maximum SUSY-QCD contributions to
      $\sigma(pp\rightarrow h\rightarrow b\,s)$, Eq.\,(\ref{eq:hqq-def}), as a
      function of \textbf{(a)}  $\deltatt$ and \textbf{(b)} value of $\bsg$.}
    \label{fig:hbs-prod-bsg-d23}
\end{figure}
Most of the loop contributions to it come from the top quark
because the bottom quark contribution is suppressed and the
squarks are rather heavy. One can clearly see this sustenance
feature in Fig.~\ref{fig:hbs-prod-gamma} in the form of a long
cross-section plateau up to around $\mA=1\TeV$, beyond which the
small $\alpha_{\rm eff}$ scenario cannot be maintained and
$\Gamma(h^0\to X)$ starts increasing and at the same time
$\sigmapphzbs$ starts decreasing. It is remarkable that this
behavior is only feasible thanks to the large radiative
corrections in the MSSM Higgs sector. When we, instead, perform
the computation using the tree-level relations for the Higgs
sector, the small $\alpha_{\rm eff}$ scenario is obviously not
possible and the enhancements/suppressions of $\sigma(pp\to h^0
b\bar{b})$/$\Gamma(h^0\to X)$ cannot take place. As a result the
FCNC rate is some 3 orders of magnitude smaller than in the
previous case.
\begin{figure}[tp]
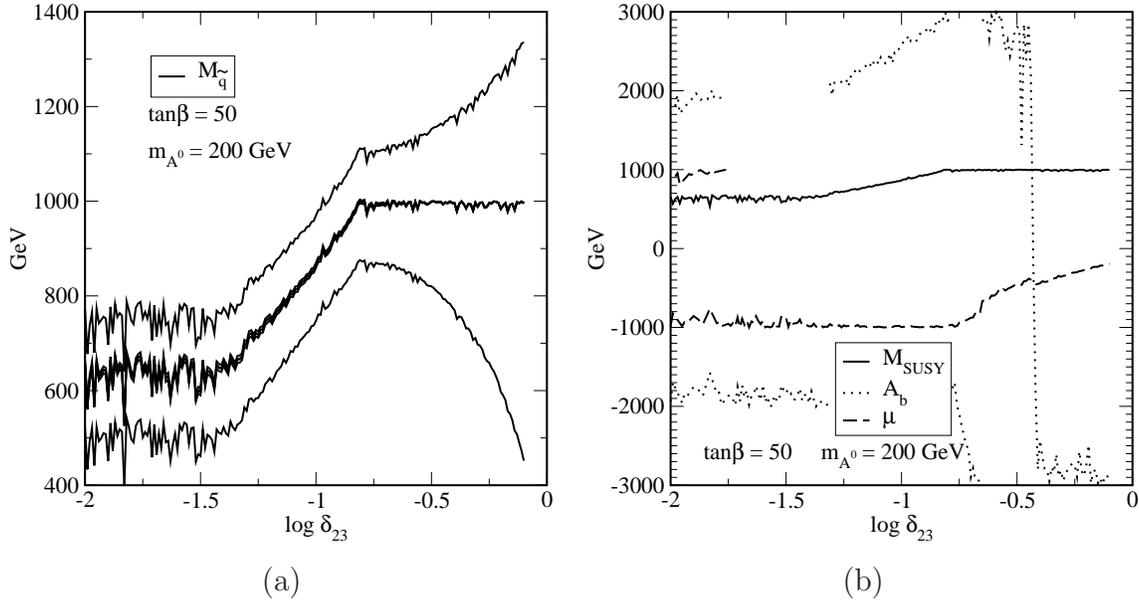

    \begin{tabular}{cc}
        \includegraphics*[height=0.45\textwidth]{hbs_msq_d23} &
        \includegraphics*[height=0.45\textwidth]{hbs_param_d23} \\
        (a) & (b)
    \end{tabular}
    \caption{Value of \textbf{(a)} down-type squark physical  masses
      ($M_{\tilde{q}}$), four of them are degenerate; \textbf{(b)}
      the
      parameters~(\ref{eq:scan-parameters}) from the maximization of
      the $h^0$ channel in Fig.~\ref{fig:hbs-prod-bsg-d23}a.}
    \label{fig:hbs-msq-param-d23}
\end{figure}

Next, we turn our attention to the FCNC mixing parameter
$\deltatt$ in Eq.~(\ref{delta23}). The value of \deltatt\ at the
cross-section maximum is not necessarily the maximum allowed
value of \deltatt\ in (\ref{eq:scan-parameters}). This is because
it is a conditioned maximum, namely a maximum obtained under the
restrictions imposed by $\bsg$, as illustrated in
Fig.~\ref{fig:hbs-prod-bsg-d23}. For further reference in
our discussion, and to better grasp some qualitative features of
our results, let us write the general form of the SUSY-QCD
contribution to $\bsg$. If we emphasize only the relevant
supersymmetric terms under consideration (obviating the powers of
the gauge couplings and other factors) we have
\begin{equation}\label{bsgamma}
B(\bsg)\sim \deltatt^2\,m_b^2(A_t-\mu\tb)^2/\msusy^4\,.
\end{equation}
Fig.~\ref{fig:hbs-prod-bsg-d23}a shows the maximum of
$\sigmapphbs$ as a function of $\deltatt$ for a fixed value of
the CP-odd Higgs boson mass $\mA=200\GeV$, whereas
Fig.~\ref{fig:hbs-prod-bsg-d23}b shows the computed value of
$\Bbsg$ corresponding to the parameter space points where each
maximum is attained. At small $\deltatt$ the SUSY-QCD
contribution to \Bbsg\ is negligible, and the experimental
restriction $\Bbsg=(2.1-4.5)\times 10^{-4}$ does not place
constraints on the other MSSM
parameters~(\ref{eq:scan-parameters}); in other words, in this
region the dependence is $\sigmapphbs\propto (\deltatt)^2$ -- the
naively expected one. Here $\deltatt\lesssim 10^{-1.5}\simeq
0.03$ and the computed $\Bbsg$ value lies well within the
experimental limit. For larger $\deltatt$, $\Bbsg$ can be
saturated at its uppermost experimentally allowed limit, and the
rest of the parameters in~(\ref{eq:scan-parameters}) must change
accordingly in order not to cross that limit. This can be
appreciated in Fig.~\ref{fig:hbs-msq-param-d23} where we show the
range of values taken by the physical down-type squark
masses\footnote{There are six different down-type squarks, but
four of them are nearly degenerate in mass in our approximation.
For the $bs$-channel, the down squarks are so heavy that the
conditions required in the last two rows of
(\ref{eq:scan-parameters}) are automatically satisfied by them in
practically all the allowed range for $\mA$. }
(Fig.~\ref{fig:hbs-msq-param-d23}a) and the lagrangian
parameters~(Fig.~\ref{fig:hbs-msq-param-d23}b) from
Eq.~(\ref{eq:scan-parameters}) that provide the maximum values of
$\sigmapphzbs$ in Fig.~\ref{fig:hbs-prod-bsg-d23}. In the small
$\deltatt$ region ($\deltatt\lesssim10^{-1.5}\simeq 0.03$) the
parameters and masses that maximize $\sigmapphbs$ are constant --
except for the statistical noise unavoidable in a Monte-Carlo
procedure. Note also that there are two possible values for the
parameters $\mu$ and $A_b$, due to the fact that (the leading
contribution of) $\sigmapphbs$ is independent of the sign of
these parameters and the Monte-Carlo procedure picks either sign
for each point with equal probability. At $\deltatt\simeq
10^{-1.5}$ the value of $\Bbsg$ becomes saturated and
$\sigmapphzbs$ reaches its maximum (cf.
Fig.~\ref{fig:hbs-prod-bsg-d23}b); however $\deltatt$ can keep
growing, yet without overshooting the $\Bbsg$ limits, because the
increasing value of $\deltatt$ is compensated by the growing
squark masses (cf. Fig.~\ref{fig:hbs-msq-param-d23}a). But this
is not all that simple, the higher range of $\deltatt$ can be
further divided in two more segments where different dynamical
features occur.  In the first range, namely $10^{-1.5} \lesssim
\deltatt \lesssim 10^{-0.75}$, the heavy Higgs boson channels
keep on increasing their FCNC rates, but not so the lightest
Higgs boson channel $\sigmapphzbs$, the reason being that for
higher squark masses we reach the region where the small
$\alpha_{\rm eff}$ scenario is feasible and hence the $h^0$
couplings become weakened.  The relevant terms of the
cross-section can roughly be written as follows (see
chapter \ref{cha:hbsSUSY_br}):
\begin{equation}\label{sigmahz}
\sigmapphzbs\sim\sigma(pp\rightarrow h^0)\times \deltatt^2
\cos^2(\beta-\alpha_{\rm eff})\, \mg^2\,\mu^2/\msusy^4\,,
\end{equation}
so that for large $\tb$ and small $\alpha_{\rm eff}$ it becomes
reduced. In the second high range of $\deltatt$, i.e. for
$\deltatt\gtrsim 10^{-0.75}\simeq 0.18$, the SUSY mass parameter
$\msusy$ has already reached its allowed maximum value specified
in~(\ref{eq:scan-parameters}), therefore other parameters have to
change to compensate for the larger $\deltatt$.
\begin{figure}[tp]
    \centering
    \includegraphics*[height=0.45\textwidth]{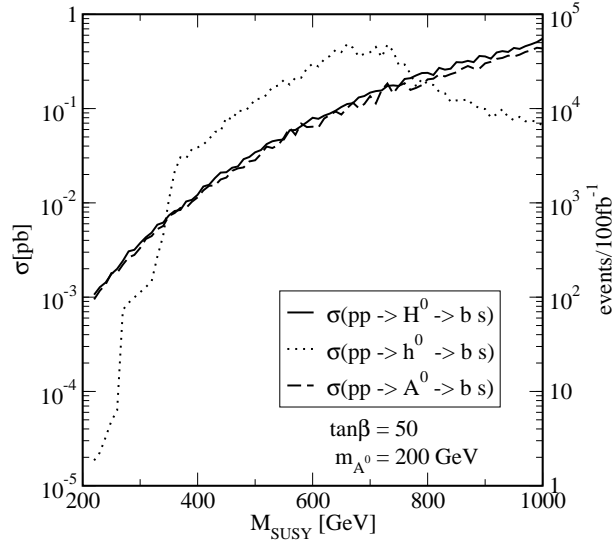}
    \caption{Maximum SUSY-QCD contributions to
      $\sigma(pp\rightarrow h\rightarrow b\,s)$, Eq.\,(\ref{eq:hqq-def}), as a
      function of \msusy.}
    \label{fig:hbs-prod-msusy}
\end{figure}
This is confirmed in Fig.~\ref{fig:hbs-msq-param-d23}b, where for
$\deltatt\gtrsim 10^{-0.75}$ the absolute value of $\mu$
decreases to preserve the $\Bbsg$ upper bound. Correspondingly,
in this region $\sigmapphzbs$ further falls down, as it is patent
in Fig\,\ref{fig:hbs-prod-bsg-d23}a. This additional feature can
also be understood from the approximate expression of the
cross-section given above. At the same time the FCNC rates for
$h=H^0,A^0$ keep further growing, but at a much lower pace. This
is because their (approximate) contribution goes like
(see \eqref{eq:approxleading}:
\begin{equation}\label{sHA}
\sigma(pp\to (H^0,A^0)\to bs)\sim\sigma(pp\rightarrow
H^0,A^0)\times \deltatt^2\, (\sin^2(\beta-\alpha_{\rm
eff}),\,1)\,\mg^2\,\mu^2/\msusy^4\,,
\end{equation}
similar to the $h^0$ case but with angular dependences on
$\alpha_{\rm eff}$ and $\beta$ which are non-suppressing in this
region (see chapter~\ref{cha:hbsSUSY_br}). Again, any further increase of
$\deltatt$ is now partially cancelled by the $\bsg$ constraint,
which demands smaller values of $\mu$. This explains the
stabilization of the FCNC rates of the $H^0$ and $A^0$ channels
in the highest $\deltatt$ range (cf.
Fig.~\ref{fig:hbs-prod-bsg-d23}a). The profile of the squark mass
curves in Fig.~\ref{fig:hbs-msq-param-d23}a implies a mixing mass
matrix with constant diagonal terms (with value $\msusy$) and
growing mixing terms ($\deltatt$).

We finish our analysis of $\sigmapphbs$ by looking at its
behavior as a function of the SUSY mass scale $\msusy$, viz. the
overall scale for the squark and gluino masses -- cf.
Eq.~(\ref{eq:scan-parameters}). Fig.~\ref{fig:hbs-prod-msusy}
shows the maximum of $\sigmapphbs$ as a function of $\msusy$ for
fixed $\mA=200\GeV$ and $\tb=50$. The interpretation of this
figure follows closely the results of the previous ones. At small
values of $\msusy$ the potentially large contribution to $\Bbsg$
has to be compensated  -- see Eq.\,(\ref{bsgamma}) -- by
small values of $\deltatt$ and/or $|\mu|$, resulting in a
(relatively) small value of $\sigmapphbs$. As $\msusy$ grows,
$\deltatt$ and $|\mu|$ can take larger values without disturbing
the restrictions from \Bbsg. The leading contribution to our FCNC
cross-sections is actually independent of the overall SUSY mass
scale $\msusy$, because for all Higgs boson channels we have
found the general behavior (leaving aside other terms mentioned
above)
\begin{equation}\label{generalsigma}
\sigmapphbs\sim\sigma(pp\rightarrow h)\times
\deltatt^2\,\mg^2\,\mu^2/\msusy^4\ \ \  (h=h^0,\,H^0,\,A^0)\,.
\end{equation}
The last factor effectively behaves as
$\deltatt^2\,\mu^2/\msusy^2$, and grows as $\deltatt^2$ for
increasing $\msusy$ at fixed ratio $\mu/\msusy$. Under the same
conditions \Bbsg\ causes no problem because it is additionally
suppressed by $m_b^2/\msusy^2$ -- cf.
Eq.\,(\ref{bsgamma}). Therefore, we are led to a sort of
``non-decoupling behavior'' of the FCNC rates with increasing
$\msusy$. In other words, we find that for the heavy neutral
Higgs bosons ($H^0,A^0$) the interesting region is (contrary to
naive expectations) the high $\msusy$ range!\, The lightest Higgs
boson ($h^0$) channel shows a similar overall behavior, but it
presents additional features because it is more tied to the
evolution of the CP-even mixing angle $\alpha$. The most
interesting region for this channel is (cf.
Fig.~\ref{fig:hbs-prod-msusy}) the central squark mass scale
$\msusy\sim 600-800\GeV$, where the small $\alpha_{\rm eff}$
scenario can take place.

From the combined analysis of
Figs.~\ref{fig:hbs-prod-ma-tb}-\ref{fig:hbs-prod-msusy} we arrive
at the following conclusions concerning the $bs$ final state:
\begin{itemize}
\item A significant event rate of FCNC Higgs boson decays $\sigmapphbs$
  is expected at the LHC, even after taking into account the limits
  on $\Bbsg$;
\item Lightest Higgs boson case, $h^0$:
  \begin{itemize}
  \item  For $\mA\lesssim300\GeV$ the rate $\sigmapphzbs$ decreases
    with $\mA$ but it is the largest in this interval, being
    produced by the combination of a large production cross-section
    $\sigma(pp\to h^0 b\bar{b})$ and a moderate $B(\hzbs)$. It
    amounts to a number of events between $\sim5\times 10^3$ and
    $\sim 12\times 10^5$ for every $100\fb^{-1}$ of integrated luminosity
    at the LHC;
  \item For $300\GeV\lesssim\mA\lesssim 850\GeV$ we expect a maximum
    of $\sim 6 \times 10^3$ events/$100\fb^{-1}$ in the small $\alpha_{\rm eff}$ scenario,
    provided by a large $B(\hzbs)$ and $\sigma(pp(gg)\to h^0)$ as the
    dominant production cross-section;
  \item For $\mA>850\GeV$ the number of events starts to decrease
    slowly;
  \item In all cases, this maximum is attained for a large value of
    $\tb\sim50$, a moderate value of
    the SUSY mass scale ($\msusy\sim 600-800\GeV$) and a \textit{low}
    value of $\deltatt\sim10^{-1.3}\sim0.05$;
  \end{itemize}
\item Heavy Higgs bosons, $H^0, A^0$:
  \begin{itemize}
  \item Although not shown in our plots, we have checked that their production
  rate $\sigma(pp\to H^0\,A^0)$ decreases fast
  with the Higgs boson mass
    (due to the decreasing of the production cross-section).
    We find a maximum FCNC rate of $\sim5\times 10^4$ events for
    $\mA\simeq 200\GeV$, and $20$ events for $\mA\simeq 1\TeV$.
  \item The maximum is produced at i) large $\tb>30$, ii) at the
  highest allowed values of the SUSY mass scale, $\msusy\sim1\TeV$,
  and iii) at a relatively large value of the FCNC mixing
  parameter, $\deltatt\sim 10^{-0.75}\sim0.18$, but not at the largest allowed
    value. The small $\alpha_{\rm eff}$ scenario plays no role in the heavy
    Higgs boson channels.
  \end{itemize}
\end{itemize}
Altogether one should expect a total maximum of some $120,000$
events/$100\fb^{-1}$.

\section{Analysis of the top-charm channel}
\label{sec:analysis-tc}

\begin{figure}[tp]
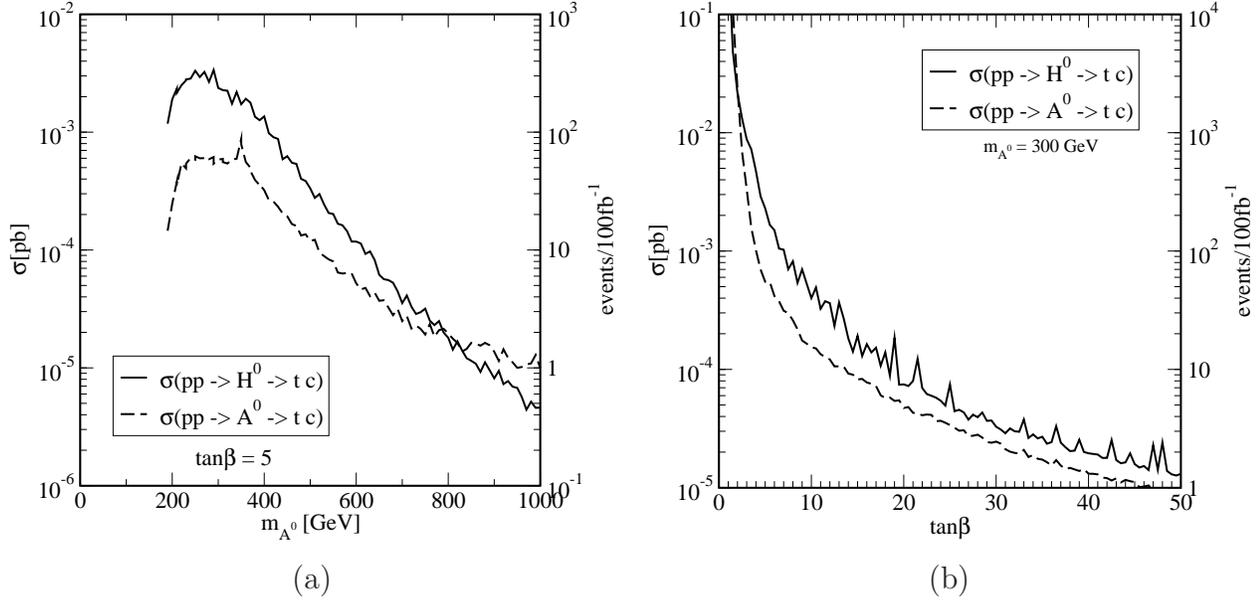

    \begin{tabular}{cc}
        \includegraphics*[height=0.45\textwidth]{htc_prod_ma} &
        \includegraphics*[height=0.45\textwidth]{htc_prod_tb} \\
        (a) & (b)
    \end{tabular}
    \caption{Maximum SUSY-QCD contributions to $\sigma(pp\rightarrow
      h\rightarrow t\,c)$, Eq.\,(\ref{eq:hqq-def}), as a function of
      \textbf{(a)} $\mA$ (at fixed $\tb$) and \textbf{(b)} $\tan\beta$ (at fixed $\mA$).}
    \label{fig:htc-prod-ma-tb}
\end{figure}

The results of the numerical scan for this channel are similar to
the $bs$ channel, so we will focus mainly on the differences.
Fig.~\ref{fig:htc-prod-ma-tb}a shows the maximum value of the
production cross-section $\sigmapphtc$, Eq.\,(\ref{eq:hqq-def}),
under study as a function of $\mA$; Fig.~\ref{fig:htc-prod-ma-tb}b
displays the cross-section as a function of $\tan\beta$.
Obviously the lightest Higgs boson ($h^0$) channel does not
appear in these plots, since in the MSSM this boson is always
lighter than the top quark.
 Looking at Fig.~\ref{fig:htc-prod-ma-tb} one can
see immediately the following: i) the dominant channel in this
case is the heavy scalar Higgs boson, $H^0$; ii) it varies between
$1$ and $300$ events/$100\fb^{-1}$; iii) the $\tan\beta$ value is
critical with preference for low values. In Table~\ref{tab:htc-maxims} we show
the numerical values of $\sigmapphtc$ together with the parameters which
maximize the production for $\tb=5$ and $\mA=300\GeV$.  We have included the
value of $B(h\to tc)$ at the maximization point of the FCNC cross-section. It
is remarkable that for the heavy CP-even Higgs boson one can reach $B(H^0\to
tc)\sim 10^{-3}$ compatible with the $\bsg$ constraint.

Let us remark that for the heavy Higgs boson channels the features
of the small $\alpha_{\rm eff}$ scenario play no significant role
because the partial widths into $b\bar{b}$ are proportional
either to $\cos^2\alpha_{\rm eff}$ (in the $H^0$ case) or to
$\sin^2\beta$ ($A^0$ case). Moreover, in the low $\tb\gtrsim 1$
region (the relevant allowed one for the $tc$ channel) the small
$\alpha_{\rm eff}$ scenario does not even have a chance to take
place.
\begin{table}
    \center
    \begin{tabular}{|c||c|c|c|}
        \hline
        $h$ &  $H^0$ & $A^0$ \\\hline\hline
        \sigmapphtc &  $2.4\times 10^{-3}\pb$ & $5.8\times 10^{-4}\pb$ \\\hline
        events/$100\fb^{-1}$ & 240 & 58  \\\hline
        $B(h\to tc)$ & $1.9\times 10^{-3} $& $5.7\times 10^{-4} $\\\hline
        $\Gamma(h\to X)$ & $0.41\GeV$ & $0.39\GeV$ \\\hline
        $\delta_{23}$ & $10^{-0.10}$ & $10^{-0.13}$ \\\hline
        $m_{\squark}$ & $880\GeV$ & $850\GeV$ \\\hline
        $A_t$ & $-2590\GeV$ & $2410\GeV$ \\\hline
        $\mu$ & $-700\GeV$ & $-930\GeV$ \\\hline
        \Bbsg & $4.13\times 10^{-4}$ & $4.47\times 10^{-4}$ \\\hline
    \end{tabular}
    \caption{Maximum value of $\sigmapphtc$ (and of the number of
      $tc$ events per $100\,fb^{-1}$) in the LHC,  for $\mA=300\GeV$ and
      $\tan\beta=5$. Shown are also the corresponding values of the
      relevant branching ratio $B(h\to tc)$ and of the total width of
      the Higgs bosons, together with the values of the SUSY
      parameters. The last row includes $B(\bsg)$.}
    \label{tab:htc-maxims}
\end{table}

\begin{figure}[tp]
    \begin{tabular}{cc}
        \includegraphics*[height=0.45\textwidth]{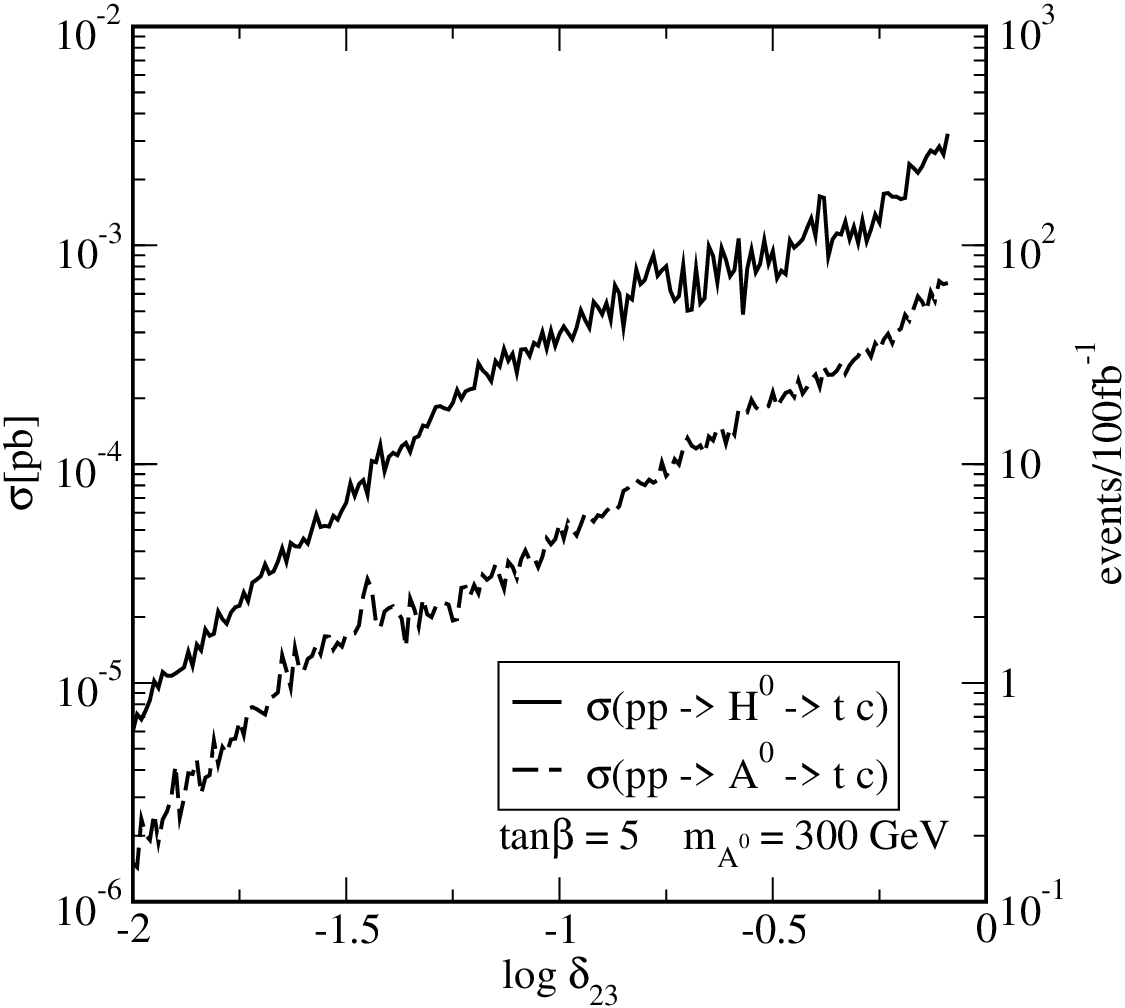} &
        \includegraphics*[height=0.45\textwidth]{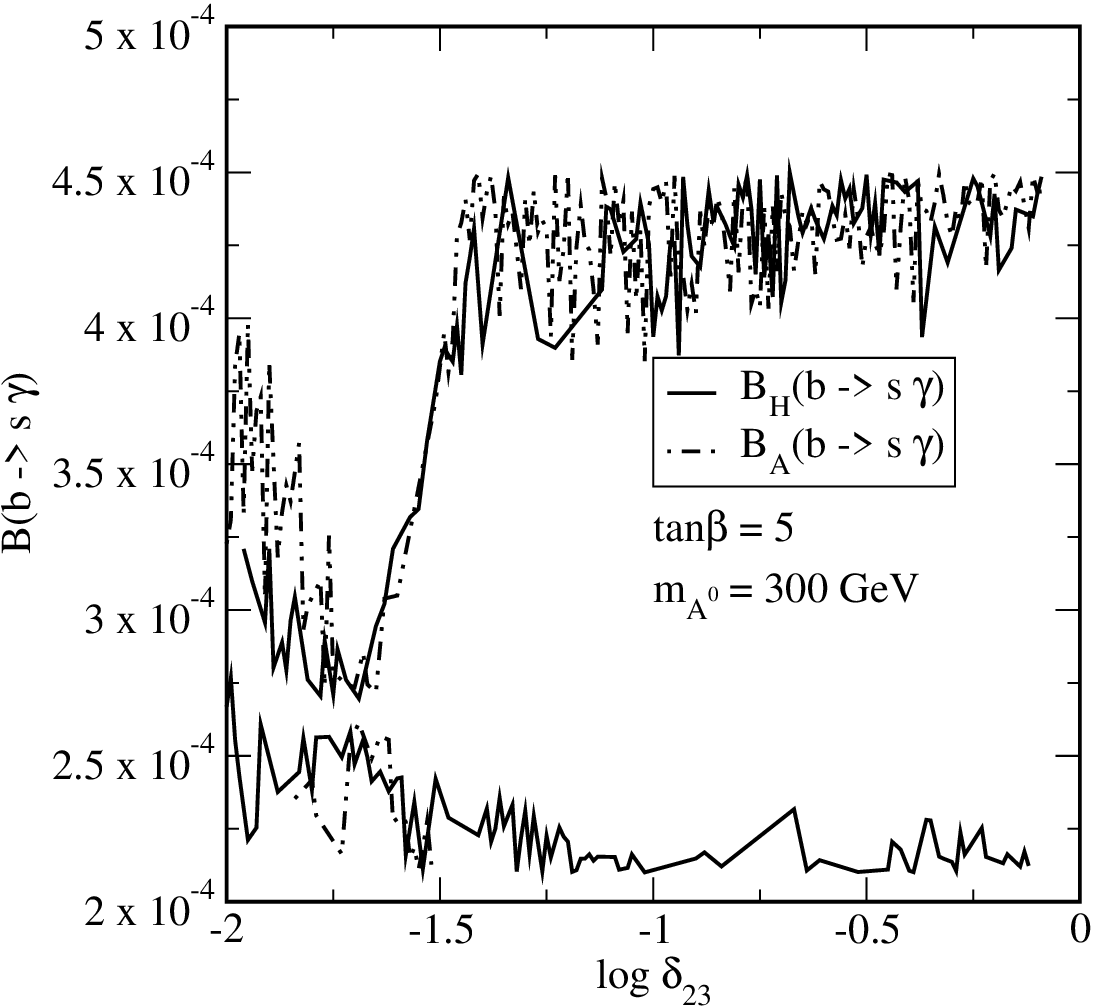} \\
        (a) & (b)
    \end{tabular}
    \caption{Maximum SUSY-QCD contributions to
      $\sigma(pp\rightarrow h\rightarrow t\,c)$, Eq.\,(\ref{eq:hqq-def}), as a
      function of \textbf{(a)} $\deltatt$ and \textbf{(b)} value of $\bsg$, for fixed $\tb$ and $\mA$.}
    \label{fig:htc-prod-bsg-d23}
\end{figure}

\begin{figure}[tp]
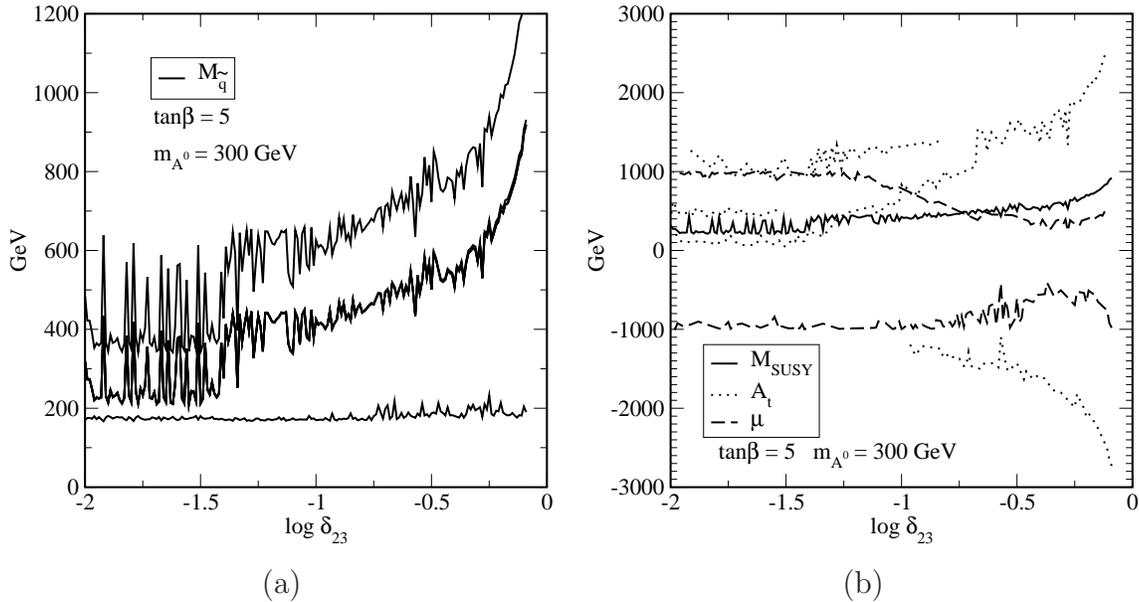

    \begin{tabular}{cc}
        \includegraphics*[height=0.45\textwidth]{htc_msq_d23} &
        \includegraphics*[height=0.45\textwidth]{htc_param_d23} \\
        (a) & (b)
    \end{tabular}
    \caption{Value of \textbf{(a)} up-type squark physical  masses ($M_{\tilde{q}}$),
    four of them are degenerate;   \textbf{(b)} the
      parameters~(\ref{eq:scan-parameters}) from the maximization of
      the $H^0$ channel in Fig.~\ref{fig:htc-prod-bsg-d23}a.}
    \label{fig:htc-msq-param-d23}
\end{figure}

\begin{figure}[tp]
    \centering
    \includegraphics*[height=0.45\textwidth]{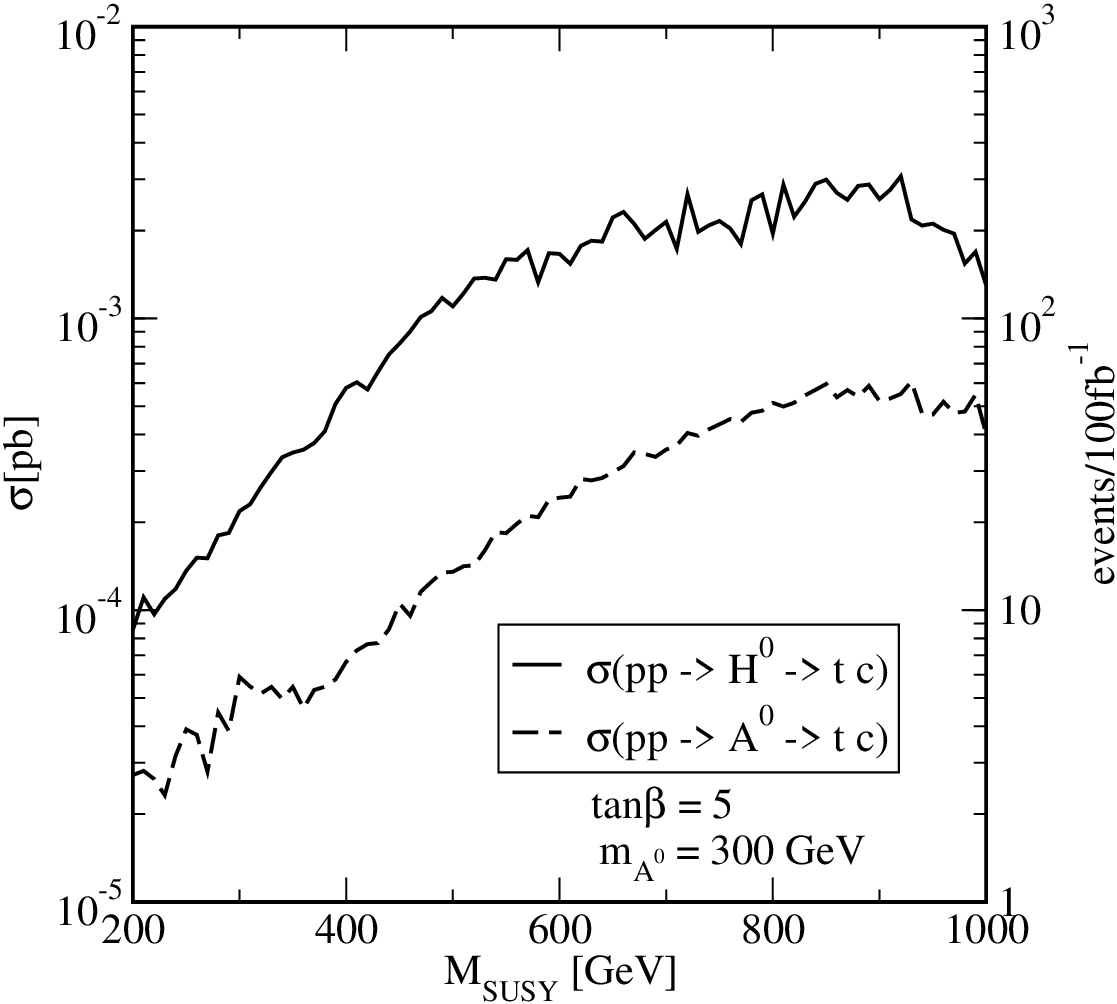}
    \caption{Maximum SUSY-QCD contributions to
      $\sigma(pp\rightarrow h\rightarrow t\,c)$, Eq.\,(\ref{eq:hqq-def}), as a
      function of \msusy.}
    \label{fig:htc-prod-msusy}
\end{figure}

We turn now our view to the role of the $\Bbsg$ restriction in
Figs.~\ref{fig:htc-prod-bsg-d23} and~\ref{fig:htc-msq-param-d23}.
Fig.~\ref{fig:htc-prod-bsg-d23} shows the maximum value of
$\sigmapphtc$ as a function of $\deltatt$ for a fixed value of
$\mA=300 \GeV$ together with the corresponding computed value of
$\Bbsg$, while Fig.~\ref{fig:htc-msq-param-d23} shows the values
of the parameters, Eq.~(\ref{eq:scan-parameters}), that realize
this maximum, together with the physical up-type squark masses.
In this case the $\Bbsg$ restriction is not as critical as in the
$bs$ channel, in part due to the fact that the SUSY-QCD
contribution to $\Bbsg$ is not enhanced at low $\tb$. For
$\deltatt\lesssim10^{-1.5}$ the value of $\Bbsg$ is well inside
the experimental limits (Fig.~\ref{fig:htc-prod-bsg-d23}b), there
is no restriction on the rest parameters
(Fig.~\ref{fig:htc-msq-param-d23}b), the squark masses remain
constant (Fig.~\ref{fig:htc-msq-param-d23}a), and the maximum
value of $\sigmapphtc$ grows here in the naively expected way
$(\deltatt)^2$ (Fig.~\ref{fig:htc-prod-bsg-d23}a).
Above this value ($\deltatt\gtrsim 10^{-1.5}$) the parameters
have to be adjusted to provide and acceptable range for
$\Bbsg$\,\footnote{The two lines appearing in this region for
$B_H(\bsg)$ mean that the maximum of $\sigmappHztc$ is attained
either by the maximum or the minimum allowed value of $\Bbsg$,
our Monte-Carlo sampling procedure picks either choice with equal
probability for each value of $\deltatt$.}. In this region the
SUSY mass $\msusy$ grows (Fig.~\ref{fig:htc-msq-param-d23}b), and
$|\mu|$ decreases, but not so fast as in the  $bs$ channel case
(Fig.~\ref{fig:hbs-msq-param-d23}b).
At the same time $A_t$ increases with increasing
$\deltatt> 10^{-1.5}$. Most of the physical squark masses grow,
but one of the up-type squarks (stop squark) can always have the
minimum allowed mass -- Eq.~(\ref{eq:scan-parameters}). In this
region the observables under study grow more slowly, since
their original $\deltatt^2$ behavior is partially
cancelled by the growing of $\msusy$ with $\deltatt$. In
Fig.~\ref{fig:htc-prod-msusy} we see, again, that $\sigmapphtc$
grows with the SUSY mass scale (although in a way less pronounced
than in the $bs$ channel), due to the relaxation of the $\bsg$
constraint for large $\msusy$. While $\sigmapphbs$ is augmented
nearly three orders of magnitude in the range
$\msusy=200-1000\GeV$ (Fig.~\ref{fig:hbs-prod-msusy}), the $tc$
channel undergoes only an increase of roughly a factor 10 in the
same parameter range.

How do the two Higgs boson channels $H^0$ and $A^0$ compare as
sources of $tc$ events? The $gg$-fusion mechanism is one of the
leading processes for Higgs boson production at relatively small
values of $\tb\gtrsim 1$ -- associated production with $b\bar{b}$
remaining still sizeable. Due to CP conservation, the squark
contributions to $gg\rightarrow A^0$ cancel out at one-loop and
only the quark contributions remain\,\cite{Spira:1997dg}. For
large squark masses the production cross-sections for $H^0$ and
$A^0$ are similar, the latter being slightly larger.  We see from
Fig.\,\ref{fig:htc-prod-ma-tb} that, despite the similarity in
production, the $H^0$ channel gives larger FCNC rates than the
$A^0$ one. The excess of FCNC events from the former can be
explained mainly from the constraints that we have imposed from
the very beginning on the squarks masses in relation to the Higgs
boson masses (see Eq.\,(\ref{eq:scan-parameters})). As we have
noted in section \ref{sec:hbsSUSY_prod:numerical-analysis}, we can have
squarks of the same chirality-type in the $H^0\,\tilde{q}
\tilde{q}^*$ vertex, whereas they must necessarily be of opposite
chirality-type in the $A^0\tilde{q} \tilde{q}^*$ case. As a
result, for small $\mA$ the squark mass constraints expressed in
(\ref{eq:scan-parameters}) allow the FCNC cross-section
maximization process to pick points near the saturation of the
mass condition $2\,M_{\tilde{q_i}}>\mH+ 50\GeV$, but not of
$M_{\tilde{q}_i}+M_{\tilde{q}_j}>\mA + 50\GeV$ for $( i\neq j)$
because only one of the up-squarks can be light (see Fig.
\ref{fig:htc-msq-param-d23}a). This produces an enhancement of
the branching ratio of $H^0\to tc$ and for this reason this FCNC
channel dominates in the relatively small $\mA$ region. However,
as soon as $\mA$ is sufficiently heavy the second mass constraint
can also be satisfied and then the two curves in
Fig.~\ref{fig:htc-prod-ma-tb} tend to converge.

From the combined analysis of
Figs.~\ref{fig:htc-prod-ma-tb}-\ref{fig:htc-prod-msusy}, we
conclude that in the case of the $H^0$ channel we expect a maximum
of $\sim 300$ events/$100\fb^{-1}$ decays into top quarks at the
LHC. This maximum is achieved for a CP-odd Higgs boson mass of
$\mA\sim300 \GeV$, and a moderately low $\tb\sim5$. This rate can
grow one order of magnitude by a lower value of $\tb\sim2$, but
decreases significantly with $\mA$. The maximum is obtained at
the largest possible value of $\deltatt$, and a moderate SUSY
mass scale $\msusy\sim600-800\GeV$, but having one of the squarks
light. While the number of events is significantly lower than the
$bs$-channel ones, the $tc$-channels offers a better opportunity
for detection, due to the much lower background.

\section{Discussion and conclusions}
\label{sec:hbsSUSY_prod:conclusions}

We have carried out a systematic study of the production rate of
FCNC processes at the LHC mediated by the decay of neutral Higgs
bosons of the MSSM: $\sigmapphqq\ (h=h^0,H^0,A^0)$ -- see
Eq.\,(\ref{eq:hqq-def}). Specifically, we have concentrated on
the FCNC production of the heavy quark pairs $qq'=bs$ and $tc$,
because they are the only ones that have a chance of being
detected. We have focused on the FCNC supersymmetric effects
stemming from the strongly interacting sector of the MSSM, namely
from the gluino-mediated flavor-changing interactions. We have
performed a maximization of the event rates in the parameter
space under a set of conditions that can be considered
``irreducible'', see Eq.\,(\ref{eq:scan-parameters}), i.e. we
cannot further shorten this minimal set (e.g. by making additional
assumptions on the relations among the parameters) without
potentially jeopardizing the conclusions of this study. Even
within this restricted parameter subspace the computer analysis
has been rather demanding. The numerical scan has been performed
using Monte Carlo techniques which we have partially
cross-checked with more conventional methods. The maximization of
the cross-sections (\ref{eq:hqq-def}) has been performed by
simultaneously computing the corresponding MSSM quantum effects
on the (relatively well-measured) low-energy FCNC decay $\bsg$
and requiring that the experimental limits on this observable are
preserved.

To summarize our results: the total number of FCNC heavy flavor
events originating from supersymmetric Higgs boson interactions at
the LHC can be large (of order $10^6$), but this does not mean
that they can be easily disentangled from the underlying
background of QCD jets where they are immersed. For example, it
is well known that the simple two-body decay $h\rightarrow
b\,\bar{b}$ is impossible to isolate due to the huge irreducible
QCD background from $b\,\bar{b}$ dijets -- a result that holds
for both the SM and the MSSM\,\cite{\LHC}. This led a long time
ago to complement the search with many other channels,
particularly $h\rightarrow\gamma\,\gamma$ which has been
identified as an excellent signature in the appropriate range.
Similarly, the FCNC Higgs boson decay channels may help to
complement the general Higgs boson search strategies, mainly
because the FCNC processes should be essentially free of QCD
background. Notwithstanding other difficulties can appear, such as
misidentification of jets. For instance, for the $bs$ final
states misidentification of $b$-quarks as $c$-quarks in
$cs$-production from charged currents may obscure the possibility
that the $bs$-events can be really attributed to Higgs boson FCNC
decays. This also applies to the $tc$ final states, where
misidentification of $b$-quarks as $c$-quarks in e.g. $tb$
production might be a source of background to the $tc$ events,
although in this case the clear-cut top quark signature should be
much more helpful (specially after an appropriate study of the
distribution of the signal versus the background).  However, to
rate the actual impact of these disturbing effects one would need
an additional study which is beyond the scope of the present
work.

An interesting (and counter-intuitive) result of our work is that
for all the Higgs boson channels $h=h^0,H^0,A^0$, the FCNC
cross-section $\sigmapphqq$ increases with growing SUSY mass
scale $\msusy$. Due to this effective ``non-decoupling'' behavior
(which is more pronounced for the $bs$ channel) the FCNC rates
are maximal when the overall squark and gluino mass scale is of
order of $\msusy\lesssim 1\TeV$ -- with the only proviso that for
the $tc$-channel a single squark should have a low mass ($\gtrsim
150\GeV$).  Moreover, we find that the two types of FCNC final
states ($bs$ and $tc$) prefer different ranges of $\tb$. The
$bs$-channel is most efficient at high $\tb>30$, whereas the
$tc$-channel works better in the regime of low $\tb<10$ (see
below). As for the mixing parameter $\deltatt$ (which is the
fundamental supersymmetric FCNC parameter of our analysis, see
its definition in (\ref{delta23})) we remark that the maximum
number of events is not always attained for the largest possible
values of it (due to the influence of the $\bsg$ constraint): in
the $bs$-channel the maximum is achieved for moderate values in
the range $0.05\lesssim\deltatt\lesssim 0.2$, whereas the
$tc$-channel prefers the maximum allowed values in our analysis
($\deltatt\lesssim 0.8$). We also remark that the naive
expectation $\sigmapphqq\sim\deltatt^2$ does not always apply.

The number of FCNC events originating from the two channels $bs$
and $tc$ is not alike, and it also depends on the particular
Higgs boson. For the $bs$ final states we have found that the
optimized value of $\sigmapphbs$ produced by our analysis is
$\sim 12\pb$. This amounts to $\sim12 \times 10^5$ events per
$\int {\cal L}dt= 100\fb^{-1}$ of data at the LHC. The most
favorable Higgs boson channel is the one corresponding to the
lightest MSSM Higgs boson, $h^0$. For this boson there are
non-trivial correlations between the two factors in
Eq.~(\ref{eq:hqq-def}), namely between the Higgs boson production
cross-section and the FCNC branching ratio. These correlations
permit an increase of the total number of FCNC events up to two
orders of magnitude in certain cases as compared to the number of
events produced by the heavy Higgs bosons $H^0$ and $A^0$, which
are essentially free of these correlations. The latter stem from
relevant quantum effects on the parameters of the Higgs boson
sector at one-loop precisely in the regions of our interest.

On the other hand, the maximum value of $\sigmapphtc$ is more
moderate, to wit: $3\times 10^{-3}\pb$, or $\sim 300$
events/$100\fb^{-1}$. For the total integrated luminosity during
the operative lifetime of the LHC, which amounts to some
$(300-400)\fb^{-1}$, we estimate that a few thousand $tc$ events
could be collected in the most optimistic conditions. This number
is of course sensitive to many MSSM parameters, but most
particularly to two: $\mA$ and $\tb$. The mass of the CP-odd
Higgs boson should not be heavier than $\mA\sim (400-500)\GeV$ if
one does not want to decrease the number of events below a few
hundred per $100\fb^{-1}$. On the other hand the number of events
is very much dependent on the particular range of $\tb$. As we
have said above, the lowest possible values are preferred for the
$tc$ channel, but the sensitivity in this range is so high that
the order of magnitude of $\sigmapphtc$ may change for different
(close) choices of $\tb$. Throughout all the analysis of the $tc$
channel we have fixed $\tb$ at an intermediate value, but the
maximum number of events per $100\fb^{-1}$ would grow from $\sim
300$ (for our standard choice $\tb=5$) up to $\sim(500,900,2000)$
if we would have chosen $\tb=(4,3,2)$ respectively. Such lower
values of $\tb$ are usually avoided in some MSSM analyses in the
literature, but as a matter of fact there are no fully
water-tight experimental bounds on $\tb$ excluding this lower
range, apart from the more incontrovertible strict lowest limit
$\tb>1$. We recall that the lower limit on $\tb$ is obtained
  indirectly from the LEP exclusion data on the light Higgs boson
  search, and is therefore very sensitive to the
  inputs used in the computation, specially the top quark
  mass~\cite{Heinemeyer:1999zf}.
Unlike the difficulties in the $bs$-channel, and in spite of the
substantially smaller number of events, we deem more feasible to
extract the $tc$ signal at the LHC, due to the presence of the
quark top, which carries a highly distinguishable signature.

At this point a comparison with the previous
chapter~\ref{cha:htc2HDM} is in order. In that work we have
studied in quite some detail the maximum FCNC production rates of
Higgs bosons decaying into $tc$ final states within the general
two-Higgs-doublet model. It was found that the maximal branching
ratio in the 2HDM takes place in the type II 2HDM (or 2HDM II),
and reads $B^{II}(h\rightarrow tc)\sim10^{-5}$, whereas in the
2HDM I it is comparatively negligible. After a detailed
computation of the event rates (including also the particular
restrictions of the $\bsg$ process, which are different in the
2HDM case as compared to the MSSM) the conclusion was that
several hundred $tc$ events could be collected at the LHC under
optimal conditions. We clearly identified which are the most
relevant Higgs boson modes for this purpose and the domains of
the 2HDM II parameter space where these events could originate
from. To make it short, our conclusion in chapter~\ref{cha:htc2HDM}
was that the $h^0$ is the most gifted decay and the ideal
situation occurs when $\tan\beta$ and $\tan\alpha$ are both
large, and also when the CP-odd state $A^0$ is much heavier than
the CP-even ones ($h^0,H^0$). Furthermore we found that in the
general 2HDM the $A^0$ state never gives any appreciable FCNC
rate into $tc$. It is easy to see that the mode $A^0\rightarrow
bs$ is not favored either (unless $\tb$ is very small and
$\mHp$ unusually light, both situations rather
unappealing).

How it compares with the MSSM case under study?  To start with, we
note that here the $A^0$ channel gives essentially the same $bs$
rate as the $H^0$ one, and that both modes can be quite relevant.
At the same time the $A^0$ rate into $tc$ is, though not dominant,
not negligible at all. Some few hundred events of this nature per
$100\fb^{-1}$ are possible. If this is not enough, in the MSSM
the most relevant $\tb$ region for the $tc$ final states is not
the highest one (as in the 2HDM II case) but just the opposite:
the lowest allowed one. This is an important difference, and one
that should help to discriminate between the 2HDM and MSSM models
in case that some $tc$ events would be unambiguously tagged at the
LHC. After all there are many high precision observables that are
highly sensitive to the preferred range of $\tb$, so that the
favorite value of $\tb$ could already be fixed from other
experiments by the time that some FCNC events could be detected.
Apart from the different correlation of the parameters in the
non-supersymmetric and supersymmetric model, in the latter case
the maximal event rates for the $tc$ mode are typically one order
of magnitude higher than the maximal event rates in the former.

The corresponding study for the 2HDM branching ratios into $bs$
final states was performed in~\cite{Arhrib:2004xu}, although in
this work the cross-section and number of events were not
computed. However, from the maximum size of the expected
branching ratios compatible with $\bsg$ (viz.
$B^{II}(h\rightarrow bs)\sim10^{-6}-10^{-5}$ in the 2HDM II) it
is already pretty obvious that the number of events can never be
competitive with the supersymmetric case where
$B^{MSSM}(h\rightarrow
bs)\sim10^{-4}-10^{-3}$\,as seen in chapter \ref{cha:hbsSUSY_br}. As for the 2HDM I
(which is insensitive to the $\bsg$ bounds) the branching ratios
can be at most of order $B^{I}(h\rightarrow bs)\sim
10^{-5}-10^{-3}$ and only so for very small values of
$\tb=0.1-0.5$ which are actually excluded in the MSSM.

Summing up and closing: the maximum number of FCNC events in the
MSSM case is larger than the highest expected rates both in the
2HDM I and II , the two kind of signatures of physics beyond the
SM being perfectly distinguishable because the relevant regions
of the parameter space are completely different. If a sample of
FCNC events of this kind could be collected, we should be able to
ascertain which is its ultimate origin.  At the end of the day if
one single thing should be emphasized is that the FCNC event rate
into $bs$ or $tc$ is so extremely tiny in the SM that if only a
dozen events of this kind could be captured under suitable
experimental conditions it would be an undeniable signature of
new physics. From what we have seen, the odds should be
heavily in favor of attributing it to a supersymmetric origin.


%% file: conclusions.tex
\chapter{Conclusions}
\label{sec:conclusions}

In this PhD. Thesis we have presented a monographic study of some rare Flavour Changing
Neutral Currents (FCNC) induced in the Two-Higgs-Doblet-Models (of both kinds
2HDM~I and II) and in the
Minimal Supersymmetric Standard Model (MSSM). In particular we have
studied the FCNC top quark decay in the 2HDM, $t\to c h$, Higgs boson
production and decay in the 2HDM, $pp\to h+X \to t\bar{c}$,
$h\to t \bar{c}$, and the Higgs boson production and decay in the MSSM,
$pp\to h+X \to {t\bar{c},b\bar{s}}$ and $h\to {t\bar{c},b\bar{s}}$,
in the context of the LHC, where $h=h^0,H^0,A^0$ are the three neutral Higgs
bosons in these extended Higgs models.

The main results obtained are the following:

\begin{itemize}
\item The non standard effects on the top quark FCNC decay into Higgs boson in
    the 2HDM~II can be very important, at most ten orders of magnitude
    larger then in the SM, bringing the branching ratios at the level of $10^{-4}$ or higher in
    some cases. Thus there is a real chance for seeing rare decays of that sort at
    the LHC. The effects of the 2HDM~I, although sizeable, do not reach the
    visible level at the LHC.

    It is useful to compare the $t\to h c$ processes with the more
    conventional rare decay $t\to g c$.
    The maximum rates for the leading FCNC processes~(\ref{Higgschannels}) and
    (\ref {gchannel}) in the 2HDM~II (resp. in the MSSM) satisfy the relations
    \begin{equation}
        B(t\rightarrow g\,c)<10^{-6}(10^{-5})<B(t\rightarrow h\,c)\sim10^{-4}\,,
        \label{summary}
    \end{equation}
    where it is understood that $h$ is $h^{0}$ or $H^{0}$, but not both, in
    the 2HDM~II; whereas $h$ is most likely $h^{0}$, but it could also be
    $H^{0}$ and $A^{0}$, in the MSSM.These decay ratios are compatible with
    their observation at the LHC if we compare them with the estimations of
    the sensitivities for $100\,fb^{-1}$ of integrated luminosity in the
    relevant colliders~\cite{\Frey,Aguilar-Saavedra:2004wm,Aguilar-Saavedra:2000aj}:
    \begin{align}
        \mathrm{\mathbf{LHC:}}B(t\rightarrow c\,X)&
        \gtrsim5\times10^{-5}\,,\notag\\
        \mathrm{\mathbf{LC:}}B(t\rightarrow c\,X) &
        \gtrsim5\times10^{-4}\,,\\
        \mathrm{\mathbf{TEV33:}}B(t\rightarrow c\,X)&
        \gtrsim5\times10^{-3}\,\,\,.\notag
    \end{align}

    The origin of the FCNC effects from the general 2HDM contributions is
    purely electroweak
    (namely it stems from enhanced trilinear Higgs boson couplings),
    in contrast to the MSSM case where they are mainly from the strong (gluino)
    sector, with the electroweak contributions one-two orders of magnitude below. This can be
    important because the MSSM strong couplings are very weakly restrained
    from experiment.

    Although the maximum ratio is similar there are different strategies to
    distinguish between the 2HDM and MSSM FCNC Higgs boson effects
    in the case that these decays would be
    detected. A possible strategy would be to look for different signatures for
    the process $t\to h c \to c b\bar{b}$. In the favorite FCNC
    region~(\ref{inputsmixing}) of the 2HDM~II, the combined decay
    $t\rightarrow h\;\,c\rightarrow cb\overline{b}$ is possible only for
    $h^{0}$ or for $H^{0}$, but not for both -- Cf. Fig.~\ref{fig:3}a --
    whereas in the MSSM, $h^{0}$ together with $H^{0}$, are highlighted for
    $110\,GeV<m_{A^{0}}<m_{t}$, with no preferred $\tan \beta $ value. And
    similarly, $t\rightarrow A^{0}\;c$ is also non-negligible for
    $m_{A^{0}}\lesssim120\,GeV$\thinspace \cite{\NP}.

\item The potential enhancement of the 2HDM contributions to the Higgs boson
    decay into top and charm quarks may reach up to ten billion times the SM
    value $B(H^{SM}\rightarrow t\,\bar{c})\sim 10^{-15}$, thereby bringing
    the maximum value of the FCNC branching ratio $B(h^{0}\rightarrow
    t\,\bar{c})$ to the level of $\sim 10^{-5}$. Adding to this the 2HDM
    contributions of the Higgs production ($pp \to H + X$) we find that the
    events $pp \to H + X \to t\bar{c} + X$ could be visible at the LHC, getting a few hundred
    events for the $h^0$ and for the $H^0$, but no for the $A^0$. Thus, theses
    events can be very useful to complement the Higgs boson detection
    strategies, like for example those based on the decay $H^{\mathrm{SM}}\to
    b\bar{b}$ (affected with a huge QCD background) or on the decay $H^{\mathrm{SM}}\to
    \gamma\gamma$ (more promising).

    There are some strategies to distinguish between the 2HDM and the
    MSSM. An obvious one is the detection of the channel $A^0\to t\bar{c}$ that
    is highly suppressed in the 2HDM but it is not so in the
    MSSM. Another possibility is
    using the fact that the MSSM contribution to FCNC processes
    is mainly from the gluino, which is
    not particularly sensitive to $\tb$, within a range of high and moderately
    values of this parameter. If, for example, a
    few of these events were observed and at the same time the best MSSM fits
    to the electroweak precision data would favor moderate values of
    $\tan\beta$, say in the range $10-20$, then it is clear that those events
    could originate from the FCNC gluino interactions but in no way within the
    context of the general 2HDM.

\item The SUSY-QCD contributions can enhance the maximum expectation for the
    FCNC decay rates $\Bhbs$, where $qq'=tc$ or $bs$, reaching the level of $10^{-3}$, in particular
    $\Bhzbsmax\sim 3\times 10^{-3}$ and $\BHAbsmax\simeq 9\times
    10^{-4}$. This corresponds to an scenario where $\Gamma(h^0\to b\bar{b})$
    is suppressed by radiative corrections to the CP-even mixing angle
    $\alpha$ (the so-called small $\alpha_{eff}$ scenario).
    When it is not the case the maximum ratios give the same result
    except for the light Higgs boson $\Bhzbsmax\simeq 1.3\times 10^{-4}$. So
    we get branching rations three to nine orders of magnitude higher than
    in the SM case, depending on the $q\,{q'}$ channel.

    The total number of FCNC heavy flavor events ($tc,bs$) originating from
    supersymmetric Higgs boson interactions at the LHC can be large (of order
    $10^6$ for $b\bar{s}$), but this does not mean that they can be easily disentangled from
    the underlying background of QCD jets where they are immersed. For the
    $bs$ final state the maximum number of events is $\sim12\times10^5$ for a
    $100\,fb^{-1}$ of integrated luminosity at the LHC, being the most
    favorable Higgs boson channel the lightest MSSM Higgs boson, $h^0$. There
    are some correlations in this channel that increase the number of events
    up to two orders of magnitude in certain cases as compared with the other
    channels $H^0,A^0$. On the other hand, for the case of $tc$ the maximum
    number of events is $\sim300\,\mathrm{events}/100\fb^{-1}$ for $\tb=5$, which amounts
    to few thousand $tc$ events during the lifetime of the LHC. This value is
    sensitive to the rest of the parameters, in particular $\mA$ and
    $\tb$. The mass of the CP-odd Higgs boson should not be heavier than
    $\mA\sim (400-500)\GeV$ if one does not want to decrease the number of
    events below a few hundred per $100\fb^{-1}$. On the other hand the number
    of events is very much dependent on the particular range of $\tb$, with a
    preferred low values, and changing orders of magnitude with tiny changes
    in $\tb$. For example for the values $\tb=(5,4,3,2)$ we get a number of
    events $\sim(300,500,900,2000)$. Although such lower values are
    usually avoided in the literature, they are not strictly excluded
    experimentally.

    We emphasize the primary differences with the 2HDM.
    In the MSSM, (1) the $A^0$ and $H^0$
    channels give essentially the same $bs$ rate, and it can be quite
    important; (2) the $A^0$ rate into $tc$ is not negligible; (3) the
    relevant range in $\tb$ is the lowest allowed one. In general the maximum
    number of FCNC events in the MSSM case is larger than the highest expected
    rates both in the 2HDM~I and~II, the two kind of signatures of physics
    beyond the SM being perfectly distinguishable because the relevant regions
    of the parameter space are completely different.

\end{itemize}

Although we have shown that the FCNC events considered in this Thesis are in
principle visible (namely, we have proven that the branching ratios and
production cross section are sizeable enough as compared to other rare, but
detected, events), it should be emphasized that the final word can only be
said after an appropriate signal versus background experimental study, which
is out of the scope of this theoretical investigation. However, we consider
highly encouraging that there is no theoretical a priori obstruction for these
processes not to be visible in the experimental setup of the LHC collider. In
particular the $t\bar{c}+\bar{t}c$ events, with the top quark in the final
state, should be very helpful and essentially free from background.

From these positive theoretical results we can assert that
the pathway to seeing new physics through FCNC decays of the top quark and
Higgs boson is thus potentially open. It is now an experimental challenge to
accomplish this program using the high luminosity super-colliders round the
corner, in particular the Large Hadron Collider at CERN, which is scheduled to
start operating within two years from now, i.e. in 2007.


%% file: func_npunts.tex
\chapter{Vertex functions} 
\label{ap:pointfun} 
 
In this appendix we briefly collect, for notational  
convenience, the basic vertex functions frequently referred  
to in the text.  
In practice we have performed the calculations using the algebraic and
numerical programs FeynArts, FeynCalc and LoopTools\cite{\FeynArts}.
The given  
formulas are exact for arbitrary internal masses and external  
on-shell momenta. Most of them are an adaptation to the  
$g_{\mu\nu}=\{+\,-\,-\,-\}$ metric of the standard formulae  
of Refs.\cite{'tHooft:1978xw,Passarino:1978jh,Axelrod:1982yc}.  
The basic one-, two- and  
three-point scalar functions are: 
\begin{eqnarray} 
    A_{0}(m)&=& \int d^{n}\tilde{q}\, \frac{1}{[q^2-m^2]},\\ 
    B_{0}(p,m_1,m_2)&=&\int d^{n}\tilde{q}\, \frac{1}{[q^2-m_1^2]\, 
    [(q+p)^2-m_2^2]}\,, 
\end{eqnarray} 
\begin{equation} 
    C_{0}(p,k,m_1,m_2,m_3)=\int d^{n}\tilde{q}\,  
    \frac{1}{[q^2-m_1^2]\,[(q+p)^2-m_2^2]\,[(q+p+k)^2-m_3^2]}\,; 
\end{equation} 
using the integration measure 
\begin{equation} 
\label{eq:muUV} 
d^n\tilde{q}\equiv {\mu }^{(4-n)}\frac{d^nq}{(2\pi )^n}\,. 
\end{equation} 
The two and three-point tensor functions needed for our calculation  
are the following 
\begin{equation} 
    [\tilde{B}_0,B_{\mu },B_{\mu \nu }](p,m_1,m_2)=\int d^{n}\tilde{q}\,  
    \frac{[q^2,q_{\mu },q_{\mu }q_{\nu}]}{[q^2-m_1^2]\,[(q+p)^2-m_2^2]}\,, 
\end{equation} 
\begin{eqnarray} 
     \lefteqn{[\tilde{C}_0,C_{\mu },C_{\mu \nu }](p,k,m_1,m_2,m_3)=}  
     \nonumber \\ 
               & &\int d^{n}\tilde{q}\, \frac{[q^2,q_{\mu},q_{\mu}  
               q_{\nu}]}{[q^2-m_1^2]\,[(q+p)^2-m_2^2]\,[(q+p+k)^2-m_3^2]}\,. 
\end{eqnarray} 
By Lorentz covariance, they can be decomposed in terms of the  
above basic scalar functions and the external momenta: 
\begin{eqnarray} 
  \tilde{B}_0 (p,m_1,m_2) &=& A_0(m_2)+m_1^2 B_0(p,m_1,m_2)\,\,,\nonumber\\ 
    B_{\mu }(p,m_1,m_2)    &=& p_{\mu } B_1(p,m_1,m_2)\,\,,\nonumber \\ 
    B_{\mu \nu }(p,m_1,m_2)&=& p_{\mu } p_{\nu }B_{21}(p,m_1,m_2)+ 
                            g_{\mu \nu }B_{22}(p,m_1,m_2)\,\,, \nonumber \\ 
    \tilde{C}_0(p,k,m_1,m_2,m_3) &=& B_0(k,m_2,m_3)+ 
                                  m_1^2 C_{0}(p,k,m_1,m_2,m_3)\,\,, \nonumber\\ 
    C_{\mu}(p,k,m_1,m_2,m_3)&=&p_{\mu} C_{11}+k_{\mu} C_{12}\,\,,\nonumber\\ 
    C_{\mu \nu }(p,k,m_1,m_2,m_3)&=& p_{\mu }p_{\nu}C_{21}+ 
                                  k_{\mu }k_{\nu}C_{22}+ 
                                  (p_{\mu}k_{\nu}+k_{\mu}p_{\nu})C_{23}+ 
                                  g_{\mu \nu }C_{24}\,\,, 
\end{eqnarray} 
where we have defined the Lorentz invariant functions: 
\begin{eqnarray} 
B_1(p,m_1,m_2)&=&\frac{1}{2p^2}[A_0(m_1)-A_0(m_2)- 
    f_1B_0(p,m_1,m_2)], \\ 
B_{21}(p,m_1,m_2)&=&\frac{1}{2p^2(n-1)}[(n-2)A_0(m_2) 
            -2m_1^2B_0(p,m_1,m_2) \nonumber \\ 
            &-& nf_1B_1(p,m_1,m_2)],\\ 
B_{22}(p,m_1,m_2)&=&\frac{1}{2(n-1)}[A_0(m_2)\!+\!2m_1^2B_0(p,m_1,m_2) 
            \!+\!f_1B_1(p,m_1,m_2)], 
\end{eqnarray} 
\begin{equation} 
     \left(\begin{array}{c} 
       C_{11}\\C_{12} 
     \end{array}\right)=Y 
     \left(\begin{array}{c} 
       B_0(p+k,m_1,m_3)-B_0(k,m_2,m_3)-f_1C_0 \\  
       B_0(p,m_1,m_2)-B_0(p+k,m_1,m_3)-f_2C_0 
     \end{array}\right)\,, 
\end{equation} 
\begin{equation} 
     \left(\begin{array}{c} 
       C_{21}\\C_{23} 
     \end{array}\right)=Y  
     \left(\begin{array}{c} 
       B_1(p+k,m_1,m_3)+B_0(k,m_2,m_3)-f_1C_{11}-2C_{24} \\  
       B_1(p,m_1,m_2)-B_1(p+k,m_1,m_3)-f_2C_{11} 
     \end{array}\right)\,, 
\end{equation} 
\begin{eqnarray} 
        C_{22}&=&\displaystyle{\frac{1}{2[p^2k^2-(pk)^2]}}  
              \{-pk[B_1(p+k,m_1,m_3)-  
              B_1(k,m_2,m_3)-f_1C_{12}]  \nonumber\\ 
              &&+p^2[-B_1(p+k,m_1,m_3)-f_2C_{12}-2C_{24}]\}\,,             
\end{eqnarray} 
\begin{equation} 
    C_{24}=\frac{1}{2(n-2)} [B_0(k,m_2,m_3)+ 
           2m_1^2C_0+f_1C_{11}+f_2C_{12}]\,,\\ 
\end{equation} 
the factors $f_{1,2}$ and the matrix $Y$, 
\begin{eqnarray*} 
   f_1 &=& p^2+m_1^2-m_2^2, \\ 
   f_2 &=& k^2+2pk+m_2^2-m_3^2\,, 
\end{eqnarray*} 
\begin{equation} 
   Y=\frac{1}{2[p^2k^2-(pk)^2]}  
   \left(\begin{array}{cc} 
        k^2 & -pk \\-pk & p^2 
   \end{array}\right)\,. 
\end{equation} 
The UV divergences for $n\rightarrow 4$ can be parametrized as 
\begin{eqnarray} 
\label{eq:DeltaUV} 
    \epsilon &=& n-4,\nonumber \\ 
      \Delta &=& \frac{2}{\epsilon }+ 
                 {\gamma }_{\scriptscriptstyle E} -\ln (4\pi)\,, 
\end{eqnarray} 
being ${\gamma }_{\scriptscriptstyle E}$ the Euler constant. 
In the end one is left with the evaluation of the scalar one-loop 
functions: 
\begin{eqnarray} 
    A_0(m) &=& \left(\frac{-i}{16{\pi }^2} \right)  
           m^2(\Delta-1+\ln \frac{m^2}{\mu ^2})\,, \\ 
    B_0(p,m_1,m_2) &=& \left(\frac{-i}{16{\pi }^2} \right)  
           \left[\Delta+\ln \frac{p^2}{{\mu }^2}-2  
           +\ln [(x_1-1)(x_2-1)] \right.\nonumber \\ 
           & & \left . +x_1\ln \frac{x_1}{x_1-1}+ 
                        x_2\ln \frac{x_2}{x_2-1}\right], \\ 
    C_0(p,k,m_1,m_2,m_3) &=& \left(\frac{-i}{16{\pi }^2} \right)  
           \frac{1}{2}\;\frac{1}{pk+p^2\xi }\Sigma\,  
\end{eqnarray} 
with 
\begin{eqnarray} 
    x_{1,2}=x_{1,2}(p,m_1,m_2) &=& \frac{1}{2}+\frac{m_1^2-m_2^2}{2p^2} 
           \pm \frac{1}{2p^2}\lambda^{1/2}(p^2,m^2_1,m^2_2), \\ 
           \lambda (x,y,z) &=& [x-(\sqrt{y}-\sqrt{z})^2] 
                               [x-(\sqrt{y}+\sqrt{z})^2]\,, \nonumber 
\end{eqnarray} 
and where $\Sigma$ is a bookkeeping device for the following alternate  
sum of twelve (complex) Spence functions: 
\begin{eqnarray} 
    \Sigma &=& 
      Sp\left(\frac{y_1}{y_1-z_{1}^{i}}\right)-Sp\left(\frac{y_1-1} 
      {y_1-z_{1}^{i}}\right)+ 
      Sp\left(\frac{y_1}{y_1-z_{2}^{i}}\right)-Sp\left(\frac{y_1-1} 
      {y_1-z_{2}^{i}}\right)  
\nonumber \\ 
  &  -&Sp\left(\frac{y_2}{y_2-z_{1}^{ii}}\right)+Sp\left(\frac{y_2-1} 
      {y_2-z_{1}^{ii}}\right)- 
      Sp\left(\frac{y_2}{y_2-z_{2}^{ii}}\right)+Sp\left(\frac{y_2-1} 
      {y_2-z_{2}^{ii}}\right) 
\nonumber \\ 
  &  +&Sp\left(\frac{y_3}{y_3-z_{1}^{iii}}\right)-Sp\left(\frac{y_3-1} 
      {y_3-z_{1}^{iii}}\right)+ 
      Sp\left(\frac{y_3}{y_3-z_{2}^{iii}}\right)-Sp\left(\frac{y_3-1} 
      {y_3-z_{2}^{iii}}\right). 
\end{eqnarray} 
The Spence function is defined as 
\begin{equation} 
     Sp(z)=-\int_{0}^{1}\frac{\ln (1-zt)}{t}\,dt\,, 
\end{equation} 
and we have set, on one hand: 
\begin{eqnarray} 
     z_{1,2}^{i}   & = & x_{1,2}(p,m_2,m_1)\,,   \nonumber \\ 
     z_{1,2}^{ii}  & = & x_{1,2}(p+k,m_3,m_1)\,, \nonumber \\ 
     z_{1,2}^{iii} & = & x_{1,2}(k,m_3,m_2)\,; 
\end{eqnarray} 
and on the other: 
\begin{equation} 
     y_1=y_0+\xi\,,\;\;\;  
     y_2=\frac{y_0}{1-\xi }\,, \;\;\;\;  
     y_3=-\frac{y_0}{\xi }\,, \;\;\;\;  
     y_0=-\frac{1}{2}\;\frac{g+h\xi }{pk+p^2\xi }\,, \;\;\;\; 
\end{equation} 
where 
\begin{equation} 
\begin{array}{ll} 
     g=-k^2+m_2^2-m_3^2\,,\;\;\; & h=-p^2-2pk-m_2^2+m_1^2\,,  
\end{array} 
\end{equation} 
and $\xi $ is a root (always real for external on-shell momenta) of  
\begin{equation} 
     p^2\xi^2+2pk\xi +k^2=0\,. 
\end{equation} 
 
Derivatives of some 2-point functions are also needed in the  
calculation of self-energies, and we use the 
following notation: 
\begin{equation} 
     \frac{\partial}{\partial p^2}B_{*}(p,m_1,m_2)\equiv  
                             B^{\prime}_{*}(p,m_1,m_2).  
\end{equation} 
We can obtain all the derivatives from the basic $B^{\prime}_0$: 
\begin{eqnarray} 
  B^{\prime}_0(p,m_1,m_2)  
   &=&      \left( \frac{-i}{16\pi^2} \right) 
            \left\{    \frac{1}{p^2} + 
            \frac{1}{\lambda^{1/2}(p^2,m^2_1,m^2_2)} 
          \right. \nonumber \\    
   & &  \times  \left.  
            \left[   x_1(x_1-1)\ln \left( \frac{x_1-1}{x_1} \right) 
                         -x_2(x_2-1)\ln \left( \frac{x_2-1}{x_2} 
                         \right)  
\right]\right\}\,, 
\end{eqnarray} 
which has a threshold for $|p|=m_1+m_2$ and a pseudo-threshold 
for $|p|=|m_1-m_2|$.

\section{Limit of heavy internal masses}
\label{sec:limitmassesgrans}

Next we will present the results when the external moments are much heavier
than the internal masses ($p^2\ll m^2$). The basic equations is:
\begin{equation}
    \begin{split}
        \label{eq:expansiopm}
        \frac{1}{(q+p)^2-m^2}&=\frac{1}{q^2-m^2}\left[1+\frac{p^2+2p\cdot
              q}{q^2-m^2}\right]^{-1}\\
        &=\frac{1}{q^2-m^2}\left[1-\frac{p^2+2p\cdot
              q}{q^2-m^2}+\left(\frac{p^2+2p\cdot
                  q}{q^2-m^2}\right)^2-\dots\right]\,,
    \end{split}
\end{equation}
which we have to substitute in each of the denominators of the
$n$-point functions. But, all the lineal terms in the expansion become
zero under the integrations $\int d^4q$, because of the denominator is a
spherically symmetric function. The others terms are integrals of functions
of $q^2$ and $p^2/(q^2-m^2)^2\ll 1$, which are negligible in this limit.

Thus the effective substitution is:
\begin{equation}
    \label{eq:substituciopm}
    \frac{1}{(q+p)^2-m^2}\rightarrow\frac{1}{q^2-m^2}\,,
\end{equation}
in the cases where $p^2\ll m^2$.

So in this limit the $n$-point functions do not depend on the moments. It
allows us to define, without ambiguities, that the $n$-point functions without
the moments are proportional to the complete $n$-point functions with the
moments equals to zero. As:
\begin{align}
    \label{eq:fucionssensemoments}
    C_{0}(0,0,m_1,m_2,m_3)&=\left(\frac{-i}{16\pi^2}\right)
    C_0(m_1,m_2,m_3)\,,\\ 
    \intertext{with}
    C_0(m_1,m_2,m_3)&=\frac{1}{m_1^2-m_2^2}F(m_1,m_2,m_3)\,,\\
    \begin{split}
        F(m_1,m_2,m_3)&=\logdos{1}{2}+\divtres{3}{3}{1}\logdos{3}{1}-\divtres{3}{3}{2}\logdos{3}{2}=\\
        &=-F(m_2,m_1,m_3)\,.
    \end{split}
\end{align}
One can see that if, for example, $m_1^2\gg m_2^2,m_3^2$, then the
function~\eqref{eq:fucionssensemoments} only depends on this heavy mass as
$C_0\propto\frac{1}{m_1^2}$. The same is true for $m_2^2\gg m_1^2,m_3^2$ and
$m_3^2\gg m_1^2,m_2^2$.

The function $C_{\mu\nu}(m_1,m_2,m_3)$ is logarithmically divergent and can
be easily written as:
\begin{equation}
    C_{\mu\nu}(m_1,m_2,m_3)=g_{\mu\nu}\left[\frac{1}{n}B_0(m_2,m_3)+\frac{1}{4}m_1^2C_0(m_1,m_2,m_3)\right]\,.
\end{equation}
But we can not substitute directly the value of $B_0$ because is divergent and
divided by $n$.
\begin{align}
    \label{eq:resultatB0pm}
    B_0(m_2,m_3)&=\Delta-1+\log m_3^2-\divtres{2}{2}{1}\logdos{3}{2}\,,\\
    \begin{split}
        \lim_{n\rightarrow4}\frac{1}{n}B_0(m_2,m_3)&=\lim_{\epsilon\rightarrow0}\inv{4}
        \Bigl(1-\frac{\epsilon}{4}\Bigr)\Bigl[\Delta+B_0(m_2,m_3)|_{\text{finite}}\Bigr]\\
        &=\inv{4}\Bigl[\Delta-\inv{2}+B_0(m_2,m_3)|_{\text{finite}}\Bigr]\,,
    \end{split}
\end{align}
where the finite part is defined as the finite part of
the~\eqref{eq:resultatB0pm}. The final result is:
\begin{align}
    C_{\mu\nu}(m_1,m_2,m_3)&=\inv{4}g_{\mu\nu}\Bigl[\Delta+\hat{C}(m_1,m_2,m_3)\bigr]\,,\\
    \intertext{where the normalized $3$-point function is:}
    \hat{C}(m_1,m_2,m_3)&=\Bigl[-\frac{3}{2}+\log{m_3^2}+\hat{E}_0(m_1,m_2,m_3)\Bigr]\,,\\
    \intertext{with}
    \hat{E}_0(m_1,m_2,m_3)&=\divtres{2}{3}{2}\logdos{3}{2}+\divtres{1}{1}{2}F(m_1,m_2,m_3)\,.\\
    \intertext{Some frequent special cases are:}
    \hat{E}_0(M,m,m)&=\frac{m^2}{m^2-M^2}-\Bigl(\frac{M^2}{m^2-M^2}\Bigr)^2\log\frac{m^2}{M^2}\,,\\
    \hat{E}_0(m,m,M)&=\log\frac{m^2}{M^2}+\hat{E}_0(M,m,m)\,.
\end{align}
The function $\tilde{C}_0(m_1,m_2,m_3)$ is very similar to the previous one,
with the result:
\begin{align}
        \tilde{C}_0(m_1,m_2,m_3)&=B_0(m_2,m_3)+m_1^2C_0(m_1,m_2,m_3)=\Delta+\hat{C}(m_1,m_2,m_3)+\inv{2}\,.
\end{align}
And the last one is $B_1(m_1,m_2)$:
\begin{equation}
    B_1(m_1,m_2)=-\inv{2}\Bigl[\Delta+\log
    m_2^2-\inv{2}-\divtres{1}{1}{2}-\Bigl(\divtres{1}{1}{2}\Bigr)^2
    \logdos{2}{1}\Bigr]\,.
\end{equation}

%% file: diag_matrix.tex
\chapter{Diagonalizing a squared mass matrix}

In this appendix we will show a group of formulas useful to diagonalize a
$2\times2$ squared mass matrix $M^{(2)}$. We will fix our convention for the
orthogonal matrixes that diagonalize $M^{(2)}$, showing the analytic
expressions for the eigenvalues and the rotations angles, and finally how
parametrize the sector in terms of the eigenvalues.

For the mass Lagrangian in the electroweak basis
\begin{align}
    \mathcal{L}_M&=\frac{1}{2}\phi^{\prime T}M^{(2)}\phi\prime
    \intertext{where}
    \phi\prime&\equiv\begin{pmatrix}\phi^\prime_1\\\phi^\prime_2\end{pmatrix}
    \intertext{are the electroweak eigenstates, and}
    M^{(2)}&\equiv\begin{pmatrix}a&c\\c&b\end{pmatrix}\label{eq:diag:m2}
\end{align}
is the symmetric squared mass matrix that mix electroweak states. We define
$R$ as the orthogonal matrix that diagonalize $M^{(2)}$:
\begin{align}
    R&=\begin{pmatrix}
        \cos\varphi&\sin\phi\varphi\\-\sin\phi\varphi&\cos\varphi\end{pmatrix}\\
    RM^{(2)}R^T&\equiv\mathrm{diag}\{m^2_1,m^2_2\}
    \intertext{The new mass eigenstates will be:}
    \phi&\equiv\begin{pmatrix}\phi_1\\\phi_2\end{pmatrix}=R\phi^\prime
    \intertext{with the mass eigenstates the Lagrangian can be written as}
    \mathcal{L}_M&=\frac{1}{2}m^2_1\phi^2_1+\frac{1}{2}m^2_2\phi^2_2
\end{align}

The mass eigenvalues can be written in terms of the matrix elements as:
\begin{align}
    m^2_{1,2}=\frac{1}{2}\left[a+b\mp\sqrt{(a-b)^2+4c^2}\right]
    \label{eq:diag:m12}
\end{align}
where $m^2_1(m^2_2)$ is the lightest (heaviest) particle and in general we can
said that is a monotonic decreasing (increasing) function with the mixing term
$|c|$. The mixing angle that diagonalize \eqref{eq:diag:m2} can be extracted
from the relation:
\begin{align}
    \tan(2\varphi)=\frac{2c}{a-b}
\end{align}
to know the correct quadrant of $\varphi$ is more useful this pair of
relations:
\begin{align}
    \sin(2\varphi)=\frac{2c}{\sqrt{(a-b)^2+4c^2}},\qquad\qquad
    \cos(2\varphi)=\frac{a-b}{\sqrt{(a-b)^2+4c^2}}.
\end{align}
From \eqref{eq:diag:m12} we can extract two relations:
\begin{align}
    m^2_1+m^2_2&=a+b\\
    -m^2_1+m^2_2&=\sqrt{(a-b)^2+4c^2}
    \intertext{and from the invariant of the determinant:}
    m^2_1m^2_2=ab-c^2.
\end{align}

These relations can be very useful to reparametrize the mass sector. We have
put everything in terms of the three degrees of freedom of the matrix,
i.e. ${a,b,c}$. We can choose other sets to represent the degrees of freedom
and give the rest in terms of these. Next we will present some examples of
these reparametrizations.

\begin{itemize}
\item Parameters ${m_1,m_2,\varphi}$
    \begin{align}
        a&=\frac{1}{2}\left[m^2_1(1-\cos2\varphi)+m^2_2(1+\cos2\varphi)\right]\\
        b&=\frac{1}{2}\left[m^2_1(1+\cos2\varphi)+m^2_2(1-\cos2\varphi)\right]\\
        c&=-\frac{1}{2}\left[(m^2_1-m^2_2)\sin2\varphi\right]
    \end{align}
\item Parameters ${m_1,a,c}$
    \begin{align}
         b&=\frac{c^2}{a-m^2_1}+m^2_1\\
         m^2_2&=\frac{c^2}{a-m^2_1}+a\\
        \tan2\varphi&=\frac{2c}{a-m^2_1-\frac{c^2}{a-m^2_1}}
    \end{align}
\item Parameters ${m_1,m_2,c}$
    \begin{align}
        a&=\frac{1}{2}\left[m^2_1+m^2_1\pm\sqrt{(m^2_1-m^2_2)^2-4c^2}\right]\\
        b&=\frac{1}{2}\left[m^2_1+m^2_1\mp\sqrt{(m^2_1-m^2_2)^2-4c^2}\right]\\
        \tan2\varphi&=\pm\frac{2c}{\sqrt{(m^2_1-m^2_2)^2-4c^2}}
    \end{align}
\end{itemize}